\documentclass[
aps,
pra,                                                                                                                                                                                                   
twocolumn,
showpacs,
superscriptaddress,
floatfix
]{revtex4-1}
\usepackage[dvipdfmx]{graphicx}
\usepackage{epsf}
\usepackage{natbib}
\usepackage{dcolumn}
\usepackage{bm}
\usepackage{ulem}
\usepackage{amsmath}
\usepackage{amssymb}
\usepackage[dvipsnames]{color} 
\usepackage[
colorlinks=true,
citecolor=blue,
urlcolor=blue,
setpagesize=false]{hyperref}
\newcommand{\bs}   {\boldsymbol}
\newcommand{\mb}   {\mathbf}
\newcommand{\mr}   {\mathrm}

\newcommand{\mcal} {\mathcal}
\newcommand{\imag} {\mathrm{i}}
\newcommand{\dd}   {\mathrm{d}}

\newcommand{\up}   {\uparrow}
\newcommand{\dn}   {\downarrow}
\newcommand{\s}    {\sigma}
\newcommand{\w}    {\omega}

\newcommand{\eps}  {\epsilon}
\newcommand{\Tr}  {{\rm Tr}}

\begin{document}
\title{
  Emergence of massless Dirac quasiparticles in correlated hydrogenated graphene with broken sublattice symmetry 
}
\author{Kazuhiro Seki}
\affiliation{Computational Condensed Matter Physics Laboratory, RIKEN, Wako, Saitama 351-0198, Japan}
\affiliation{Computational Materials Science Research Team, RIKEN Advanced Institute for Computational Science (AICS),  Kobe Hyogo 650-0047,  Japan}
\author{Tomonori Shirakawa}
\affiliation{Computational Quantum Matter Research Team, RIKEN Center for Emergent Matter Science (CEMS), Wako, Saitama 351-0198, Japan}
\author{Qinfang Zhang}
\affiliation{Key Laboratory for Advanced Technology in Environmental Protection of Jiangsu Province, Yancheng Institute of technology,Yancheng 224051, China} 
\author{Tao Li}
\affiliation{Department of Physics, Renmin University of China, Beijing, 100872, China}
\author{Seiji Yunoki}
\affiliation{Computational Condensed Matter Physics Laboratory, RIKEN, Wako, Saitama 351-0198, Japan}
\affiliation{Computational Materials Science Research Team, RIKEN Advanced Institute for Computational Science (AICS),  Kobe Hyogo 650-0047,  Japan}
\affiliation{Computational Quantum Matter Research Team, RIKEN Center for Emergent Matter Science (CEMS), Wako, Saitama 351-0198, Japan}

\begin{abstract}
  Using the variational cluster approximation (VCA) and the cluster perturbation theory, 
  we study the finite temperature phase diagram of a half-depleted periodic Anderson model 
  on the honeycomb lattice at half filling 
  for a model of graphone, i.e., single-side hydrogenated graphene. 
  The ground state of this model is found to be ferromagnetic (FM) semi-metal. 
  The origin of this FM state is attributed to the instability of a flat band located at the Fermi energy in the 
  noninteracting limit and is smoothly connected to the 
  Lieb-Mattis type ferromagnetism. 
  The spin wave dispersion in the FM state is linear in momentum at zero temperature 
  but becomes quadratic at finite temperatures, implying that the FM state is fragile against thermal fluctuations. 
  Indeed, our VCA calculations find that the paramagnetic (PM) state dominates the finite temperature 
  phase diagram.  
  More surprisingly, we find 
  that massless Dirac quasiparticles with the linear energy dispersion emerge at the Fermi energy 
  upon introducing the electron correlation $U$ at the impurity sites  
  in the PM phase. 
  The Dirac Fermi velocity is found to be highly correlated to the 
  quasiparticle weight of the emergent massless Dirac quasiparticles at the Fermi energy and monotonically 
  increases with $U$. 
  These unexpected massless Dirac quasiparticles are also examined with the Hubbard-I approximation and 
  the origin is discussed in terms of 
  the spectral weight redistribution involving a large energy scale of $U$. 
  Considering an effective quasiparticle Hamiltonian which reproduces the single-particle excitations 
  obtained by the Hubbard-I approximation, 
  we argue that the massless Dirac quasiparticles are protected by the electron correlation. 
  Our finding is therefore 
  the first example of the emergence of massless Dirac quasiparticles due to 
  dynamical electron correlations without breaking any spatial symmetry. 
  The experimental implications of our results for graphone as well as a graphene sheet on transition metal 
  substrates are also briefly discussed. 
\end{abstract}

\date{\today}
\maketitle

\section{Introduction} 
Graphene~\cite{Novoselov2005} has been 
one of the most actively studied research subjects in current condensed 
matter physics~\cite{Neto2009RMP}. 
Although its unique electronic property is characterized already in a single-particle level, namely,
with the linear electronic energy dispersion, i.e., the Dirac cone dispersion, at the Fermi energy ($E_{\rm F}$)~\cite{Wallace1947},
many-body effects on graphene have also attracted much attention~\cite{Kotov2012}. 
For example, 
tremendous efforts have been devoted on investigating whether 
a spin liquid state can exist in the half-filled Hubbard model on the honeycomb 
lattice~\cite{Meng2010,Sorella2012,Assaad2013}, 
one of the simplest models for graphene, 
and the nature of metal-insulator transition~\cite{Toldin2015,Otsuka2015}. 

The research has been also extended to graphene derived 
systems, e.g., a series of hydrogenated graphene~\cite{Sofo2007,Elias2009,Ray2014,Peng2014}.
A first-principles calculation based on density functional theory (DFT) has predicted that 
the single-side hydrogenated graphene, called graphone, 
becomes a ferromagnetic (FM) 
semiconductor with a small indirect gap~\cite{Zhou2009}. 
Other DFT based study 
has suggested that the single-side hydrogenated and fluorinated graphenes are both candidates for 
quantum spin liquid~\cite{Rudenko2013}. 
Possible increase of the spin-orbit coupling due to the $sp^3$ lattice distortion has been also discussed~\cite{Neto2009,Gmitra2013}. 

Many-body effects on the hydrogenated graphene, however, have not been explored so far. 
On the one hand, the isolated graphene, described by, e.g., the single-band Hubbard 
model on the honeycomb lattice~\cite{Schuler2013}, remains semi-metallic 
even when a moderate amount of electron interactions are introduced before the antiferromagnetic instability 
sets in~\cite{Meng2010,Sorella2012,Assaad2013,Sorella1992}, and thus the correlation effect in semi-metallic 
phase is merely 
renormalization~\cite{Otsuka2015}. 
On the other hand, the electron correlation in the hydrogen atoms should be treated in a many-body way, 
as suggested in the Heitler-London description for the chemical bonding of a hydrogen molecule~\cite{HL}.
It is also noteworthy that many-body effects on hydrogen atoms in metal can induce a 
Kondo-like effect and make a drastic correction in the single-particle excitation spectrum 
as compared with that obtained by DFT calculations~\cite{Eder1997}. 

Here, 
we employ the variational cluster approximation (VCA)~\cite{Potthoff2003} and 
the cluster perturbation theory (CPT)~\cite{Senechal2000} 
to investigate many-body effects on graphone
by considering a half-depleted periodic Anderson model on the honeycomb lattice at half filling. 
We find that the ground state of this model is FM semi-metallic. 
The FM state is attributed to the instability of a flat band located at $E_{\rm F}$ 
in the noninteracting limit and smoothly connected to the Lieb-Mattis type ferromagnetism. 
The liner spin wave analysis of an effective spin model 
for the periodic Anderson model in the strong coupling limit 
finds that the spin wave excitations in the FM state exhibits 
the linear dispersion in momentum at zero temperature, 
while the spin wave dispersion becomes quadratic at finite temperatures, 
implying that the FM state is fragile against thermal fluctuations 
and can be stable only at zero temperature.

Our VCA calculations indeed find that the finite temperature 
phase diagram is dominated by a paramagnetic (PM) state. 
Most significantly, we find that 
massless Dirac quasiparticles emerge at $E_{\rm F}$ 
upon introducing the electron correlation in the PM phase. 
The Dirac Fermi velocity of 
the emergent massless Dirac quasiparticles is found to 
be highly correlated to the 
quasiparticle weight at the Fermi energy 
and increase monotonically with the electron correlation. 
We show that the emergence of the massless Dirac quasiparticles 
is well captured by the simple Hubbard-I approximation and 
can be understood as the result of 
the spectral weight redistribution involving a large energy scale of the electron correlation. 
By considering an effective Hamiltonian for the quasiparticles, we discuss the chiral symmetry 
of the single-particle excitations in the PM phase within the Hubbard-I approximation 
and argue that the emergent massless Dirac quasiparticles are protected by the electron correlation. 
The massless Dirac quasiparticles found here in the PM phase are therefore 
in sharp contrast to massless Dirac dispersions generated by band engineering with breaking a crystalline 
symmetry, and represent the first example of the emergence of massless Dirac quasiparticles induced by dynamical electron 
correlations. 

The rest of this paper is organized as follows. Sec.~\ref{Sec:model} introduces 
the periodic Anderson model studied here 
and explains briefly the numerical methods, i.e., the VCA 
and the CPT. These numerical methods are employed to obtain the finite temperature phase diagram 
and examine the single-particle excitations in Sec.~\ref{sec:results}. 
The results are compared with those obtained analytically 
using the mean-field theory in Sec.~\ref{sec:mf} and the Hubbard-I approximation in Sec.~\ref{sec:H-I}. 
An effective Hamiltonian for the quasiparticles is constructed on the basis of the Hubbard-I 
approximate analysis and the chiral symmetry of the quasiparticle 
excitations is discussed in Sec.~\ref{sec:H-I_chiral}. The implications of our results for experiments 
are also briefly discussed in Sec.~\ref{sec:dis} before summarizing the paper in Sec.~\ref{sec:sum}. 

In addition, five appendices are provided to supplement the main text. 
The stability of the FM state is examined with the liner spin wave theory in Appendix~\ref{appsec:sw}. 
Lieb's theorem is applied to the periodic Anderson model in Appendix~\ref{appsec:lieb}. 
This analysis, together with the numerically exact diagonalization study of small clusters 
in Appendix~\ref{appsec:ed}, reveals that the FM ground state found here is 
smoothly connected to the Lieb-Mattis type ferromagnetism. 
The single-particle excitations for several 
limiting cases are also examined within the Hubbard-I approximation in Appendix~\ref{appsec:HubI}. 
Finally, the Brillouin-Wigner perturbation theory is applied to the effective quasiparticle 
Hamiltonian in Appendix~\ref{appsec:BW}.

\section{Model and Methods}\label{Sec:model}

In this section, we first introduce a periodic Anderson model as one of the simplest models 
for graphone, i.e., single-side hydrogenated graphene, and summarize the electron band structure 
of this model in the noninteracting limit. 
Next, we briefly explain the finite temperature VCA and CPT to treat the electron correlation effect  
on the finite temperature phase diagram and the single-particle excitations 
beyond the single-particle approximation. 

\subsection{Periodic Anderson model}

\begin{figure}[!htb]
  \begin{center}
    \includegraphics[clip,width=7.2cm]{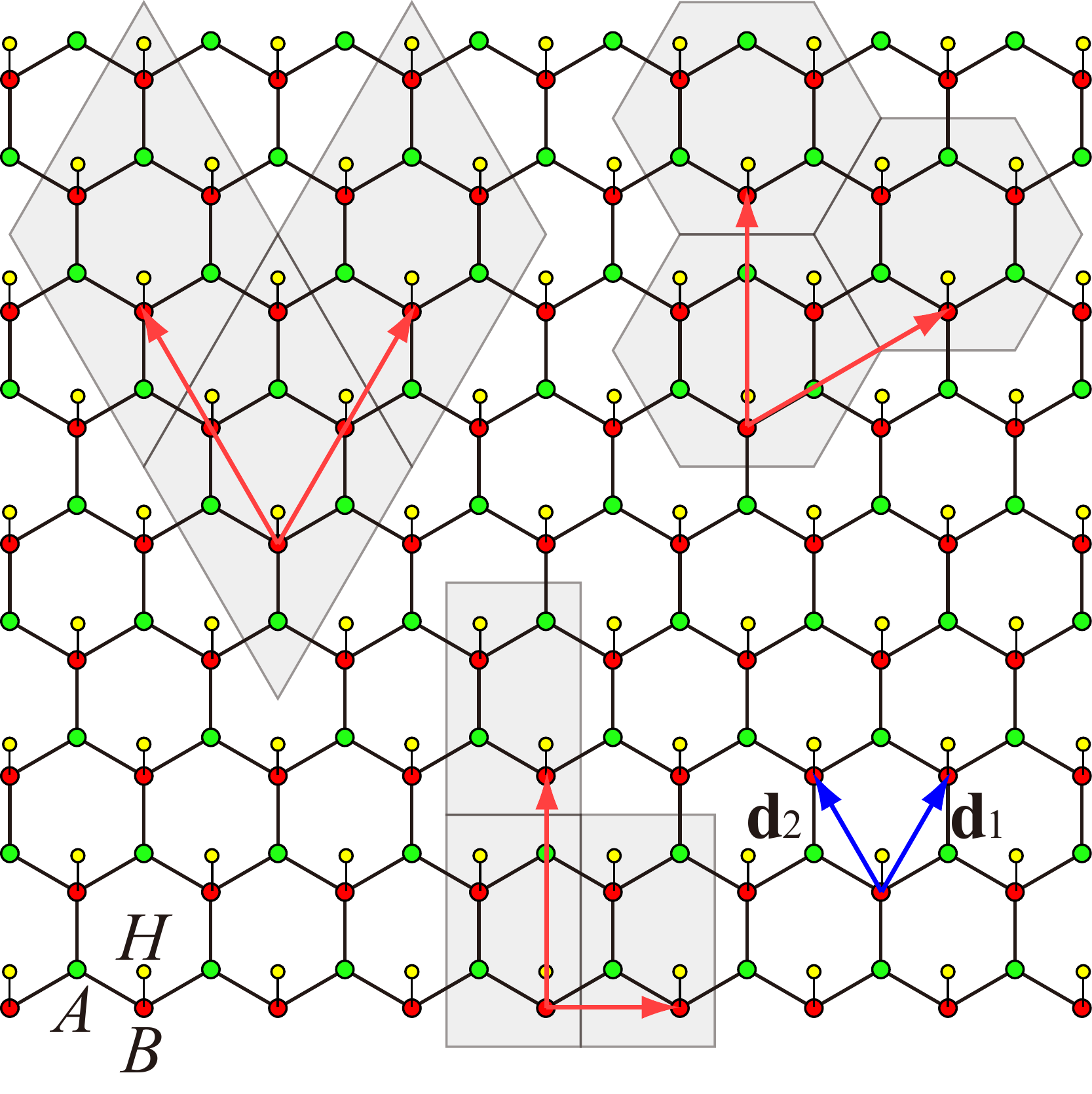}
  \end{center}
  \caption{(color online)
    Schematic honeycomb lattice structure on which the half-depleted periodic Anderson model is defined 
    for a model of graphone, i.e., single-side hydrogenated graphene. 
    Green and red circles denote the carbon conduction sites on $A$ and $B$ sublattices of the honeycomb 
    lattice, respectively, and yellow circles indicate the hydrogen impurity sites. 
    The primitive translational vectors of the honeycomb lattice, 
    $\mb{d}_1= \left(  \frac{1}{2}, \frac{\sqrt{3}}{2} \right)a$ and $\mb{d}_2= \left( -\frac{1}{2}, \frac{\sqrt{3}}{2} \right)a$, 
    are denoted by blue arrows, 
    where $a$ is the lattice constant between the next nearest-neighboring carbon sites. 
    Clusters considered in the VCA and the CPT (grey shaded regions) include  
    a 12-site cluster containing 4 unit cells (upper left), 
    a 9-site  cluster containing 3 unit cells (upper right), and 
    a 6-site cluster containing 2 unit cells (lower center). 
    The orange arrows indicate primitive translational vectors for each cluster. 
    The 6-site and 9-site clusters  
    are used for finite temperature calculations and 
    the 12-site cluster is used for zero temperature calculations.
  }
  \label{lattice}
\end{figure}

We consider a half-depleted periodic Anderson model on the honeycomb lattice defined as  
\begin{equation}
  \mcal{H}   = \mcal{H}_0 + \eps_H \sum_{i}\sum_\s n_{\mb{r}_i \s H} +  U     \sum_{i}   n_{\mb{r}_i \up H} n_{\mb{r}_i \dn H},  
  \label{eq:ham}
\end{equation}
where 
\begin{eqnarray}
  \mcal{H}_0 &=& -t \sum_{ i} \sum_{ \s} \sum_{ {\bm\delta}}\left ( c_{\mb{r}_i \s A}^\dag c_{\mb{r}_i+{\bm\delta} \s B}  
     + {\rm h.c.} \right) \nonumber \\
  & +& t_{sp} \sum_{i} \sum_{\s}   \left(     c_{\mb{r}_i \s B}^\dag c_{\mb{r}_i \s H} + \mr{h.c.} \right),
  \label{eq.Ham}
\end{eqnarray}
$c^\dag_{\mb{r}_i \s \alpha}$ 
is the electron creation operator 
with spin ${\s\,(=\up,\dn)}$ and orbital $\alpha$ (= $A$, $B$, and $H$) in the $i$-th 
unit cell locating at $\mb{r}_i$, and 
$n_{\mb{r}_i\s \alpha} = c^\dag_{\mb{r}_i\s \alpha} c_{\mb{r}_i \s \alpha}$.
Here, orbital $A$ ($B$) denotes carbon $p_z$ orbital on $A$ ($B$) sublattice of the honeycomb lattice and 
orbital $H$ indicates hydrogen $s$ orbital (see Fig.~\ref{lattice}). 
The conduction band of the periodic Anderson model is described by the first term of Eq.~(\ref{eq.Ham}), 
where the hopping integral $t$ is finite only between the nearest-neighboring carbon sites, indicated 
by the sum over ${\bm\delta}=(0,0)$, $\mb{d}_1 $, and $\mb{d}_2 $ 
with $\mb{d}_1 = \left(  1/2, \sqrt{3}/2 \right)a$ and $\mb{d}_2 = \left( -1/2, \sqrt{3}/2 \right)a$ being 
the primitive translational vectors of the honeycomb lattice 
($a$: the lattice constant between the next nearest-neighboring carbon sites). 
The hybridization between the conduction band in the graphene plane and the 
hydrogen ``impurity'' sites is denoted by $t_{sp}$, where each hydrogen impurity site 
is linked only with the carbon site on $B$ sublattice, as shown in Fig.~\ref{lattice}. 
The on-site potential energy and the on-site Coulomb repulsion at the hydrogen 
impurity sites are denoted by $\eps_H$ and $U$, respectively. 

The periodic Anderson model $\mcal H$ described in Eq.~(\ref{eq:ham}) is the simplest model for graphone, 
implicitly assuming that the hopping integral $t$ in the conduction band 
should be considered as the renormalized one due to the electron correlation in the carbon sites. 
In the following, we set the electron density $n$ to be one for any $U$ at all temperatures 
by imposing the particle-hole symmetry with $\eps_H = -U/2$, and thus  
the local electron density in each site is exactly one. 
We also set $a=\hbar = k_B= 1$.

\subsection{Noninteracting limit}\label{Sec:nonint}

\begin{figure}[!htb]
    \begin{center}
      \includegraphics[clip,width=6.8cm]{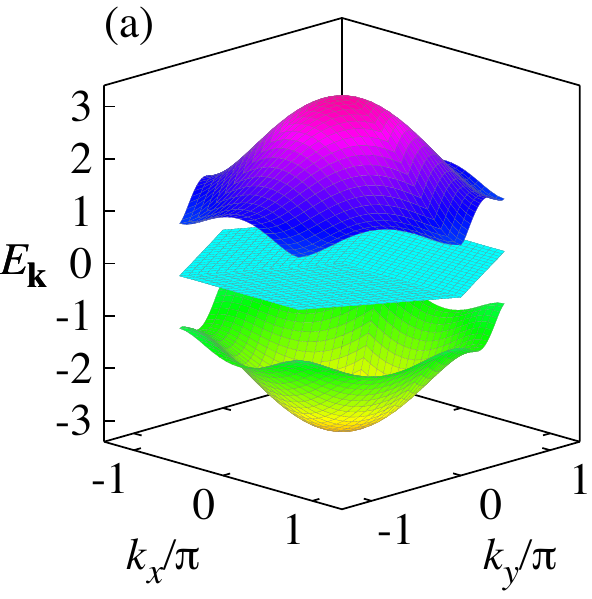}\\
      \includegraphics[clip,width=4.8cm]{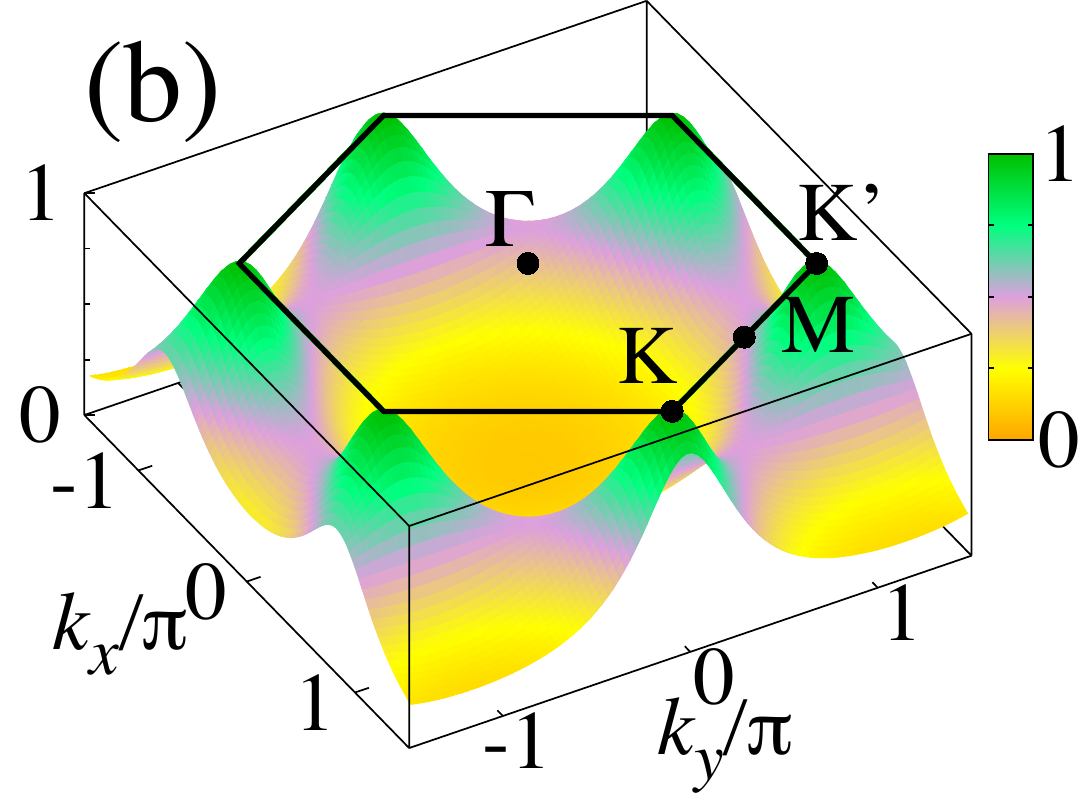}
    \end{center}
  \caption{(color online)
    (a) The energy band structure $E_{\mb k}$ (in unit of $t$) and 
    (b) $\langle\phi_{{\mb {k}}\sigma}^{\rm flat}|c^\dag_{{\mb{k}}\sigma A}c_{{\mb {k}}\sigma A}|\phi_{{\mb {k}}\sigma}^{\rm flat}\rangle$ 
    for the noninteracting case with $t_{sp}/t=1$, where $|\phi_{\mb{k}\sigma}^{\rm flat}\rangle$ is the eigenstate of the flat 
    band with $E_{\mb k}=0$ at momentum $\mb k=(k_x,k_y)$. 
    The first Brillouin zone is indicated by the hexagon with solid lines in (b), where high symmetric momenta are 
    also denoted by $\Gamma$: $(0,0)$, $K$: $\frac{4\pi}{3a}(1,0)$, $K'$: $\frac{4\pi}{3a}(\frac{1}{2},\frac{\sqrt{3}}{2})$, 
    and $M$: $\frac{4\pi}{3a}(\frac{3}{4},\frac{\sqrt{3}}{4})$.
    }
    \label{disp_nonint}
\end{figure}

In the noninteracting limit with $\eps_H =0$, the Hamiltonian leads in the momentum space 
\begin{equation}
  \mcal{H}_{0}=\sum_{\mb{k}, \s} 
  \mb{c}^\dag_{\mb{k}\s}
  \left(
  \begin{array}{ccc}
    0                 &  \gamma_{\mb{k}}  & 0      \\
    \gamma^*_{\mb{k}}  &      0           & t_{sp}  \\
    0                 & t_{sp}           & 0 
  \end{array}
  \right)
  \mb{c}_{\mb{k}\s},
  \label{h_free}
\end{equation}
where 
$\mb{c}^\dag_{\mb{k}\s} = ( c_{\mb{k} \s A}^\dag \,  c_{\mb{k} \s  B}^\dag \,   c_{\mb{k} \s H}^\dag )$ 
is the Fourier transform of the real space creation operators and 
\begin{equation}
\gamma_\mb{k} = -t \left(1 + e^{i \mb{k}\cdot \mb{d_1}} + e^{i \mb{k} \cdot \mb{d_2}} \right).
\label{eq:gamma_k}
\end{equation} 
The characteristic features are summarized as follows (see also Fig.~\ref{disp_nonint}): 
(i) The Dirac cone dispersions which are present for the pure graphene model are now absent, 
(ii) instead, the massive Dirac dispersions, described as $E_{\mb{k}} = \pm \sqrt{|\gamma_{\mb{k}}|^2 + t_{sp}^2}$, 
appear near $K$ and $K'$ points, 
and (iii) in addition there exists the flat band at $E_{\rm F}$, i.e., $E_{\mb{k}}=0$, which is composed of $A$ and $H$ 
orbitals (solely of $A$ orbital at $K$ and $K'$ points), but not $B$ orbital.  
Note also that the flat band is exactly half-filled at $n=1$. 

The features (i), (ii) and (iii) are 
understood by noticing that ${\cal H}_0$ satisfies the Lieb's condition on a bipartite lattice 
with no hopping between the same sublattices~\cite{Lieb}. Following Lieb's argument, 
${\cal H}_0$ has $|A|+|H|-|B|$ ($|\alpha|$: the number of $\alpha$ orbitals) zero eigenvalues, forming the flat band, 
and the wave functions with zero eigenvalues are contributed only from $A$ and $H$ orbitals. 
The simplified tight binding model considered here already captures the main characteristic features 
obtained by spin-unpolarized DFT calculations, 
including the almost dispersionless band near $E_{\rm F}$ 
with almost zero weight of hydrogen orbitals around $K$ and $K'$ points~\cite{Zhou2009,Rudenko2013}. 

\subsection{Variational cluster approximation}

We employ the VCA~\cite{Potthoff2003} to investigate a possible symmetry broken magnetic ordered state. 
Here, we introduce, as a variational parameter, a uniform field $h'$ on the hydrogen impurity 
sites~\cite{Horiuchi2008} described as 
\begin{equation}
  \mcal{H}_{h'} = h' \sum_i \left(n_{\mb{r}_i\up H} - n_{\mb{r}_i\dn H} \right). 
\end{equation}
The reference system $\mcal{H}_{\rm ref}$ considered is thus composed of $\mcal{H}_{h'}$ and 
a collection of disconnected 
finite size clusters, as shown in Fig.~\ref{disp_nonint} (a), where 
each cluster is described by $\mcal{H}$ but with no hopping terms between clusters, 
the corresponding Hamiltonian being denoted as $\mcal{H}_{\rm c}$. 
Hence, the reference system is described as $\mcal{H}_{\rm ref}=\mcal{H}_{\rm c}+\mcal{H}_{h'}$. 

The VCA evaluates as a function of $h'$ the grand potential functional 
\begin{equation}\label{eq.Omega-num}
  \mcal{F} =
  \frac{1}{L_\mr{c}}\mcal{F}' - \frac{T}{NL_{\mr{c}} }  
  \sum_{\s} 
  \sum_{\w_\nu} 
  \sum_{\mb{\tilde{k}}} 
  \ln \det
  \left[
    \bs{I}-\bs{V}_\s(\mb{\tilde{k}}) \bs{G}_\s'(\imag \w_\nu)
  \right],
\end{equation}
where $\w_\nu = (2\nu + 1) \pi T$ with an integer $\nu$ is the 
Matsubara frequency for a given temperature $T$ 
and the wave vector $\mb{\tilde{k}}$ is defined in the reduced Brillouin zone of the reference system. 
The reference system $\mcal{H}_{\rm ref}$ comprises $N$ identical clusters  
and each cluster contains $L_{\rm c}$ unit cells. 
The single-particle Green's function of 
a single cluster in $\mcal{H}_{\rm ref}$ is denoted as $\bs{G}'_\s(i\w_\nu)$. 
$\bs{V}_\s(\mb{\tilde{k}})$ is the $(3L_{\rm c}\times 3L_{\rm c})$ sub-matrix element of 
block-diagonalized $\bs{V}$ in the momentum ($\mb{\tilde{k}}$) and spin ($\s$) spaces, where $\bs{V}$ is a 
matrix representation of the one-body Hamiltonian $\mcal{H}_{\rm ref} - \mcal{H}$.  
${\bs I}$ is the $(3L_{\rm c}\times 3L_{\rm c})$ unit matrix. 
The grand potential $\mcal{F}'$ of the single cluster is readily evaluated as 
\begin{equation}
\mcal{F}' = -T \ln \sum_{s} \exp(-E_{s}/T),
\end{equation} 
where $E_s$ is the $s$-th eigenvalue of 
a single cluster in $\mcal{H}_{\rm ref}$. 
The exact diagonalization method is employed to obtain $\bs{G}_\s'(i\w_\nu)$ and $\mcal{F}'$ numerically exactly. 
The FM state is obtained when a saddle point 
$\partial \mcal{F} /\partial h'|_{h'=h^*} = 0$ with the lowest $\mcal{F}$ is 
at $h^* \not = 0$. 

\subsection{Cluster perturbation theory}

The CPT~\cite{Senechal2000} is employed to obtain the translationally invariant single-particle 
Green's function of the infinite system.
In the CPT, the single-particle Green's function $\mcal{G}^{\alpha \beta}_\sigma(\mb{k},z)$ 
of $\mcal H$ is given as 
\begin{eqnarray}\label{eq:gcpt}
  \mcal{G}^{\alpha \beta}_\sigma(\mb{k},z)  = 
  \frac{1}{L_c} 
  \sum_{i,j}
  \left(\bs{G}'^{-1}_\sigma(z) - \bs{V}_\sigma(\mb{k}) \right)_{i\alpha, j \beta}^{-1} 
  e^{-i \mb{k} \cdot (\mb{r}_i - \mb{r}_j)}, 
\end{eqnarray}
where $\alpha$ and $\beta$ are orbital indices (i.e., $A$, $B$, and $H$ orbitals) 
and the sums over $i$ and $j$ are for unit cells within a single cluster in $\mcal H_{\rm ref}$. 
The single-particle Green's function 
$\bs{G}'_\s(z)$ of the single cluster is 
obtained within the VCA, as described above. 
Note here that the momentum $\mb{k}$ and the complex frequency $z$ can take any values. 
Therefore, we can achieve arbitrarily fine resolution of $\mb{k}$ and $z$ for 
the single-particle excitations, which allows us for the detailed analysis of the spectral properties including 
the spectral weight and the Dirac Fermi velocity.

\section{Numerical results}\label{sec:results}

In this section, we first discuss the finite temperature phase diagram of the periodic Anderson model 
obtained by the VCA. Next, we examine in details the single-particle excitations in each phase of the 
phase diagram using the CPT. 

\subsection{Phase diagram}\label{sec:pd}

\begin{figure}
  \begin{center}
    \includegraphics[width=7cm]{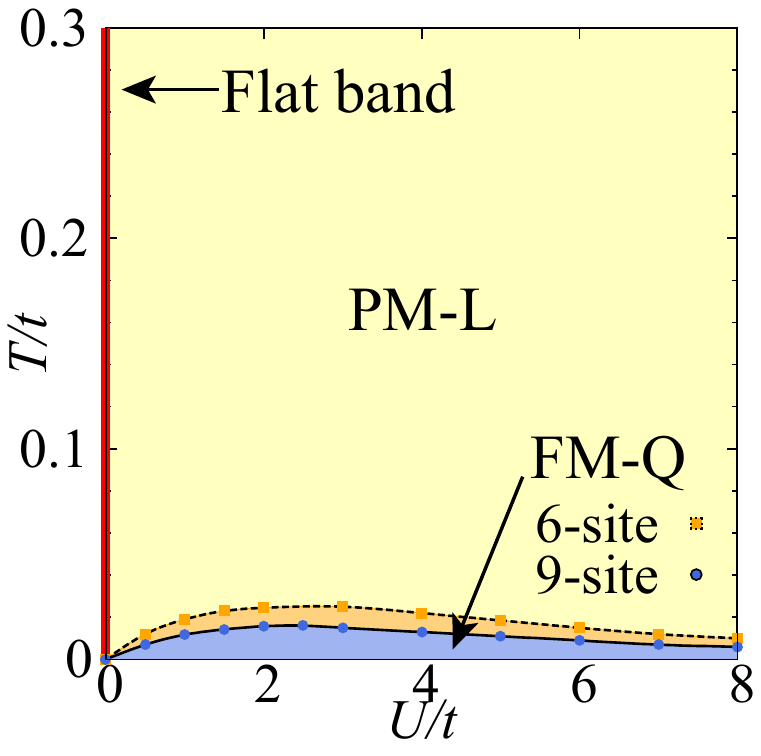}
  \end{center}
  \caption{(color online) 
    \label{fig:pd}
    Finite temperature phase diagram of the periodic Anderson model $\mcal H$ obtained by the VCA for 
    the 6-site and 9-site clusters with $t_{sp}/t=1$ at $n=1$. 
    PM-L (FM-Q) stands for a PM (FM) phase with the linear (quadratic) quasiparticle dispersion 
    around $E_{\rm F}$. 
    In the noninteracting limit with $U=0$, the flat band appears at $E_{\rm F}$ (red solid line). 
    The FM state 
    is stable in a blue (orange) shaded region below a solid (dashed) line for the 9-site (6-site) cluster. 
  }
\end{figure}

The finite temperature phase diagram obtained by the VCA 
is summarized in Fig.~\ref{fig:pd}. 
We find that the ground state is always FM 
semi-metallic for $U>0$ and the magnetic moment 
\begin{equation}
m_z=\frac{1}{NL_{\rm c}}\sum_{i}\sum_{\alpha}
\left(\langle n_{\mb{r}_i\uparrow\alpha} \rangle - \langle n_{\mb{r}_i\downarrow\alpha} \rangle\right)
\end{equation} 
is exactly one, where $\langle\cdots\rangle$ implies the thermal average. 
Thus, strictly speaking, the ground state is ferrimagnetic~\cite{seki2015}.
As discussed above in Sec.~\ref{Sec:nonint}, in the noninteracting limit with $U=0$, 
the flat band exists exactly at $E_{\rm F}$ and is half-filled [see Fig.~\ref{disp_nonint}(a)]. 
Therefore, the system is unstable against FM order upon introducing $U$. 
We assign the origin of this FM state to be flat-band ferromagnetism~\cite{Titvinidze}. 
It should be also noted that, in the strong coupling limit 
where an electron in each hydrogen impurity site is completely localized, 
the Ruderman-Kittel-Kasuya-Yoshida (RKKY) 
interaction~\cite{RKKY} between these localized spins is FM~\cite{Saremi2007} 
(see also Appendix~\ref{appsec:rkky}), which naturally induces the FM ground state. 

As discussed in Appendix~\ref{appsec:lieb}, 
Lieb's theorem itself~\cite{Lieb} does not guarantee the uniqueness of the ground state of $\mcal H$. 
This is simply because there is no on-site interaction $U_C$ for the carbon conduction sites in $\mcal H$. 
However, numerically exactly diagonalizing small clusters, we find in Appendix~\ref{appsec:ed} 
that in the parameter region studied here the ground state for $U_C\ne0$ is smoothly connected to 
the {\it non-degenerate} ground state for $U_C=0$ (apart from the trivial spin degeneracy), implying 
that the ground state of $\mcal H$ is the Lieb-Mattis type ferromagnetism 
on a bipartite lattice with total spin $S=(|A|+|H|-|B|)/2$~\cite{LM}. 

With increasing the temperature, 
however, the FM state is thermally destroyed and a PM state becomes stable. 
Notice here that the finite FM critical temperature $T_{\rm C}$ found in the VCA is 
due to a mean-field like treatment of the electron correlation beyond the size of clusters. 
Indeed, increasing the size of a cluster from 6 sites to 9 sites,  
we find that $T_{\rm C}$ decreases for all values of $U/t$, as shown in Fig.~\ref{fig:pd}. 
In Appendix~\ref{appsec:sw}, we also analyze with the linear spin wave theory an effective spin Hamiltonian 
for the periodic Anderson model in the strong coupling limit, and find that the spin wave dispersion 
$\Omega_{\mb q}$ with momentum ${\mb q}$ around $\Gamma$ point is proportional 
to $|{\mb q}|$ at $T=0$ but $|{\mb q}|^2$ at finite temperatures 
(see Fig.~\ref{fig:spinwave} and \ref{fig:spinwaveFT}), 
implying that the FM order is stable only 
at $T=0$, as is expected from Mermin-Wagner theorem~\cite{Mermin1966} 
Therefore, the finite $T_{\rm C}$ obtained in the VCA should be regarded as a temperature where the short 
range FM correlations are developed over the size of a cluster, and 
the finite temperature phase diagram is dominated by the PM phase. 

As shown below, it is more surprising to find in the PM phase that 
massless Dirac quasiparticles 
emerge at $K$ and $K'$ points with the Dirac points exactly at $E_{\rm F}$. 

\subsection{Single-particle excitations}
The single-particle excitation spectrum 
\begin{equation}
  A^{\alpha\beta}_{\s}(\mb{k},\w) =-\frac{1}{\pi} \Im  \mcal{G}^{\alpha \beta}_\s (\mb{k},\w + i \eta) 
\end{equation}
can be easily obtained from the single-particle Green's function $\mcal{G}^{\alpha \beta}_\s({\mb k},z)$ 
calculated using the CPT in Eq.~(\ref{eq:gcpt}). Here, $\w$ is the real frequency and $\eta$ is 
real positive infinitesimal for Lorentzian broadening of the spectrum. 

\subsubsection{FM ground state}\label{sec:fm}

\begin{figure*}
  \includegraphics[width=4.3cm,angle=0]{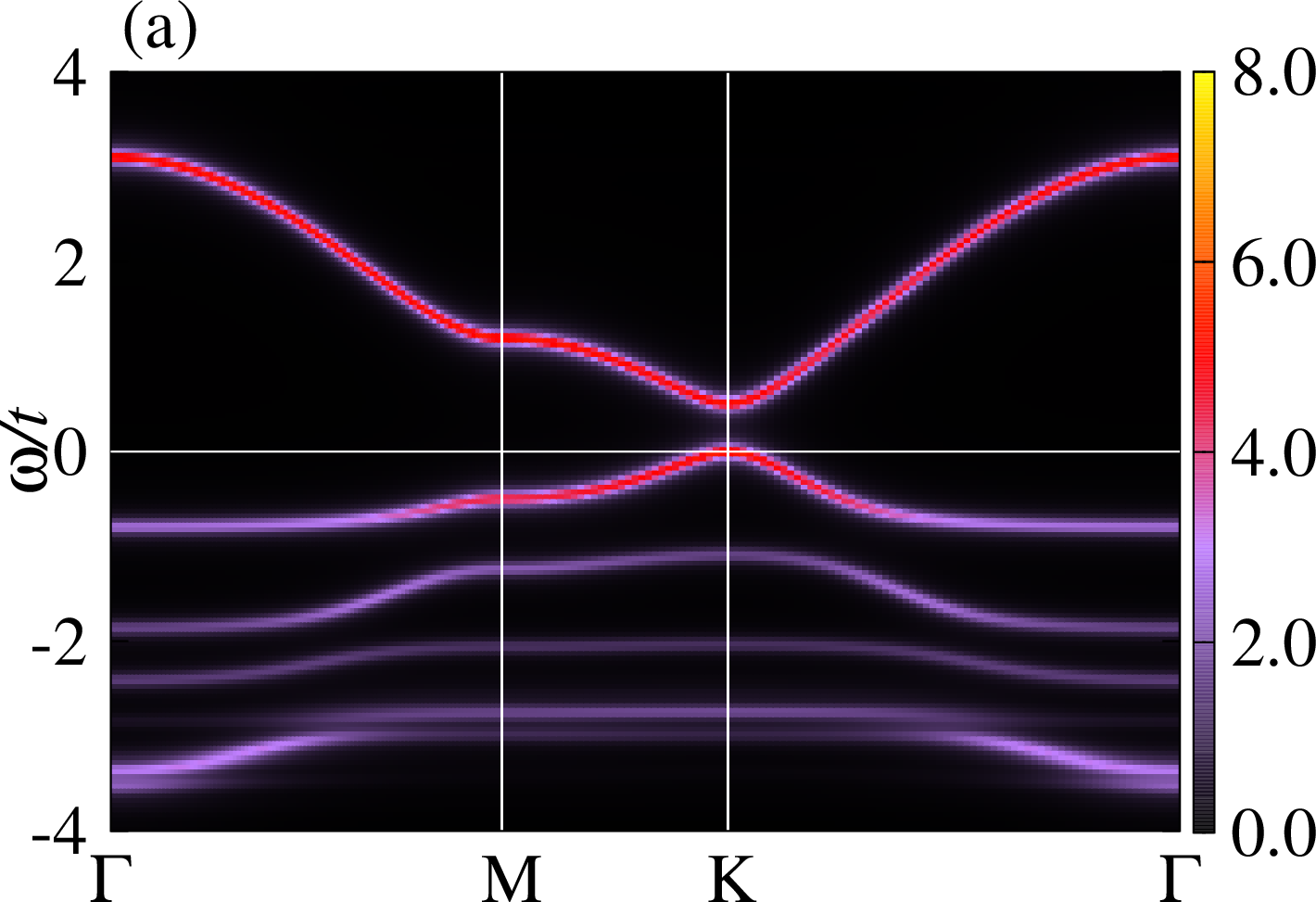}
  \includegraphics[width=4.3cm,angle=0]{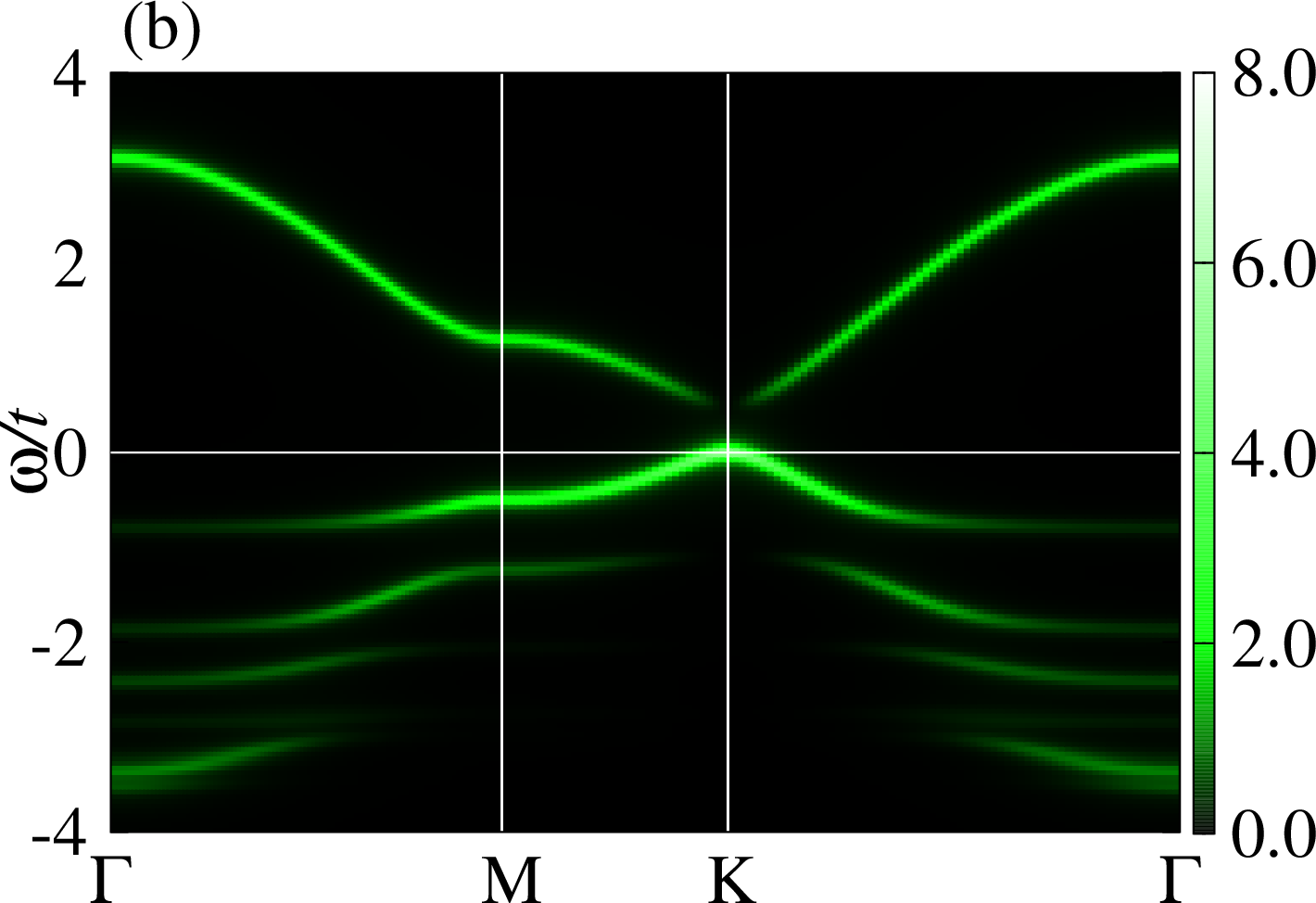}
  \includegraphics[width=4.3cm,angle=0]{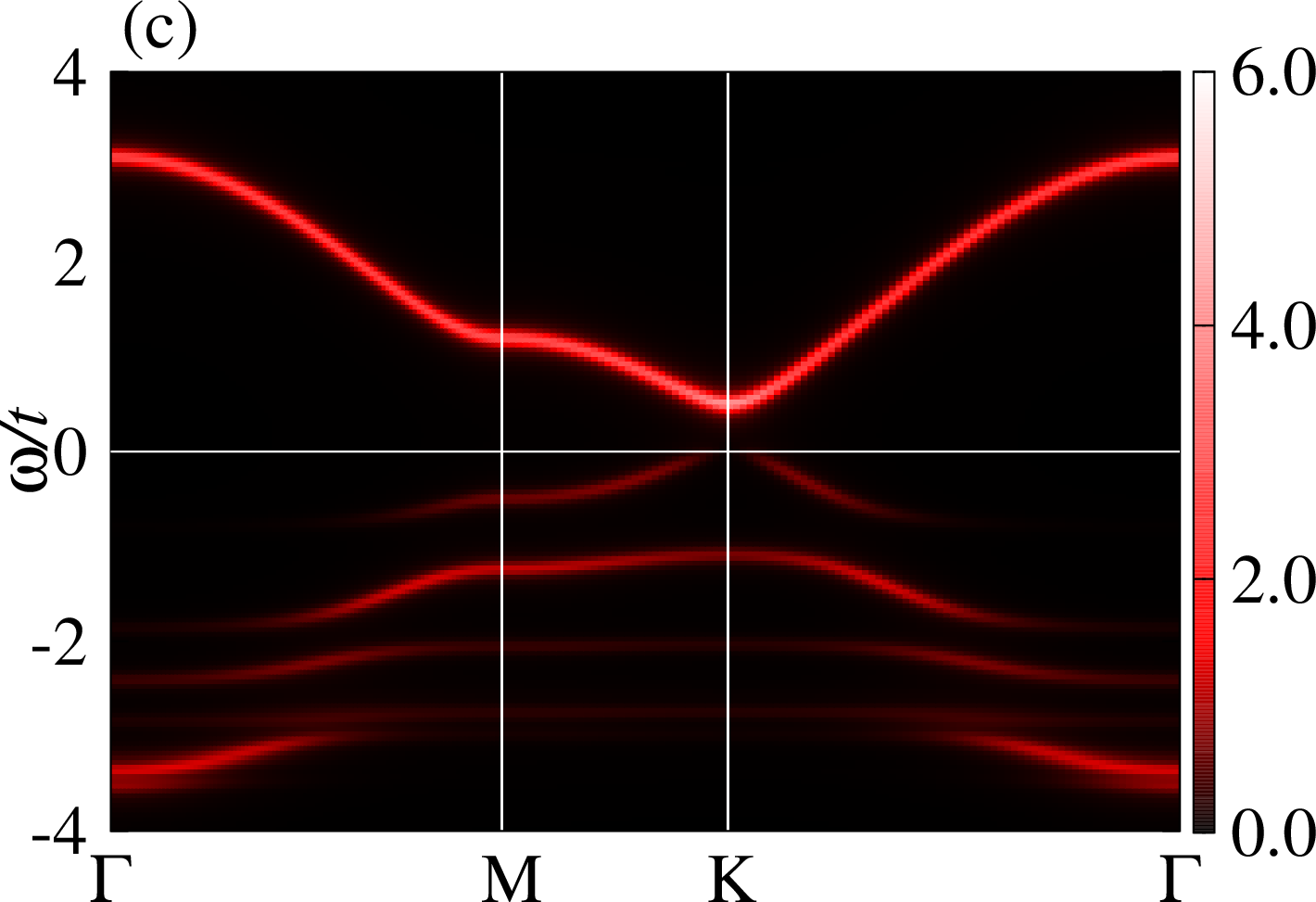}
  \includegraphics[width=4.3cm,angle=0]{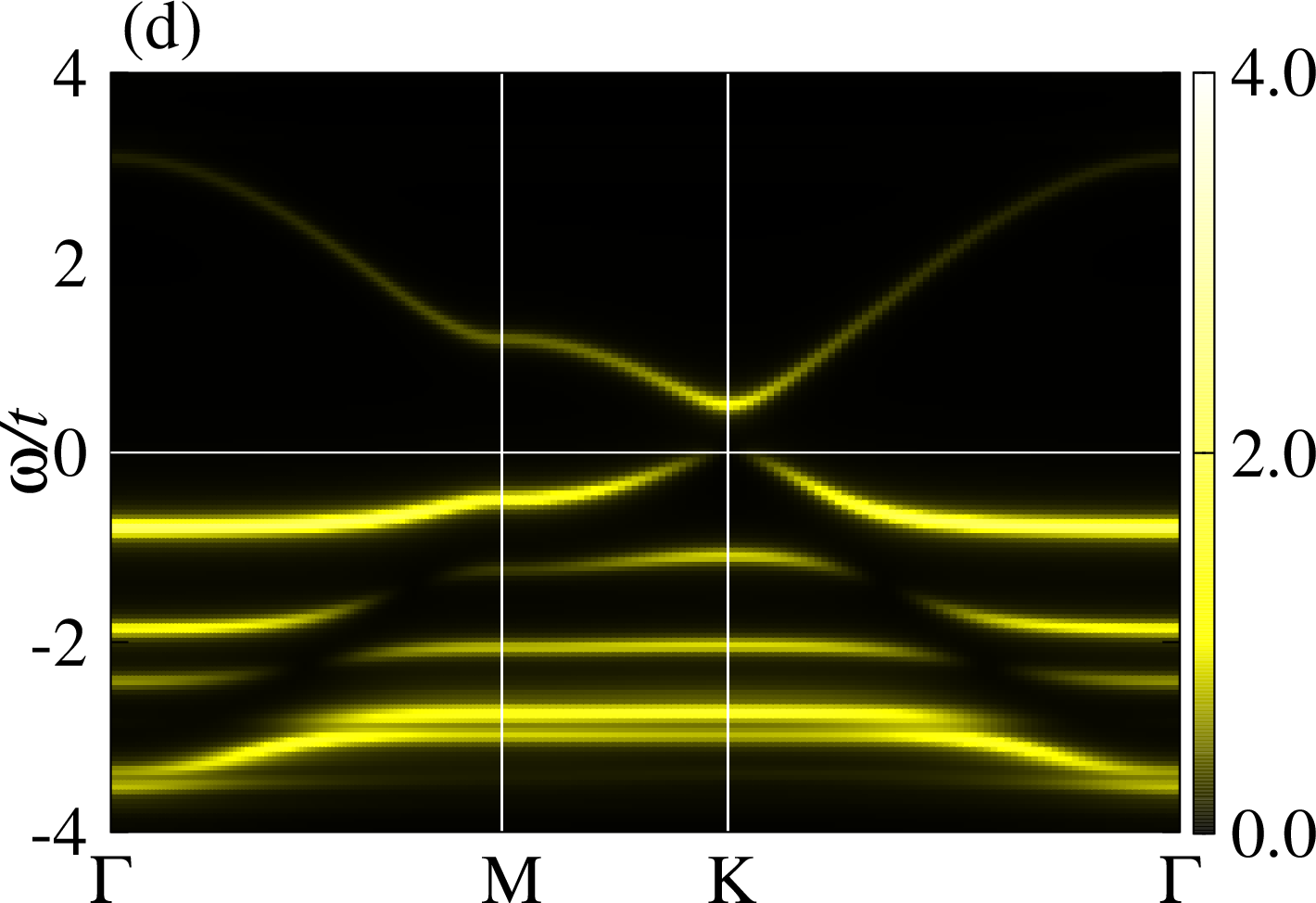}\\
  \includegraphics[width=4.3cm,angle=0]{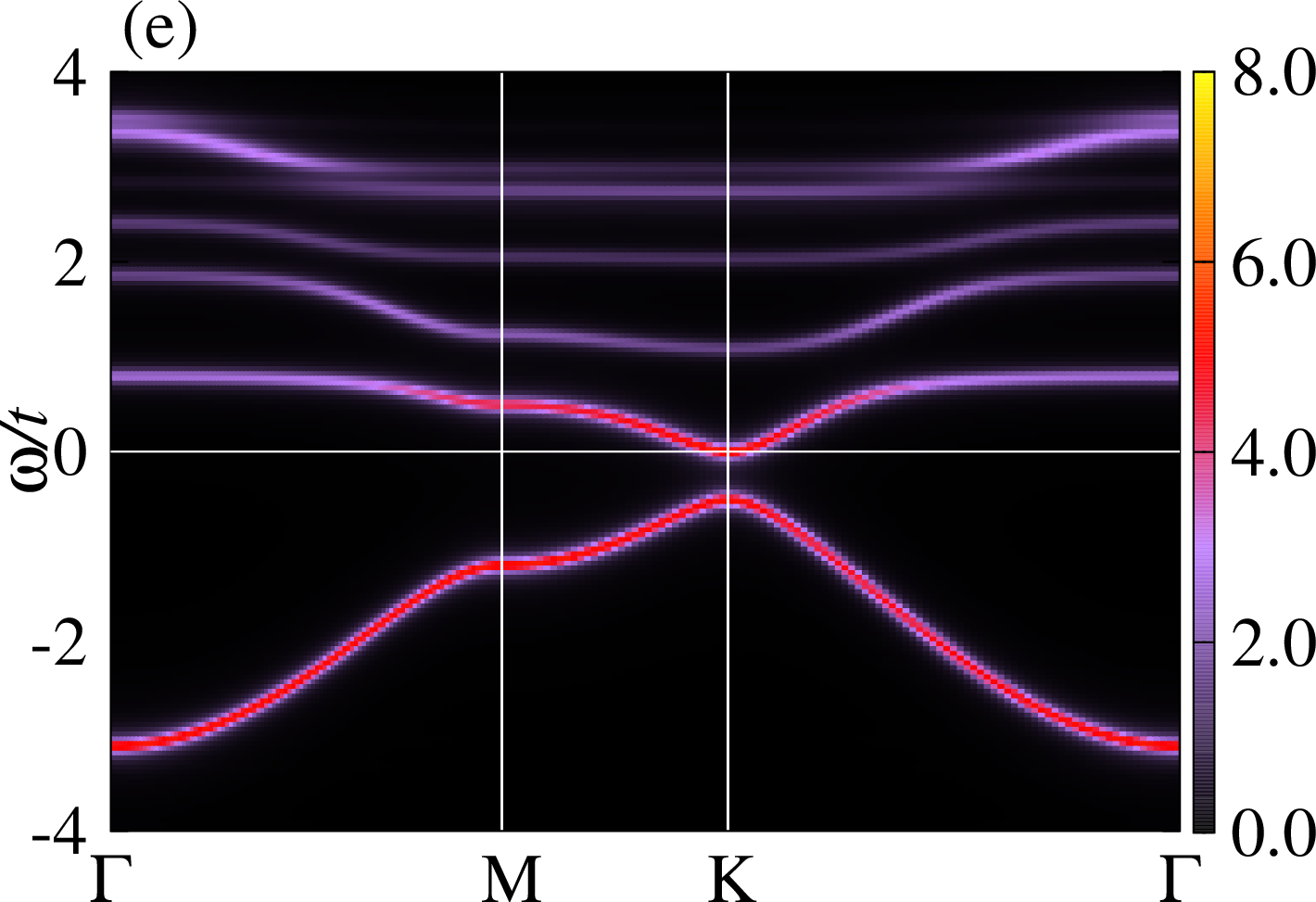}
  \includegraphics[width=4.3cm,angle=0]{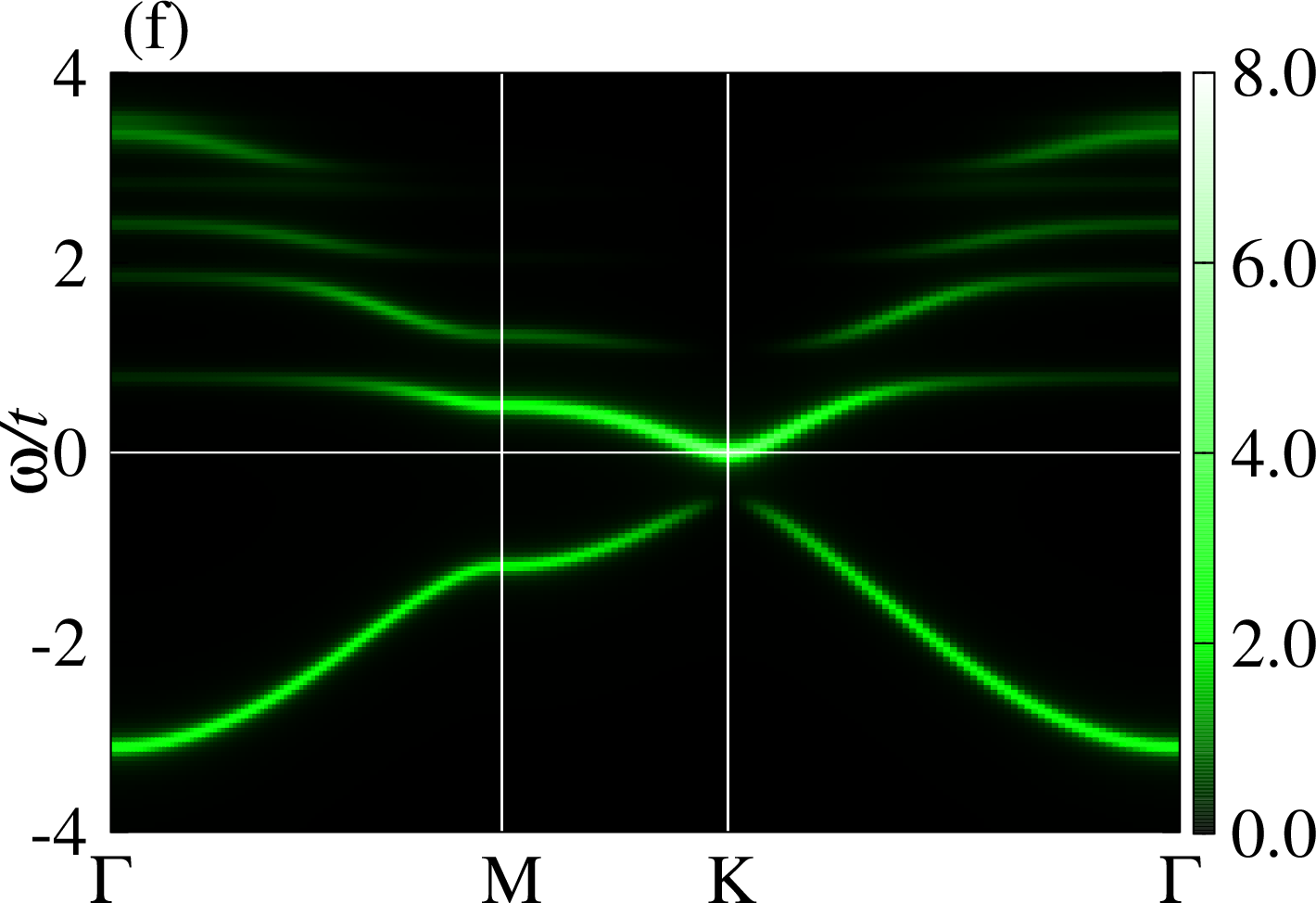}
  \includegraphics[width=4.3cm,angle=0]{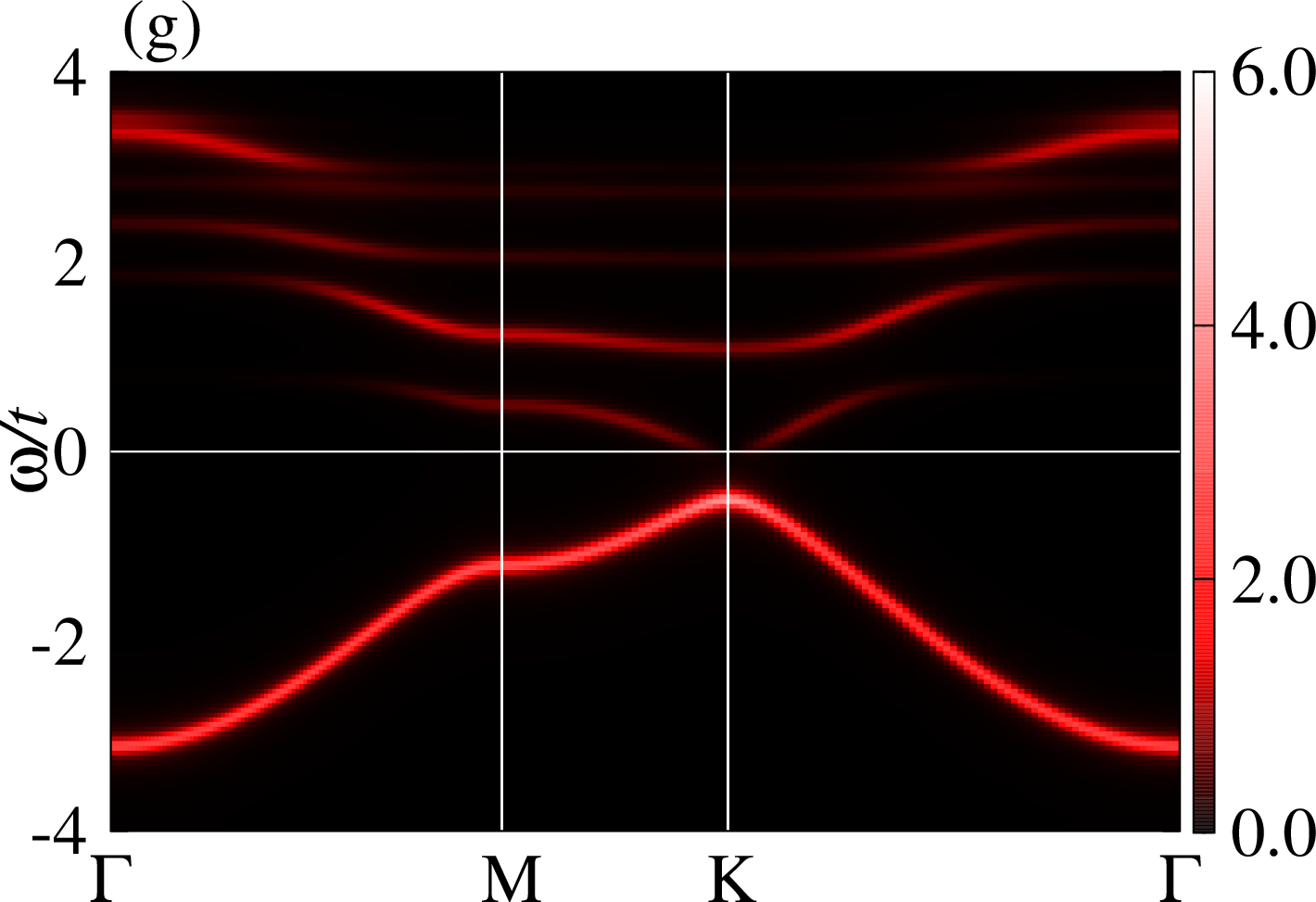}
  \includegraphics[width=4.3cm,angle=0]{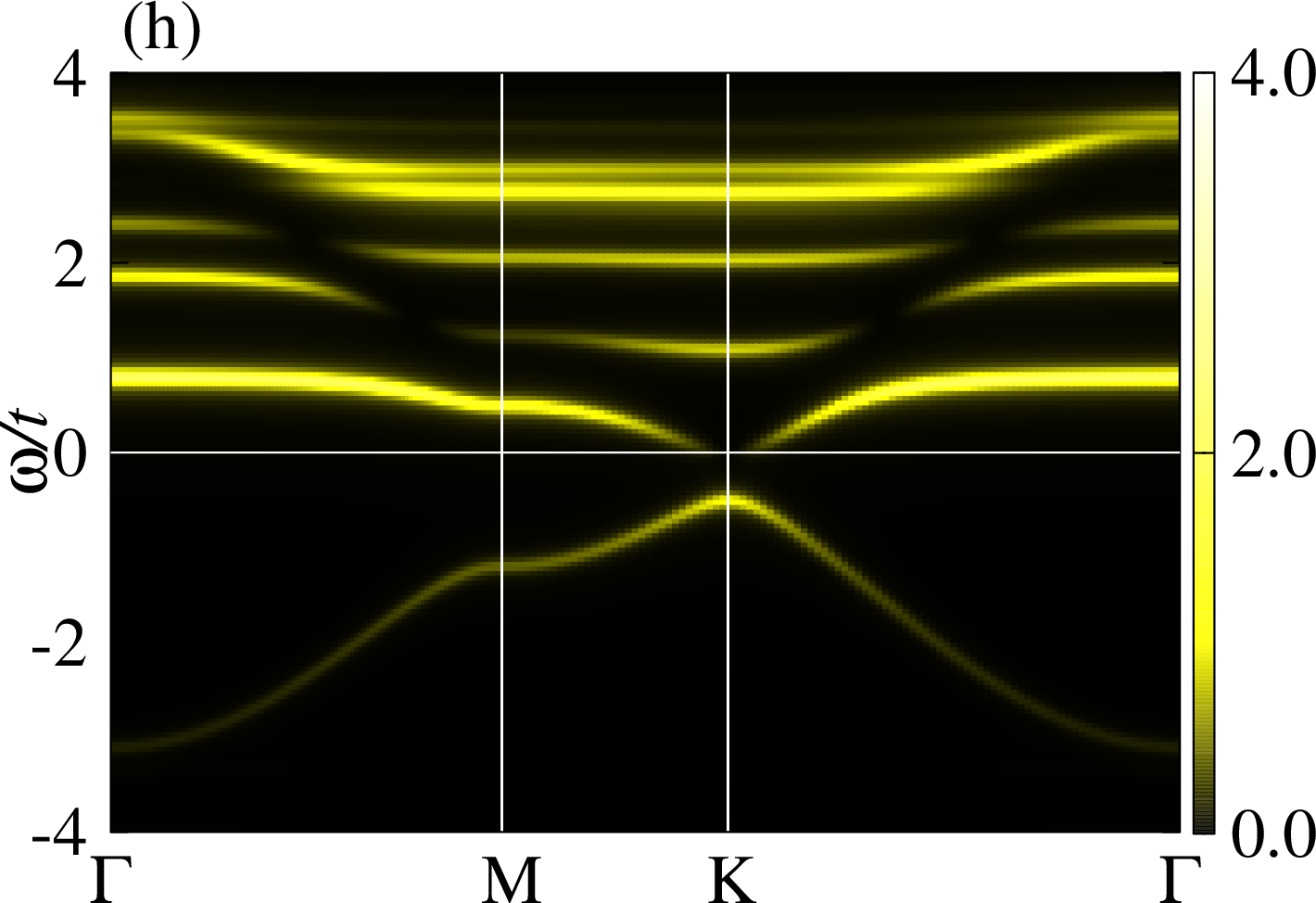}
  \caption{(color online) 
    The CPT results of the single-particle excitation spectra for (a)--(d) up and 
    (e)--(h) down electrons in the FM state at $T=0$. 
    The calculations are for $U/t = 4$ and $t_{sp}/t=1$ using the 9-site cluster (containing 3 unit cells). 
    (a) and (e): $A_\sigma(\mb{k},\w) = \sum_\alpha A_\sigma^{\alpha\alpha}(\mb{k},\w)$, 
    (b) and (f): $A^{AA}_\sigma(\mb{k},\w)$, 
    (c) and (g): $A^{BB}_\sigma(\mb{k},\w)$, and 
    (d) and (h): $A^{HH}_\sigma(\mb{k},\w)$. 
    A Lorentzian broadening of $\eta/t = 0.05$ is used. 
    The spectral intensity is
    indicated by a color bar in each figure. 
    Notice that the different intensity scales are used for different figures. 
    The Fermi energy $E_{\rm F}$ is located at $\w=0$. 
  }
  \label{fig:akw_fm}
\end{figure*}

\begin{figure*}
  \begin{center}
    \includegraphics[width=16cm]{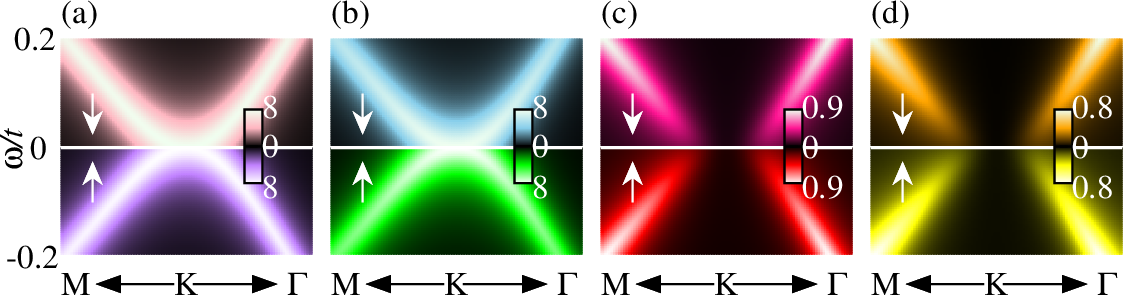}
  \end{center}
  \caption{(color online) 
    Same as Fig.~\ref{fig:akw_fm} but the enlarged scale at the vicinity of $K$ point near the Fermi energy 
    $E_{\rm F}$ (white lines). 
    (a) $A_\sigma(\mb{k},\w) = \sum_\alpha A_\sigma^{\alpha\alpha}(\mb{k},\w)$, 
    (b) $A^{AA}_\sigma(\mb{k},\w)$, 
    (c) $A^{BB}_\sigma(\mb{k},\w)$, and 
    (d) $A^{HH}_\sigma(\mb{k},\w)$. 
    Notice that the spectra $A^{\alpha\alpha}_\uparrow(\mb{k},\w)$ and 
    $A^{\alpha\alpha}_\downarrow(\mb{k},\w)$ for up and down spins, respectively, are 
    plotted with different colors below and above $E_{\rm F}$ at $\omega=0$. 
    The region of momenta taken in the horizontal axis is $0.2\pi$ in the $K$-$M$ ($K$-$\Gamma$) 
    direction from $K$. 
  }
  \label{fig:akw_fm_en}
\end{figure*}

Figure~\ref{fig:akw_fm} shows the typical results of the single-particle excitation spectrum  
in the FM ground state for $U/t=4$ at $T=0$. 
The enlarged spectrum close to the Fermi energy $E_{\rm F}$ around $K$ point is also shown 
in Fig.~\ref{fig:akw_fm_en}. 
It is clearly observed in Fig.~\ref{fig:akw_fm_en} that (i) the low-energy single-particle excitations around 
$E_{\rm F}$ display the quadratic energy dispersion in momentum, indicating massive quasiparticle 
excitations, and (ii) the lowest single-particle excitations around $K$ point (and also $K'$ point) are 
composed mostly of $A$ orbital (the lowest excitations at $K$ and $K'$ points are solely due to $A$ orbital 
and their spectral weights are independent of $U$). 
The latter is the remnant to the noninteracting case shown in Fig.~\ref{disp_nonint}(b).

It is also noticeable in Fig.~\ref{fig:akw_fm} that (iii) the lowest single-particle excitations among the same 
spin have a finite gap $\Delta_{\rm c}$, suggesting that the spin conserved charge excitations are gapped, 
while (iv) the spin excitation gap should be zero since 
the  lowest single-particle excitation gap $\Delta_{\rm s}$ among the opposite spins is zero, 
as shown in Fig.~\ref{fig:akw_fm_en}. 
We also examine the $U$ dependence of the single-particle excitation spectrum 
and find that (v) although $\Delta_{\rm c}$ ($\Delta_{\rm s}$) 
monotonically increases (remains zero) with increasing $U$, 
the effective mass $m^*$ simply deceases, where $m^*$ is inversely 
proportional to the curvature of the quadratic energy dispersion for the lowest single-particle excitations. 
As discussed in Sec.~\ref{sec:mf}, these features (i)--(v) can be understood within a simple 
mean-field theory. 

\subsubsection{PM state}\label{sec:CPT_pm}

\begin{figure*}
  \begin{center}
    \includegraphics[width=16cm]{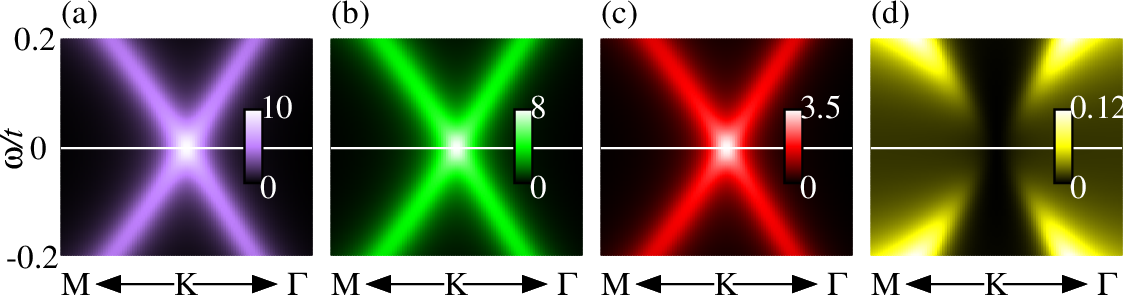}
  \end{center}
  \caption{(color online) 
    The CPT results of the single-particle excitation spectra in the PM state for $U/t = 4$ and $t_{sp}/t=1$ 
    obtained using the 9-site cluster (containing 3 unit cells) 
    at $T/t=0.025$ around $K$ point near the Fermi energy $E_{\rm F}$ (white lines). 
    (a) $A_\sigma(\mb{k},\w) = \sum_\alpha A_\sigma^{\alpha\alpha}(\mb{k},\w)$, 
    (b) $A^{AA}_\sigma(\mb{k},\w)$, 
    (c) $A^{BB}_\sigma(\mb{k},\w)$, and 
    (d) $A^{HH}_\sigma(\mb{k},\w)$. 
    Here, we only show the spectra $A^{\alpha\alpha}_\uparrow(\mb{k},\w)$ for up electrons, 
    which are exactly the same as $A^{\alpha\alpha}_\downarrow(\mb{k},\w)$. 
    A Lorentzian broadening of $\eta/t = 0.05$ is used. 
    The spectral intensity is indicated by a color bar in each figure. 
    Notice that the different intensity scales are used for different figures. 
    The region of momenta taken in the horizontal axis is $0.2\pi$ in the $K$-$M$ ($K$-$\Gamma$) 
    direction from $K$. 
  }\label{fig:akw_pm_en}
\end{figure*}

\begin{figure*}
\includegraphics[width=4.3cm,angle=0]{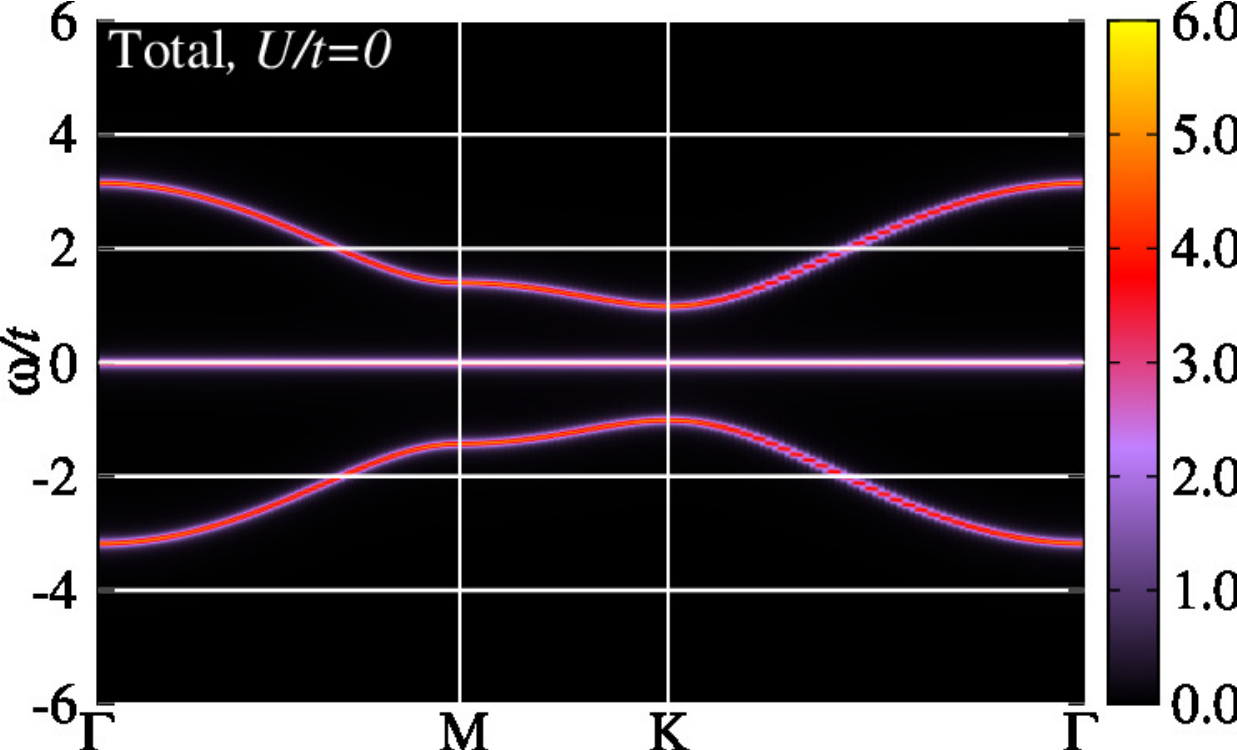}
\includegraphics[width=4.3cm,angle=0]{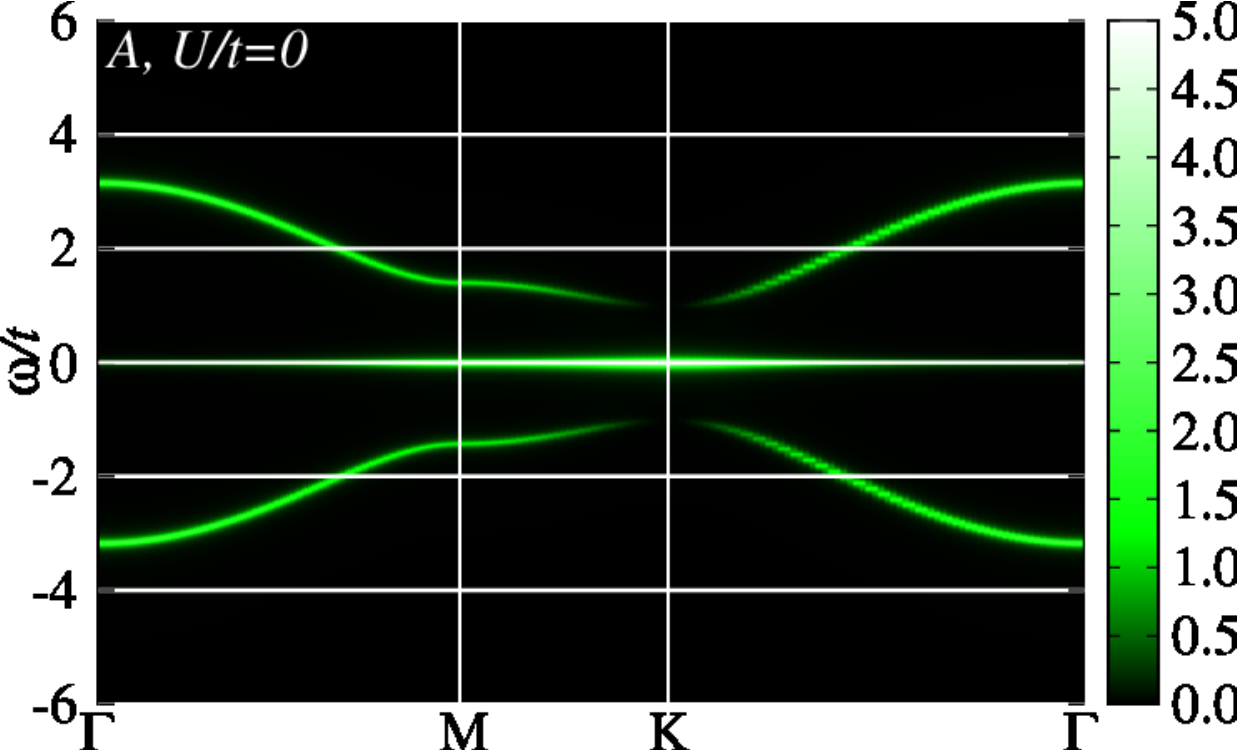}
\includegraphics[width=4.3cm,angle=0]{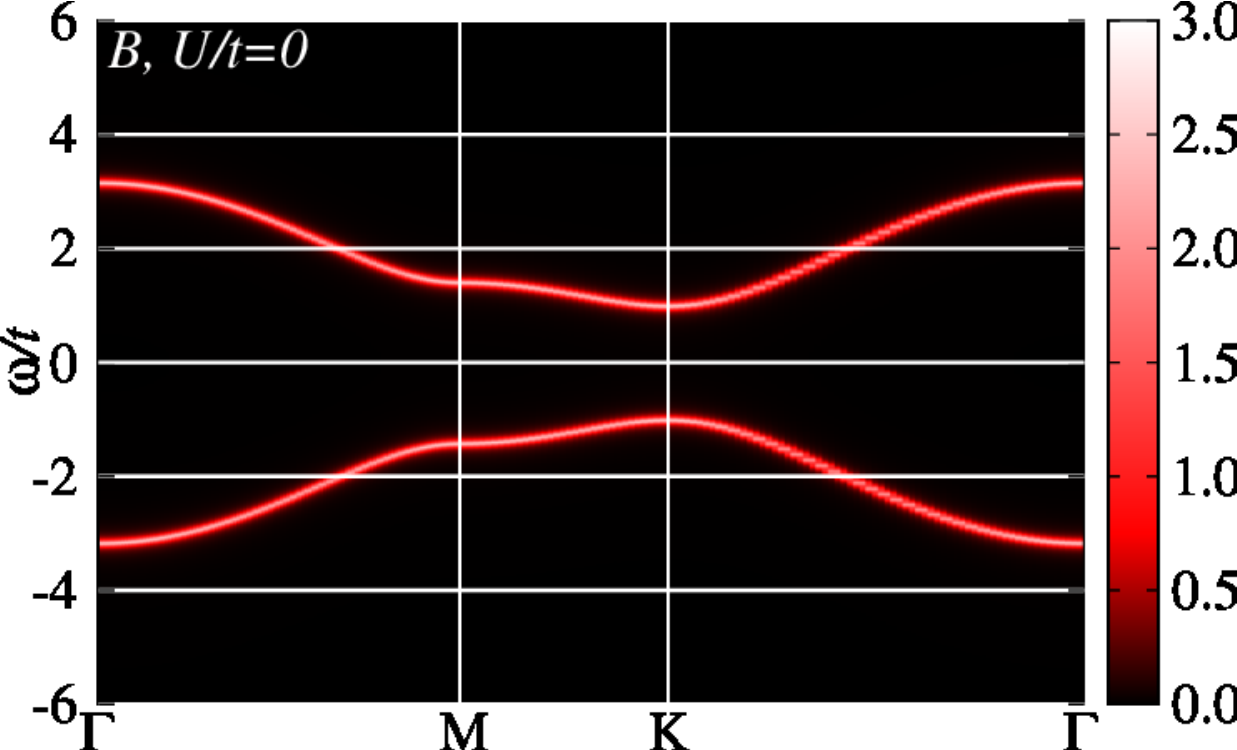}
\includegraphics[width=4.3cm,angle=0]{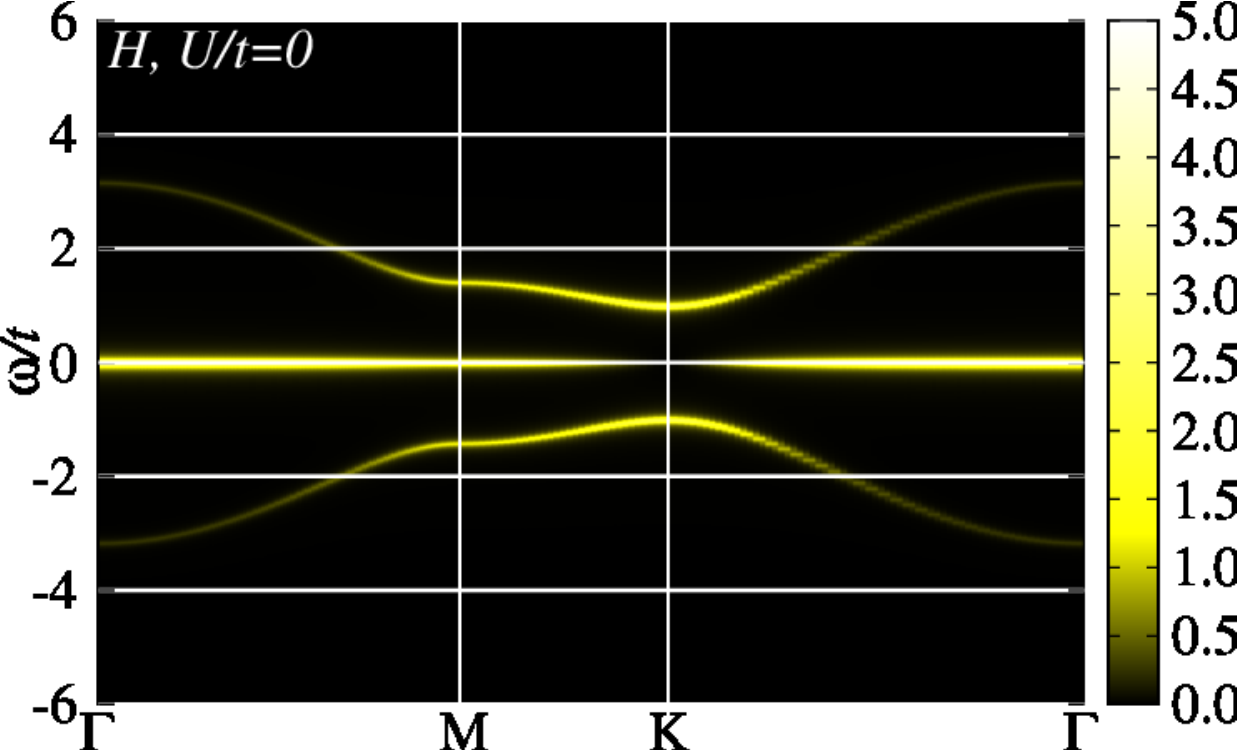}\\
\includegraphics[width=4.3cm,angle=0]{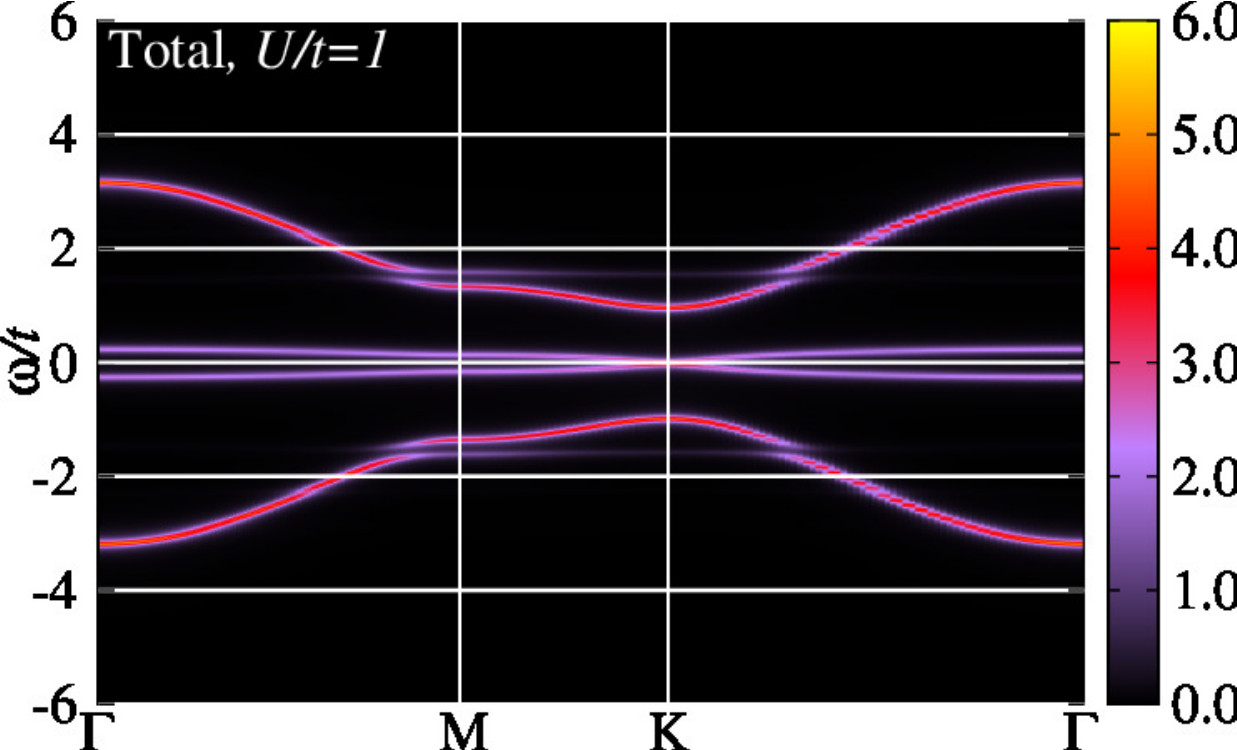}
\includegraphics[width=4.3cm,angle=0]{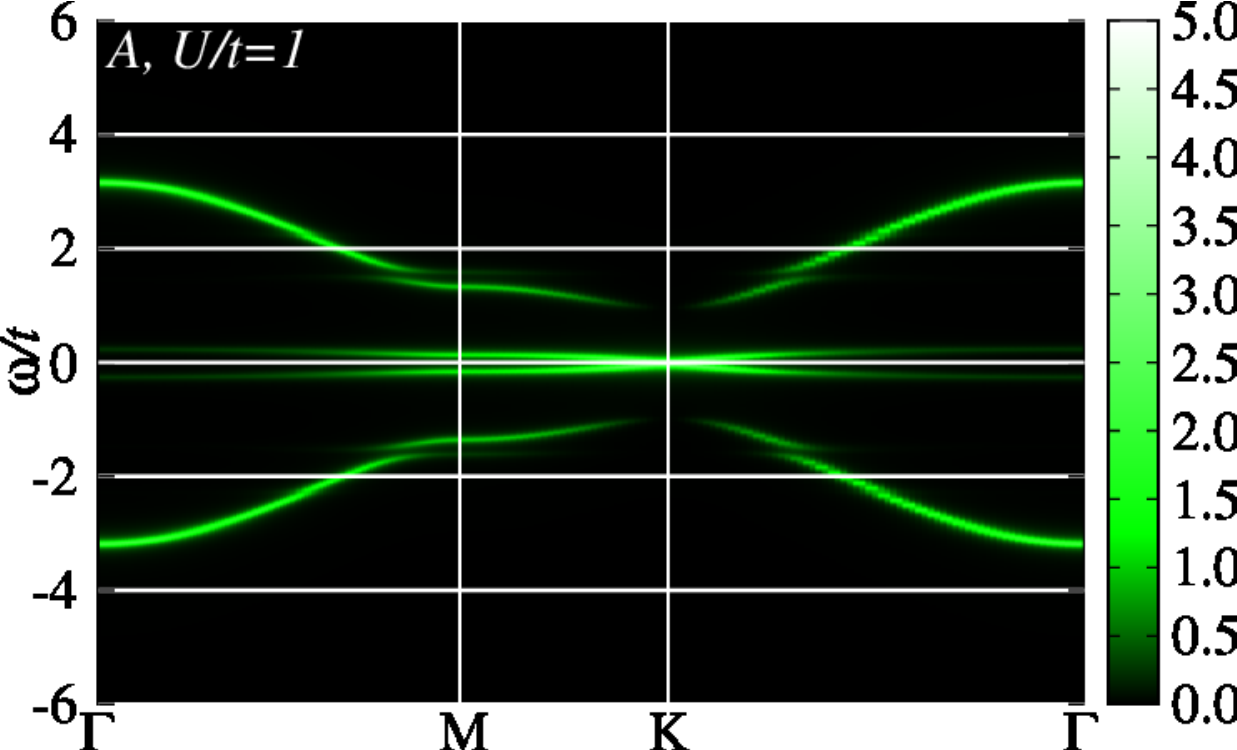}
\includegraphics[width=4.3cm,angle=0]{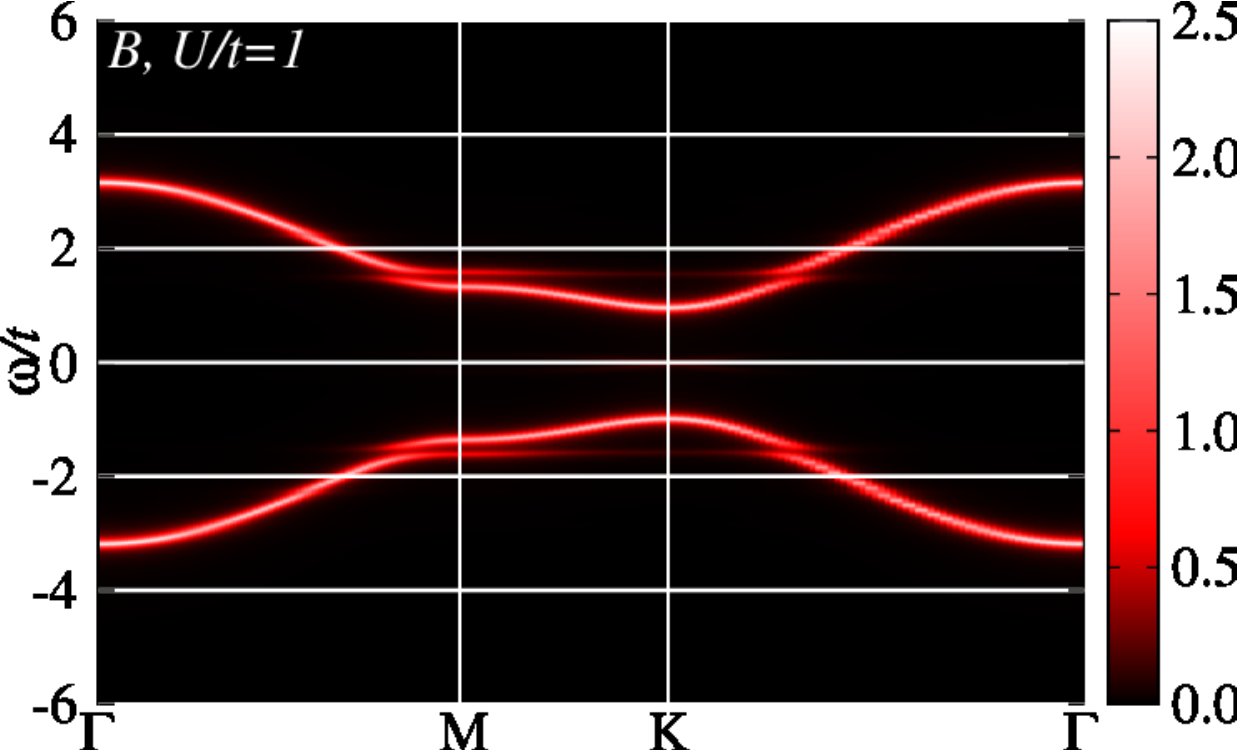}
\includegraphics[width=4.3cm,angle=0]{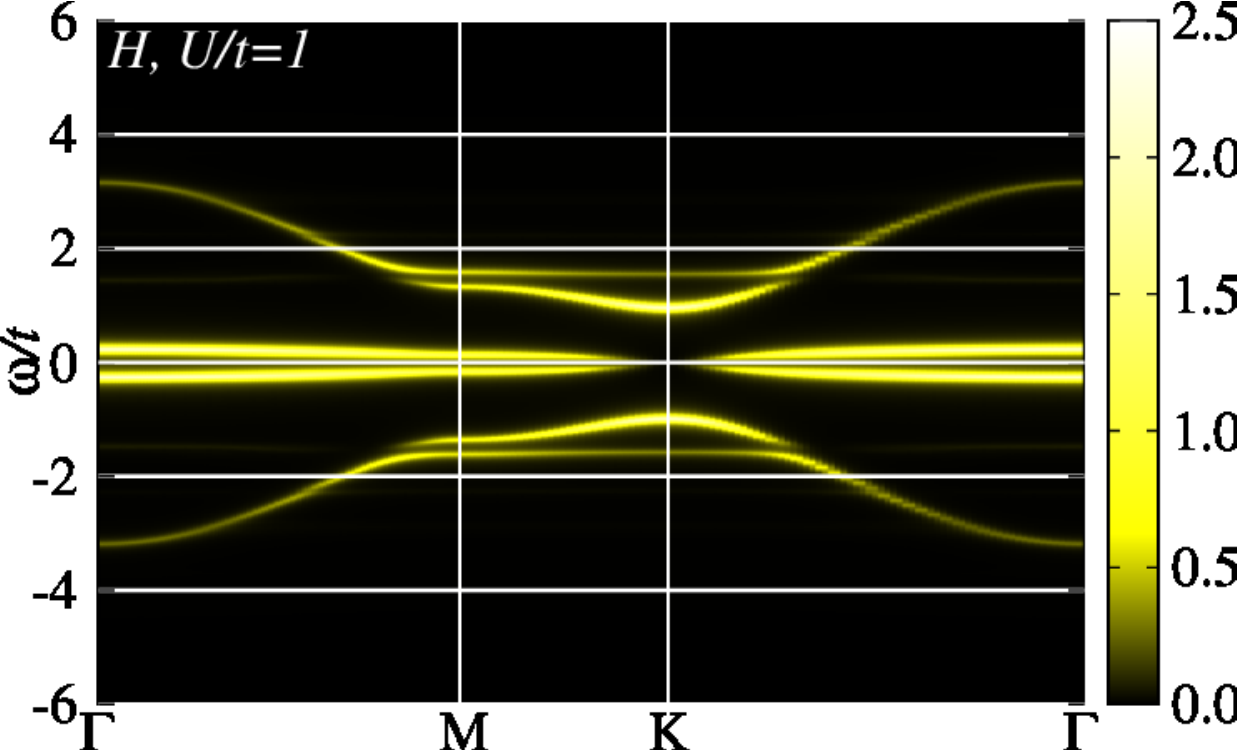}\\
\includegraphics[width=4.3cm,angle=0]{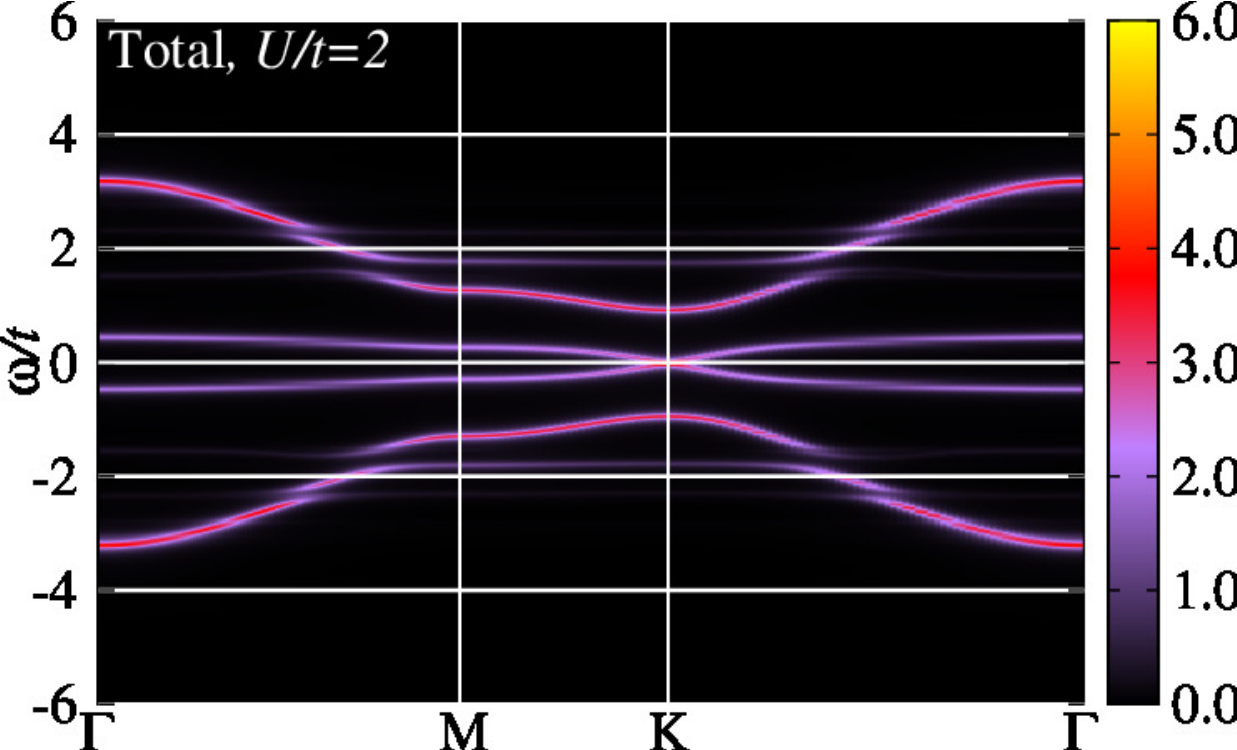}
\includegraphics[width=4.3cm,angle=0]{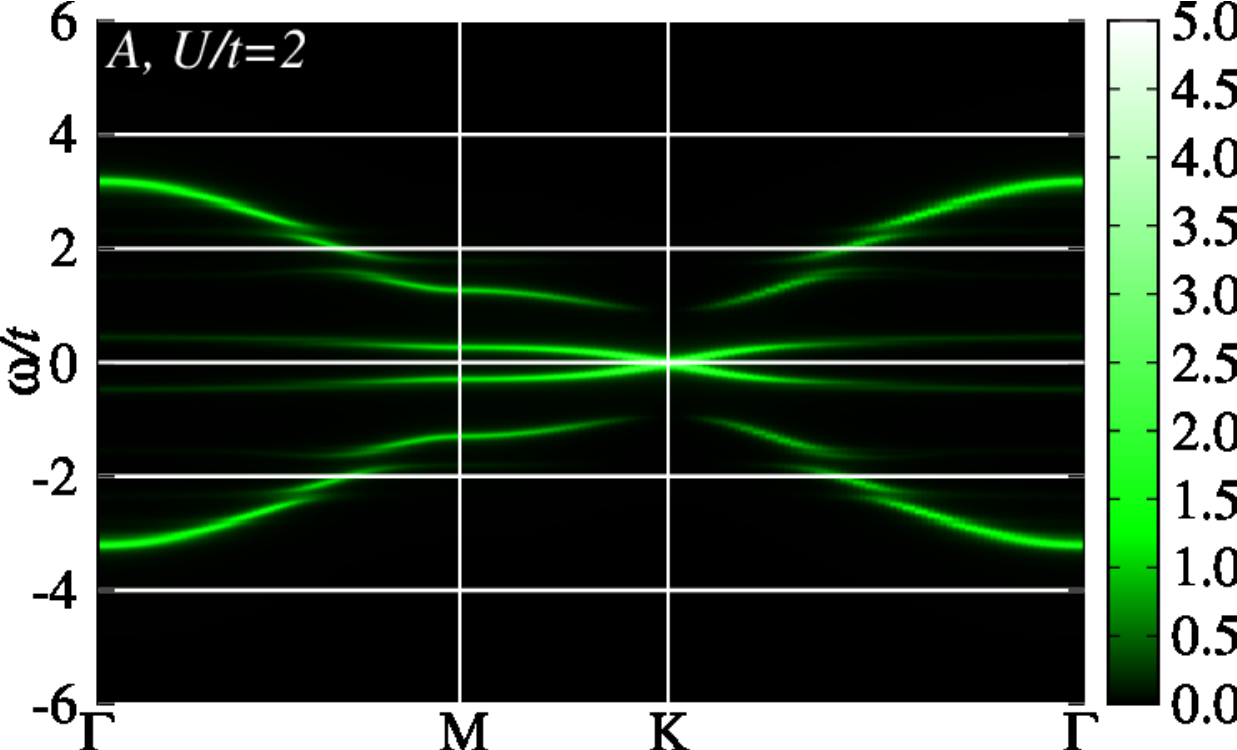}
\includegraphics[width=4.3cm,angle=0]{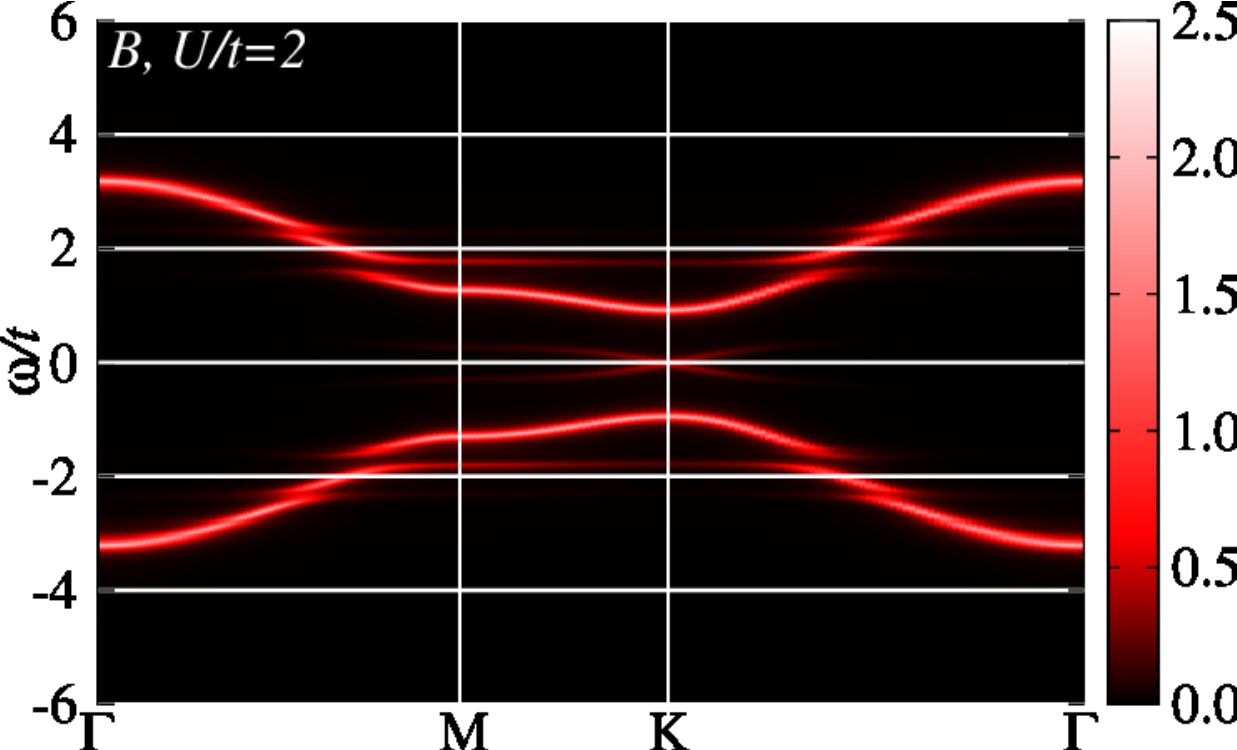}
\includegraphics[width=4.3cm,angle=0]{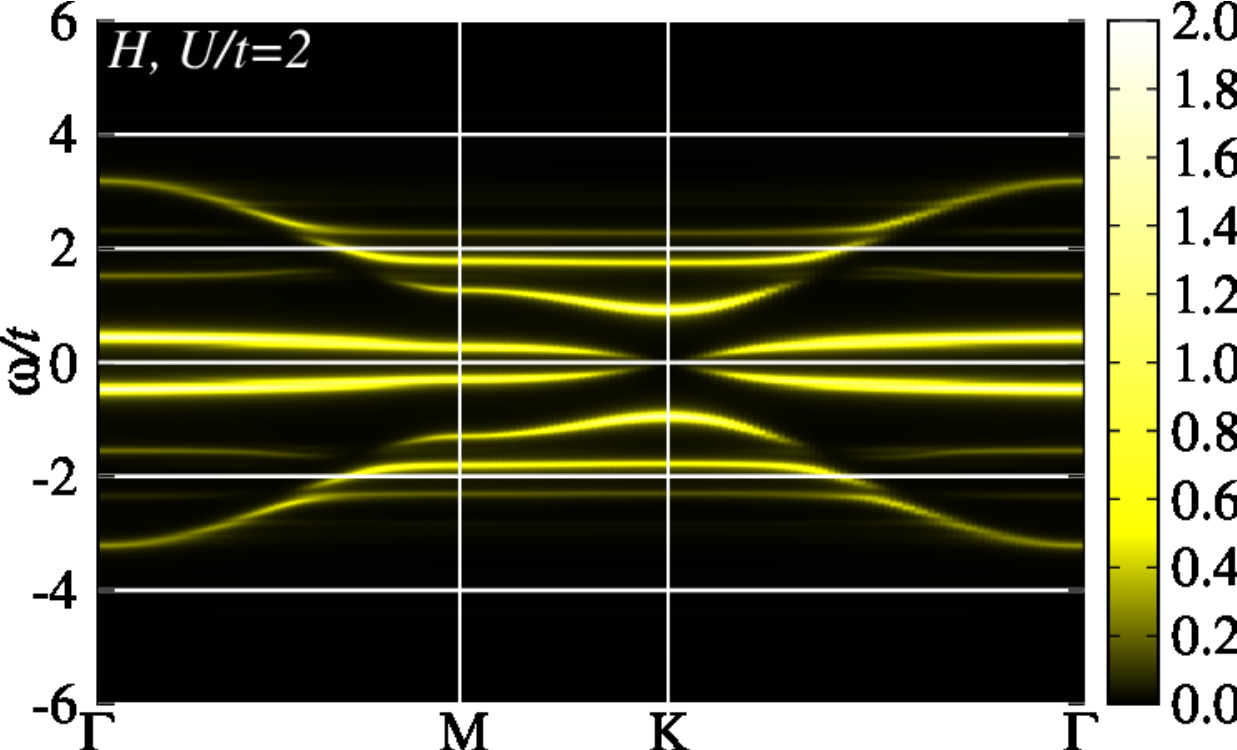}\\
\includegraphics[width=4.3cm,angle=0]{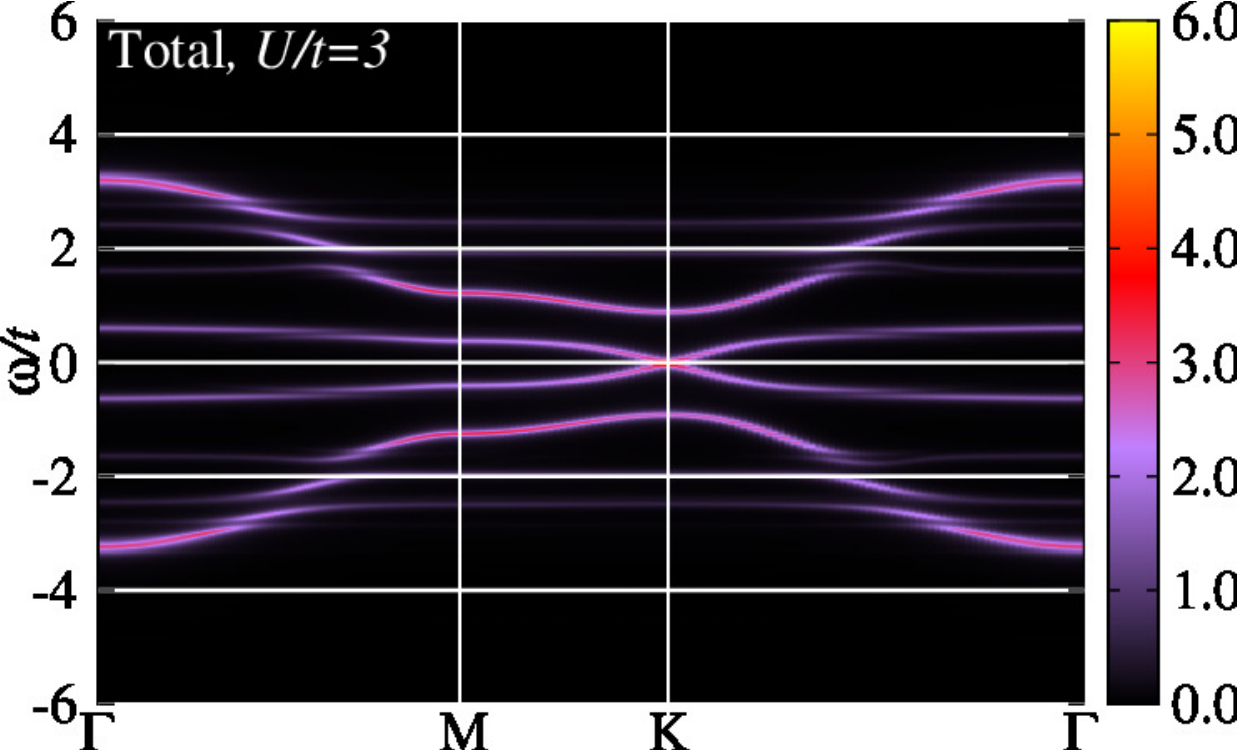}
\includegraphics[width=4.3cm,angle=0]{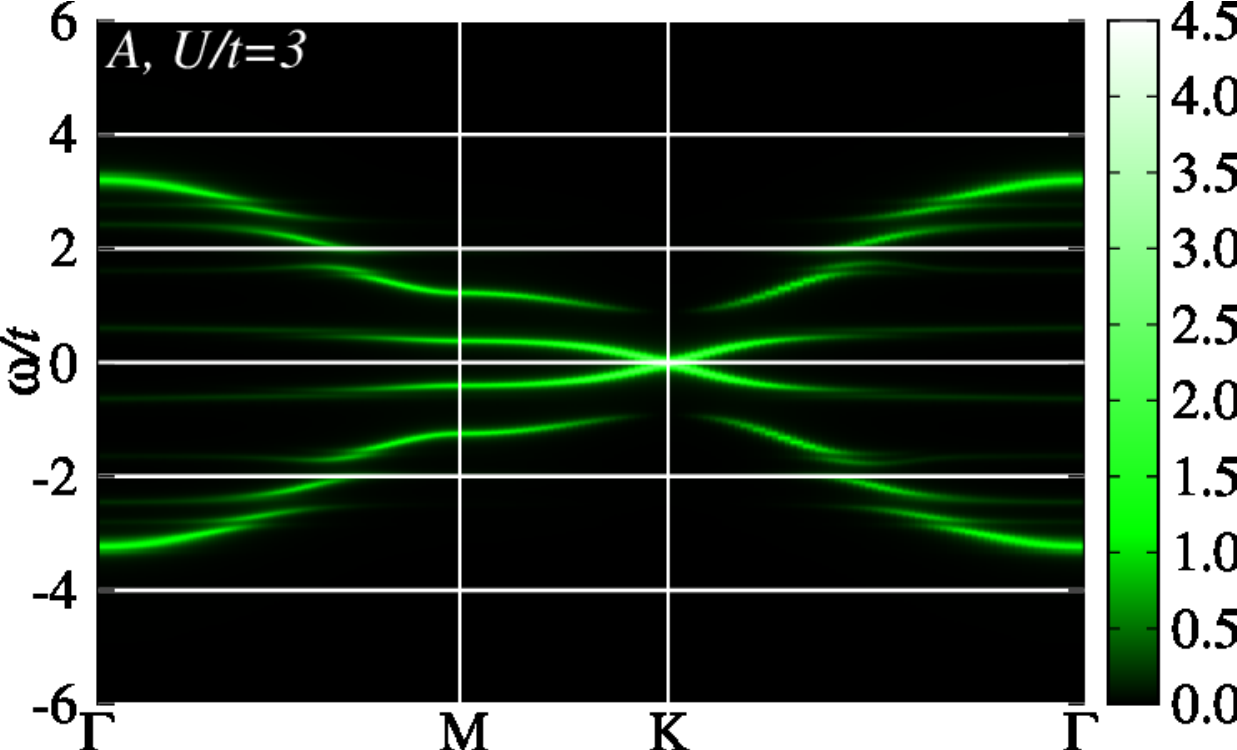}
\includegraphics[width=4.3cm,angle=0]{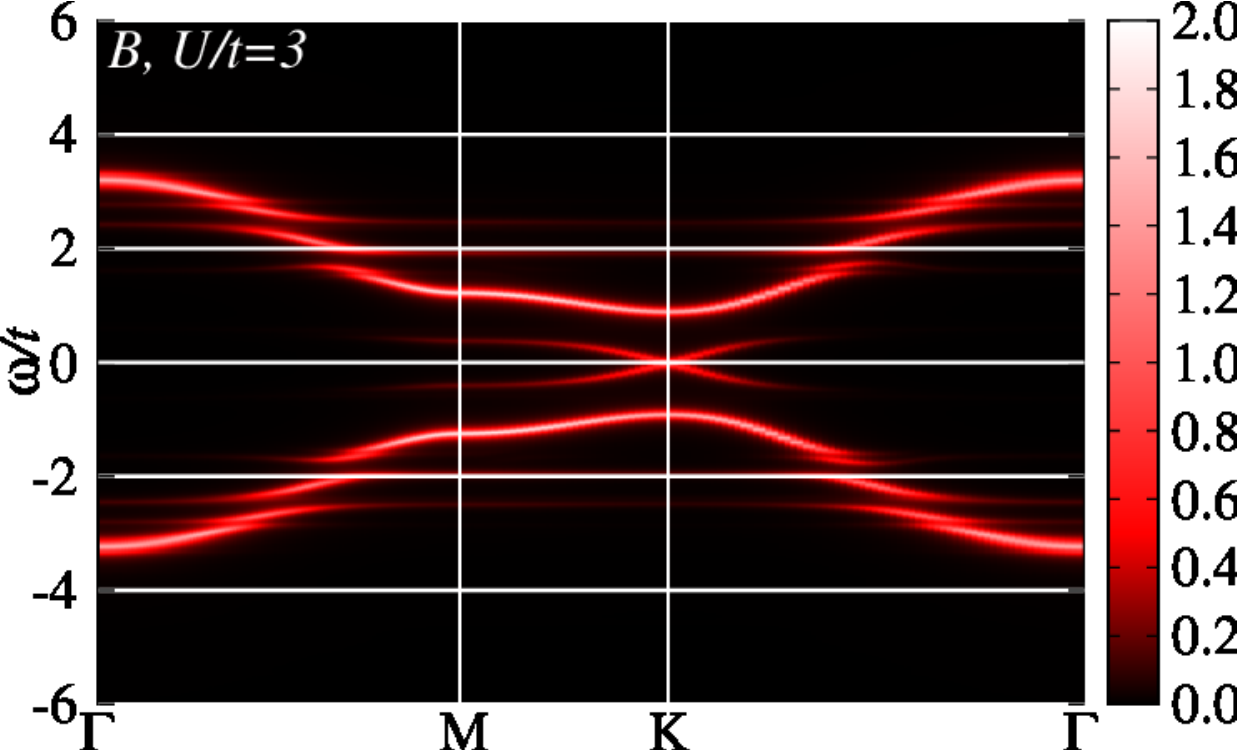}
\includegraphics[width=4.3cm,angle=0]{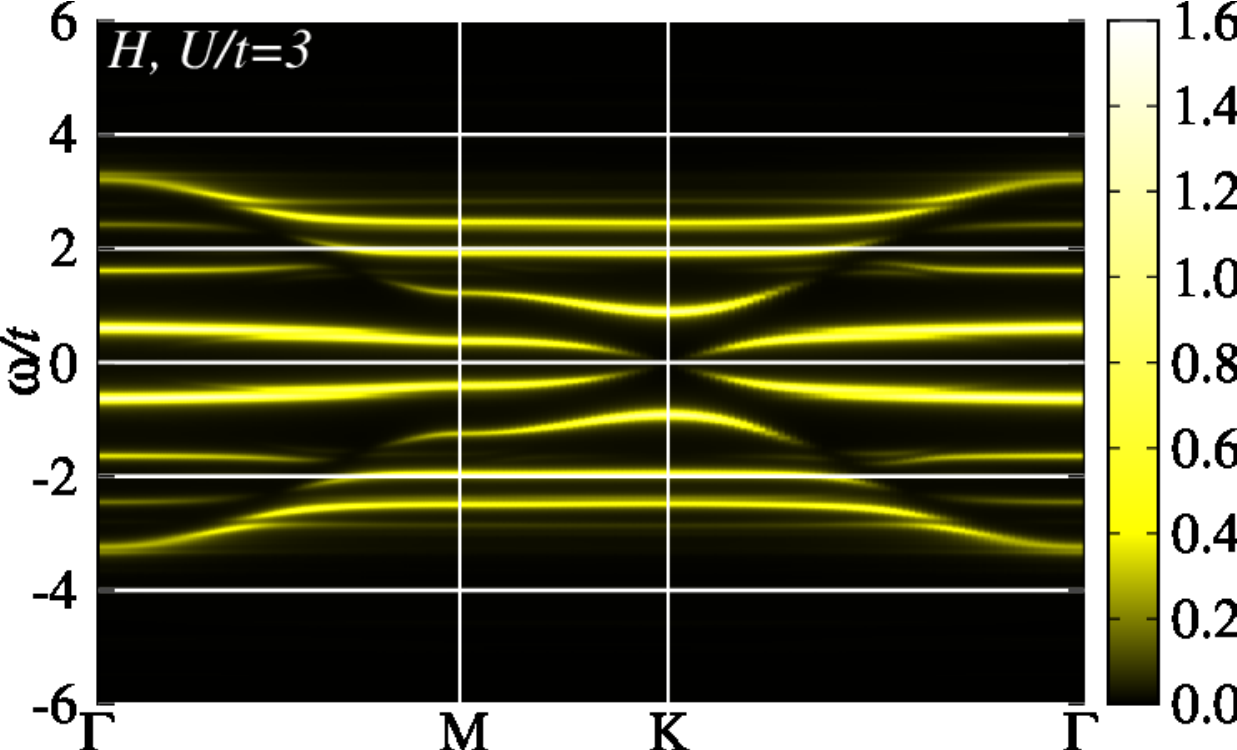}\\
\includegraphics[width=4.3cm,angle=0]{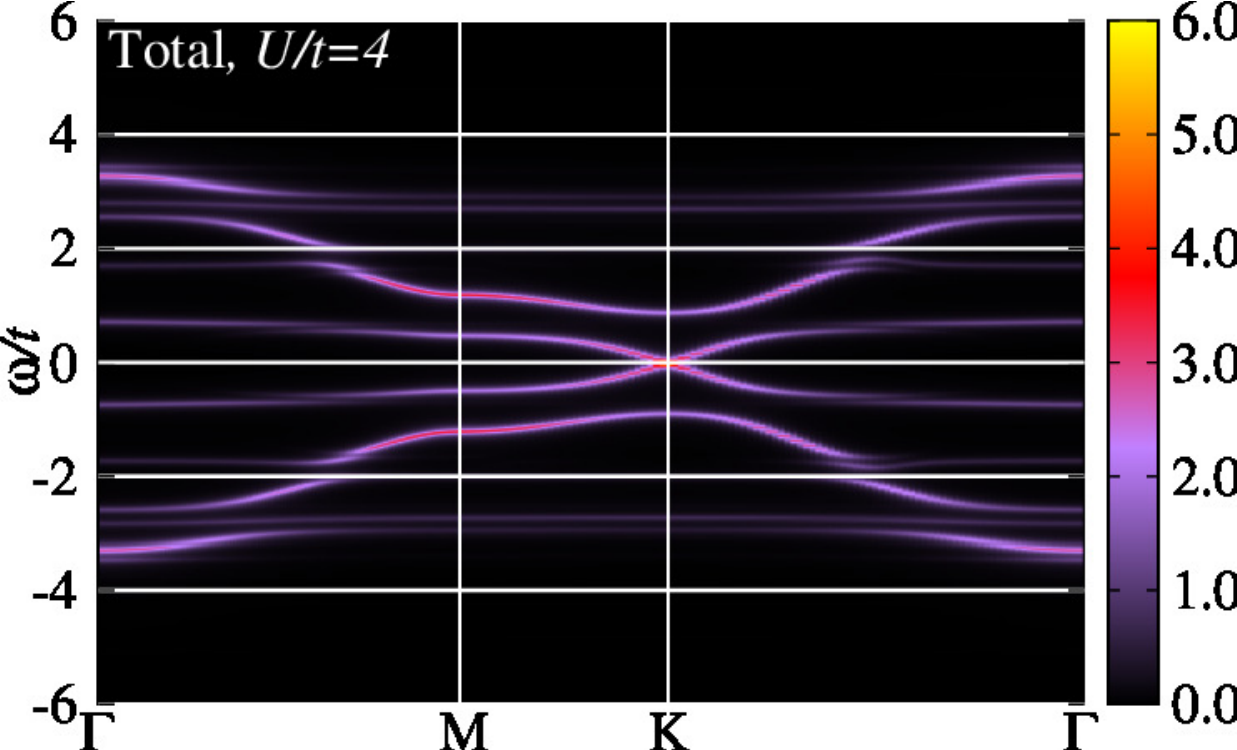}
\includegraphics[width=4.3cm,angle=0]{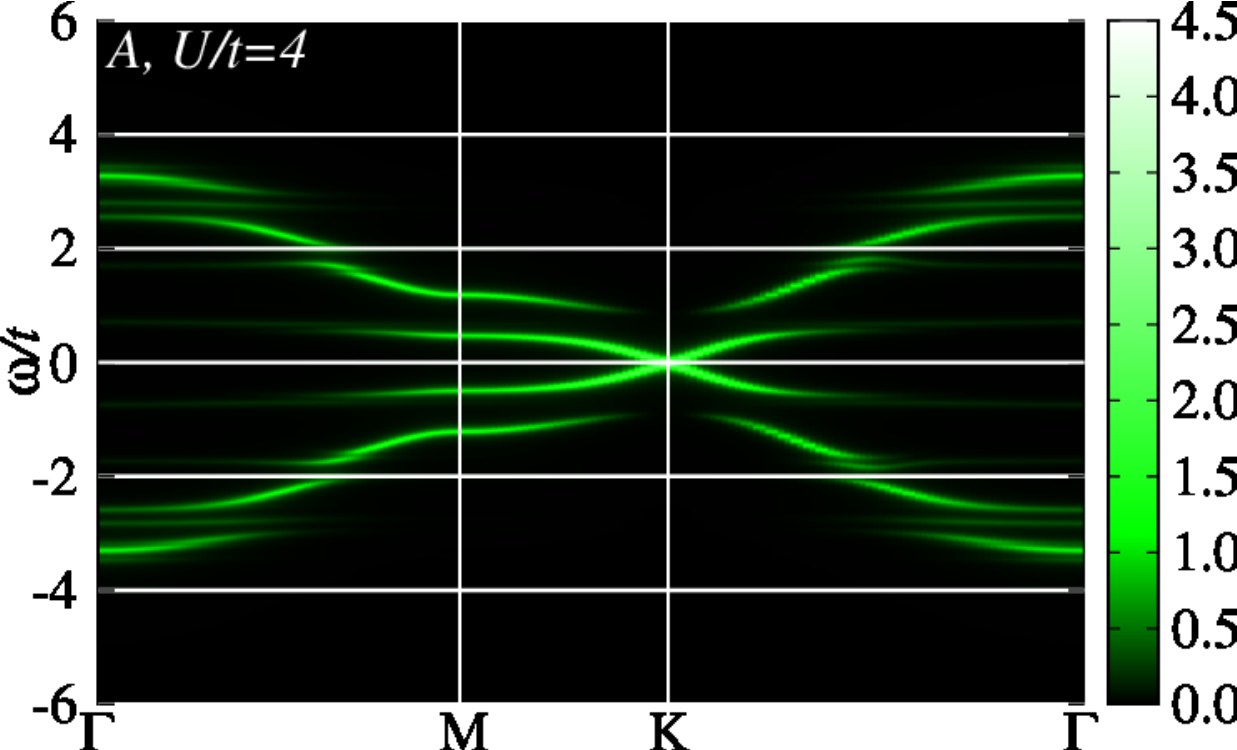}
\includegraphics[width=4.3cm,angle=0]{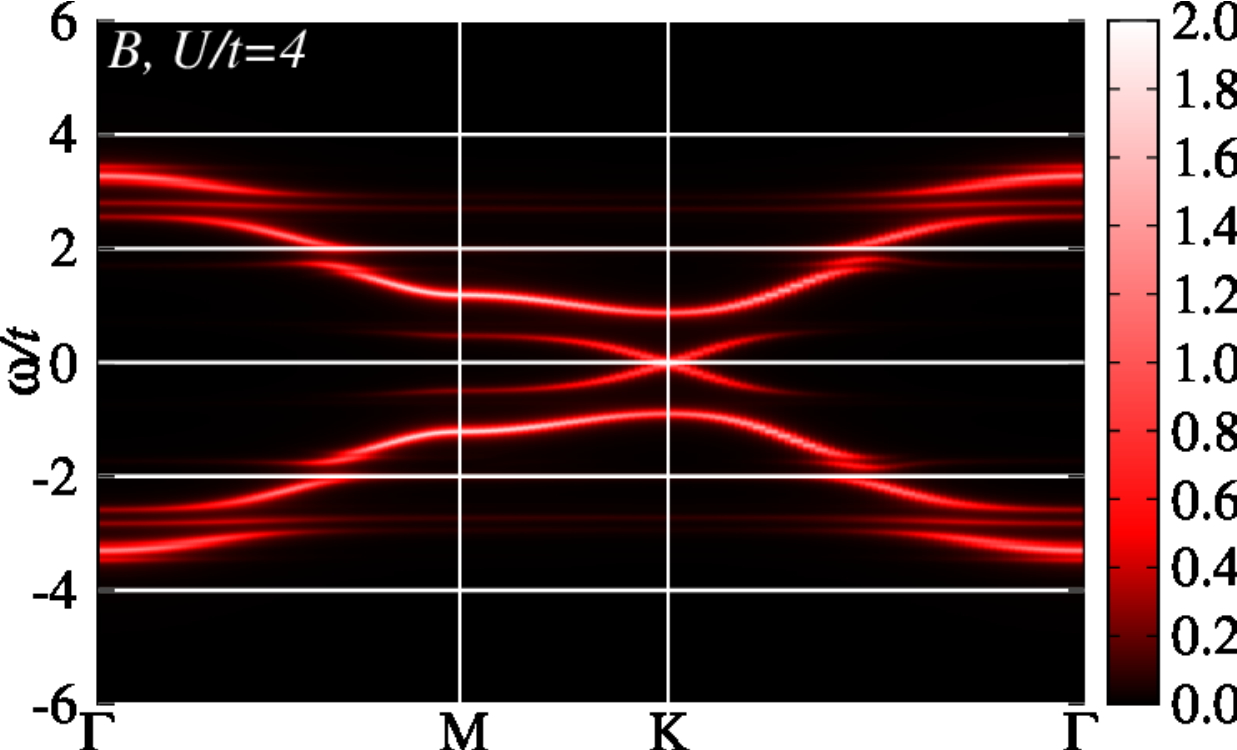}
\includegraphics[width=4.3cm,angle=0]{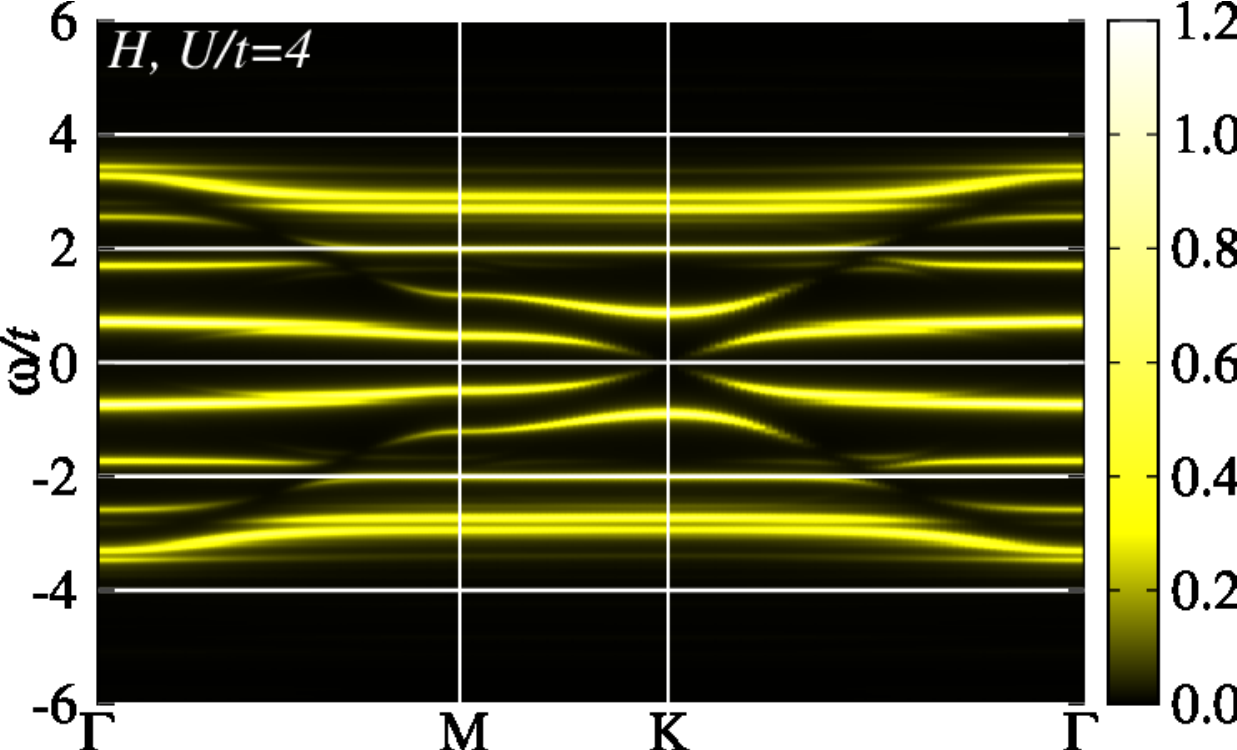}\\
\includegraphics[width=4.3cm,angle=0]{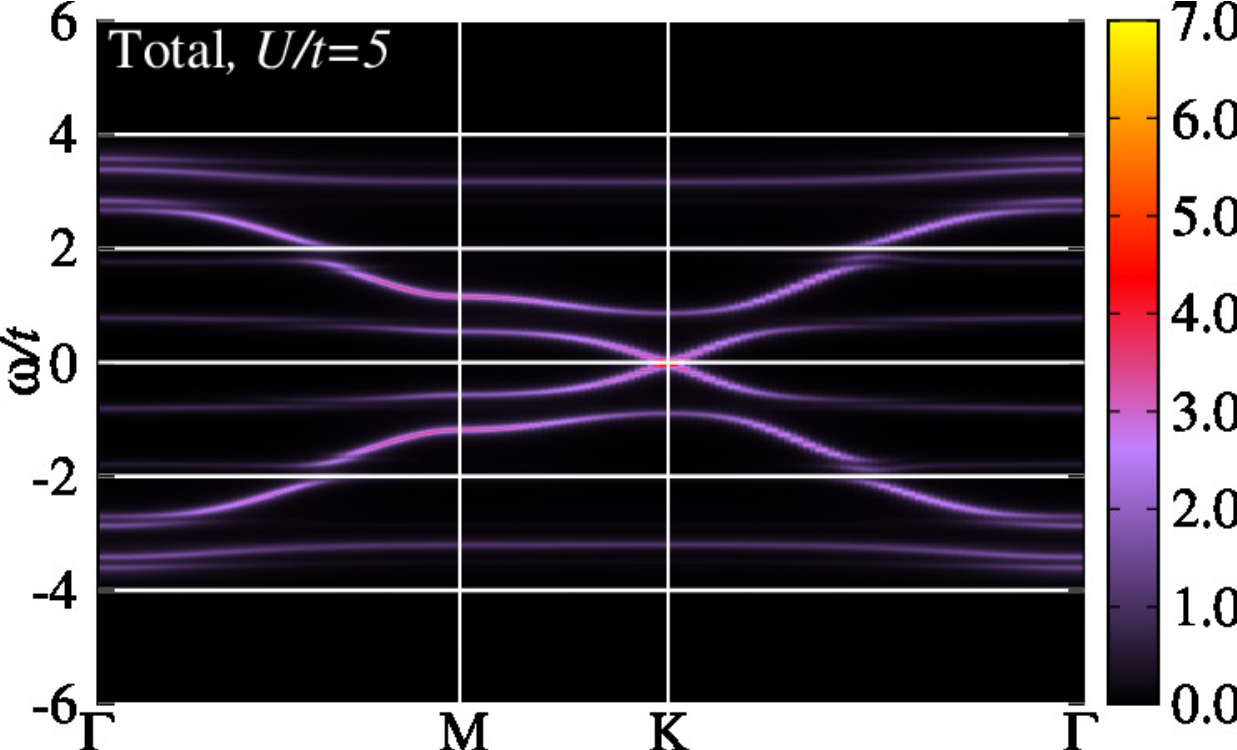}
\includegraphics[width=4.3cm,angle=0]{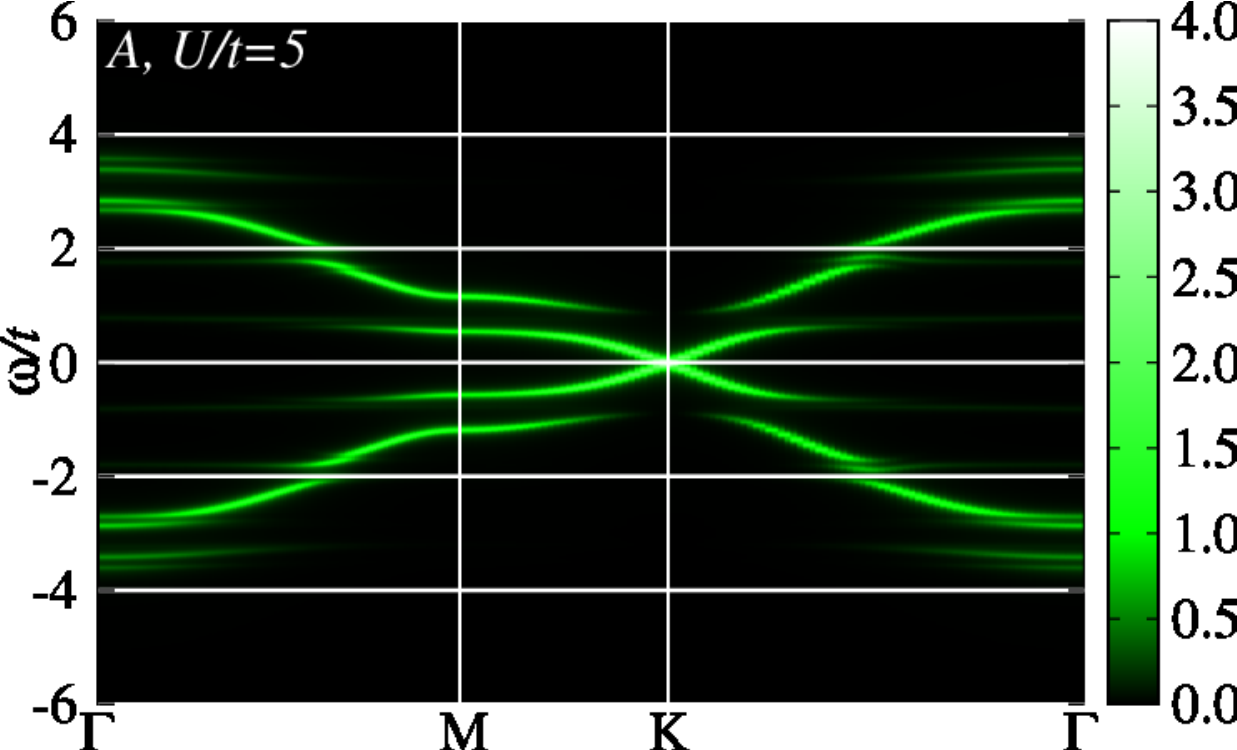}
\includegraphics[width=4.3cm,angle=0]{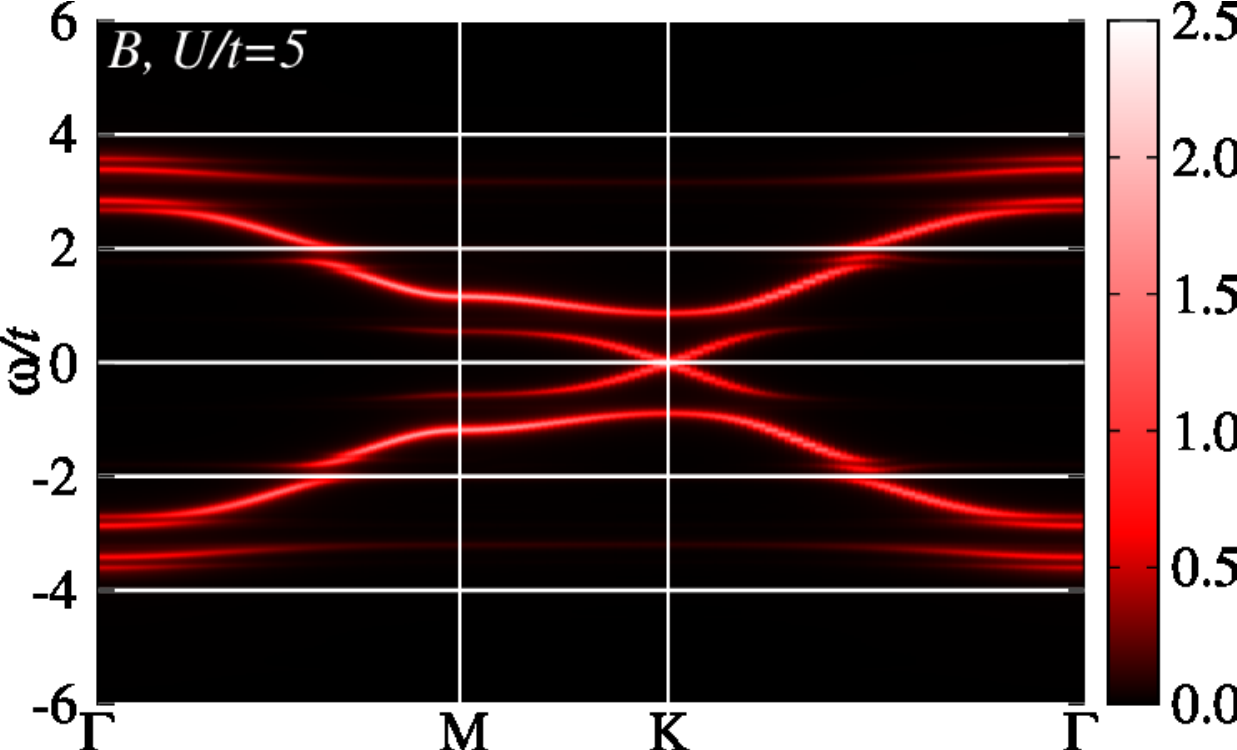}
\includegraphics[width=4.3cm,angle=0]{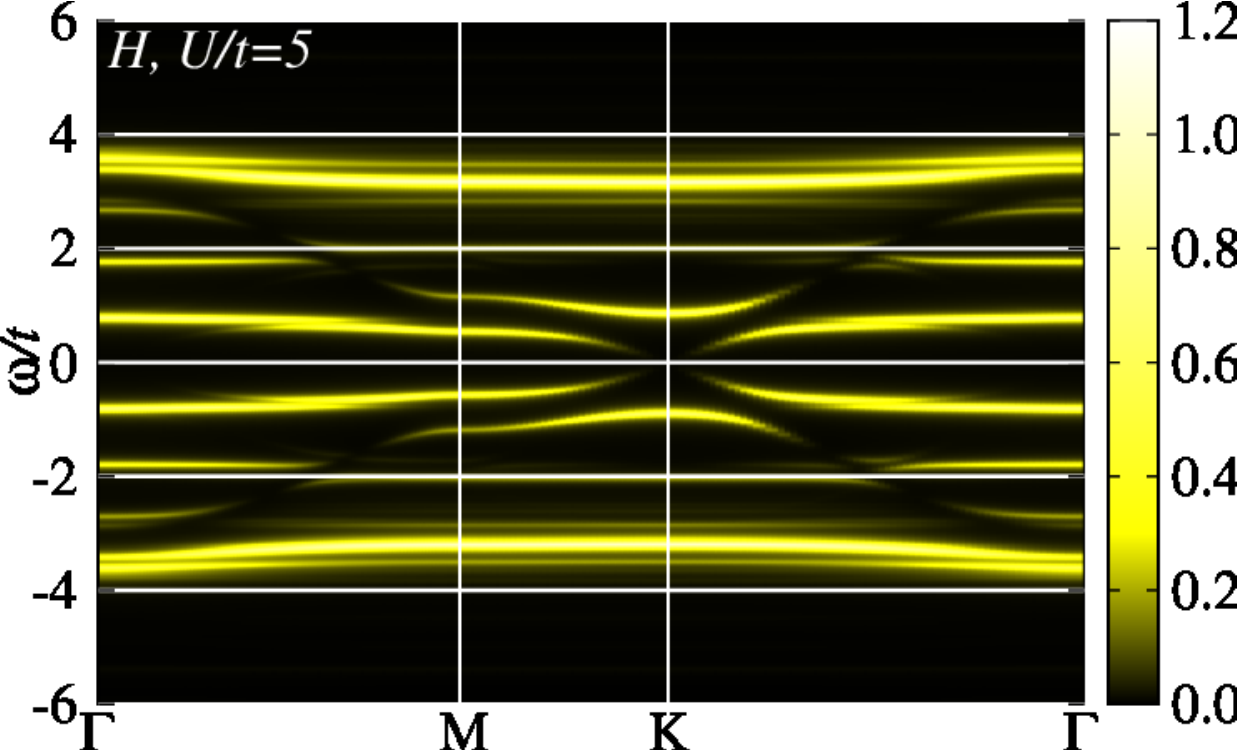}\\
\includegraphics[width=4.3cm,angle=0]{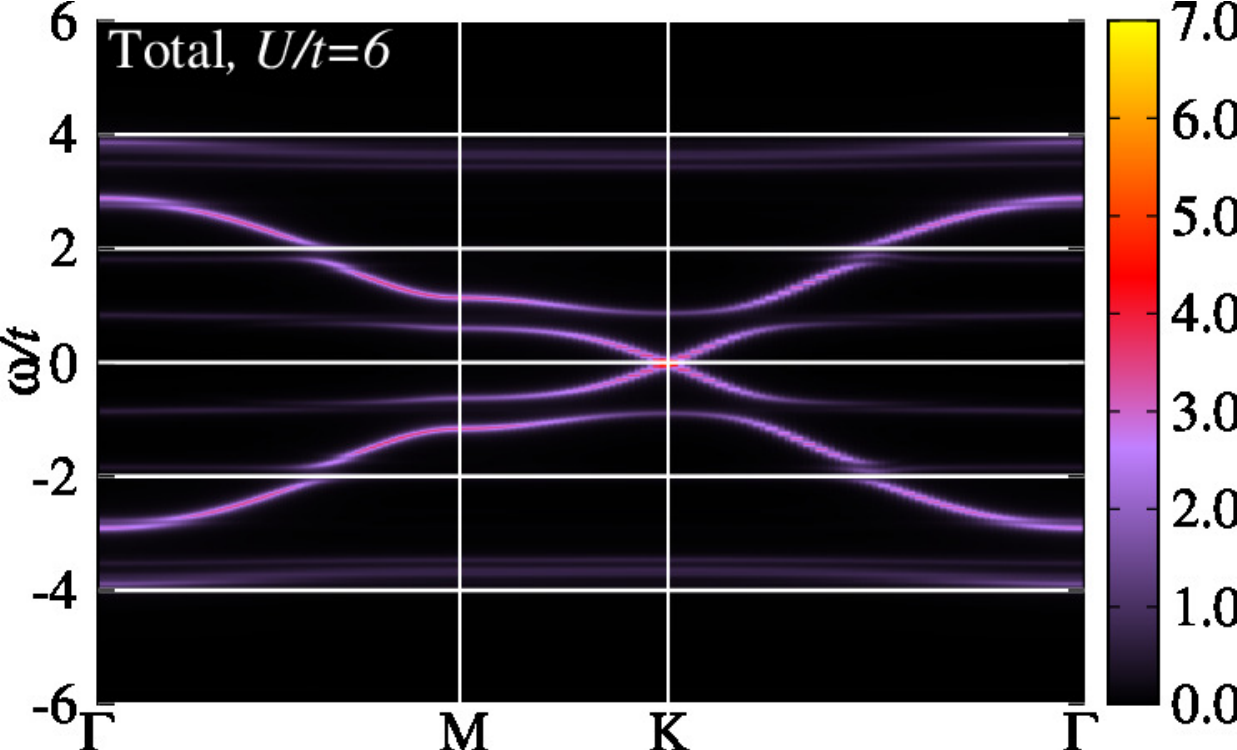}
\includegraphics[width=4.3cm,angle=0]{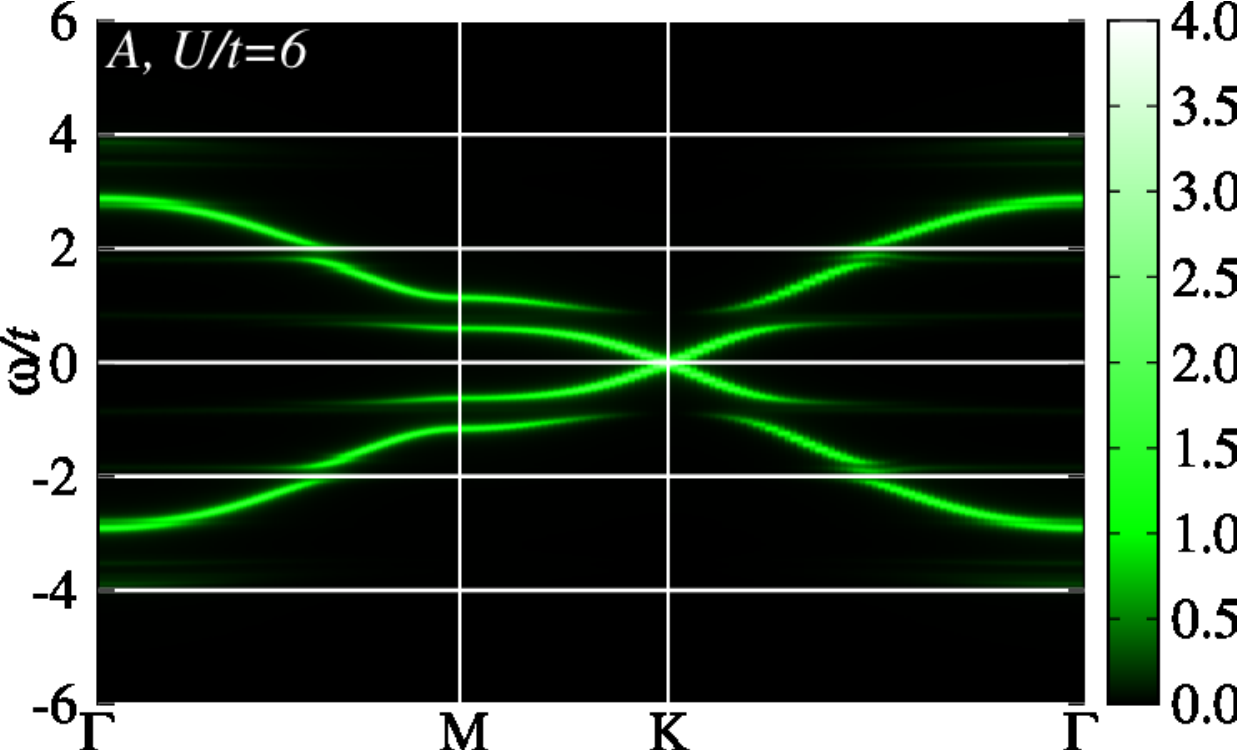}
\includegraphics[width=4.3cm,angle=0]{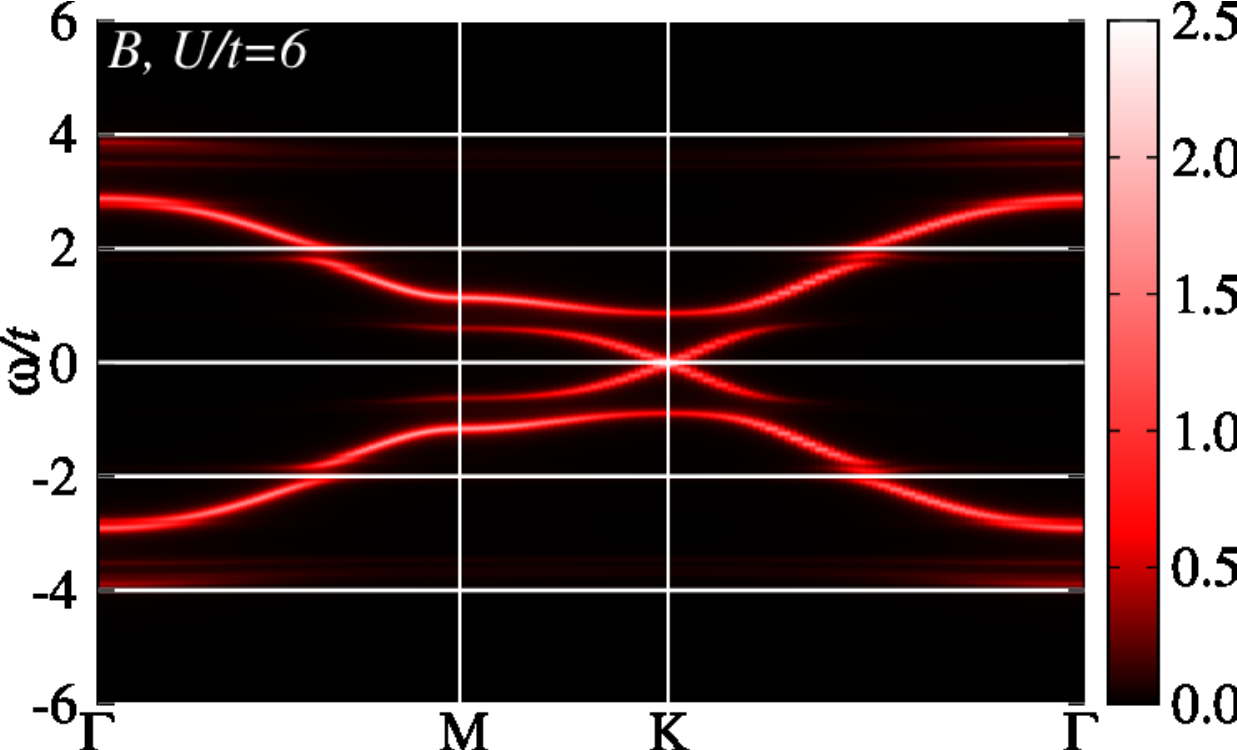}
\includegraphics[width=4.3cm,angle=0]{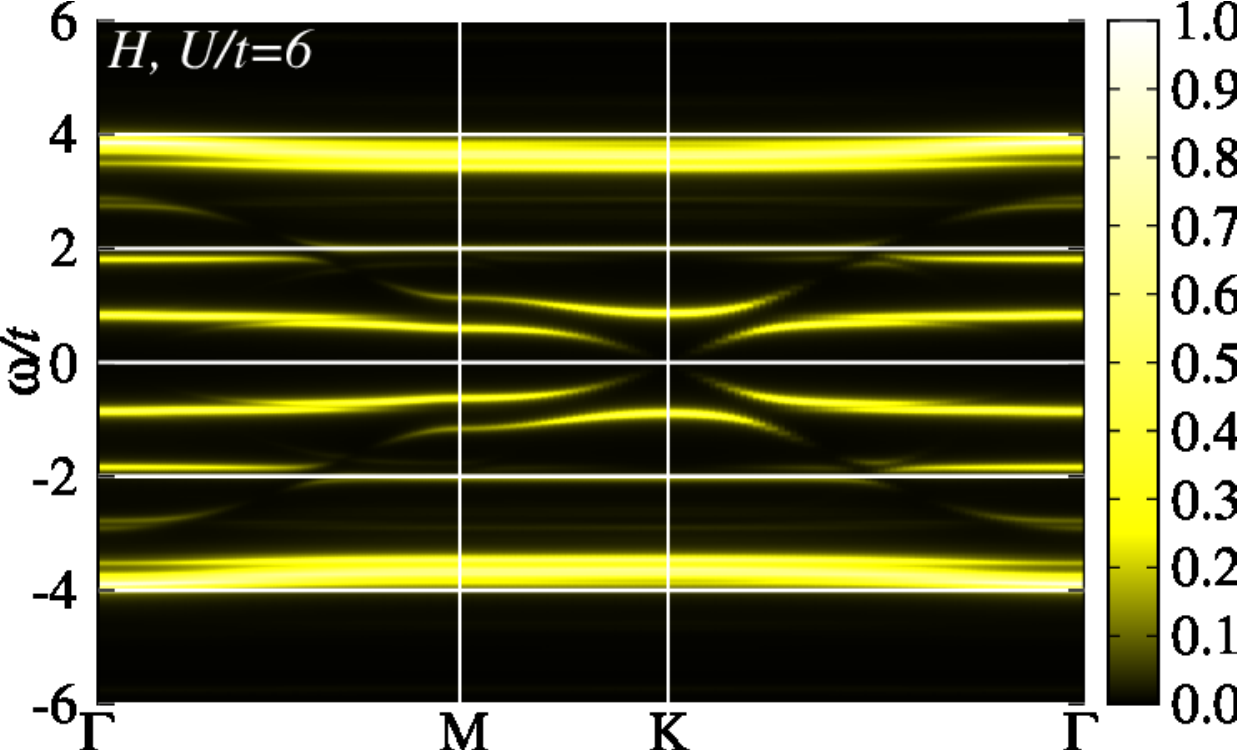}\\
\includegraphics[width=4.3cm,angle=0]{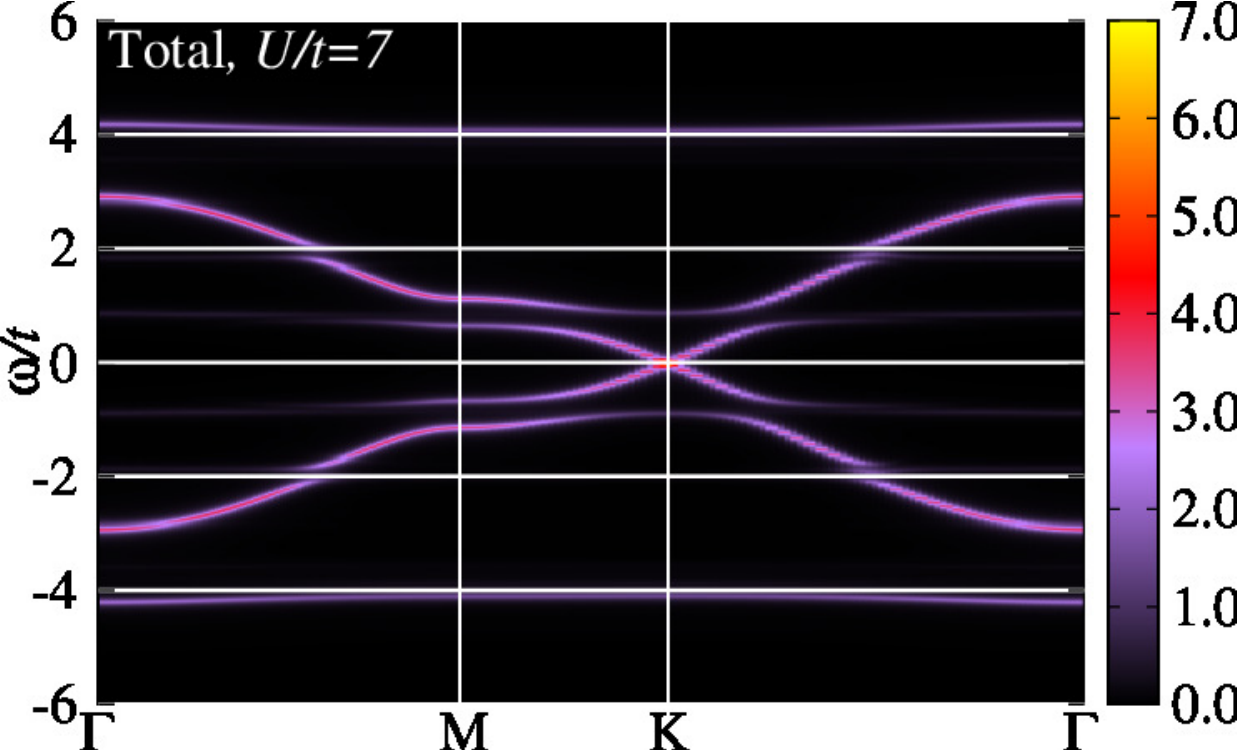}
\includegraphics[width=4.3cm,angle=0]{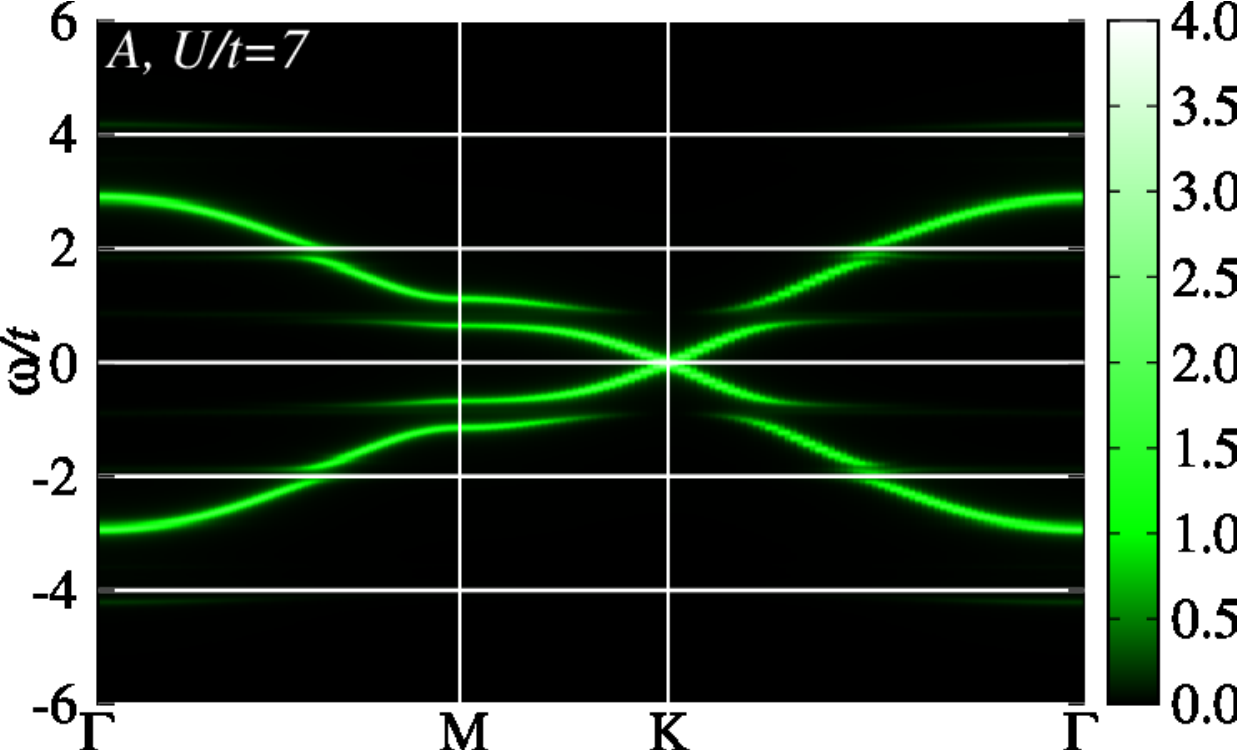}
\includegraphics[width=4.3cm,angle=0]{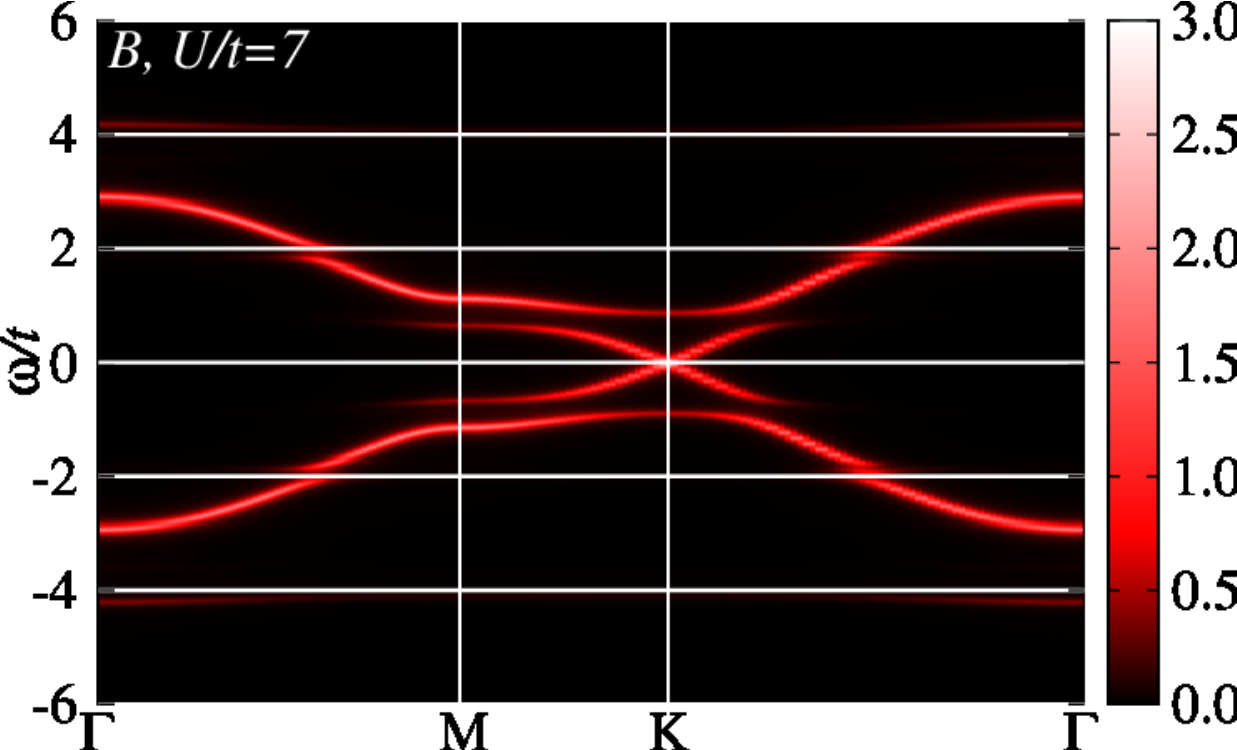}
\includegraphics[width=4.3cm,angle=0]{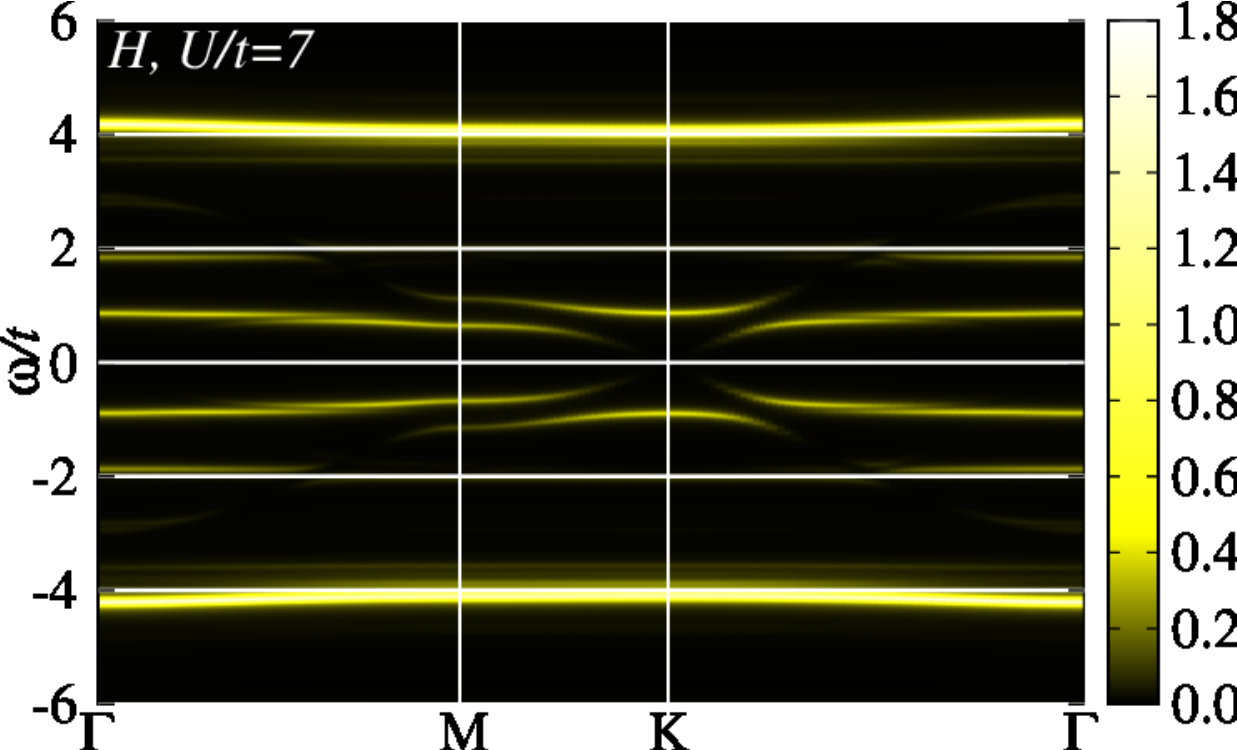}
\caption{(color online) 
  The CPT results of the single-particle excitation spectra 
  for $U/t=0,1, \cdots, 7$ (from top to bottom panels) and $t_{sp}/t=1$ in the PM state 
  at $T=0$ obtained using 9-site cluster (containing 3 unit cells). 
  $A_\sigma(\mb{ k},\w)=\sum_\alpha A^{\alpha\alpha}_\sigma(\mb{ k},\w)$, 
  $A^{AA}_\sigma(\mb{ k},\w)$, $A^{BB}_\sigma(\mb{ k},\w)$, and $A^{HH}_\sigma(\mb{ k},\w)$ are shown 
  from left to right panels. 
  Here, we only show the spectra for up electrons, 
  which are exactly the same as $A^{\alpha\alpha}_\downarrow(\mb{k},\w)$.
  A Lorentzian broadening of $\eta/t=0.05$ is used.
  The spectral intensity is indicated by a color bar in each figure.
  Note that the different intensity scales are used for different figures.
  The Fermi energy $E_{\rm F}$ is located at $\w=0$. 
} 
\label{fig:akw_pm}
\end{figure*}

\begin{figure}
  \includegraphics[width=6.0cm]{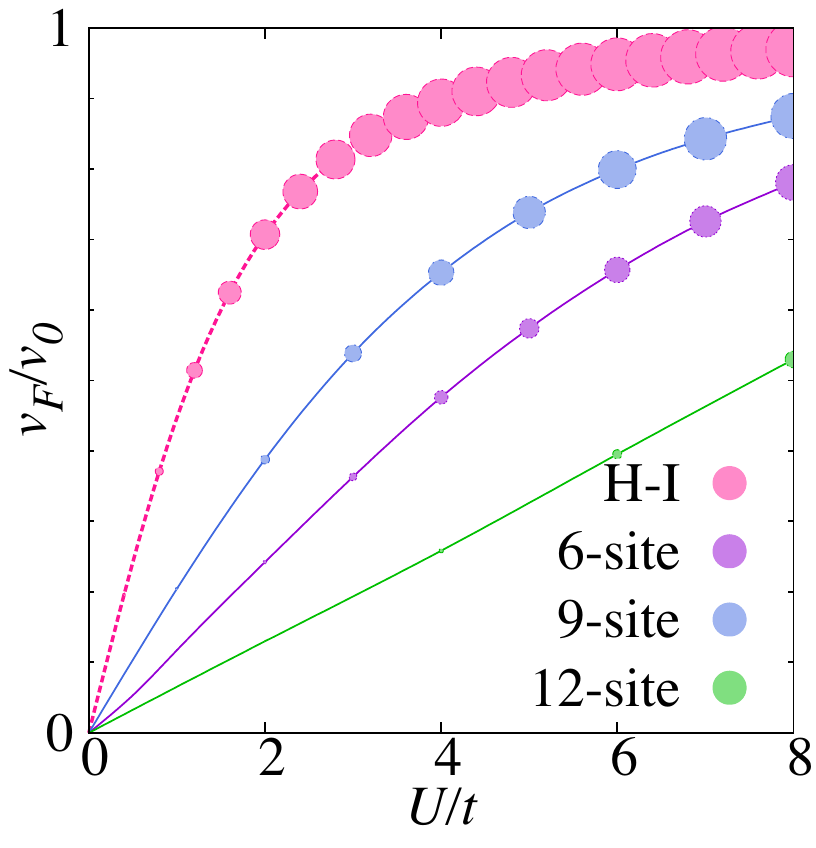}
  \caption{(color online) 
    $U$ dependence of the Dirac Fermi velocity $v_{\rm F}$ calculated using 
    the CPT and the Hubbard-I (H-I) approximation. 
    The CPT calculations are done for the 6-site cluster (containing 2 unit cells) at 
    $T=0.025t$ in the PM phase, and for the 9-site cluster (containing 3 unit cells) and 
    the 12-site cluster (containing 4 unit cells) at $T=0$ where the PM state is assumed. 
    Both calculations are for $t_{sp}/t=1$ at $n=1$. 
    The size of dots is proportional to the spectral weight $\rho_K^{BB}$ for $B$ orbital at the Fermi energy 
    and $v_0=\sqrt{3}t/2$ is the Dirac Fermi velocity of the pure graphene model. 
  } 
  \label{fig:vf_vs_u}
\end{figure}

In sharp contrast to the results for the FM state, we find that the PM state exhibits the linear energy dispersions 
near $E_{\rm F}$. As shown in Fig.~\ref{fig:akw_pm_en} for $U/t=4$, we can clearly observe that 
the massless Dirac quasiparticle excitations emerge with the Dirac point exactly at $E_{\rm F}$.  
The $U$ dependence of the single-particle excitations for the PM state is summarized in 
Fig.~\ref{fig:akw_pm}. 
The massless Dirac quasiparticle excitations always exist in the PM state 
as long as $U$ is finite. 
It is also interesting to notice in Fig.~\ref{fig:akw_pm} that the Dirac Fermi velocity $v_{\rm F}$, 
i.e., the slope of the linear energy dispersion at $E_{\rm F}$, monotonically 
increases with increasing $U$ and approaches to the Dirac Fermi velocity $v_0=\sqrt{3}t/2$ of the pure graphene 
model in the limit of $U/t\to\infty$ (see also Fig.~\ref{fig:vf_vs_u}). 

While the simplest mean-field theory and the DFT calculations can 
reproduce qualitatively the FM state with the quadratic energy dispersion, 
they fail to capture the Dirac-like quasiparticle excitations 
with the linear energy dispersion in the PM state~\cite{Zhou2009}. 
The inability of describing the massless Dirac quasiparticles in the single-particle approximations 
immediately implies that the dynamical correlation effect is responsible for the 
emergent massless Dirac quasiparticles, which will be discussed more.

We should also notice in Fig.~\ref{fig:akw_pm_en} and Fig.~\ref{fig:akw_pm}
that the emergent massless Dirac quasiparticles 
are composed of $A$ and $B$ orbitals, but not $H$ orbital: 
while the contribution of $A$ ($H$) orbital to the low energy spectral weight near 
$K$ point (and also $K'$ point) remains large (vanishing) with varying $U$, the contribution of $B$ orbital 
is small but finite for small $U$ and gradually increases as the massless Dirac quasiparticles becomes 
more visible in the single-particle excitation spectrum for large $U$. 

To be more quantitative, we also evaluate the spectral weight for $\alpha$ orbital 
at the Dirac point with $\w=0$, 
\begin{equation}\label{eq:sw}
  \rho^{\alpha\alpha}_K := -  \lim_{\eta \rightarrow 0^+} \frac{\eta}{\pi} \Im
      {\cal G}^{\alpha\alpha}_\sigma (\mb{k}=K, i \eta), 
\end{equation} 
where $0^+$ is positive infinitesimal, and find that 
$\rho^{AA}_K=1$ and $\rho^{HH}_K=0$ (the same also at $K'$ point), irrespectively of the value of $U$, 
whereas $\rho^{BB}_K$ increases monotonically from zero with increasing $U$, as shown in Fig.~\ref{fig:vf_vs_u}. 
This is understood by recalling that in the noninteracting limit, $A$ orbital is completely decoupled from $H$ orbital 
at $K$ and $K'$ points where $\gamma_{\mb k}=0$ [see Eq.~(\ref{h_free})]. 
Therefore, even when the interaction $U$ on the hydrogen impurity sites 
is turned on, the contribution from $A$ orbital to the spectral weight 
at $K$ and $K'$ points remains the same. 

\begin{figure*}
  \includegraphics[width=4.3cm,angle=0]{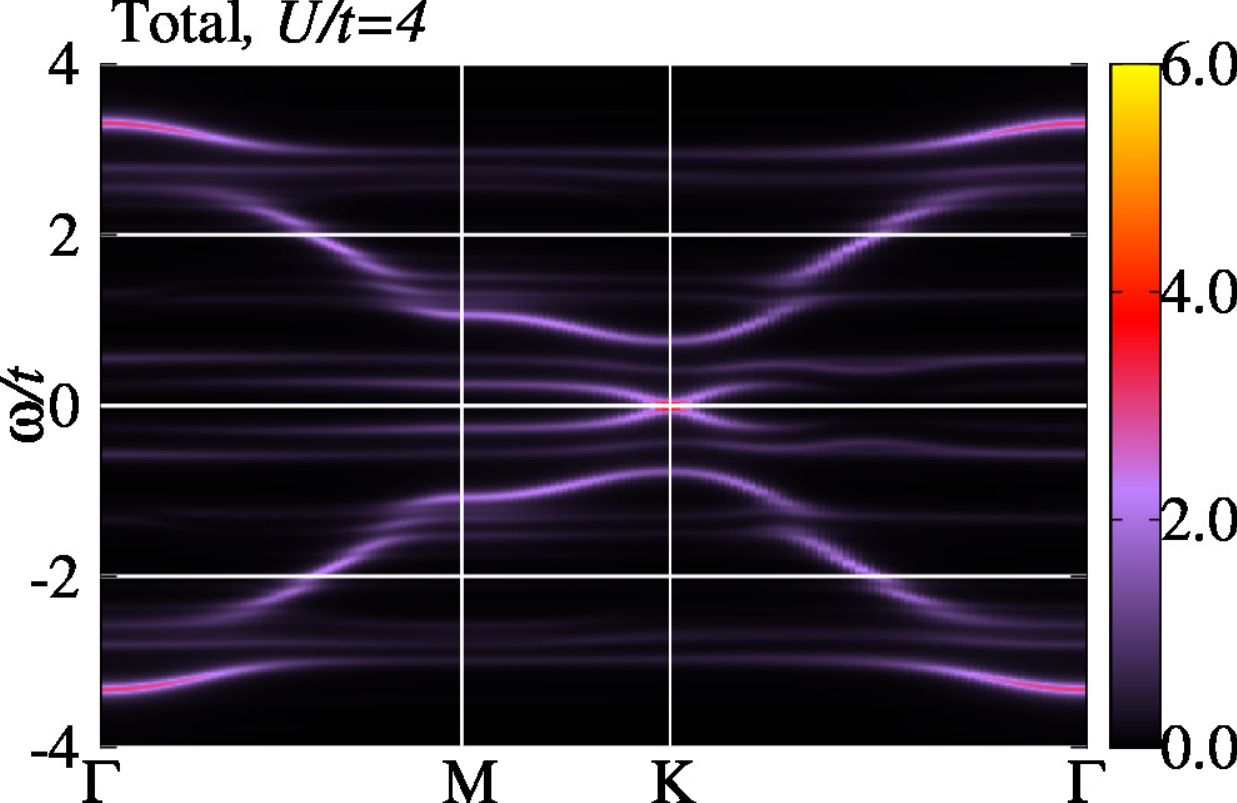}
  \includegraphics[width=4.3cm,angle=0]{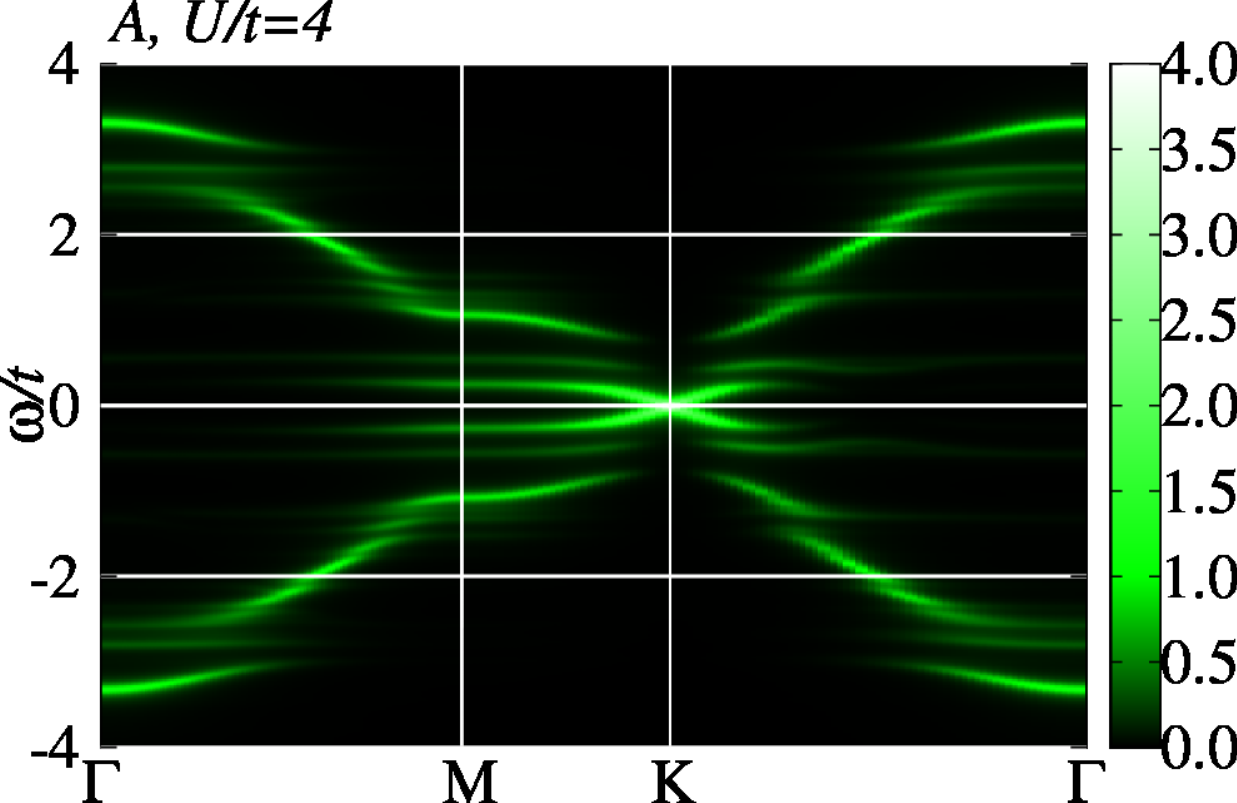}
  \includegraphics[width=4.3cm,angle=0]{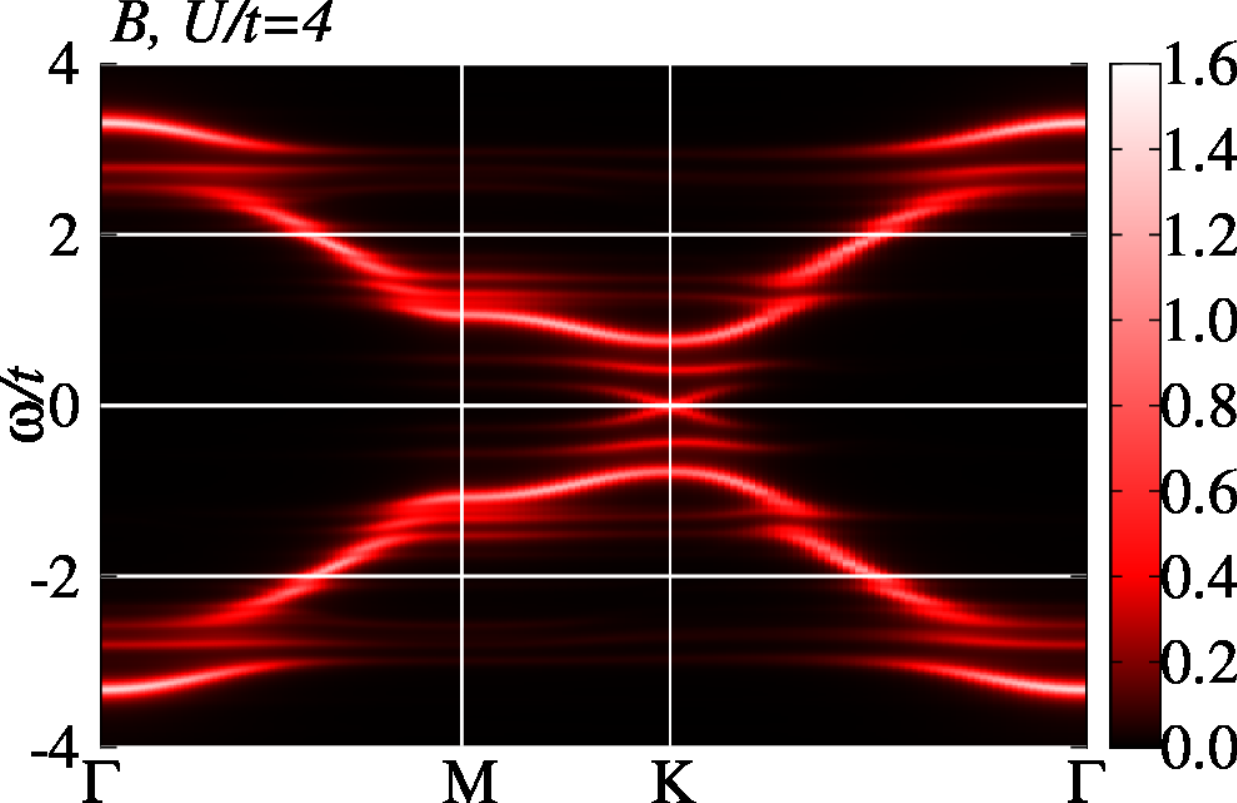}
  \includegraphics[width=4.3cm,angle=0]{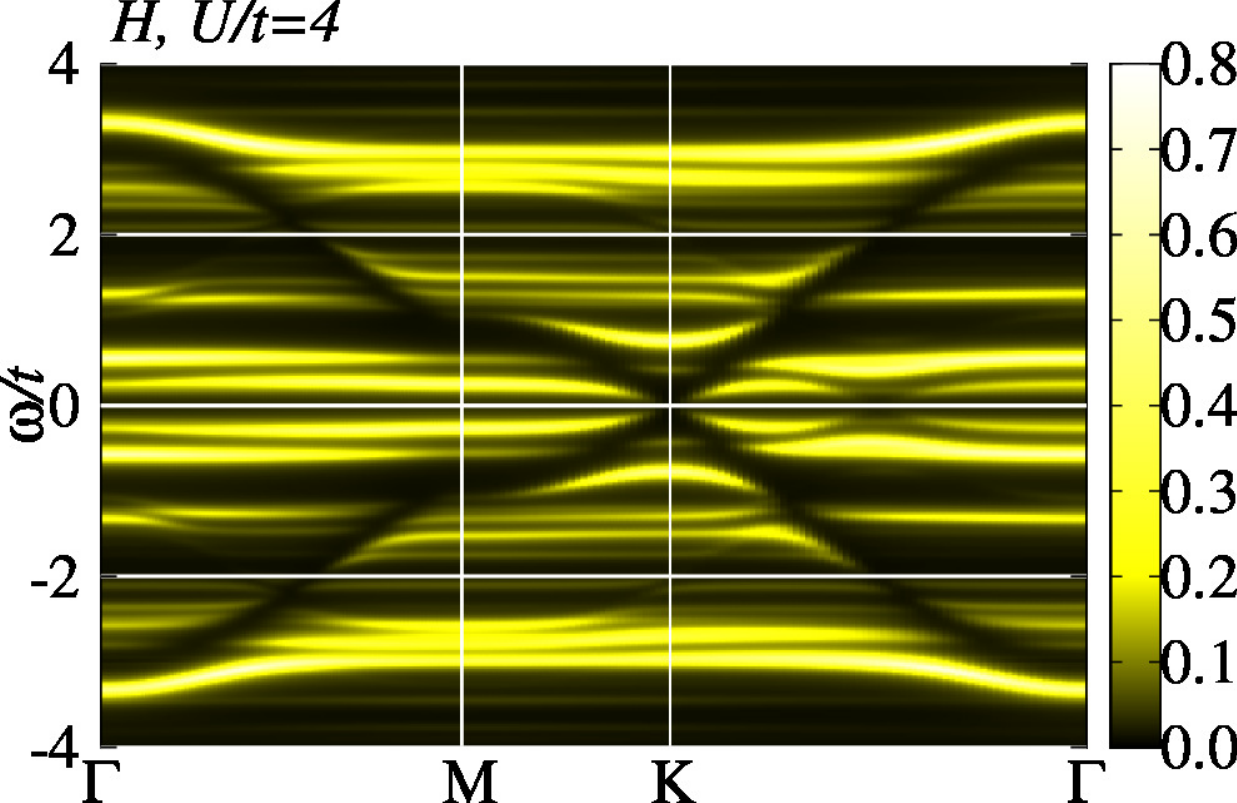}\\
  \includegraphics[width=4.3cm,angle=0]{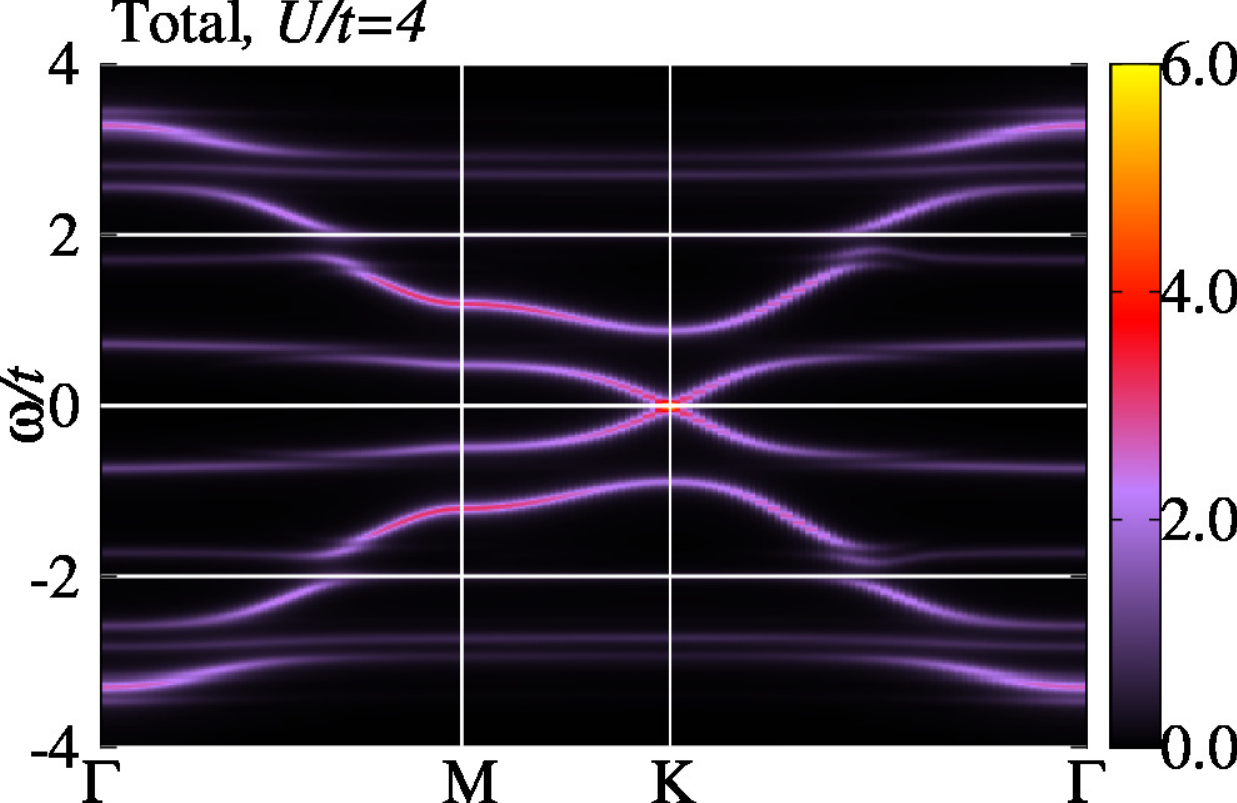}
  \includegraphics[width=4.3cm,angle=0]{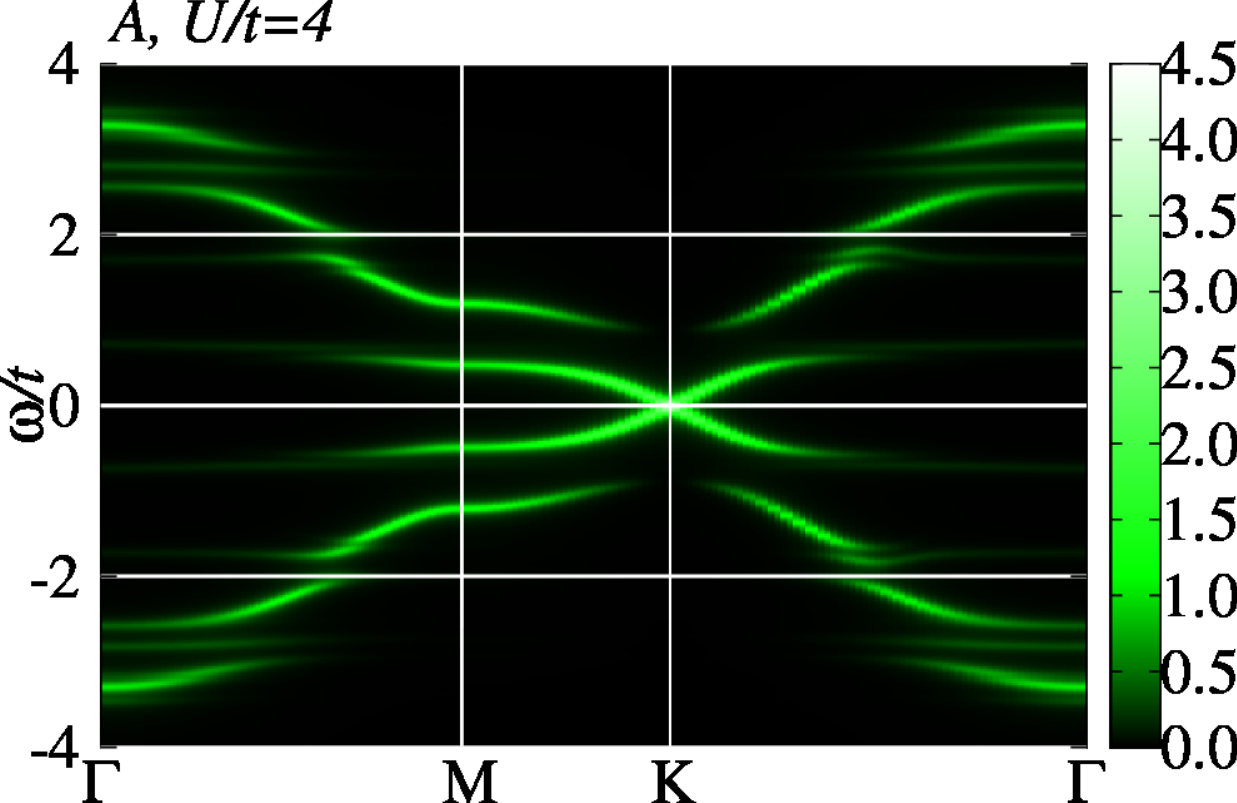}
  \includegraphics[width=4.3cm,angle=0]{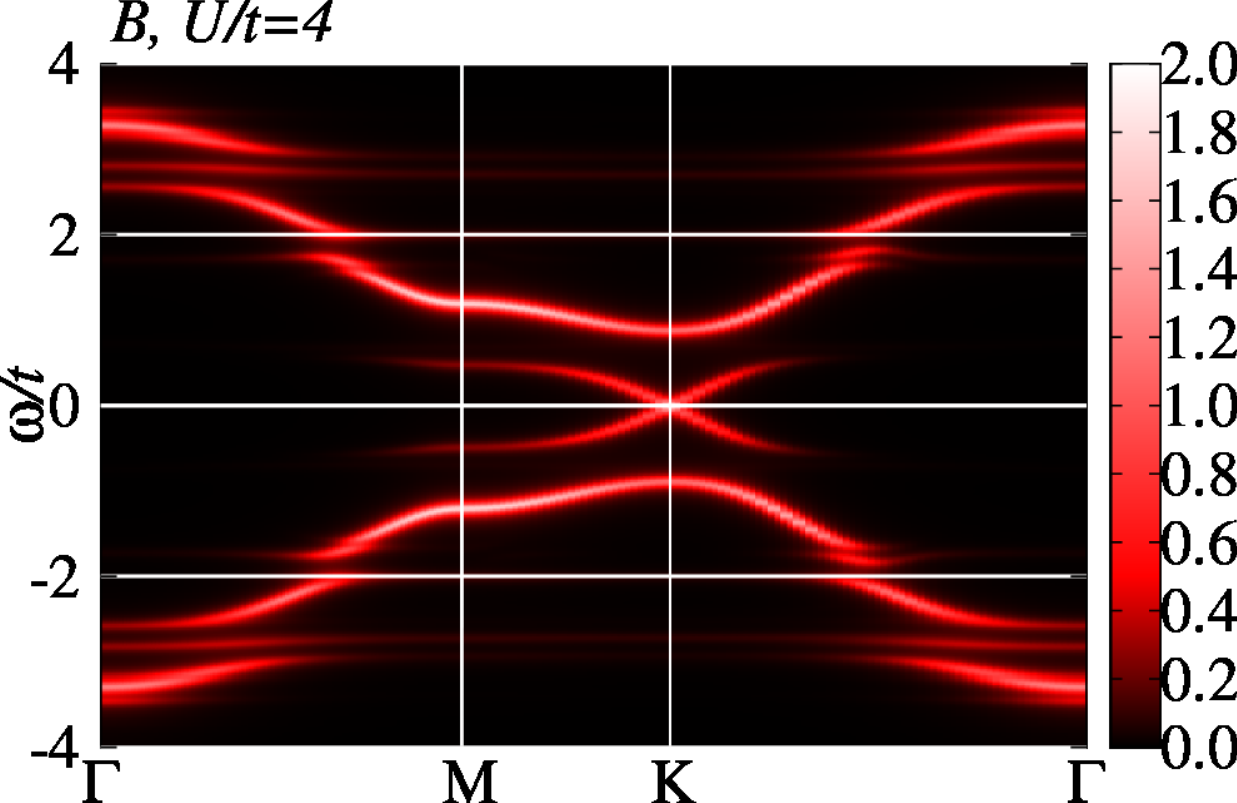}
  \includegraphics[width=4.3cm,angle=0]{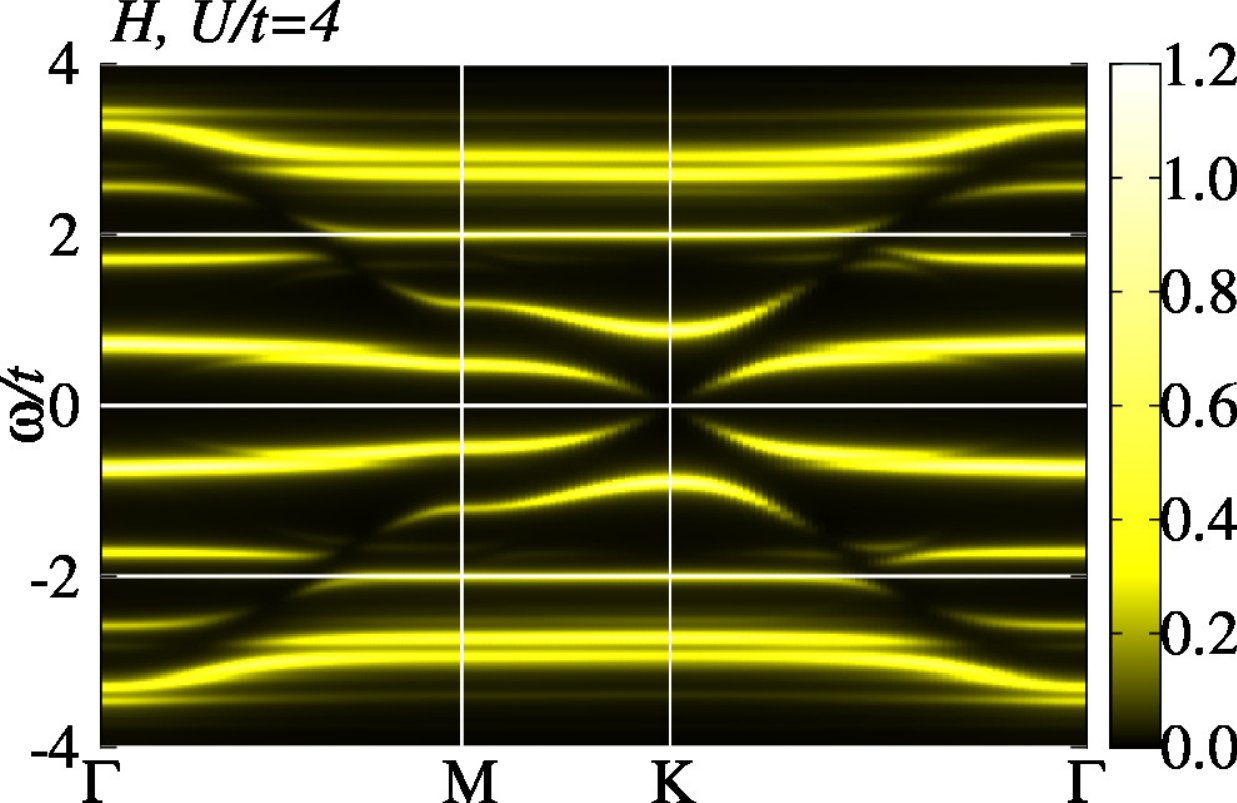}\\
  \includegraphics[width=4.3cm,angle=0]{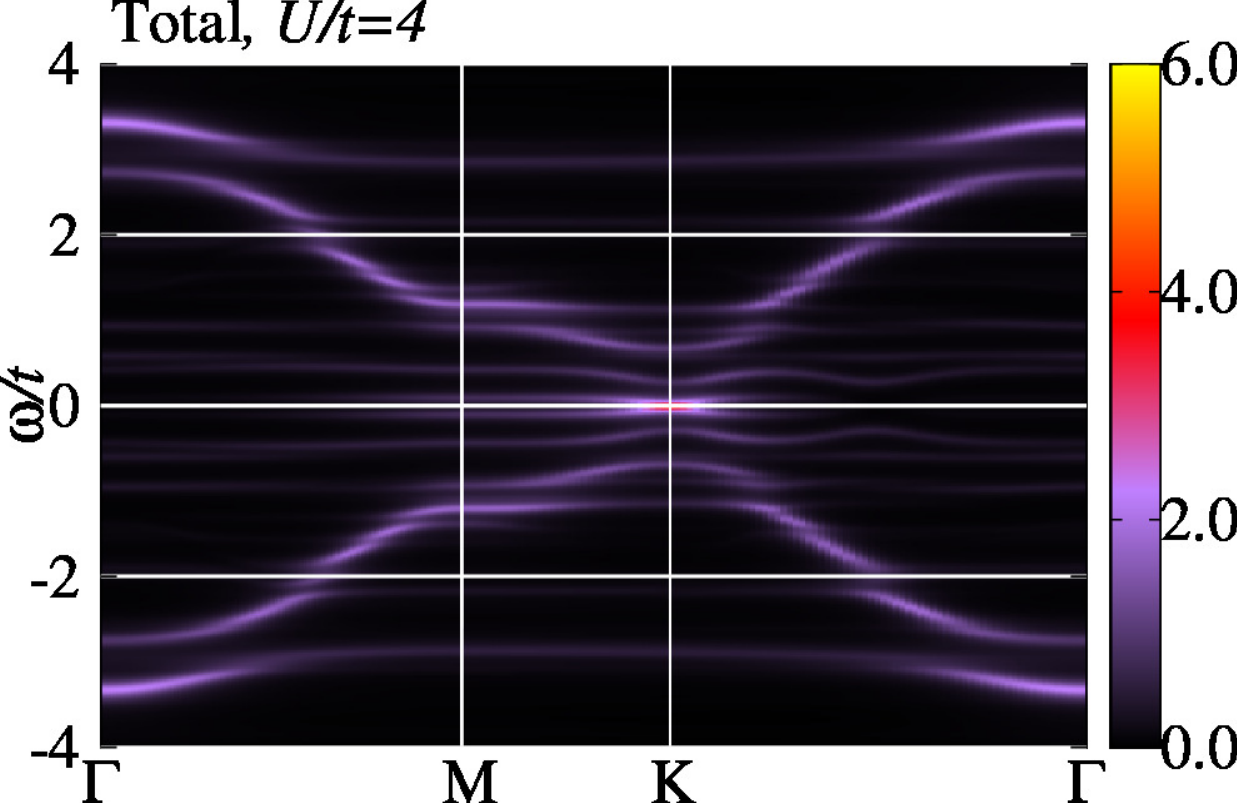}
  \includegraphics[width=4.3cm,angle=0]{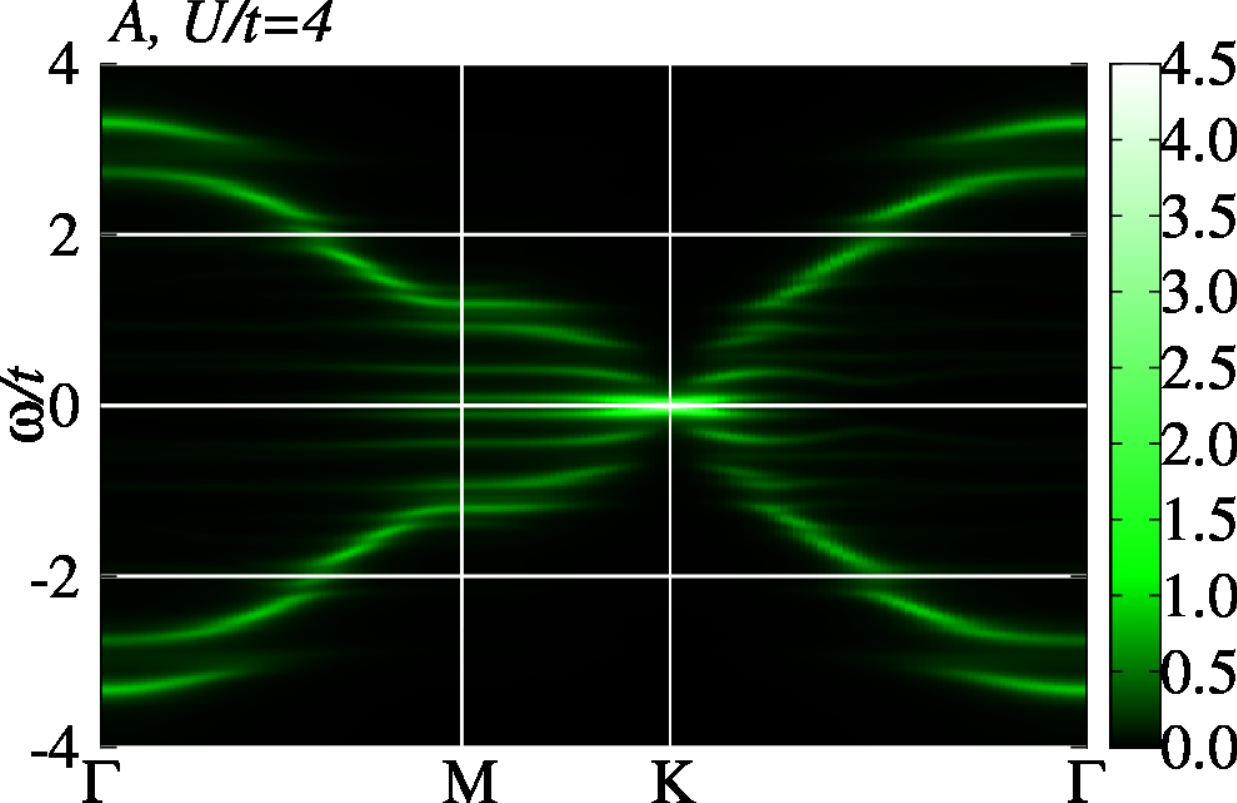}
  \includegraphics[width=4.3cm,angle=0]{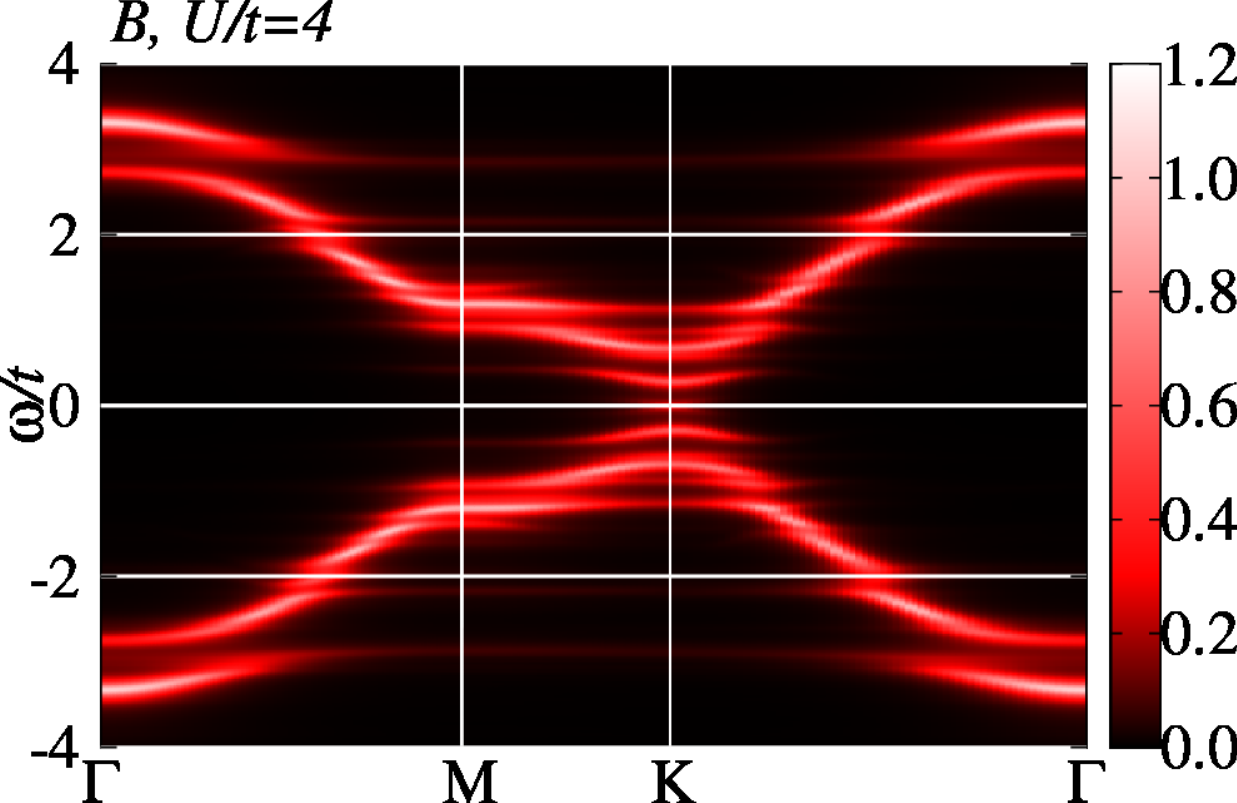}
  \includegraphics[width=4.3cm,angle=0]{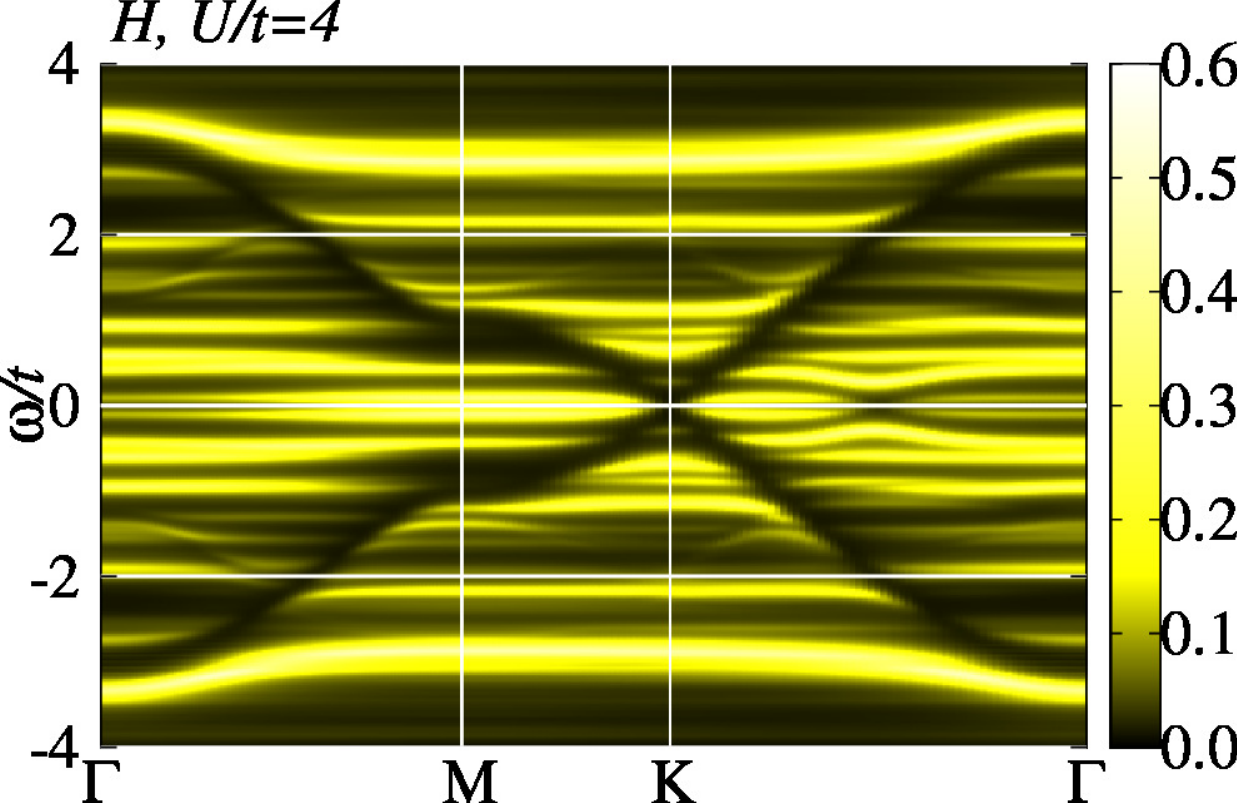}
  \caption{(color online) 
    The CPT results of the single-particle excitation spectra in the PM state for $U/t=4$ and $t_{sp}/t=1$ obtained using 
    the 6-site cluster (containing 2 unit cells) at $T=0.025t$, 
    the 9-site cluster (containing 3 unit cells) at $T=0$, and 
    the 12-site cluster (containing 4 unit cells) at $T=0$ (from top to bottom panels). 
    $A_\sigma(\mb{k},\w)=\sum_\alpha A^{\alpha\alpha}_\sigma(\mb{k},\w)$, 
    $A^{AA}_\sigma(\mb{k},\w)$, $A^{BB}_\sigma(\mb{k},\w)$, and $A^{HH}_\sigma(\mb{k},\w)$ are shown 
    from left to right panels. 
    A Lorentzian broadening of $\eta/t=0.05$ is used.
    The spectral intensity is indicated by a color bar in each figure.
    Note that the different intensity scales are used for different figures.
    The Fermi energy $E_{\rm F}$ is located at $\w=0$. 
    Although the emergent massless Dirac quasiparticles for the 12-site calculations (bottom panels) 
    are not as clear as the other cases, it becomes apparent in the enlarged scale near $E_{\rm F}$, 
    as shown in Fig.~\ref{fig:akw_pm_fs_en}. 
  } 
  \label{fig:akw_pm_fs}
\end{figure*}

\begin{figure*}
  \includegraphics[width=4.3cm,angle=0]{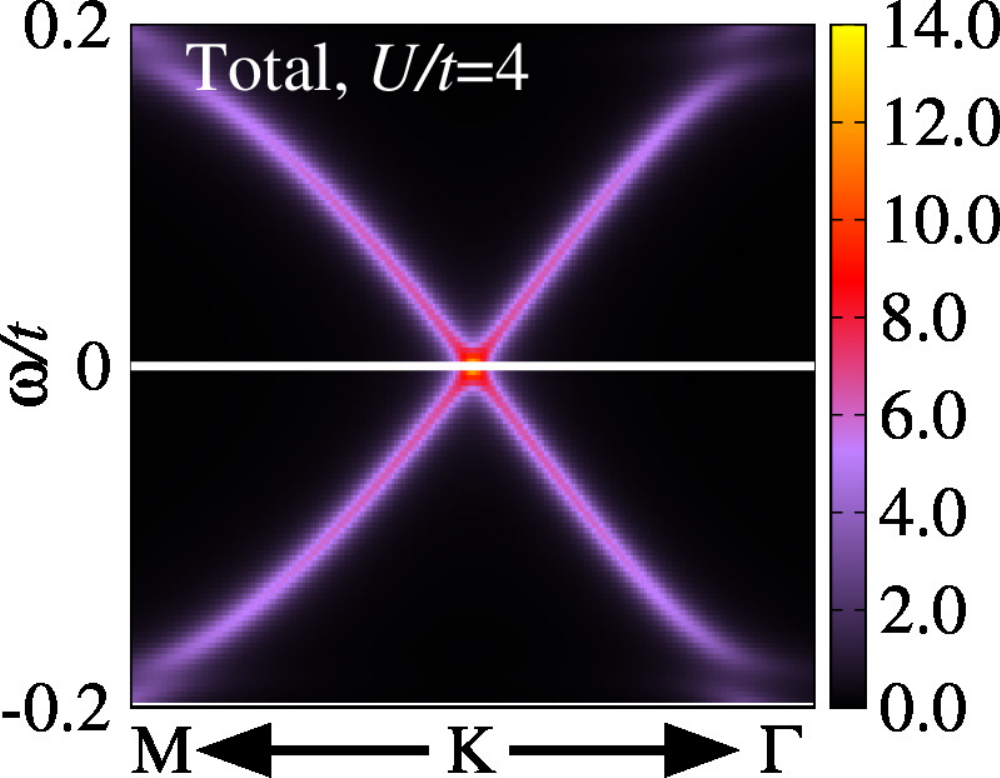}
  \includegraphics[width=4.3cm,angle=0]{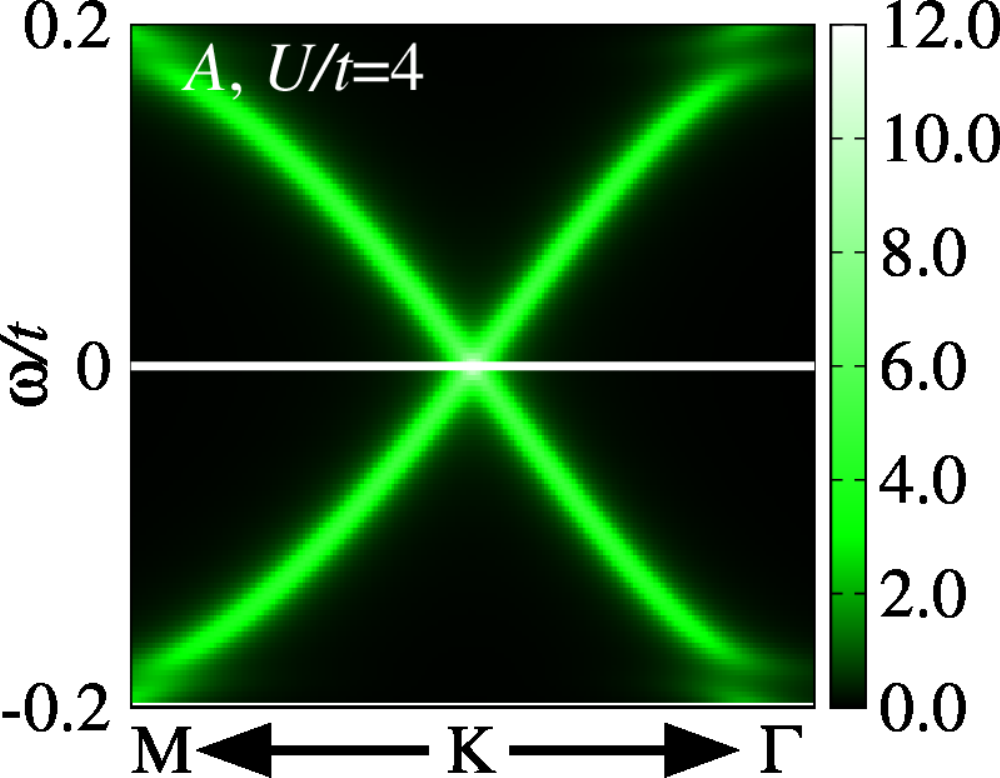}
  \includegraphics[width=4.3cm,angle=0]{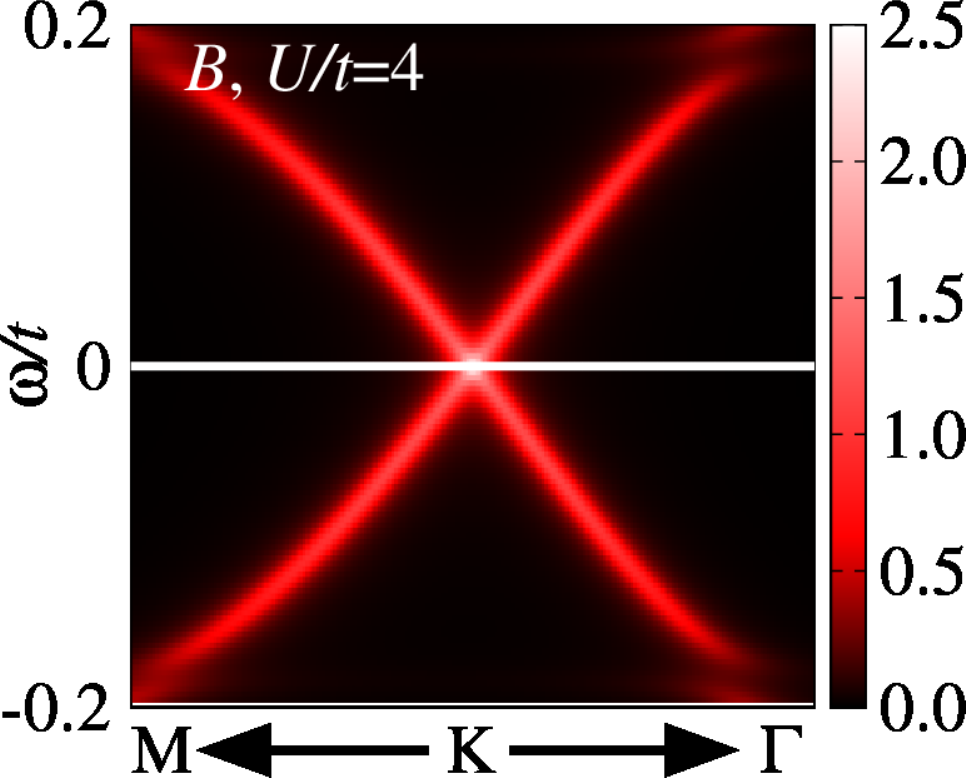}
  \includegraphics[width=4.3cm,angle=0]{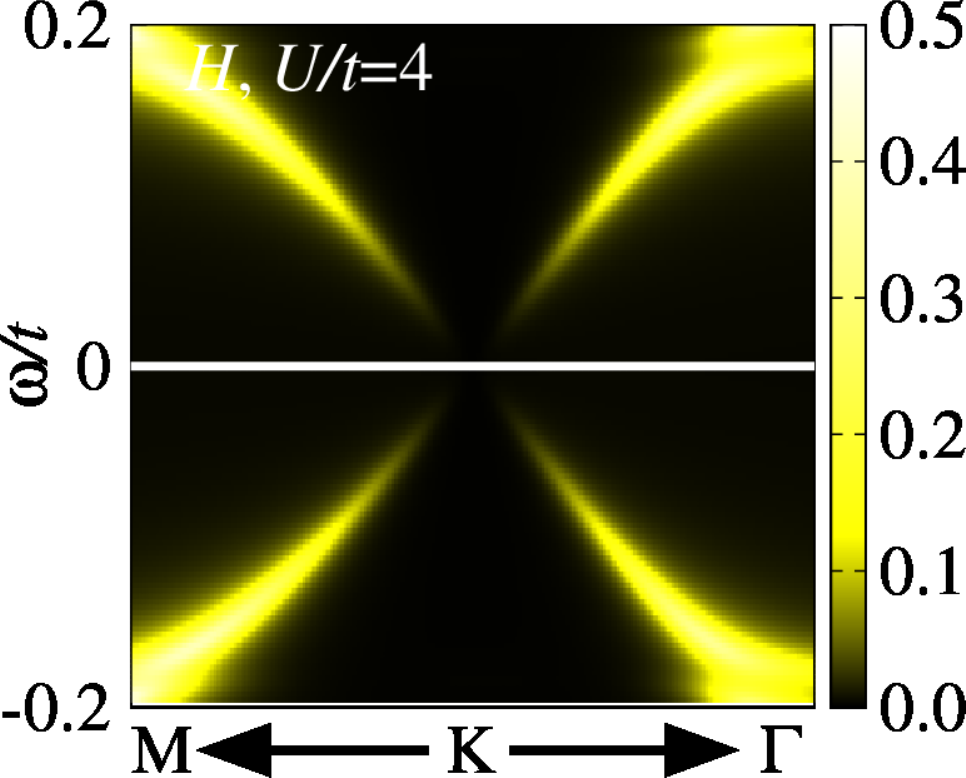}\\
  \includegraphics[width=4.3cm,angle=0]{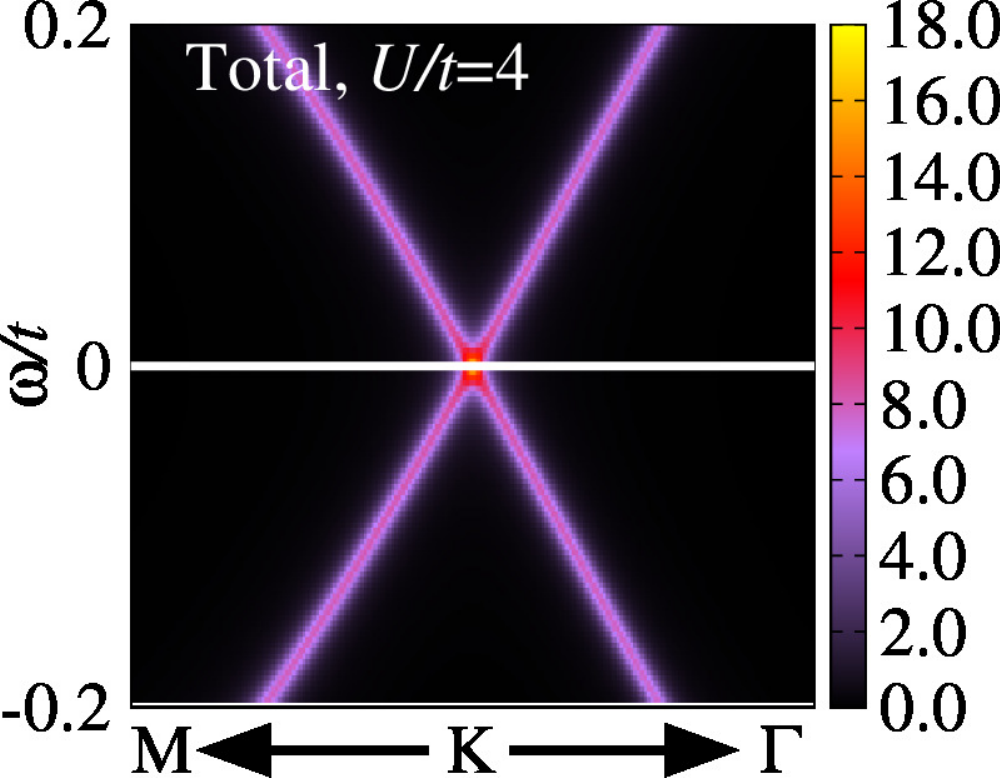}
  \includegraphics[width=4.3cm,angle=0]{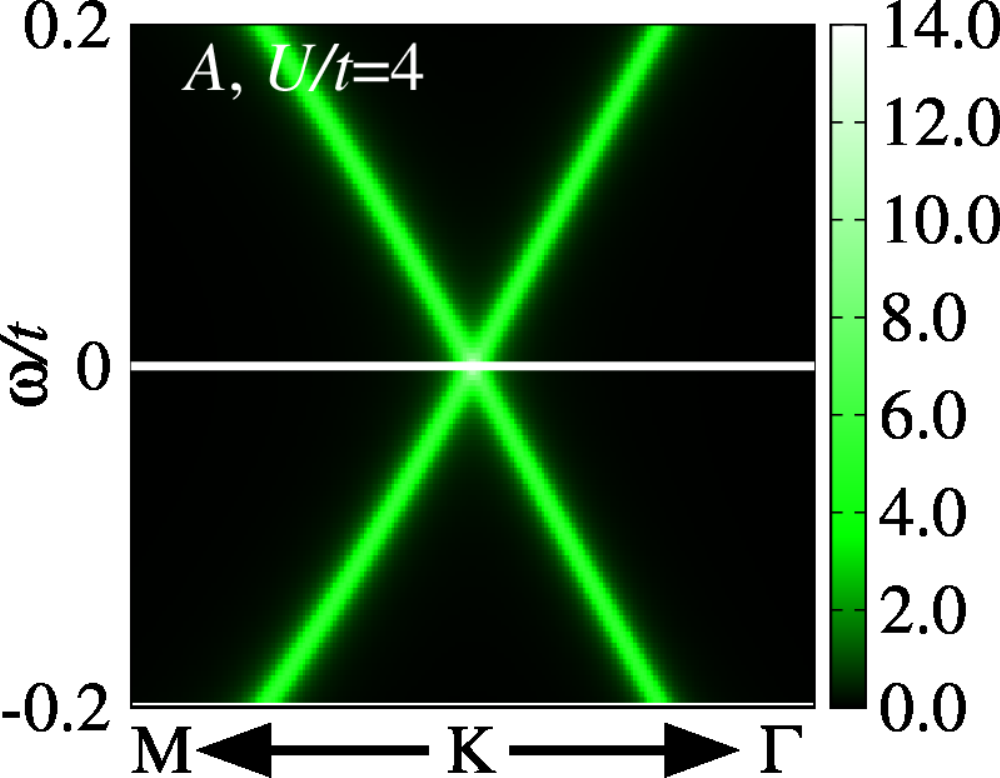}
  \includegraphics[width=4.3cm,angle=0]{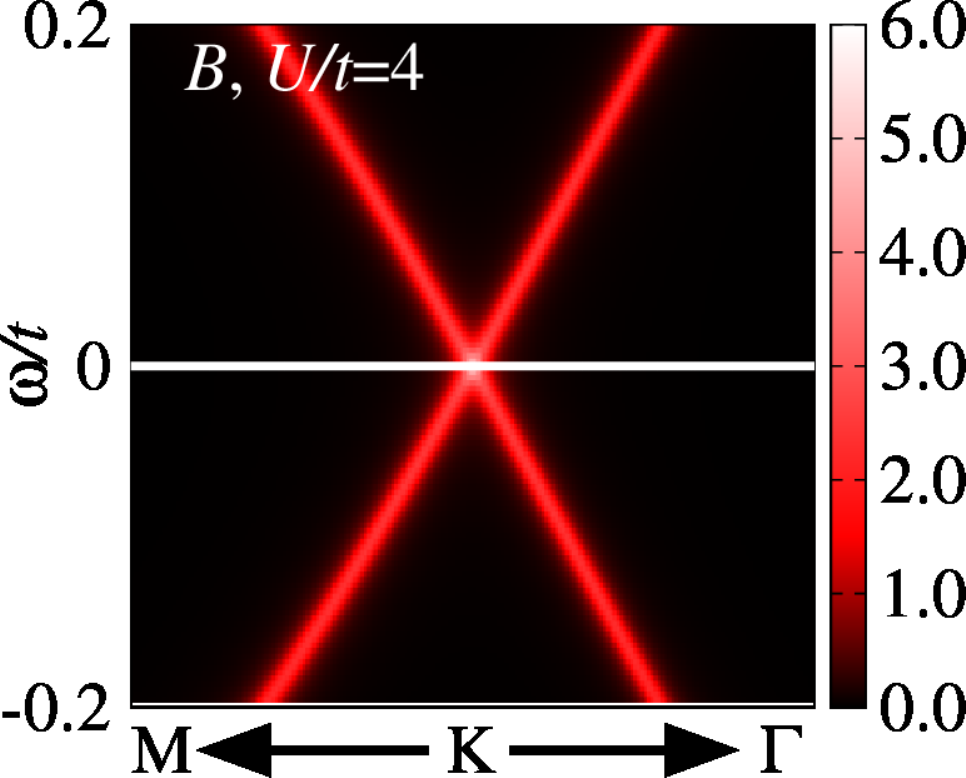}
  \includegraphics[width=4.3cm,angle=0]{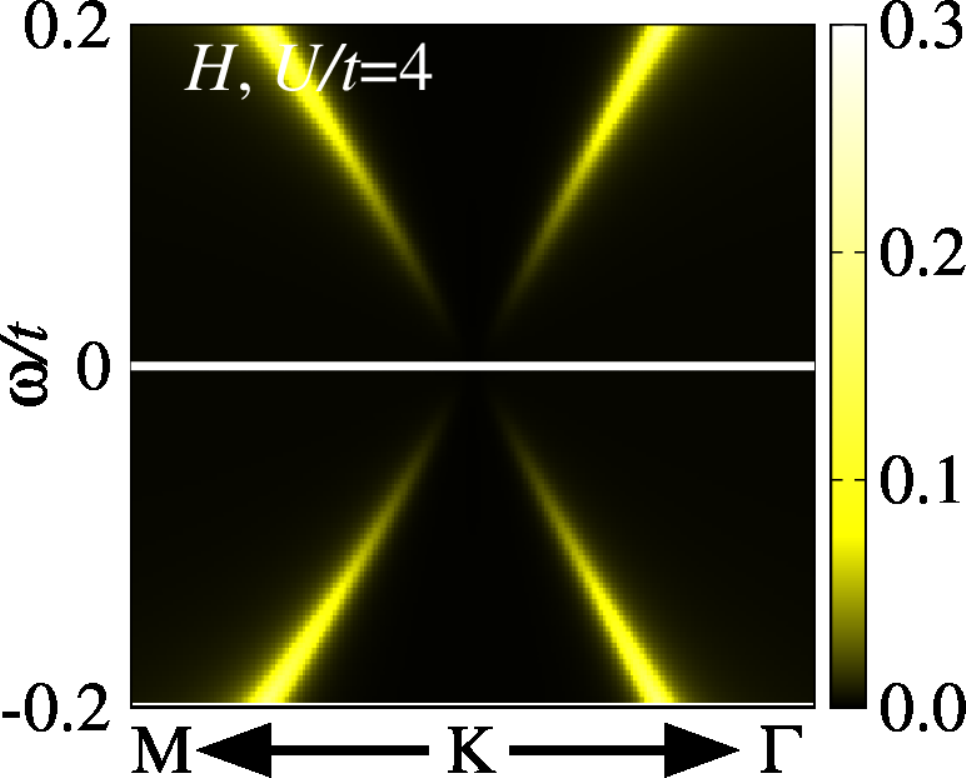}\\
  \includegraphics[width=4.3cm,angle=0]{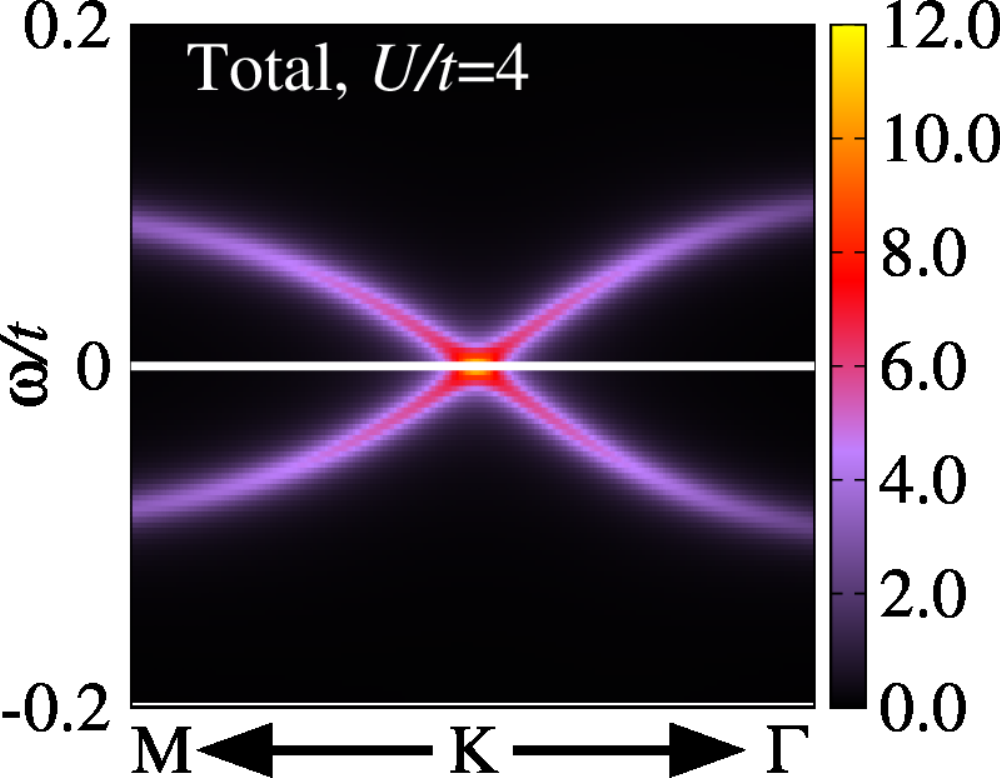}
  \includegraphics[width=4.3cm,angle=0]{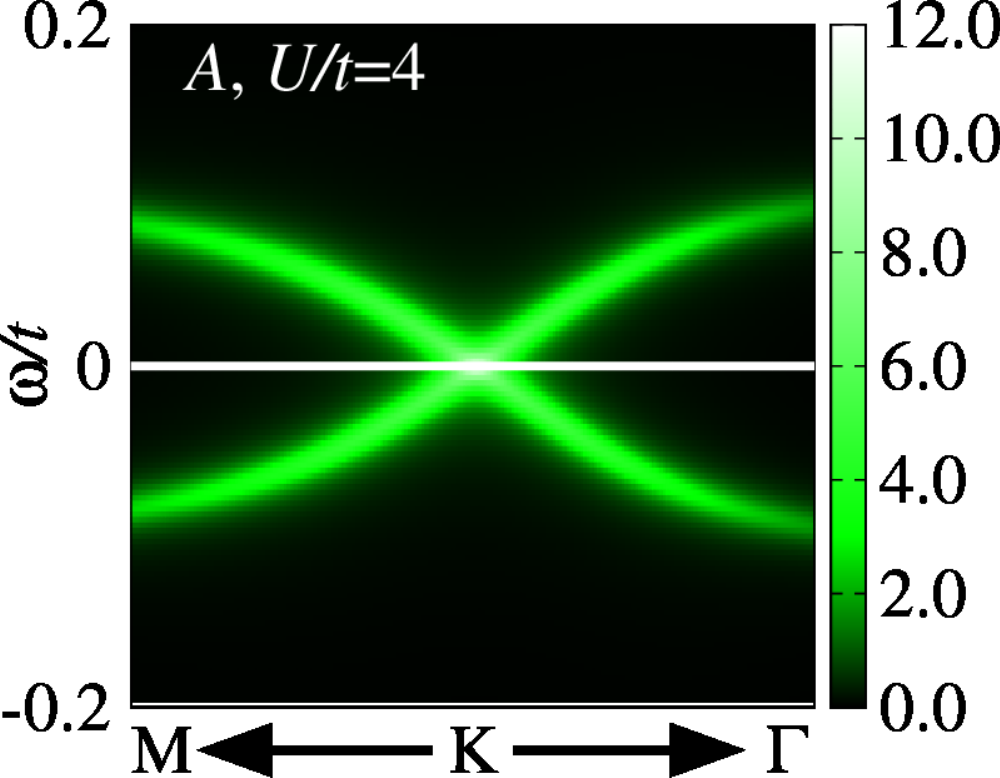}
  \includegraphics[width=4.3cm,angle=0]{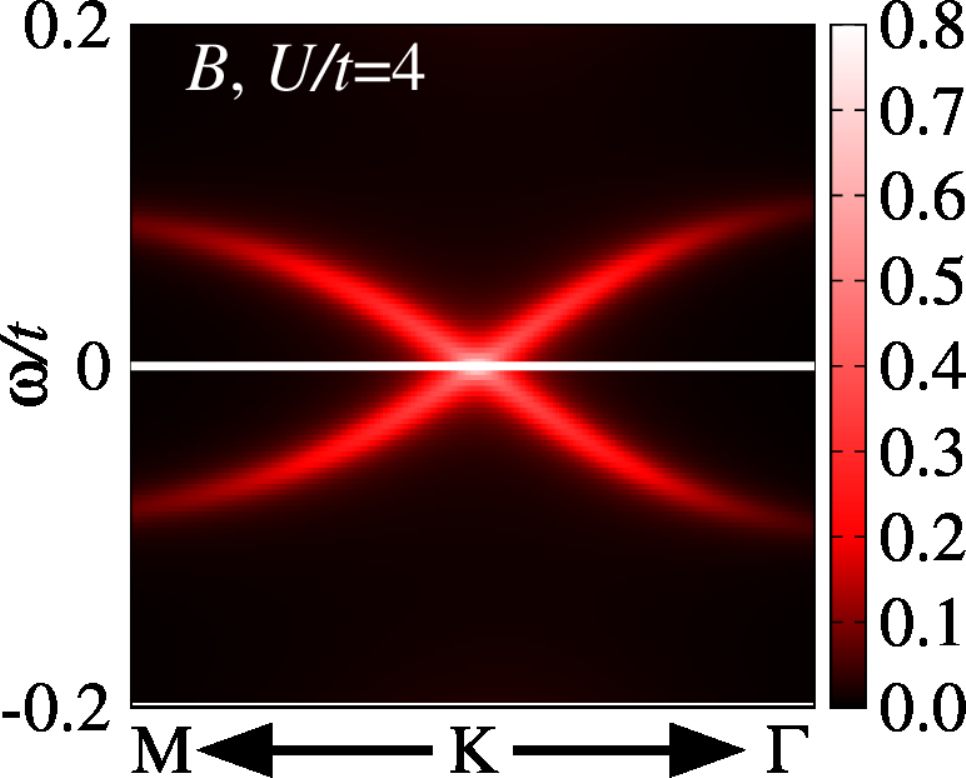}
  \includegraphics[width=4.3cm,angle=0]{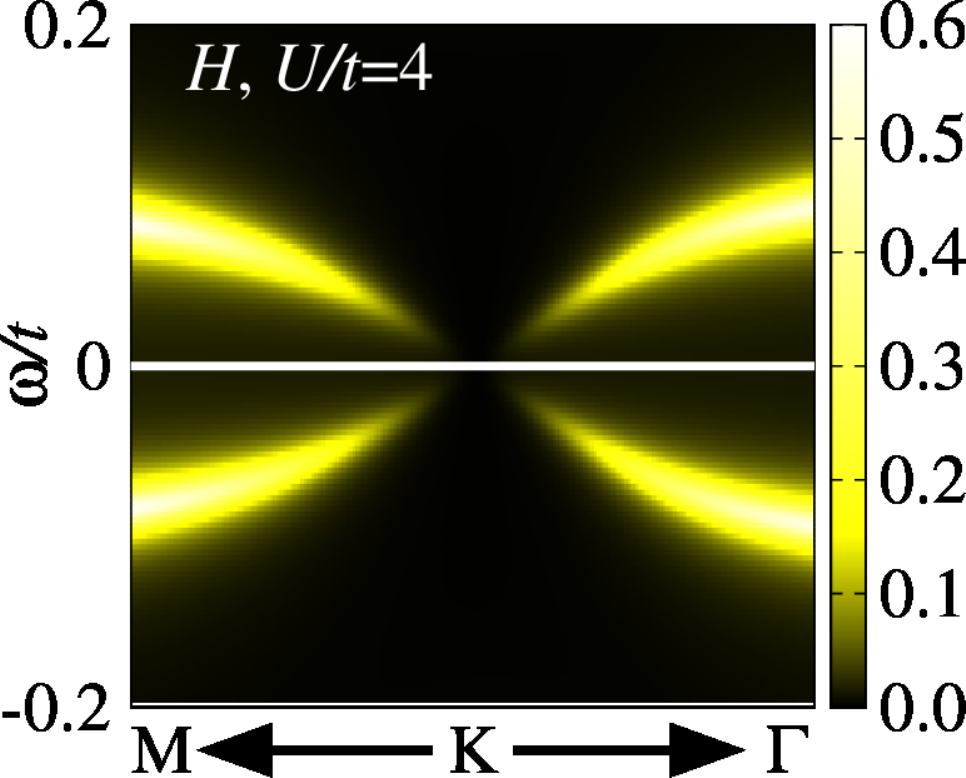}
  \caption{(color online) 
    Same as Fig.~\ref{fig:akw_pm_fs} but the enlarged scale at the vicinity of $K$ point near 
    the Fermi energy $E_{\rm F}$ (white lines). 
    A Lorentzian broadening of $\eta/t=0.01$ is used.
    The region of momenta taken in the horizontal axis is $0.2\pi$ in the $K$-$M$ ($K$-$\Gamma$) 
    direction from $K$. 
  } 
  \label{fig:akw_pm_fs_en}
\end{figure*}

We also examine the finite size effects on the single-particle excitations in the PM state using three different 
clusters and find no qualitative difference. 
Namely, as shown in Fig.~\ref{fig:akw_pm_fs}, we still find the emergent massless 
Dirac quasiparticles at $K$ and $K'$ points with the Dirac points exactly at $E_{\rm F}$ 
and the same characteristic features of 
their spectral weights. 
Although the emergent massless Dirac quasiparticles are not clear for the 12-site calculations 
shown in the bottom panels of Fig.~\ref{fig:akw_pm_fs}, 
it is indeed apparent in the enlarged scale near $E_{\rm F}$ shown in Fig.~\ref{fig:akw_pm_fs_en}. 
Therefore, the emergence of the massless Dirac quasiparticles is not subjected to the finite size effects 
of the clusters. 

\begin{figure}
  \includegraphics[width=6.0cm]{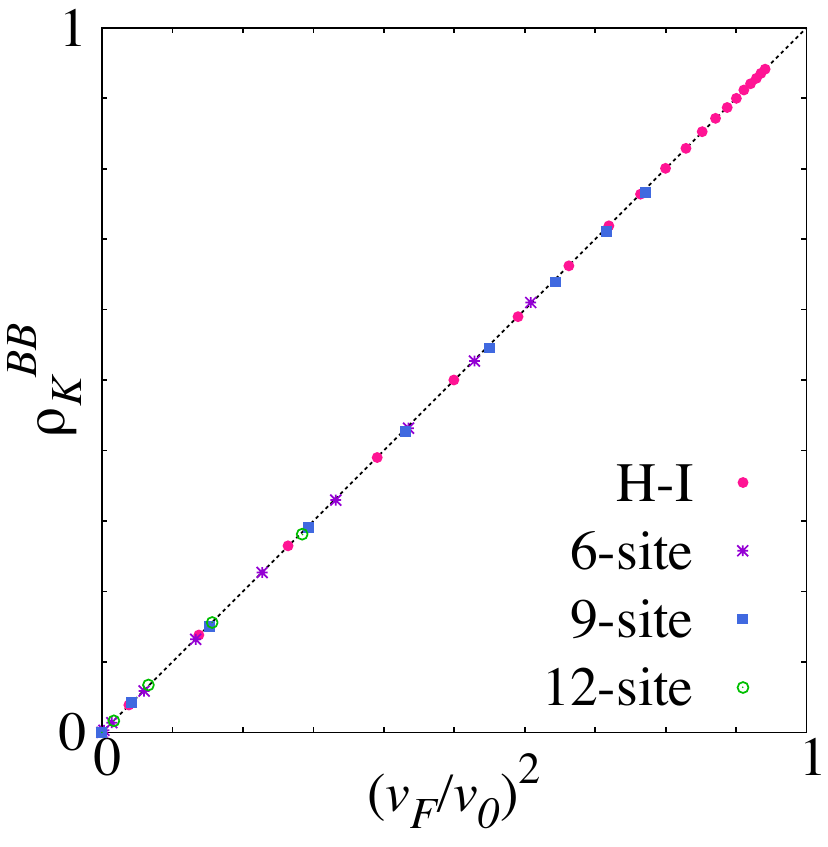}
  \caption{(color online) 
    The spectral weight $\rho_K^{BB}$ for $B$ orbital 
    at the Fermi energy $E_{\rm F}$ and momentum ${\mb k}=K$ versus $v_{\rm F}^2$ calculated using 
    the CPT and the Hubbard-I (H-I) approximation. 
    The CPT calculations are done for the 6-site cluster (containing 2 unit cells) at 
    $T=0.025t$ in the PM phase, and for the 9-site cluster (containing 3 unit cells) and 
    the 12-site cluster (containing 4 unit cells) at $T=0$ where the PM state is assumed.  
    Both calculations are for $t_{sp}/t=1$ at $n=1$ with various values of 
    $U$ shown in Fig.~\ref{fig:vf_vs_u}. 
    Here, $v_0=\sqrt{3}t/2$ is the Dirac Fermi velocity of the pure graphene model. 
  } 
  \label{fig:vf_vs_sw}
\end{figure}

On the other hand, we find in Fig.~\ref{fig:akw_pm_fs} and Fig.~\ref{fig:akw_pm_fs_en} 
that the Dirac Fermi velocity $v_{\rm F}$ depends quantitatively on the cluster 
size used. The $U$ dependence of $v_{\rm F}$ for three different clusters is summarized in 
Fig.~\ref{fig:vf_vs_u}. 
Although the value itself depends on the cluster size, the qualitative behavior of $v_{\rm F}$ is the same: 
$v_{\rm F}$ monotonically increases with increasing $U$ for all clusters 
used. It is also interesting to note that, irrespectively of the cluster sizes, the spectral weight $\rho^{BB}_K$ 
for $B$ orbital at the Dirac point is found to be 
proportional to $v_{\rm F}^2$, i.e., 
\begin{equation}\label{eq.rhobb}
\rho^{BB}_K=\left(\frac{v_{\rm F}}{v_0}\right)^2,
\end{equation}
as shown in Fig.~\ref{fig:vf_vs_sw}. 
This universal behavior is intuitively understood by 
assuming that the electron annihilation operator is renormalized with the renormalization factor 
$z_A$ ($z_B$) for $A$ ($B$) orbital 
at the Dirac point $k_{\rm F}$, i.e., $c_{k_{\rm F}\sigma A (B)}\to z_{A (B)} c_{k_{\rm F}\sigma A (B)}$ 
with $z_A\approx1$. 
Due to this renormalization, $\rho^{BB}_K\sim z_B^2$ while $v_{\rm F}/v_0\sim z_B$ 
because $t$ is renormalized to $tz_Az_B$.
The simple and yet significant universal relation in Eq.~(\ref{eq.rhobb})  
concisely expresses the inevitable involvement of $B$ orbital in the low-energy excitations 
for the emergent massless Dirac quasiparticles.

Finally, we summarize how the massless Dirac quasiparticles emerge and evolve with increasing $U$ 
and how different orbitals contribute to the formation of the massless Dirac quasiparticles. 
As shown in Fig.~\ref{fig:akw_pm}, the spectral weight $A^{BB}_\sigma(\mb{k},\w)$ for $B$ orbital 
around the Fermi energy is exactly zero when $U=0$ since $B$ orbital does not contribute to the flat band 
formation in the noninteracting limit (see also Fig.~\ref{disp_nonint}). However, 
with increasing $U$, $A^{BB}_\sigma(\mb{k},\w)$ around $K$ and $K'$ points near the Fermi energy 
gradually increases to form the massless Dirac quasiparticles. On the other hand, 
the $A$ orbital component of the spectral weight $A^{AA}_\sigma(\mb{k},\w)$ at the Fermi energy is already finite 
even when $U=0$, since $A$ orbital contributes to the formation of the flat band in the noninteracting limit, 
and as $U$ increases it develops into the low energy excitations with the linear energy dispersions 
at $K$ and $K'$ points. 
As shown in Sec.~\ref{sec:H-I}, these features are in good agreement with those obtained in the Hubbard-I 
approximation.

The contribution of $H$ orbital is different. First, we notice in Fig.~\ref{fig:akw_pm} that the 
spectral weight $A^{HH}_\sigma(\mb{k},\w)$ for $H$ orbital displays almost dispersionless spectra, 
indicating that $H$ orbital is rather localized in real space, even when $U$ is small. 
Furthermore, as clearly observed in Fig.~\ref{fig:akw_pm} and Fig.~\ref{fig:akw_pm_fs}, 
$A^{HH}_\sigma(\mb{k},\w)$ exhibits a ``dark spectral" region where no spectral intensity exists and 
the ``dispersion" of the dark spectral region very much resembles the energy dispersion of the conduction 
band described by the pure graphene model, i.e., $\w=\pm|\gamma_{\mb{k}}|$, 
suggesting that 
\begin{equation}
A^{HH}_\sigma(\mb{k},\w = \pm |\gamma_{\mb{k}}|) = 0.
\end{equation} 
This is exactly the case for the Hubbard-I approximation, as discussed in Sec.~\ref{sec:H-I}, 
because $G_{\sigma}^{HH} (\mb{k}, \w=\pm |\gamma_{\mb{k}}|) = 0$ in Eq.~(\ref{eq.Gkw}) 
(see also Fig.~\ref{fig.akw.h1}).  
This implies that $H$ orbital is ``level repulsive", i.e., dynamically decoupled to the conduction band composed of 
$A$ and $B$ orbitals, and thus
$H$ orbital does not contribute to the formation of the emergent massless Dirac quasiparticles. 

\section{Analytical results}\label{sec:analytical}

In this section, the periodic Anderson model is analyzed using 
the simplest mean-filed theory for the FM state and the Hubbard-I approximation for the 
PM state. We also construct an effective Hamiltonian to describe the single-particle 
excitations in the PM state and discuss the chiral symmetry of the quasiparticles as well as 
the origin of the emergent massless Dirac quasiparticles. 

\subsection{Mean-field approximation}\label{sec:mf}
We first consider the simplest mean-field theory for the FM state to show that the main characteristic features 
of the single-particle excitations obtained by the CPT in Sec.~\ref{sec:fm} can be reproduced by the single-particle 
approximation. 
Applying the mean-field decoupling to the on-site Coulomb term for the hydrogen impurity sites,
\begin{eqnarray}
n_{\mb{r}_i\uparrow H} n_{\mb{r}_i\downarrow H} &\approx& n_{\mb{r}_i\uparrow H} \left< n_{\mb{r}_i\downarrow H} \right> + n_{\mb{r}_i\downarrow H} \left< n_{\mb{r}_i\uparrow H} \right> \nonumber \\
&-& \left< n_{\mb{r}_i\uparrow H} \right> \left< n_{\mb{r}_i\downarrow H} \right>,
\end{eqnarray}
the mean-field Hamiltonian $\mathcal{H}_{\rm MF}$ for the periodic Anderson model with $\eps_H = -U/2$
is given as 
\begin{eqnarray}
\mathcal{H}_{\rm MF} &=& \sum_{\bf k}\sum_\s  
\mb{c}^\dag_{\mb{k}\s}
\left(
\begin{array}{ccc}
0 & \gamma_{\bf k} & 0 \\
\gamma_{\bf k}^{\ast} & 0 & t_{sp} \\
0 & t_{sp} & \Delta_{\bar\sigma} \\
\end{array}
\right) 
\mb{c}_{\mb{k}\s}
\nonumber \\
&-& U \sum_i \left< n_{\mb{r}_i\uparrow H} \right>\left< n_{\mb{r}_i\downarrow H} \right>,
\label{eq:mf}
\end{eqnarray}
where
\begin{eqnarray}
\Delta_{\sigma} = U \left( \left< n_{\mb{r}_i \sigma H} \right> - \frac{1}{2} \right)  
\end{eqnarray}
and $\bar\s$ denotes the opposite spin of $\s$. 
We assume that $\left< n_{\mb{r}_i\sigma H} \right>$ is site independent. 

\begin{figure*}
  \includegraphics[width=4.3cm,angle=0]{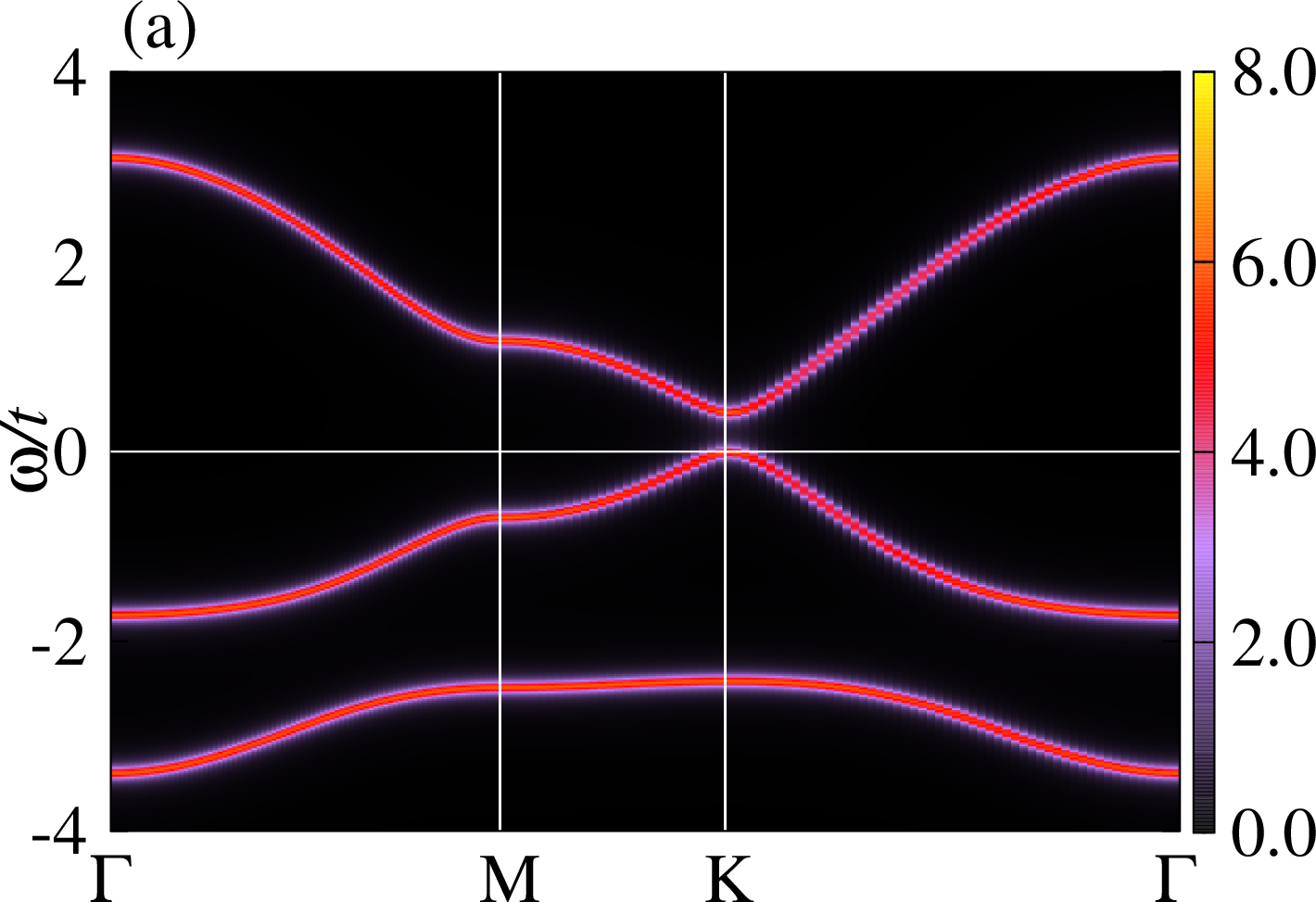}
  \includegraphics[width=4.3cm,angle=0]{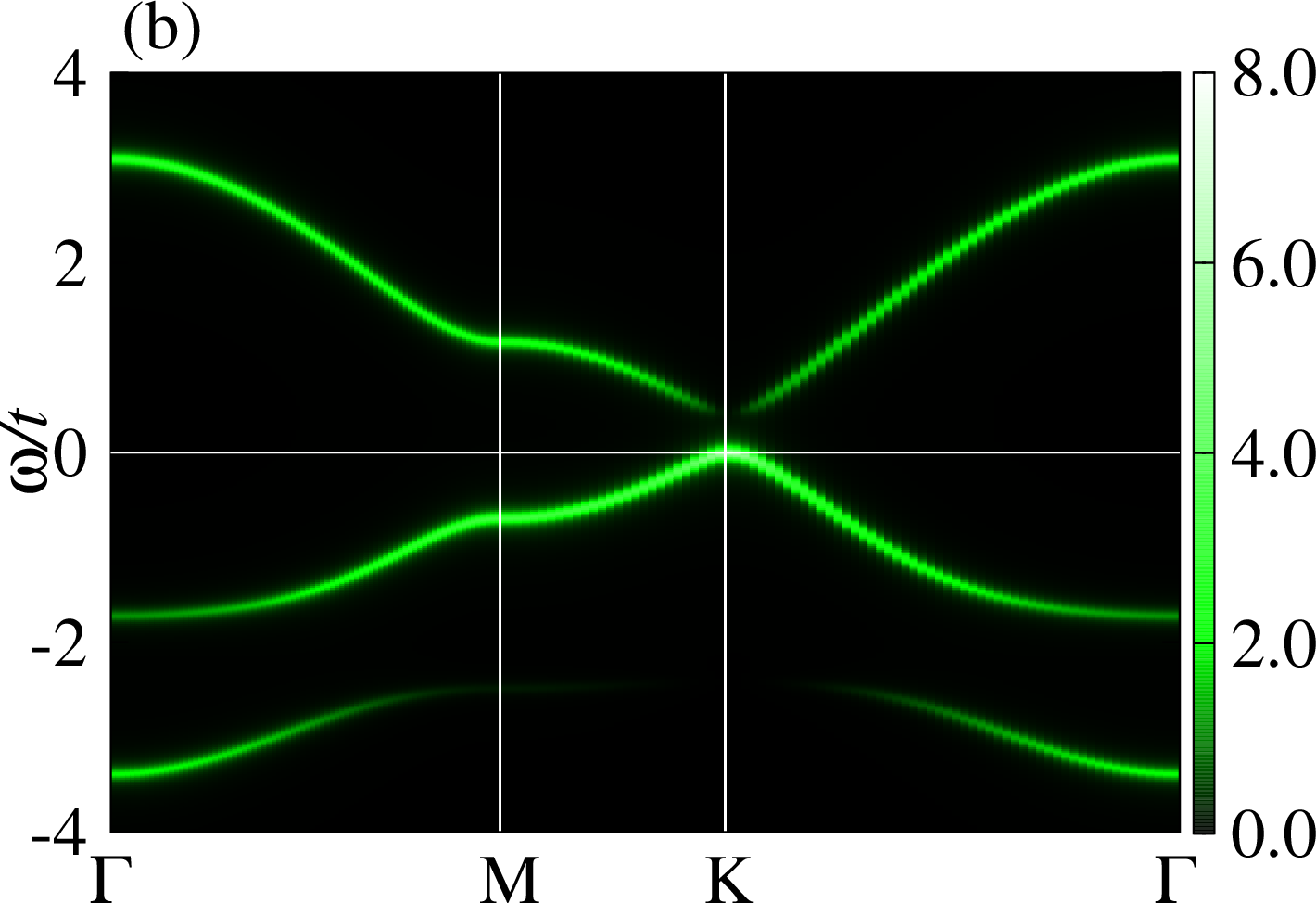}
  \includegraphics[width=4.3cm,angle=0]{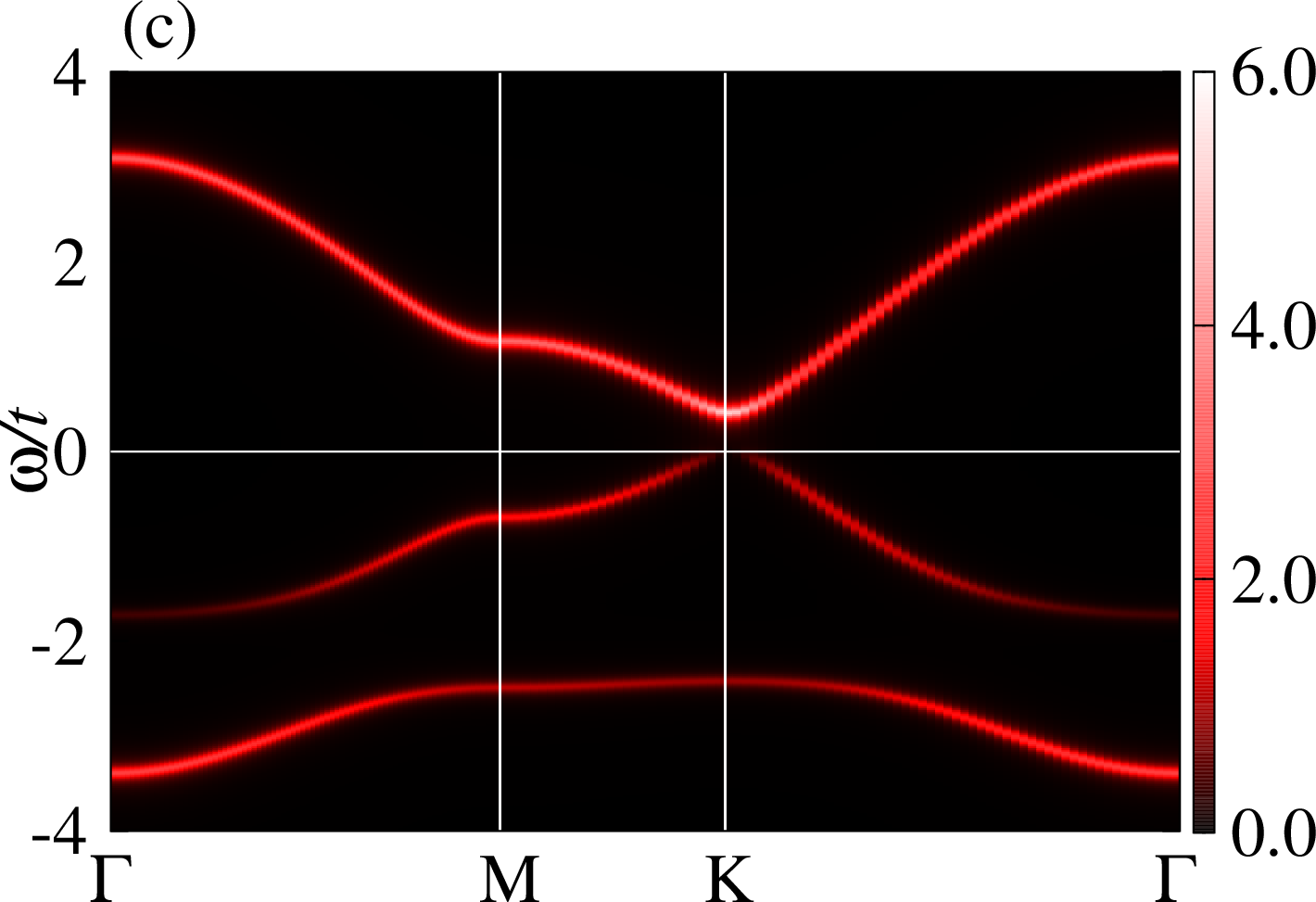}
  \includegraphics[width=4.3cm,angle=0]{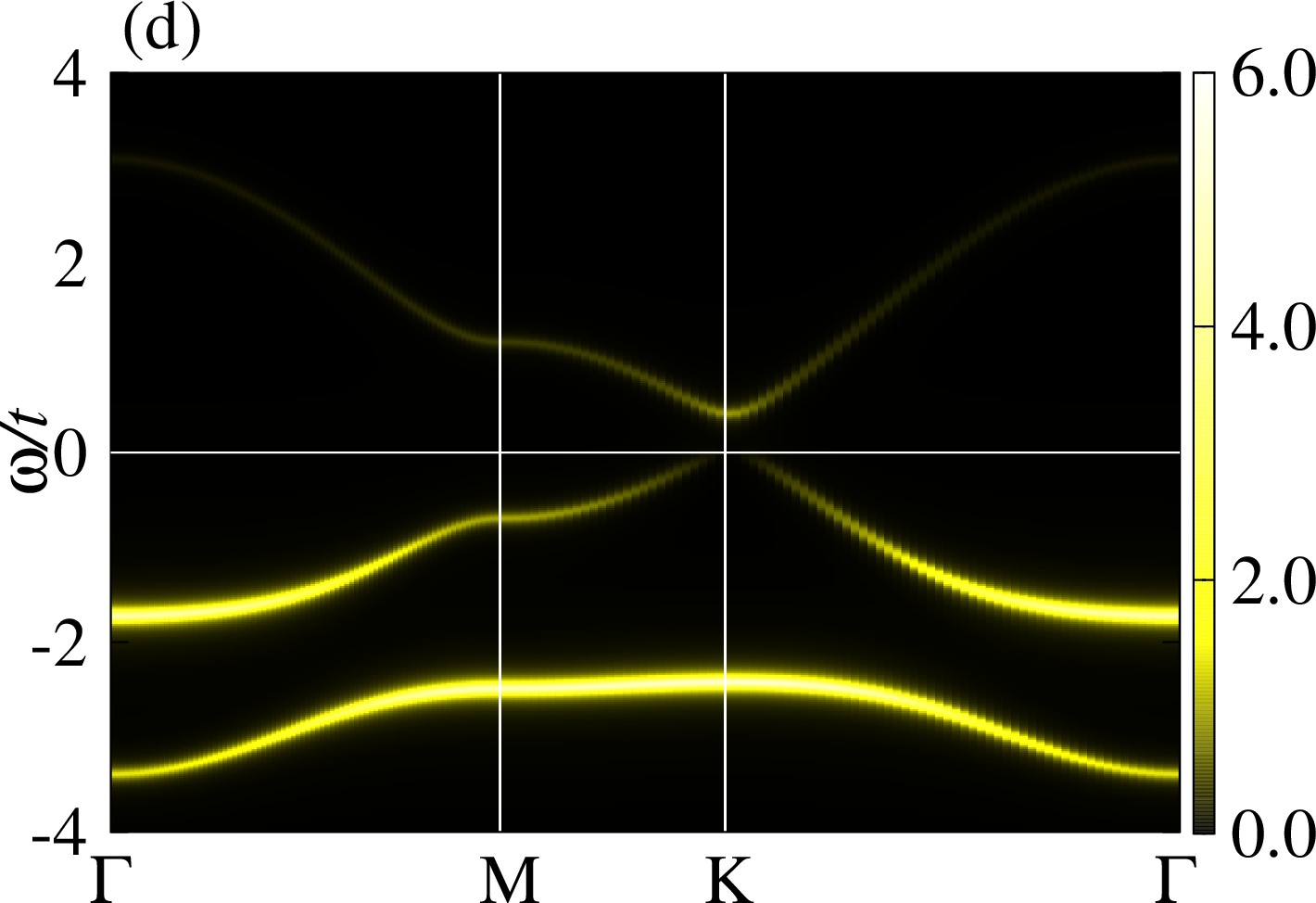}\\
  \includegraphics[width=4.3cm,angle=0]{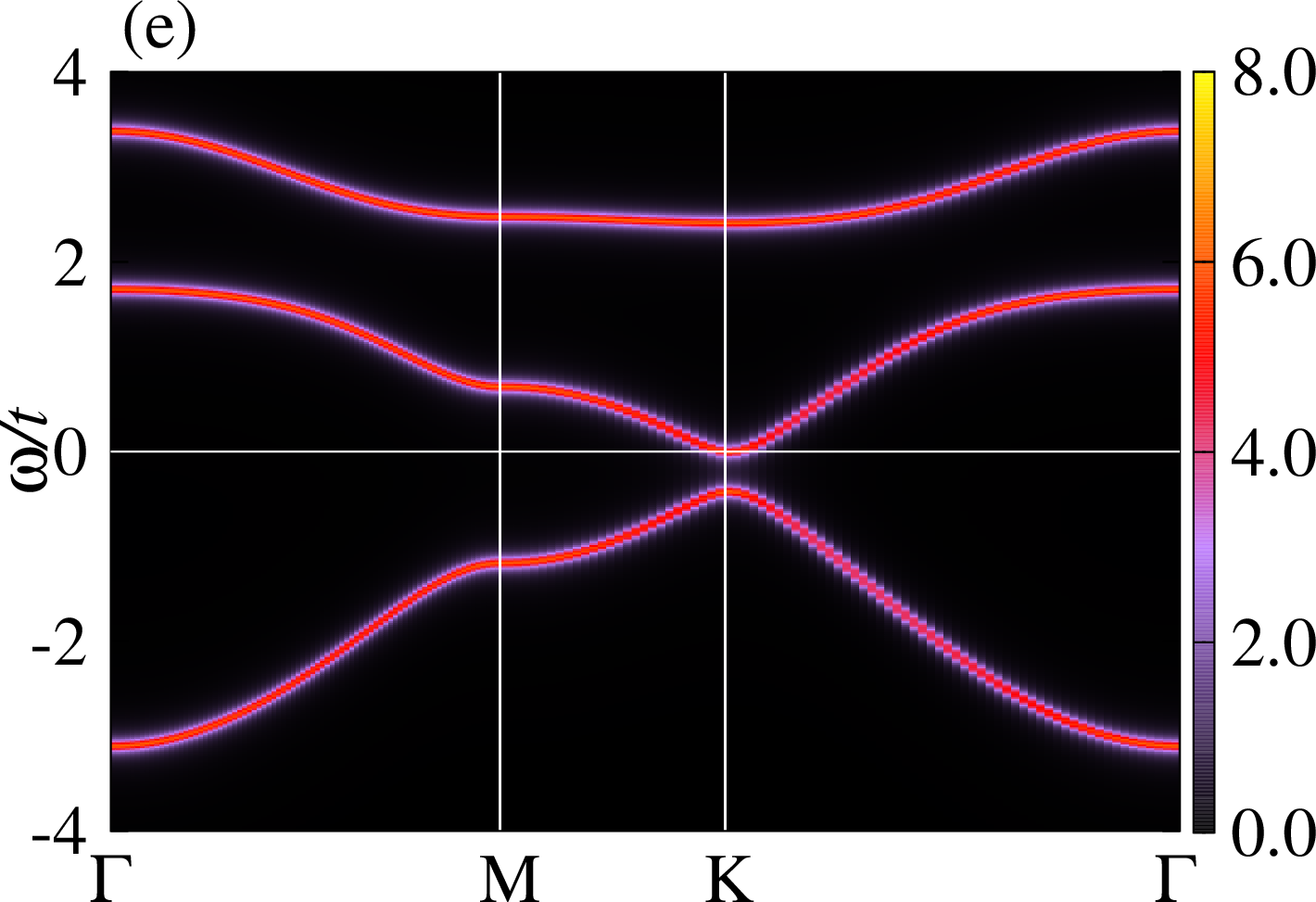}
  \includegraphics[width=4.3cm,angle=0]{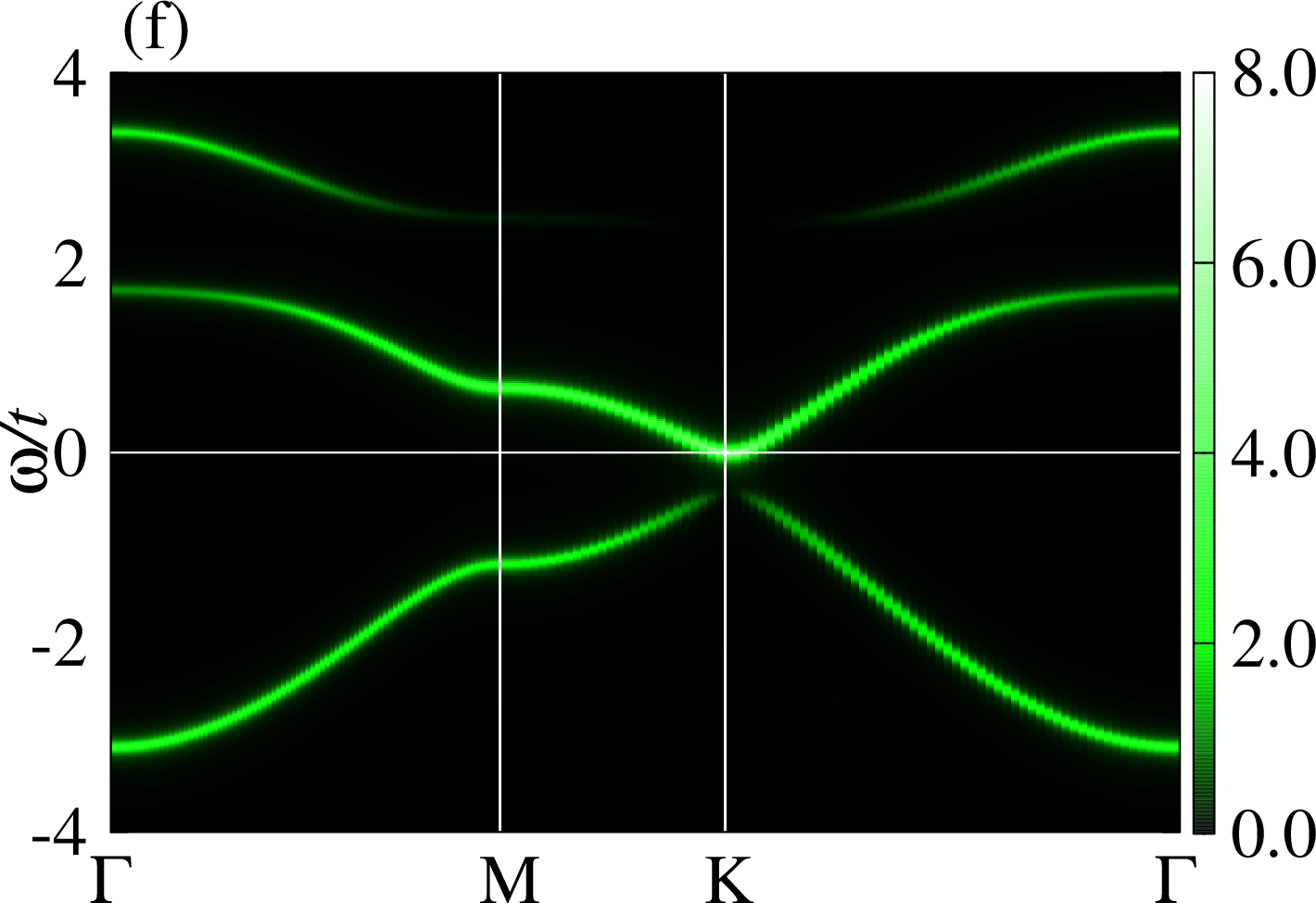}
  \includegraphics[width=4.3cm,angle=0]{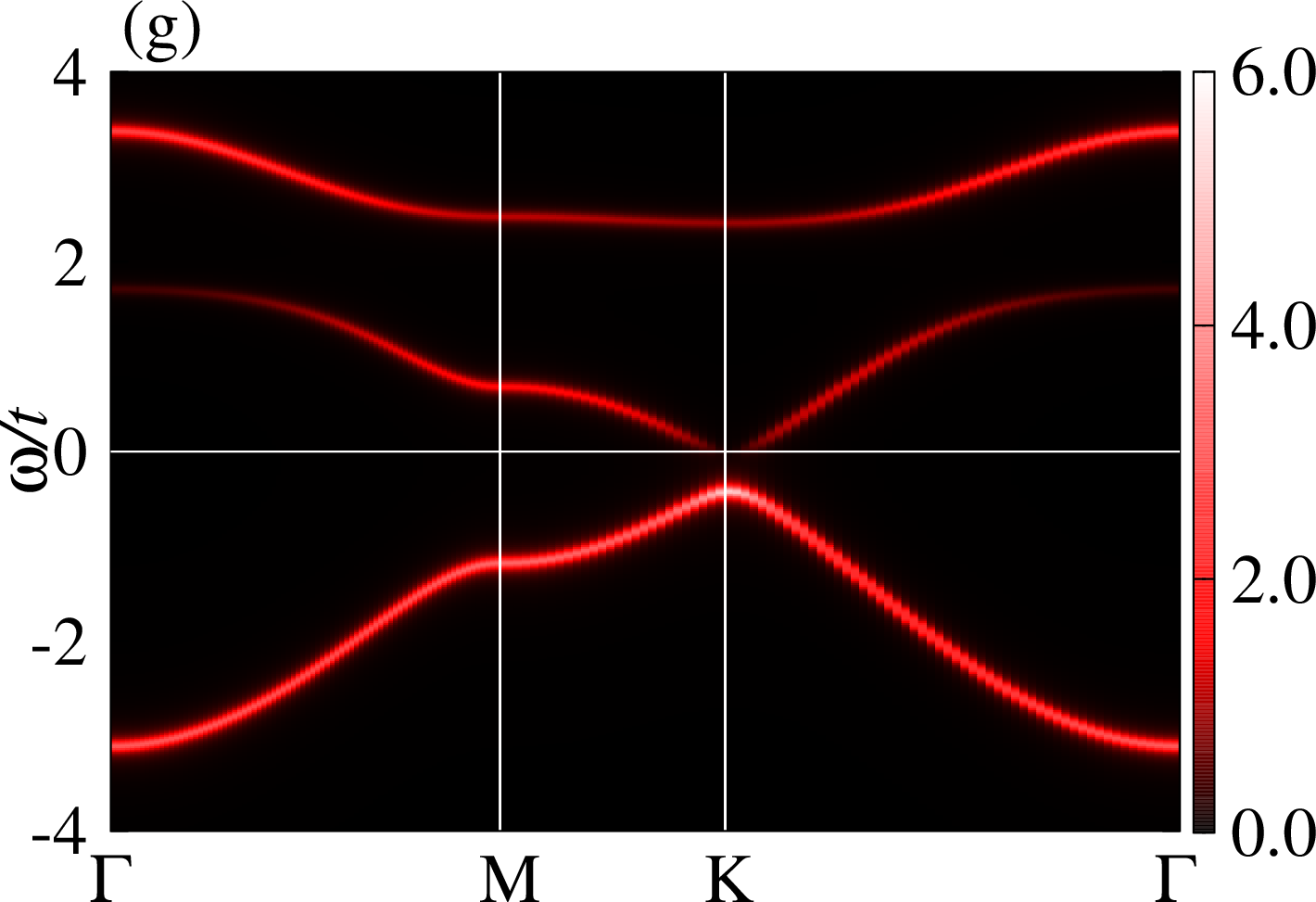}
  \includegraphics[width=4.3cm,angle=0]{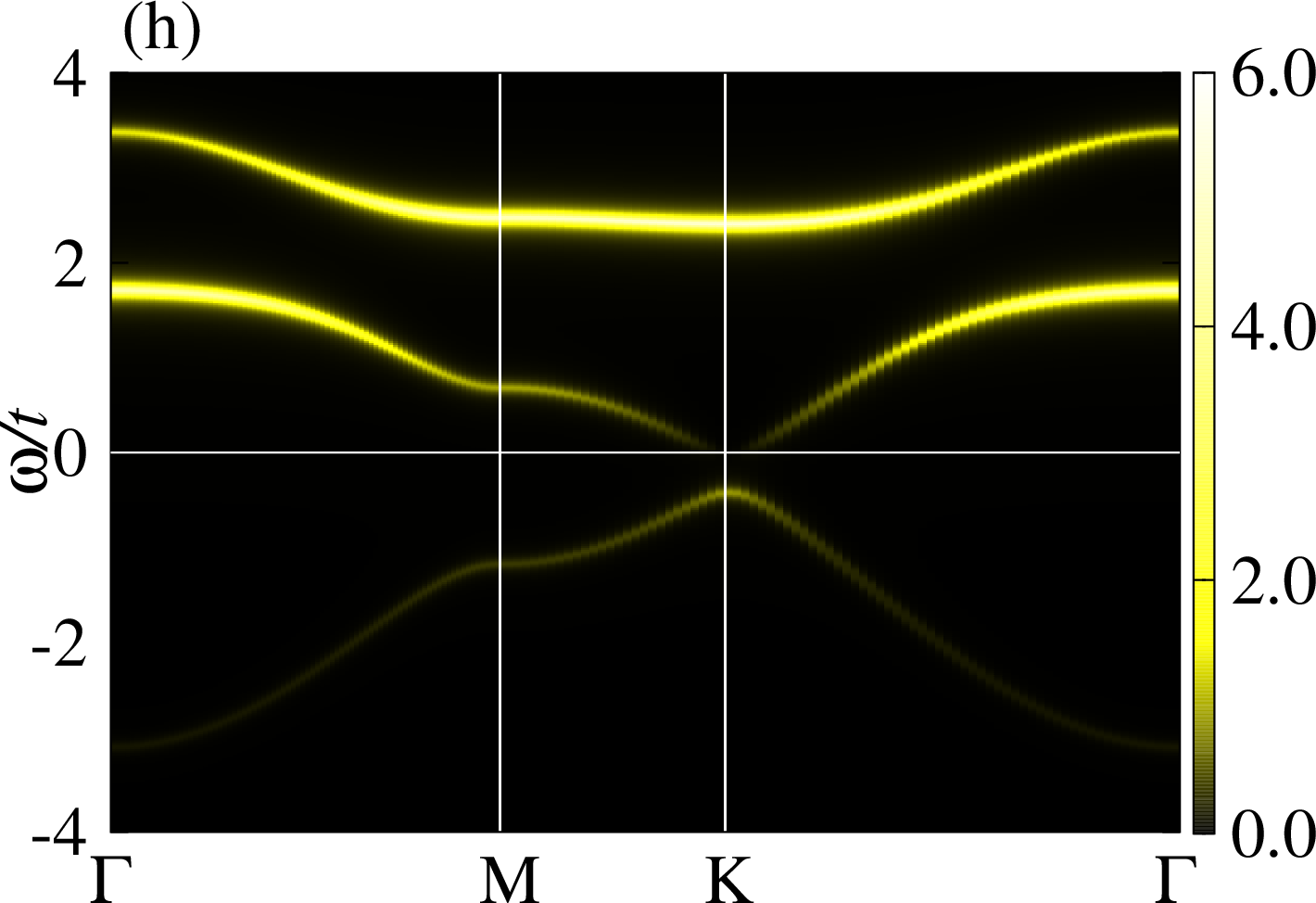}
  \caption{(color online) 
    The mean-field results of the single-particle excitation spectra for (a)--(d) up and 
    (e)--(h) down electrons in the FM state at $T=0$. 
    The calculations are for $t_{sp}/t=1$ and $\Delta/t = 2$ at $n=1$. 
    (a) and (e): $A_\sigma(\mb{k},\w) = \sum_\alpha A_\sigma^{\alpha\alpha}(\mb{k},\w)$, 
    (b) and (f): $A^{AA}_\sigma(\mb{k},\w)$, 
    (c) and (g): $A^{BB}_\sigma(\mb{k},\w)$, and 
    (d) and (h): $A^{HH}_\sigma(\mb{k},\w)$. 
    A Lorentzian broadening of $\eta/t = 0.05$ is used. 
    The spectral intensity is
    indicated by a color bar in each figure. 
    Notice that the different intensity scales are used for different figures. 
    The Fermi energy $E_{\rm F}$ is located at $\w=0$. 
  }
  \label{fig:mfband}
\end{figure*}

Assuming the FM ansatz, i.e.,  
\begin{eqnarray}
\Delta_{\uparrow} = - \Delta_{\downarrow} = \Delta,
\end{eqnarray}
we can easily obtain the 
single-particle excitation spectrum of ${\mathcal H}_{\rm MF}$ 
for a given $\Delta$. 
A typical example of the single-particle excitation spectrum for the FM state is shown in Fig.~\ref{fig:mfband}. 
The main features are summarized as follows. 
First, for any $t_{sp}$ and $\Delta > 0$, the Fermi energy locates 
at the top (bottom) of the middle band for up (down) electrons, thus indicating that 
the total magnetic moment $2S$ is exactly 
$|H| + |A| - |B|$, independently of the value of $\Delta\,(> 0)$. 
Second, the top of the middle band for up electrons and 
the bottom of the middle band for down electrons 
touch exactly at the Fermi energy and momentum ${\bf k}=K$ and $K^{\prime}$. 
This degeneracy is easily understood because $\gamma_{\bf k}=0$ at 
$K$ and $K'$ points, and therefore one of the eigenvalues of 
$3\times 3$ matrices in Eq.~(\ref{eq:mf}) for each spin component must be zero at these momenta. 
This also indicates that the energy dispersion is quadratic near the Fermi energy. 
It should be also noticed in Fig.~\ref{fig:mfband} that the low-energy excitations close to the 
Fermi energy is mostly 
composed of $A$ orbital and indeed only $A$ orbital contributes to the spectral weight at the Fermi energy. 
These results are qualitatively the same as those obtained using the CPT in Fig.~\ref{fig:akw_fm} 
and Fig.~\ref{fig:akw_fm_en}. 

\subsection{Hubbard-I approximation}\label{sec:H-I}

It is highly interesting to examine the single-particle excitations in the PM state 
using the Hubbard-I approximation~\cite{Hubbard1963} 
since this is the simplest approximation to 
treat dynamical electron correlations with no spatial fluctuations.

\subsubsection{Self-energy}

Within the Hubbard-I approximation~\cite{Hubbard1963}, the self-energy $\Sigma_{\s}(\w)$ of the single-particle 
Green's function for the hydrogen impurity site with spin $\s$ is given as
\begin{equation}
\Sigma_{\s}(\w) = U \langle n_{\mb{r}_i\bar\s H}\rangle + U^2
\frac{\langle n_{\mb{r}_i\bar\s H}\rangle \left(1-\langle n_{\mb{r}_i\bar\s H}\rangle \right)}
     {\w - \eps_H - U                     \left(1-\langle n_{\mb{r}_i\bar\s H}\rangle \right)}.
\end{equation}
Assuming the PM state at half-filling, i.e., $\langle n_{\mb{r}_i\up H}\rangle = \langle n_{\mb{r}_i\dn H}\rangle = 1/2$ and $\eps_{H}=-U/2$, the self-energy is 
\begin{equation}
\Sigma_\sigma(\w) = \frac{U}{2} + \frac{U^2}{4\w}.
\end{equation}

\subsubsection{Dispersion relation}\label{sec:HI_disp}

\begin{figure*}
  \begin{center}
    \includegraphics[clip,width=5.8cm]{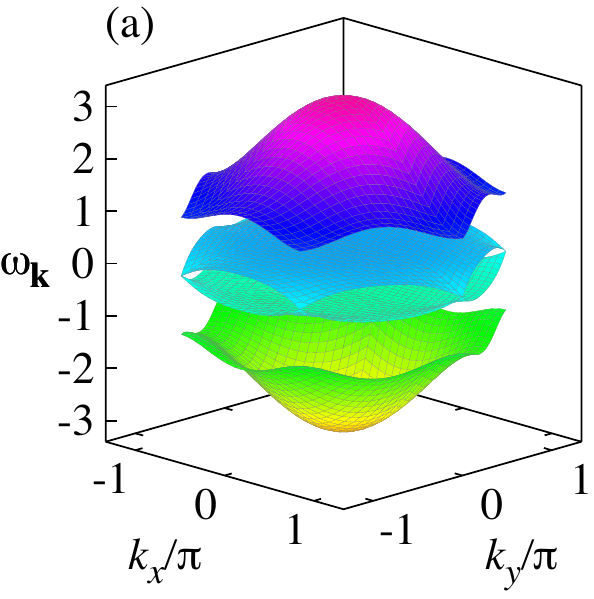}
    \includegraphics[clip,width=5.8cm]{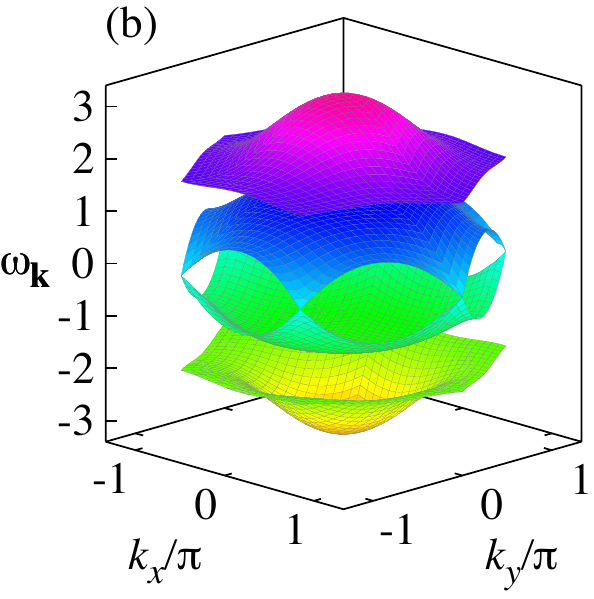}
    \includegraphics[clip,width=5.8cm]{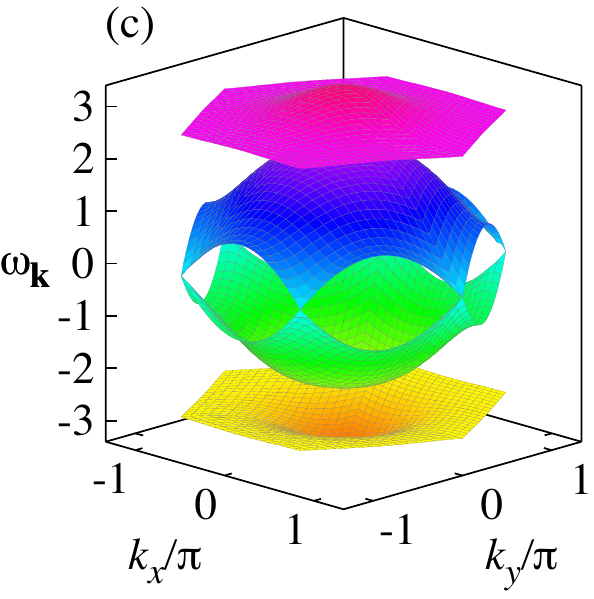}
    \caption{(color online)
      Single-particle excitation dispersion $\w_{\mb k}$ (in unit of $t$) 
      for $U/t=1$ (a), $3$ (b), and $5$ (c) with $t_{sp}/t=1$ obtained by the Hubbard-I approximation. 
      Note that the inner (outer) two bands correspond to $\w_{\mb{k}} = \pm \w_{-,\mb{k}}$ ($\pm \w_{+,\mb{k}}$). 
      The Fermi energy $E_{\rm F}$ is located at $\w_{\mb k}=0$. 
      The inner two bands display the massless Dirac dispersions at $K$ and $K'$ points 
      with the Dirac points exactly at $E_{\rm F}$, and the outer two bands correspond to the upper and lower 
      Hubbard bands.   
      \label{fig:HubI_disp}
    }
  \end{center}
\end{figure*}

Once the self energy $\Sigma_\sigma(\w)$ is obtained, the inverse of the interacting single-particle Green's function 
$\bs{G}_\sigma(\mb{k},\w)$ for spin $\sigma$ and momentum $\mb k$ is simply given as 
\begin{eqnarray}
\bs{G}_\sigma^{-1}(\mb{k},\w) &=& \bs{G}^{-1}_{0\sigma}(\mb{k},\w) - \bs{\Sigma}_\sigma(\w) \nonumber \\
&=&
\left(
\begin{array}{ccc}
  \w                & -\gamma_{\mb{k}}  & 0      \\
 -\gamma^*_{\mb{k}}  &      \w          & -t_{sp}  \\
   0                & -t_{sp}               & \w - \eps_H - \Sigma_\sigma(\w) 
\end{array}
\right),
\label{eq.Ginv}
\end{eqnarray}
where $\bs{G}_{0\sigma}(\mb{k},\w)$ is the noninteracting single-particle Green's function. 
In this matrix representation, the bases for the first, second, and third column and row correspond to 
$A$, $B$, and $H$ orbitals, respectively. 
The particle-hole symmetry is guaranteed by setting 
the on-site energy of the hydrogen impurity site to be $\eps_H = -\frac{U}{2}$. 
The dispersion relation of the single-particle excitations is obtained as the poles of the single-particle 
Green's function. 
Thus, by solving the following equation  
\begin{eqnarray}
\det \bs{G}_{\s}^{-1} (\w,\mb{k}) &=& 
\frac{1}{\w} \left\{ 
\w^4 - \left(\frac{U^2}{4} + |\gamma_{\mb{k}}|^2 + t_{sp}^2  \right) \w^2 \right. \nonumber \\
&&\quad\quad\quad + \left. \frac{U^2}{4} |\gamma_{\mb{k}}|^2 
\right\} = 0,
\label{eq.det0}
\end{eqnarray}
we find that there exist four poles at 
$\w_{1,{\mb k}} = \w_{+,{\mb k}}$, $\w_{2,{\mb k}} = \w_{-,{\mb k}}$, $\w_{3,{\mb k}} = -\w_{-,{\mb k}}$, 
and $\w_{4,{\mb k}} = - \w_{+,{\mb k}}$, where
\begin{widetext}
\begin{equation}
\w_{\pm,\mb{k}}^2 = \frac{1}{2} \left\{ \left(\frac{U^2}{4} + |\gamma_{\mb{k}}|^2 + t_{sp}^2  \right)  \pm 
\sqrt{ 
\left(\frac{U^2}{4} + |\gamma_{\mb{k}}|^2 + t_{sp}^2 + U|\gamma_{\mb{k}}| \right)   
\left(\frac{U^2}{4} + |\gamma_{\mb{k}}|^2 + t_{sp}^2 - U|\gamma_{\mb{k}}| \right) }
\right\}
\label{eq.poles}
\end{equation}
\end{widetext}
and $\w_{\pm,\mb{k}} \ge 0$, i.e., $\w_{1,\mb k}\ge\w_{2,\mb k}\ge0\ge\w_{3,\mb k}\ge\w_{4,\mb k}$. 
The dispersion relation for various $U$ is shown in Fig.~\ref{fig:HubI_disp}. 
It is interesting to notice in Fig.~\ref{fig:HubI_disp} that the inner two bands with $\nu=2$ and $3$ 
exhibit the massless Dirac dispersions 
at $K$ and $K'$ points with the Dirac 
points exactly at the Fermi energy as soon as a finite $U$ is turned on, thus  
in good qualitative agreement with the results obtained by the CPT in Sec.~\ref{sec:CPT_pm}. 
It should be also noted that the outer two bands with $\nu=1$ and $4$ shown in Fig.~\ref{fig:HubI_disp} 
correspond to the upper and lower Hubbard bands, respectively. 

\subsubsection{Dirac Fermi velocity}
Let us define 
 $\bs{\kappa}=\mb k - {\mb k}_{K}$ 
($\bs{\kappa}=\mb k - {\mb k}_{K'}$) 
and $\kappa = |\bs{\kappa}|$,  
where $\mb k_{K}$ ($\mb k_{K'}$) is the momentum 
at $K$ ($K'$) point.  
By expanding $|\gamma_{\mb{k}}| \simeq v_0 \kappa \ll 1$ 
around $K$ ($K'$) point, 
we obtain the massless Dirac quasiparticle dispersion 
\begin{equation}
  \pm \w_{-,\mb{k}} \simeq \pm v_{\rm F} \kappa,
\end{equation}
where $v_{\rm F}$ is the Dirac Fermi velocity 
\begin{equation}
  v_{\rm F} = \frac{U}{\sqrt{U^2 + 4t_{sp}^2}} v_0.
  \label{eq.vf}
\end{equation}
Here, $v_0 = \sqrt{3}t/2$ is the Dirac Fermi velocity of the pure graphene model.

In the small $U$ limit (i.e., $U \ll |t_{sp}|$), the Dirac Fermi velocity increases linearly with $U$,
\begin{equation}
  v_{\rm F} \simeq \frac{U}{2|t_{sp}|} v_0,
  \label{vf_s}
\end{equation}
while in the large $U$ limit (i.e., $U \gg |t_{sp}|$), the Dirac Fermi velocity is approximated as 
\begin{equation}
  v_{\rm F} \simeq \frac{U}{U + J_{\mr{K}}/4} v_0,
  \label{vf_l}
\end{equation}
where $J_{\mr{K}} = 8t_{sp}^2/U$ is the Kondo coupling between a localized spin on the hydrogen 
impurity site and 
a conduction electron on the carbon site [see Eq.~(\ref{H_KL})]. 
As shown in Fig.~\ref{fig:vf_vs_u}, we find that $v_{\rm F}$ monotonically increases 
from zero with increasing 
$U$ and reaches to $v_0$ in the limit of $U/t\to \infty$. Therefore, 
$v_{\rm F}$ calculated by the Hubbard-I approximation 
is qualitatively compared with the result obtained by the CPT.  

\subsubsection{Spectral representation}

Simply inverting the $3 \times 3$ matrix in Eq.~(\ref{eq.Ginv}), we can obtain 
the single-particle Green's function 
\begin{widetext}
\begin{equation}
  \bs{G}_\sigma(\mb{k},\w) = \frac{1}{\det \bs{G}_\sigma^{-1}(\mb{k},\w)} \\
  \left(
    \begin{array}{ccc}
      \w^2 - \frac{1}{4} \left( U^2 + 4t_{sp}^2 \right) &
      \gamma_{\mb{k}} \left(\w - \frac{U^2}{4\w} \right) &
      t_{sp} \gamma_{\mb{k}} \\
      \gamma_{\mb{k}}^* \left(\w - \frac{U^2}{4\w} \right) &
      \w^2 - \frac{U^2}{4} &
      t_{sp} \w \\
      t_{sp} \gamma_{\mb{k}}^{*} &
      t_{sp} \w  &
      \w^2 - |\gamma_{\mb{k}}|^2 
    \end{array}
  \right),
  \label{eq.Gkw}
\end{equation}
\end{widetext}
where the determinant is readily evaluated using Eqs.~(\ref{eq.det0}) and (\ref{eq.poles}) as 
\begin{equation}
\det \bs{G}_\sigma^{-1}(\mb{k},\w) = \frac{1}{\w} \prod_{\nu=1}^{4} (\w - \w_{\nu,\mb{k}}).
\end{equation} 

The spectral representation of the single-particle Green's function is thus 
\begin{equation}
  \bs{G}_\sigma(\mb{k},\w) = \sum_{\nu = 1}^{4} \frac{\bs{\rho}_{\nu,\mb{k}}}{\w - \w_{\nu,\mb{k}}},
  \label{eq.spectral}
\end{equation}
where $\bs{\rho}_{\nu,\mb{k}}$ is a $3 \times 3$ matrix and its element 
$\left (\bs{\rho}_{\nu,\mb{k}}\right)_{\alpha\beta}= \rho^{\alpha \beta}_{\nu,\mb{k}} $ is defined as 
\begin{equation}
  \rho^{\alpha \beta}_{\nu,\mb{k}} = \lim_{\w \rightarrow \w_{\nu,\mb{k}}} (\w - \w_{\nu,\mb{k}}) G_\sigma^{\alpha \beta} (\mb{k},\w)
\end{equation}
with $\alpha$ and $\beta=A,B$, and $H$. 
Directly calculating $\rho^{\alpha \beta}_{\nu,\mb{k}}$, we obtain the explicit form of $\bs{\rho}_{\nu,\mb{k}}$, i.e., 
\begin{widetext}
\begin{equation}
  \bs{\rho}_{1/4,\mb{k}} = \frac{1}{2(\w_{+,\mb{k}}^2 - \w_{-,\mb{k}}^2)}
  \left(
    \begin{array}{ccc}
      \w_{+,\mb{k}}^2 - \frac{1}{4} \left(U^2 + 4t_{sp}^2 \right)               &
      \pm \gamma_{\mb{k}}  \left( \w_{+,\mb{k}} - \frac{U^2}{4 \w_{+,\mb{k}}} \right)    & 
      t_{sp} \gamma_{\mb{k}}  \\
      \pm \gamma_{\mb{k}}^* \left( \w_{+,\mb{k}} - \frac{U^2}{4 \w_{+,\mb{k}}} \right)    &
      \w_{+,\mb{k}}^2 - \frac{U^2}{4}   &
      \pm   t_{sp}   \w_{+,\mb{k}}     \\
      t_{sp} \gamma_{\mb{k}}^*        &
      \pm   t_{sp}   \w_{+,\mb{k}}    &
      \w_{+,\mb{k}}^2 - |\gamma_{\mb{k}}|^2              
    \end{array}
  \right)
  \label{eq.rho14}
\end{equation}
\end{widetext}
for the outer two bands with $\nu=1$ and $4$, 
where plus and minus signs in the right hand side correspond to 
$\nu=1$ and $4$, respectively, and 
\begin{widetext}
\begin{equation}
\bs{\rho}_{2/3,\mb{k}} = \frac{-1}{2(\w_{+,\mb{k}}^2 - \w_{-,\mb{k}}^2)}
\left(
\begin{array}{ccc}
 \w_{-,\mb{k}}^2 - \frac{1}{4} \left(U^2 + 4t_{sp}^2 \right)               &
 \pm \gamma_{\mb{k}}  \left( \w_{-,\mb{k}} - \frac{U^2}{4 \w_{-,\mb{k}}} \right)    & 
 t_{sp} \gamma_{\mb{k}}  \\
 \pm \gamma_{\mb{k}}^* \left( \w_{-,\mb{k}} - \frac{U^2}{4 \w_{-,\mb{k}}} \right)    &
 \w_{-,\mb{k}}^2 - \frac{U^2}{4}   &
 \pm  t_{sp}   \w_{-,\mb{k}}     \\
 t_{sp} \gamma_{\mb{k}}^*    &
 \pm  t_{sp}   \w_{-,\mb{k}}    &
 \w_{-,\mb{k}}^2 - |\gamma_{\mb{k}}|^2              
\end{array}
\right)
\label{eq.rho23}
\end{equation}
\end{widetext}
for the inner two bands with $\nu=2$ and $3$, showing the emergent massless Dirac quasiparticles, 
where plus and minus signs in the right hand side correspond to 
$\nu=2$ and $3$, respectively.  
It is now easy to directly confirm that the spectral weights fulfill the sum rule  
\begin{equation}
  \sum_{\nu=1}^{4} \rho_{\nu,\mb{k}}^{\alpha \beta} = \delta_{\alpha \beta}. 
  \label{sumrule}
\end{equation}

The single-particle Green's function $\bs{G}_\sigma(\mb{k},\w)$ in the Hubbard-I approximation is 
thus evaluated using Eq.~(\ref{eq.spectral}) with the excitation energy dispersions $\w_{\nu,\mb{k}}$ 
in Eq.~(\ref{eq.poles}) and the spectral weights $\bs{\rho}_{\nu,\mb{k}}$ in 
Eqs.~(\ref{eq.rho14}) and (\ref{eq.rho23}). 
The excitation energy dispersions $\w_{\nu,\mb{k}}$ and the spectral weights $\bs{\rho}_{\nu,\mb{k}}$ for 
several limiting cases are studied in Appendix~\ref{appsec:HubI}. Among these limiting cases, 
it is rather interesting to note that the single-particle excitations in 
the strong coupling limit with $U\to\infty$ are exactly the same 
as those in the decoupling limit with $t_{sp}\to0$, i.e., both showing the 
massless Dirac energy dispersion with the Dirac Fermi velocity $v_0$. 

\subsubsection{Density of states}

\begin{figure*}
  \begin{center}
    \includegraphics[width=5.6cm]{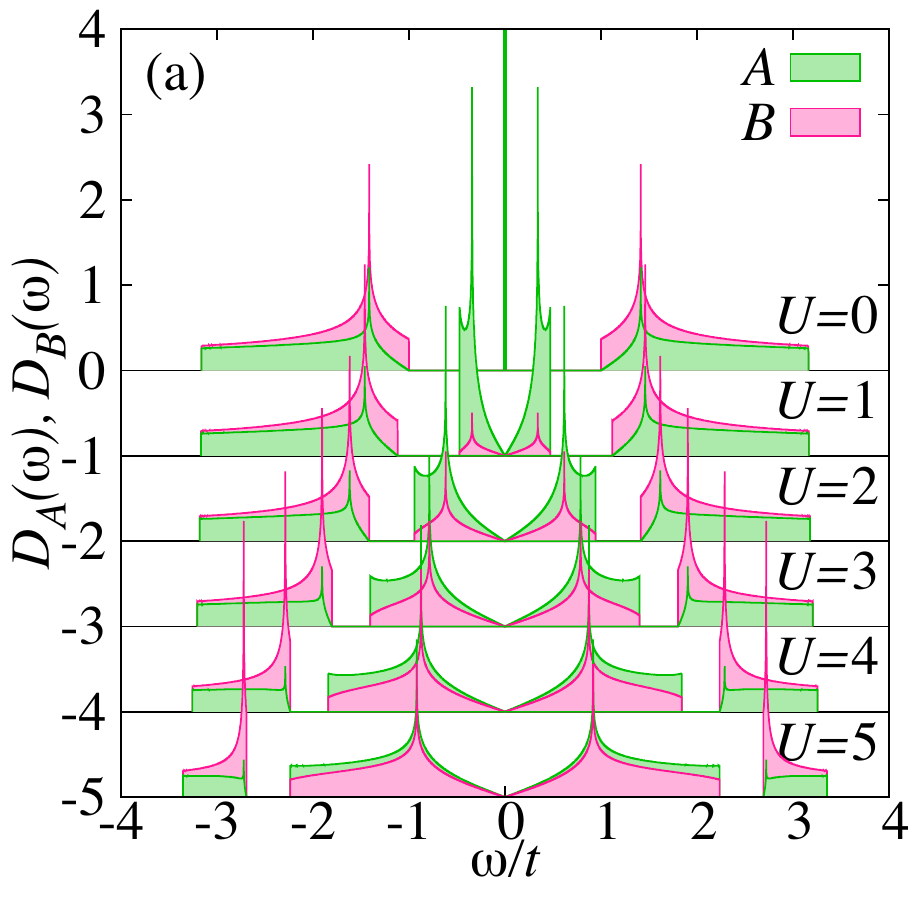}
    \includegraphics[width=5.6cm]{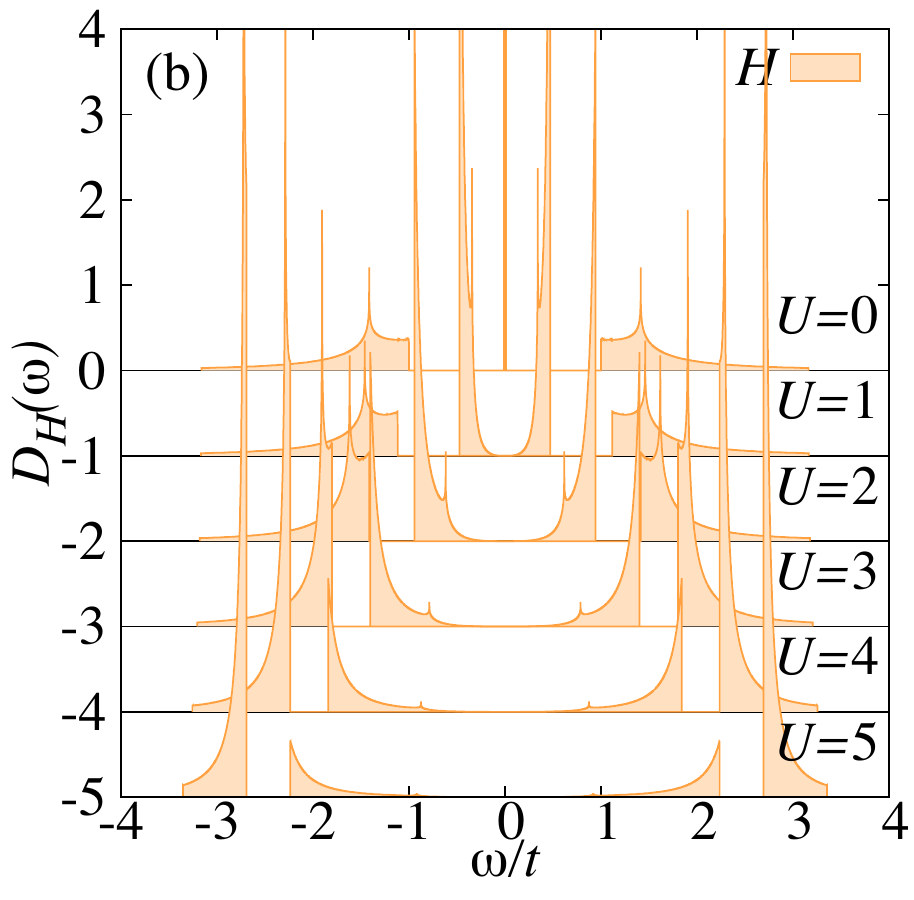}
    \includegraphics[width=5.6cm]{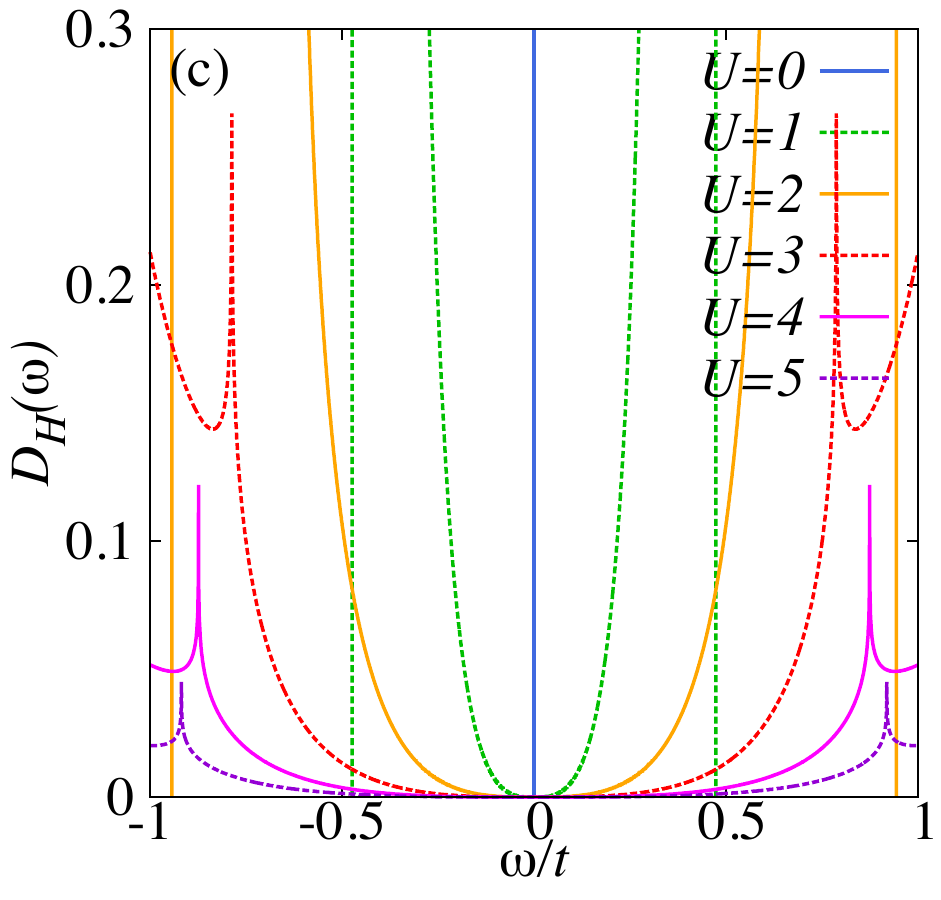}
    \caption{(color online) 
      Orbital resolved density of states (a) $D_{A/B}(\w) $ and (b) $D_H(\w)$ for various values of $U$ 
      (indicated in the figures) with $t_{sp}/t=1$ obtained by the Hubbard-I 
      approximation. 
      Notice that the results are shifted by $U$ for clarity. 
      (c) The enlarged plot of $D_H(\w)$ in (b) near the Fermi energy. 
       The diverging density of states due to the flat band for $U=0$ is 
       represented by the vertical line at $\omega=0$. 
      The Fermi energy $E_{\rm F}$ is located at $\omega=0$. 
      \label{fig:HubIdos}      
    }
  \end{center}
\end{figure*}

The density of states (DOS) $D_\alpha(\omega)$ projected onto $\alpha$ orbital is evaluated as 
\begin{eqnarray}
  D_\alpha (\w) 
  &=& -\frac{1}{\pi N_a} \sum_{\mb{k}} \sum_\sigma \lim_{\eta \rightarrow 0^+} \Im G_\sigma^{\alpha \alpha}(\mb{k},\w+ \imag \eta) \nonumber\\
  &=& \frac{2}{N_a} \sum_{\nu=1}^4 \sum_{\mb{k}} \rho^{\alpha \alpha}_{\nu,\mb{k}} \delta (\w - \w_{\nu, \mb{k}} ), 
  \label{dos_a}
\end{eqnarray}
where 
$N_a$ is the number of unit cells and  
no magnetic order is assumed in the last equation. 
Figure~\ref{fig:HubIdos} shows the evolution of $D_\alpha(\omega)$  
obtained within the Hubbard-I approximation. 
It is clearly observed in Fig.~\ref{fig:HubIdos} that the significant redistribution of the spectral weight occurs 
with increasing $U$. 

The characteristic features of the spectral weight redistribution are summarized as follows. 
The flat band which appears in the noninteracting limit 
(see Sec.~\ref{Sec:nonint}) causes a delta function peak at $\w=0$ in $D_A(\omega)$ and $D_H(\omega)$, 
but not in $D_B(\omega)$. However, 
once the Coulomb interaction $U$ is introduced, the flat band in $D_A(\omega)$ splits into two bands around the 
Fermi energy to form a ``Dirac band" with the massless Dirac quasiparticle dispersion 
(see Fig.~\ref{fig:HubI_disp} and also Fig.~\ref{fig.akw.h1}). 
It is also noticed in Fig.~\ref{fig:HubIdos}(a) that the high energy spectral weight in $D_A(\omega)$ 
for $|\omega| \agt \Delta_c$,  where 
\begin{equation}
\Delta_c=\sqrt{\left(\frac{U}{2}\right)^2+t_{sp}^2} = \frac{1}{2}\frac{v_0}{v_{\rm F}}U 
\label{eq:delta_c}
\end{equation}
is the lower (upper) bound of the upper (lower) Hubbard band at $K$ and $K'$ points, 
is transferred to the low energy region to participate in the formation of the massless Dirac quasiparticles.  
Simultaneously, the spectral weight $D_B(\omega)$ for $B$ orbital in the low energy 
region $|\omega|< \Delta_c$ 
is gradually transferred from the upper and lower Hubbard bands located in the high energy region  
$|\omega| \agt \Delta_c$, and the contribution of 
$B$ orbital to the ``Dirac band" becomes as significant as that of $A$ orbital for larger values of $U$. 
Indeed, $D_A (\w) \approx D_B (\w)$ in the ``Dirac band" for $|\w|<\Delta_c$ when $U/t\agt 5$. 
On the other hand, as shown in Figs.~\ref{fig:HubIdos}(b) and \ref{fig:HubIdos}(c), the spectral weight 
$D_H(\omega)$ for $H$ orbital loses its intensity in the low energy region near the Fermi energy 
and the large spectral weights are piled up in rather narrow high energy regions, exhibiting 
a typical localized incoherent feature.  
This spectral weight redistribution enhances the coherent hybridization between $A$ and $B$ orbitals 
in the low energy region. Therefore, the participation of $B$ orbital 
together with the disengagement of $H$ orbital in the low energy excitations  
is essential to form the massless Dirac quasiparticles near the Fermi energy.

\begin{figure*}
  \begin{center}
    \includegraphics[clip,width=6.cm]{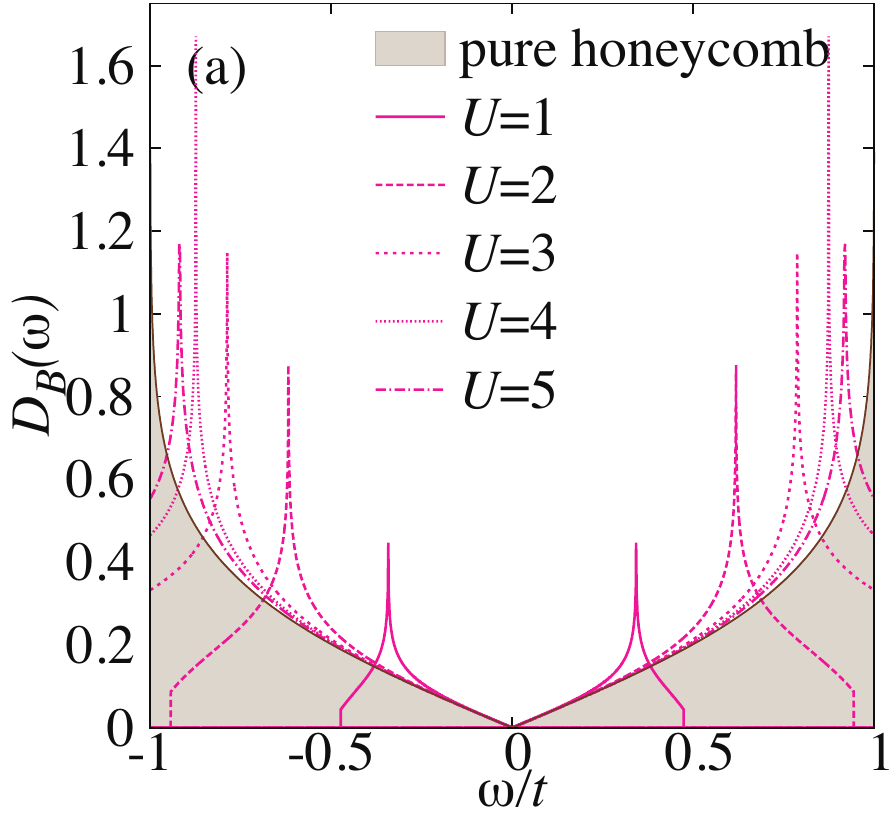}
    \includegraphics[clip,width=6.cm]{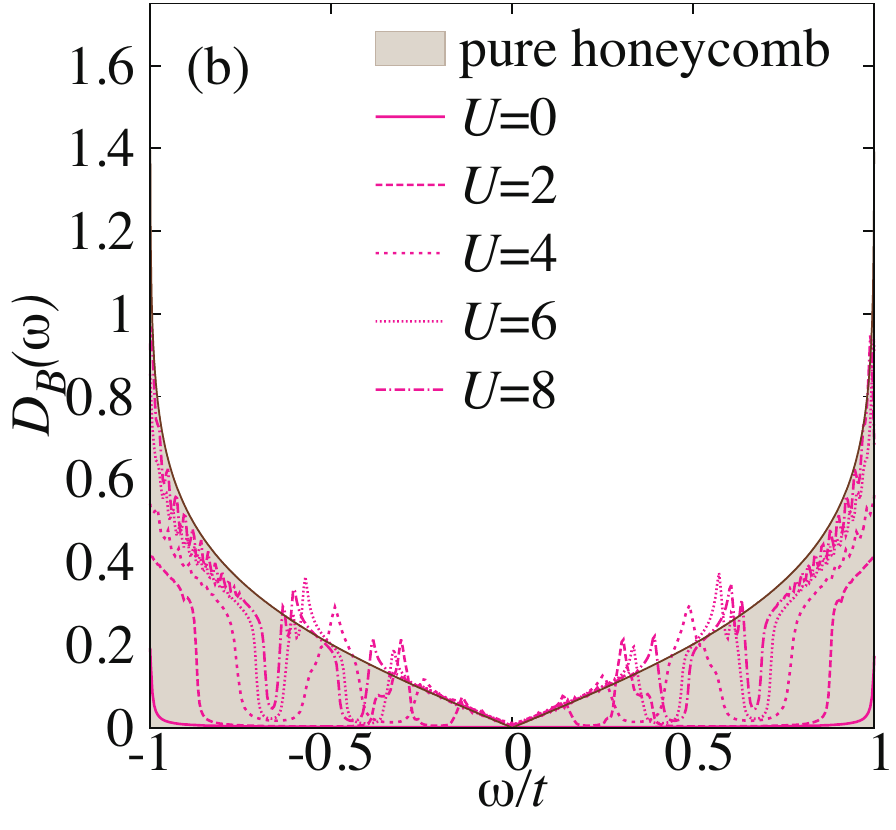}
    \includegraphics[width=5.7cm]{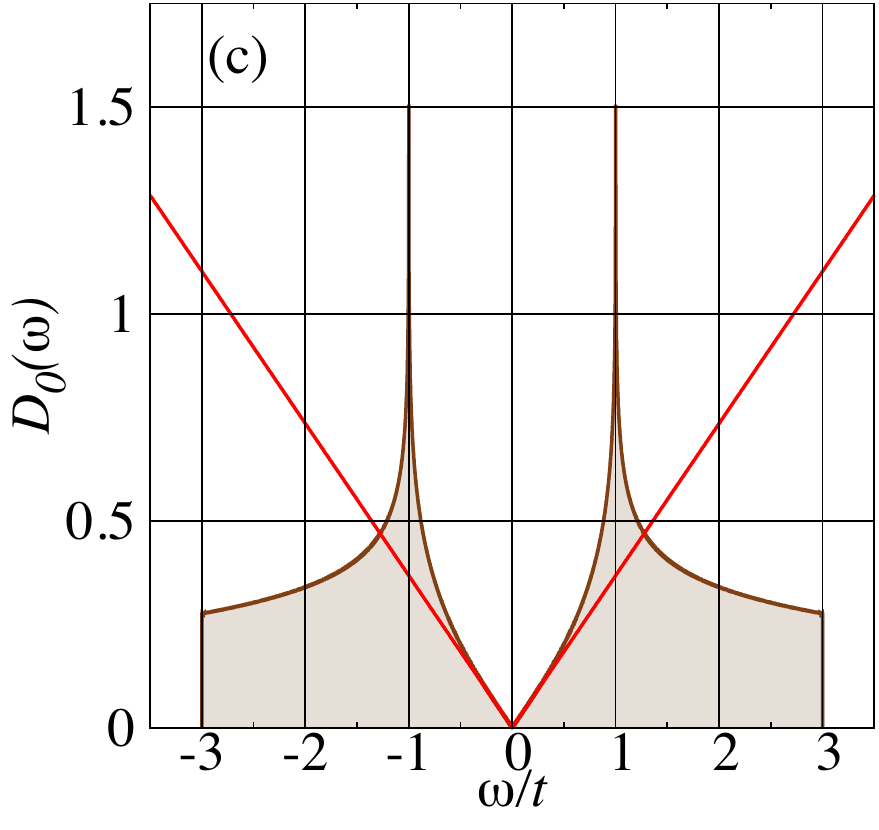}
    \caption{(color online)
      Density of states $D_B(\w)$ for $B$ orbital near the Fermi energy for various values of $U$ 
      (indicated in the figures) with $t_{sp}/t=1$ obtained by (a) the Hubbard-I approximation and (b) the CPT. 
      The CPT calculations are for the PM state at $T/t=0.025$ using the 6-site cluster (containing 2 unit cells) 
      with a Lorentzian broadening of $\eta/t = 0.005$. 
      For comparison, density of states $D_0(\w)$ per orbital for the pure graphene model 
      is plotted in (c) and also indicated by a shaded region in (a) and (b). 
      Red straight lines ($\frac{2\sqrt{3}}{3\pi} \frac{|\w|}{t^2}$)
      in (c) represent the initial slope of $D_0(\w)$ around the Fermi energy $E_{\rm F}$ at $\omega=0$. 
      \label{fig:dos_aa}
    }
  \end{center}
\end{figure*}

It is indeed noticed in Fig.~\ref{fig:HubIdos}(a) that $D_A(\w)$ and $D_B(\w)$ both exhibit the linearly 
vanishing density of states near the Fermi energy, a characteristic feature of the massless Dirac dispersion.  
It is well known that the Van Hove singularity appears in the DOS at 
$\w=\pm t$, i.e., $\w=\pm2v_0/\sqrt{3}$, for the pure graphene 
model, as shown in Fig.~\ref{fig:dos_aa}(c). 
Similarly, we find in Fig.~\ref{fig:HubIdos} that the Van Hove singularity appears exactly 
at $\w/t = \pm v_{\rm F}/v_0$, i.e., $\w=\pm2v_{\rm F}/\sqrt{3}$, even for the periodic Anderson model,   
indicating that the low energy band for $|\w|<\Delta_c$ 
can in fact be regarded as an effective pure graphene band with the renormalized Dirac Fermi velocity 
$v_{\rm F}$.

As shown in Fig.~\ref{fig:dos_aa}(a), we also find that the slope of $D_B(\w)$ around the Fermi energy is 
independent of $U/t$ and is identical to that for the pure graphene model. 
The density of states $D_0(\w)$ per orbital for the pure graphene model 
is shown in Fig.~\ref{fig:dos_aa}(c) and can be evaluated as 
\begin{eqnarray}
  D_0(\w) 
  &=& -\frac{1}{\pi N_a} \sum_{\mb k}  \sum_{\s} \lim_{\eta \rightarrow 0^+} \Im G_{0}^{\alpha \alpha} (\mb{k},\w + \imag \eta) \nonumber \\
  &=& 
  \frac{1}{N_a} \sum_{\mb{k}} 
  \left\{
    \delta(\w - |\gamma_{\mb{k}}|) + \delta(\w + |\gamma_{\mb{k}}|) 
  \right\},
  \label{eq:d0}
\end{eqnarray}  
where
\begin{equation}
  G_0^{\alpha \alpha}(\mb{k},\w) = \frac{1}{2} \left(\frac{1}{\w - |\gamma_{\mb{k}}|} + \frac{1}{\w + |\gamma_{\mb{k}}|} \right)
\end{equation} 
is the diagonal element of the noninteracting single-particle Green's function with orbital $\alpha\,(= A$ and $B$) 
and spin $\s\,(=\up$ and $\dn)$ for the pure graphene model.  
Indeed, one can find within the Hubbard-I approximation 
that the DOS for $B$ orbital near the Fermi energy is  
\begin{equation} 
  D_B(\w) 
    =\frac{\sqrt{3}}{2\pi}\frac{|\w|}{v_0^2}
    =\frac{2\sqrt{3}}{3\pi}\frac{|\w|}{t^2}
\end{equation}
for $|\w|\sim0$, exactly the same slope of the linearly increasing DOS for the pure graphene 
model~\cite{Neto2009RMP} and independent of the value of $U$. 
As shown in Fig.~\ref{fig:dos_aa}(b), the same results are also found in the CPT calculations for the PM state.

We can now show that the spectral weight $\rho^{BB}_{\nu,\mb{k}}$ for $B$ orbital 
at the Dirac points, i.e., at $K$ (and also $K'$) point  
on the Fermi energy, is related to the Dirac Fermi velocity $v_{\rm F}$ via  
\begin{equation} 
  \rho^{BB}_{2/3, K(K')} = 
  \frac{1}{2} \left( \frac{v_{\rm F}}{v_0} \right)^2.
  \label{eq.vsq}
\end{equation}
From Eq.~(\ref{dos_a}), the slope of the DOS near the Fermi energy 
is evaluated as  
\begin{eqnarray} 
  \lim_{\w \to 0} \frac{D_B(\w)}{\w} 
  &=& 
   \lim_{\w \to 0}\frac{4}{\w V_{\rm BZ}} \int_{0}^{\Lambda}  \dd \kappa \, \kappa  \int_0^{2\pi} \dd \theta \nonumber \\
  &\times& \left[ 
    \rho_{2, K }^{BB}\, \delta \left(\w - v_{\rm F} \kappa \right)  + 
    \rho_{3, K }^{BB}\, \delta \left(\w + v_{\rm F} \kappa \right) \right]  \nonumber \\
  &=& 
  \left\{
  \begin{array}{ll}
    \displaystyle
    \frac{\sqrt{3}}{\pi} \frac{\rho^{BB}_{2, K} }{v_{\rm F}^2} & (\w \to 0^+), \\  
    \displaystyle
    -\frac{\sqrt{3}}{\pi} \frac{\rho^{BB}_{3, K} }{v_{\rm F}^2} & (\w \to 0^-),
  \end{array}
  \right.
  \label{dos_w}
\end{eqnarray} 
where
the factor $4$ on the right hand side of the first line 
accounts for the contributions to the DOS of the four emergent Dirac cones with 
the low-energy linear energy dispersions in the neighborhoods of $K$ and $K'$ points, including the 
spin degeneracy. $V_{\rm BZ}= 8 \pi^2 /\sqrt{3}$ is the volume of the Brillouin zone, 
$\Lambda$ is a positive cut-off momentum within which the low energy dispersion is approximated 
linear in momentum around $K$ and $K'$ points, 
and $0^{+(-)}$ is positive (negative) infinitesimal. 
We have also used that 
\begin{equation}
  \frac{1}{N_a}\sum_{\mb k} \cdots = 
  \frac{1}{V_{\rm BZ}}\int \dd^2 k \cdots  . 
\end{equation}
Since we can show form Eq.~(\ref{eq:d0}) that $\rho^{BB}_{2, K} = \rho^{BB}_{3, K} = 1/2$ 
and $v_{\rm F} = v_{0}$ in 
Eq.~(\ref{dos_w}) for the pure graphene model, the fact that the slope of $D_B(\w)$ 
near the Fermi energy is the same as that of $D_0(\w)$ 
naturally leads to Eq.~(\ref{eq.vsq}).  
This is indeed derived analytically below in Eq.~(\ref{eq:rho_BB}).

\subsubsection{Spectral weight at the Dirac points}

Although the density of states $D_\alpha(\w)$ vanishes at $\w=0$, the spectral weight $\bs{\rho}_{\nu,\mb{k}}$ 
itself is finite even at $\w=0$ for ${\mb k}=$~$K$ and $K'$. 
Here, we derive the analytical expression of the spectral weight at $K$ point where the massless Dirac 
dispersions emerge. 
Because $\gamma_{\mb{k}=K}=0$, the spectral weight at $K$ point is determined only by $U$ and $t_{sp}$. 
Indeed, the poles of the single-particle Green's function in Eq.~(\ref{eq.poles}) are located at 
$\w_{1/4,K} =\pm\Delta_c $ 
and $\w_{2/3,K} = 0$. 
Therefore, from Eqs.~(\ref{eq.rho14}) and (\ref{eq.rho23}), 
the spectral weights $\rho^{\alpha\alpha}_{\nu,K}$ at $K$ point are given as 
\begin{eqnarray}
\rho_{1/4,K}^{AA} &=& \rho_{2/3,K}^{HH} = 0, \label{eq:rho_2A}\\ 
\rho_{2/3,K}^{AA} &=& \rho_{1/4,K}^{HH} = \frac{1}{2}, \label{eq:rho_1A}\\ 
\rho_{1/4,K}^{BB} &=& \frac{1}{2}\left\{1-\left(\frac{v_{\rm F}}{v_0}\right)^2\right\}, \label{eq:rho_1B}\\ 
\label{rKBB}
\rho_{2/3,K}^{BB} &=& \frac{1}{2}\left(\frac{v_{\rm F}}{v_0}\right)^2. \label{eq:rho_BB} 
\end{eqnarray}
The spectral weight at the Dirac point, corresponding to $\nu=2$ and $3$, is indeed finite 
even though the density of states is zero at the Fermi energy.

Since the spectral weight $\rho_{\nu,K}^{\alpha\alpha}$ considered here is related to the 
spectral weight $\rho_K^{\alpha\alpha}$ of the single-particle excitation spectrum 
$A^{\alpha\alpha}_\s({\mb k}=K,\w=0)$, defined in Eq.~(\ref{eq:sw}), 
as 
\begin{equation}
\rho_K^{\alpha\alpha} = \rho_{2,K}^{\alpha\alpha} + \rho_{3,K}^{\alpha\alpha}, 
\end{equation}
we find that $\rho^{BB}_K$ is highly correlated to $v_{\rm F}$, i.e., $\rho_K^{BB} = (v_{\rm F}/v_0)^2$, 
exactly the same relation found by the CPT for the PM state in Fig.~\ref{fig:vf_vs_sw}, and   
monotonically increases with increasing $U$ as $v_{\rm F}$ also monotonically increases 
(see Fig.~\ref{fig:vf_vs_u}). 
On the other hand, we find that $\rho_K^{AA}=1$ and $\rho_K^{HH}=0$, irrespectively 
of the value of $U$. This is also in good agreement with that obtained by the CPT for the PM state. 
These results therefore suggest that the involvement of $B$ orbital in 
the low-energy excitations, 
which is absent in the noninteracting limit, 
is essential to form the emergent massless Dirac quasiparticles. 

Now, using the spectral weights $\rho^{\alpha\alpha}_{\nu,K}$ in Eqs.~(\ref{eq:rho_2A})--(\ref{eq:rho_BB}), 
we can readily obtain within the Hubbard-I approximation the single-particle Green's function at $K$ point as 
\begin{eqnarray}
  G_{\s}^{AA}(K,\w) &=& \frac{1}{\w}, \label{GAAK} \\
  G_{\s}^{BB}(K,\w) &=& \left(\frac{v_{\rm F}}{v_0}\right)^2 \frac{1}{\w} 
  + \frac{1}{2} \left\{ 1 - \left(\frac{v_{\rm F}}{v_0}\right)^2 \right\} \nonumber \\ 
  &\times&
  \left( 
    \frac{1}{\w - \Delta_c} + \frac{1}{\w + \Delta_c}
  \right), \label{GBBK} \\
  G_{\s}^{HH}(K,\w) &=& \frac{1}{2} 
  \left( 
    \frac{1}{\w - \Delta_c} + \frac{1}{\w + \Delta_c}
  \right). \label{GHHK}
\end{eqnarray}
From these analytical forms, we can find several characteristic features of the single-particle excitations. 
First, $G_{\s}^{AA}(K,\w)$ does not depend on $U$ and it remains in the same form as in the noninteracting case. 
Namely, it has a single pole at zero energy ($\w=0$) and its spectral weight is one.  
Second, $G_{\s}^{BB}(K,\w)$ has a pole at $\w=0$  
with its  spectral weight proportional to the square of the Dirac Fermi velocity $v_{\rm F}$, i.e., 
$\left(\frac{v_{\rm F}}{v_0}\right)^2 = \frac{U^2}{U^2 + 4t_{sp}^2}$.
The other two poles are located at $\w = \pm\Delta_c  $
and their spectral weights are both 
$\frac{1}{2} \left\{1 - \left(\frac{v_{\rm F}}{v_0}\right)^2 \right\}$.
Therefore, as $U$ increases, the spectral weight is transferred 
from the high energy poles at $\w = \pm\Delta_c $ in the upper and lower Hubbard bands 
to the zero energy one in the Dirac band. 
Third, $G_{\s}^{HH}(K,\w)$ has no poles at $\w=0$ for any finite value of $U$, but at 
$\w= \pm\Delta_c$. 

\subsubsection{Single-particle excitation spectrum}

\begin{figure*}
  \includegraphics[width=4.3cm,angle=0]{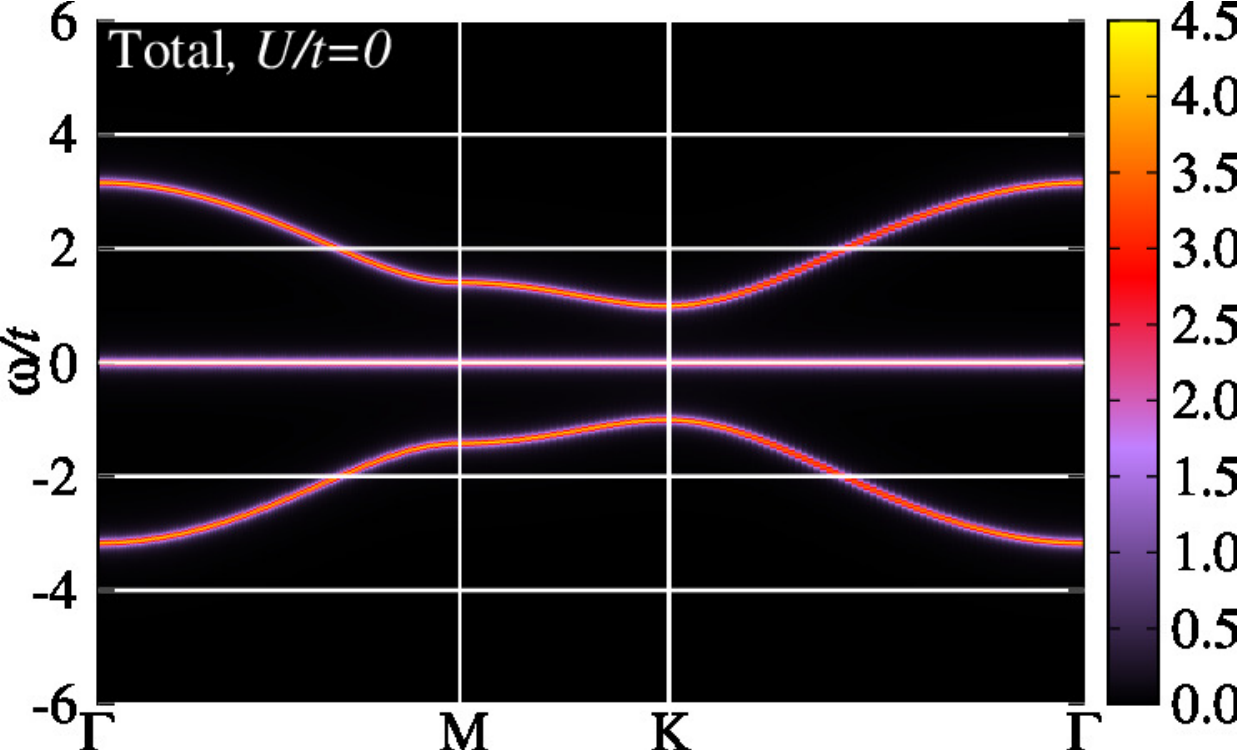}
  \includegraphics[width=4.3cm,angle=0]{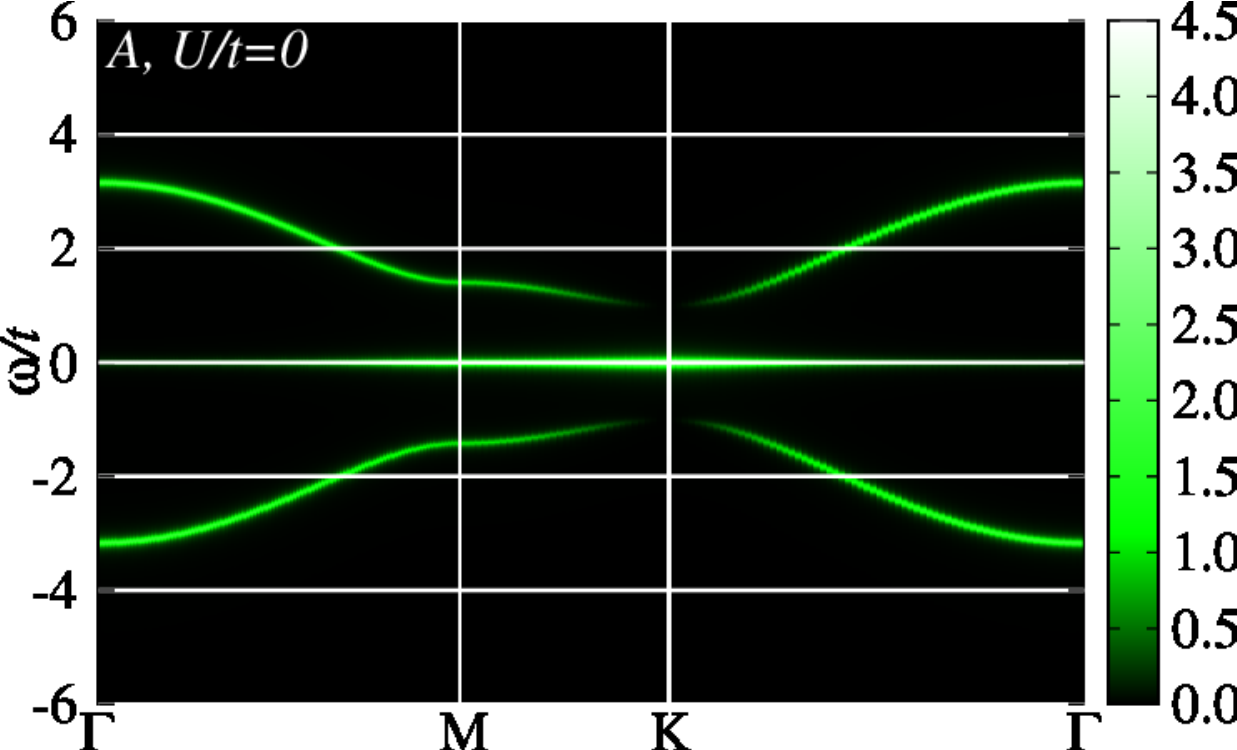}
  \includegraphics[width=4.3cm,angle=0]{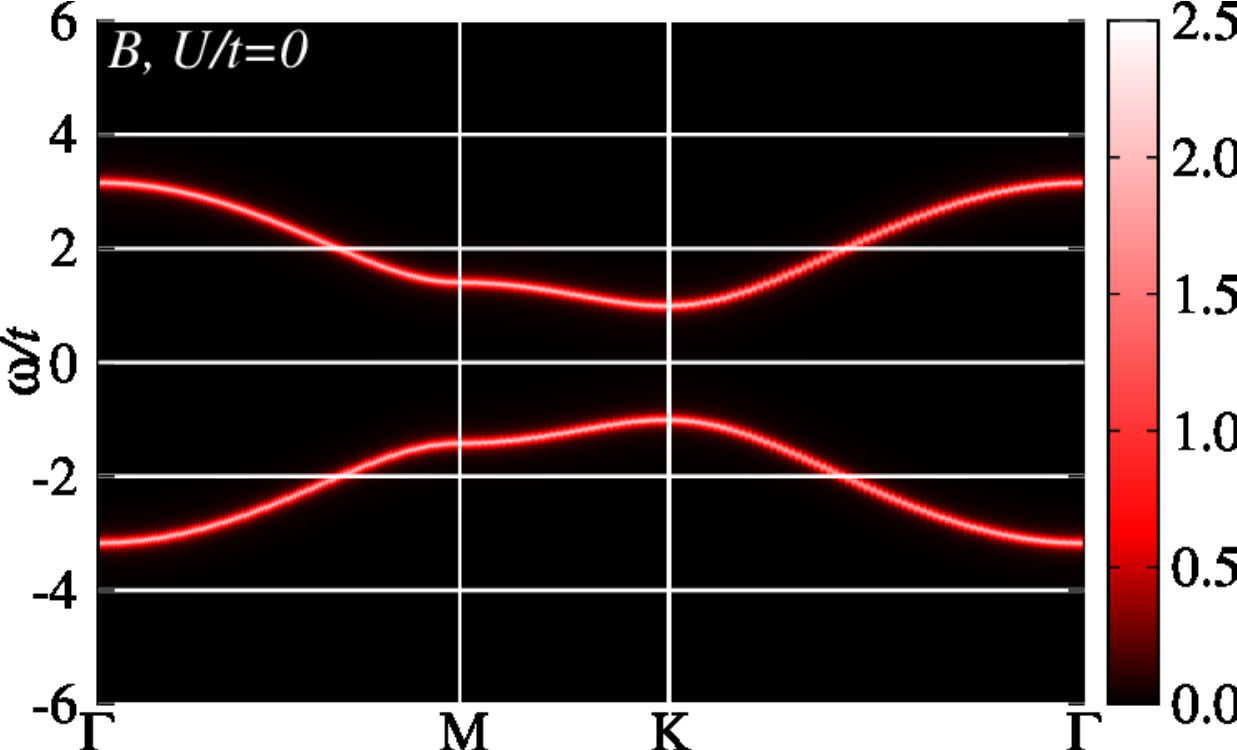}
  \includegraphics[width=4.3cm,angle=0]{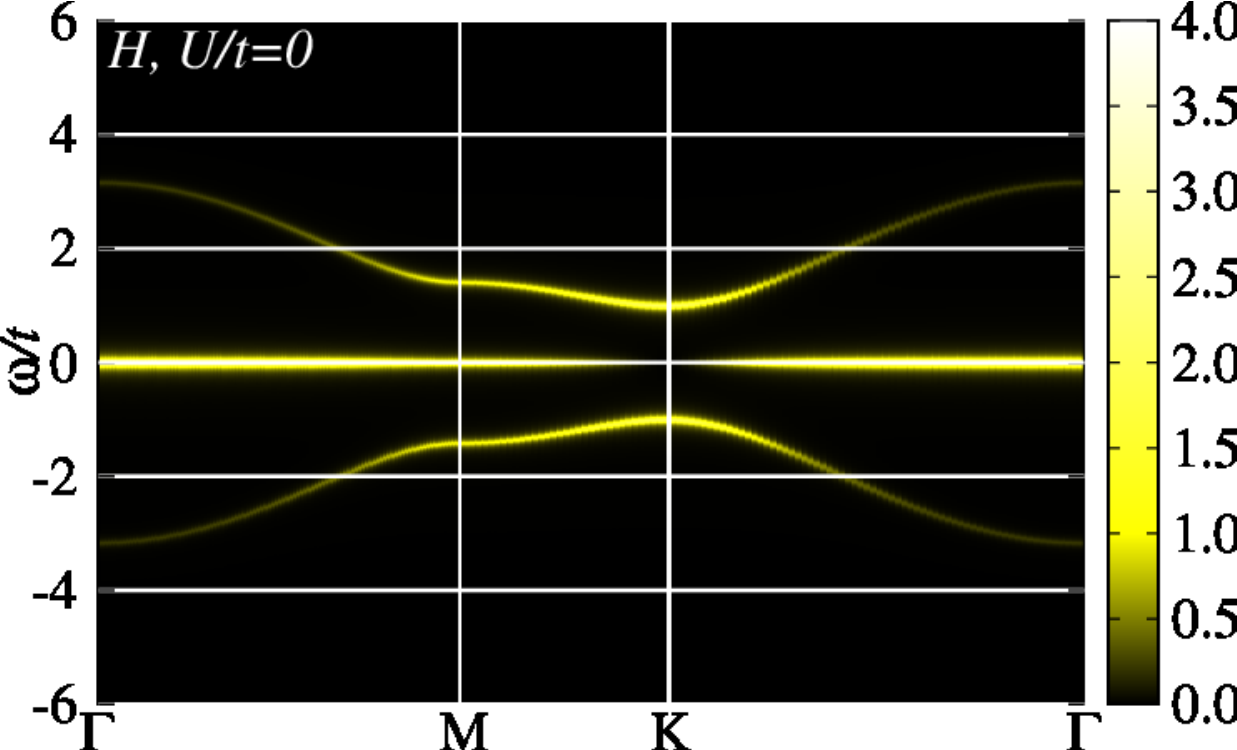}\\
  \includegraphics[width=4.3cm,angle=0]{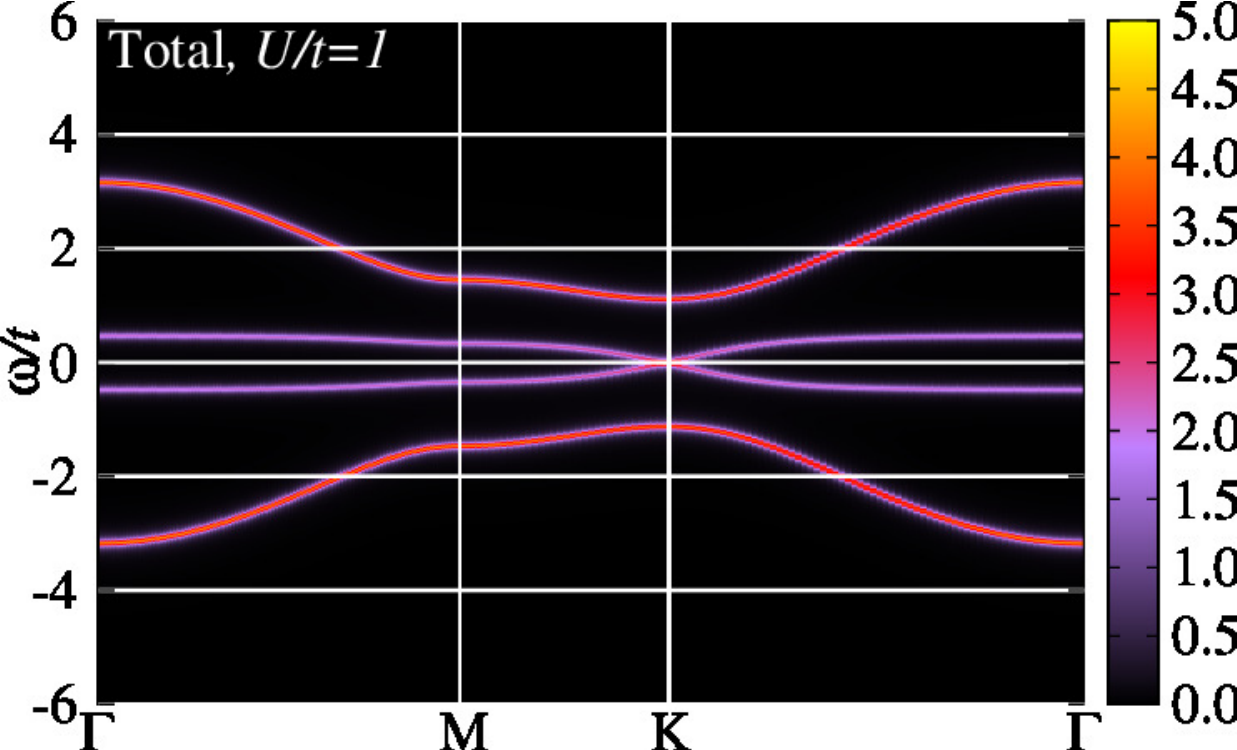}
  \includegraphics[width=4.3cm,angle=0]{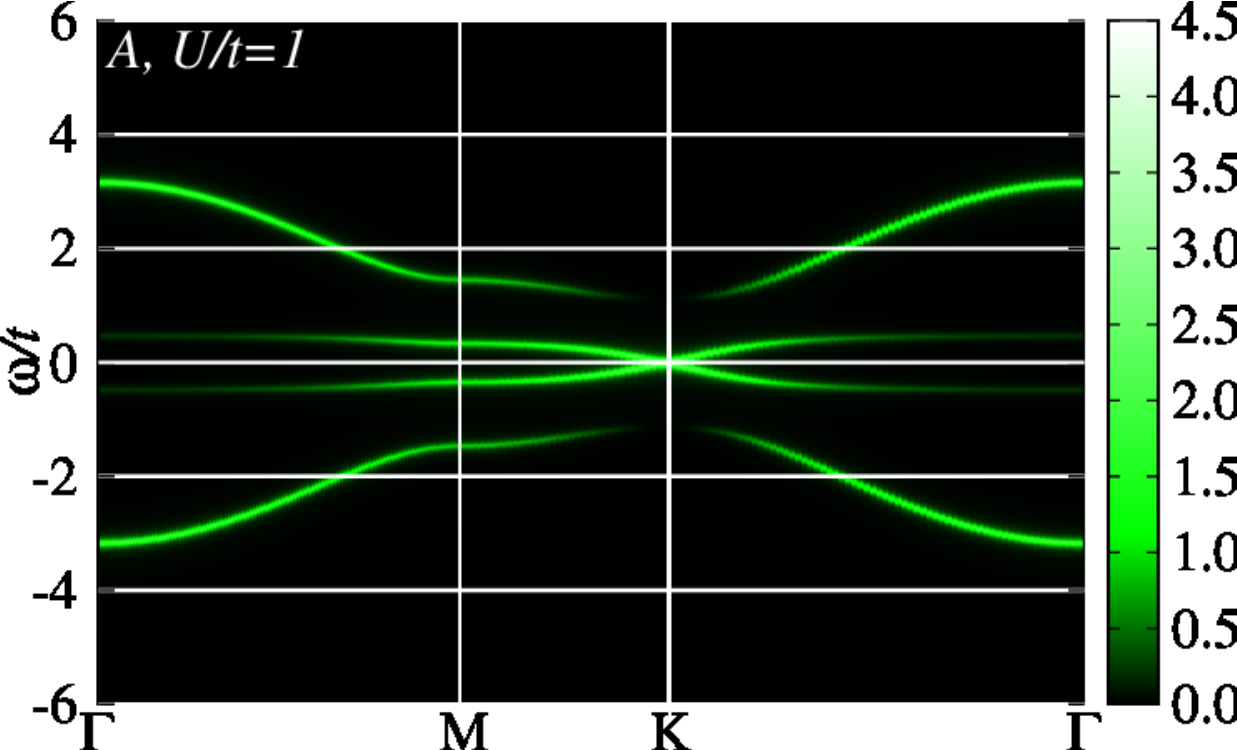}
  \includegraphics[width=4.3cm,angle=0]{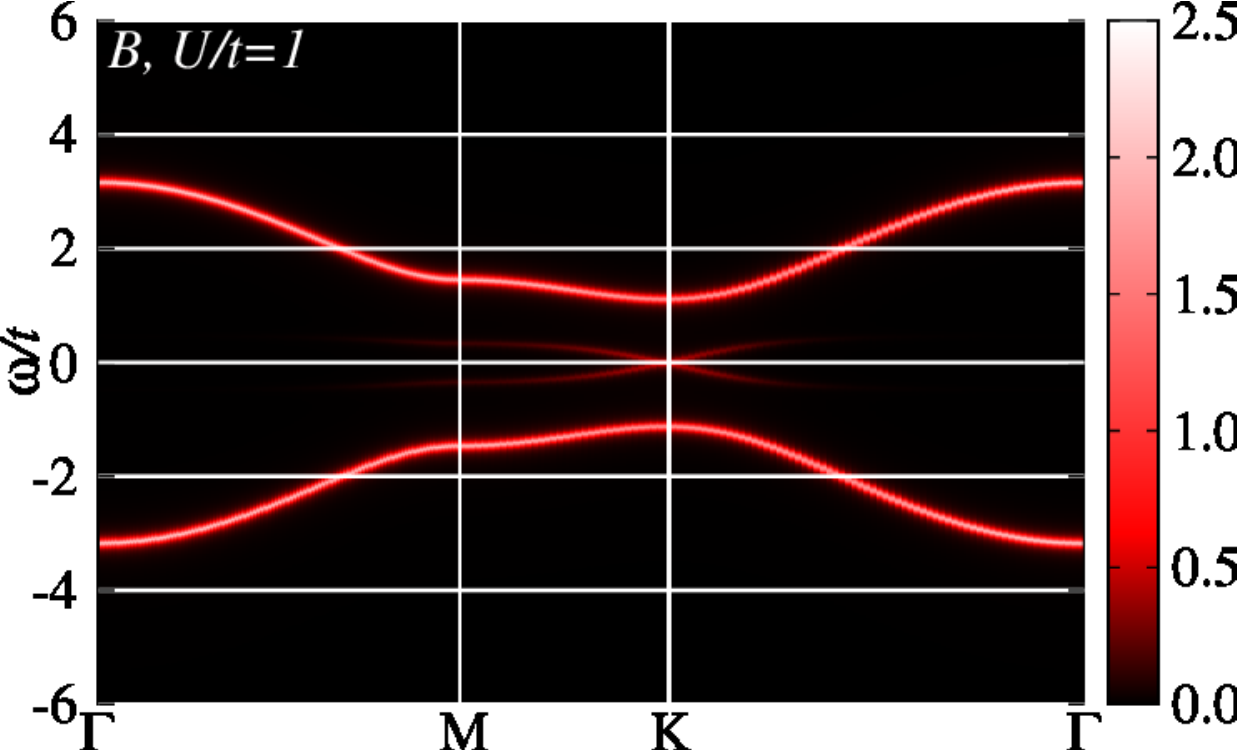}
  \includegraphics[width=4.3cm,angle=0]{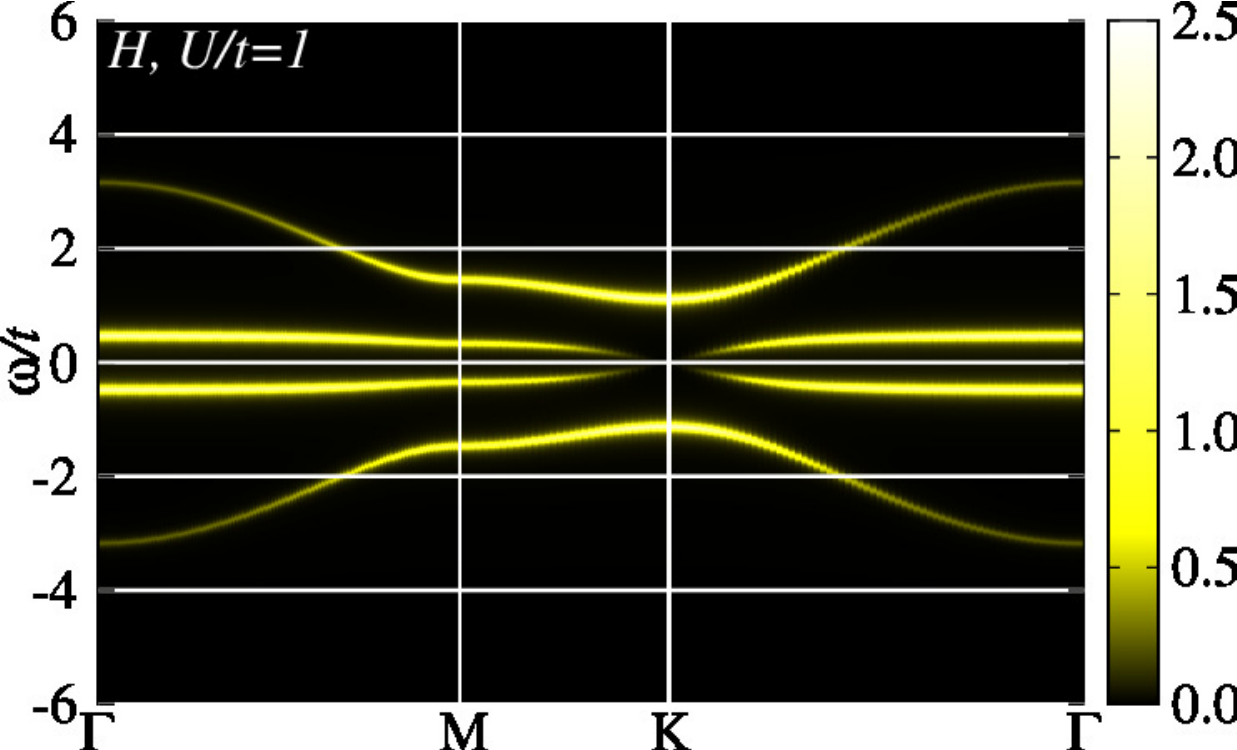}\\
  \includegraphics[width=4.3cm,angle=0]{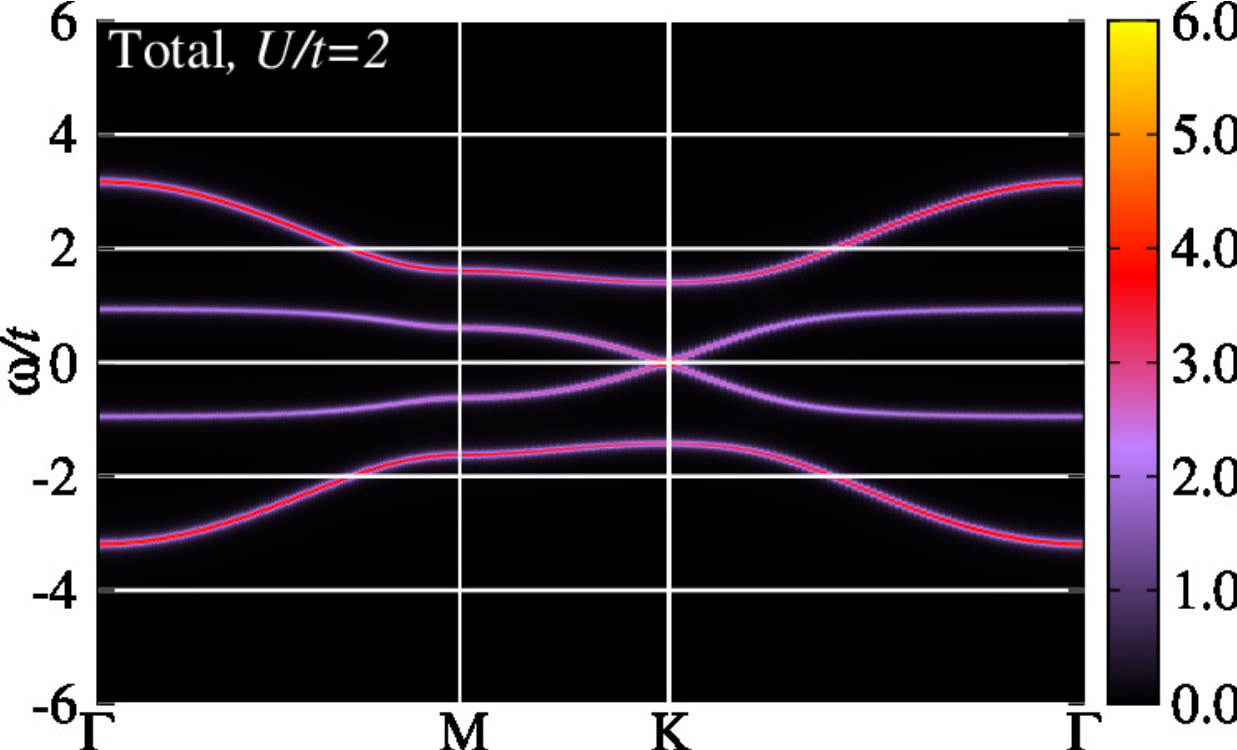}
  \includegraphics[width=4.3cm,angle=0]{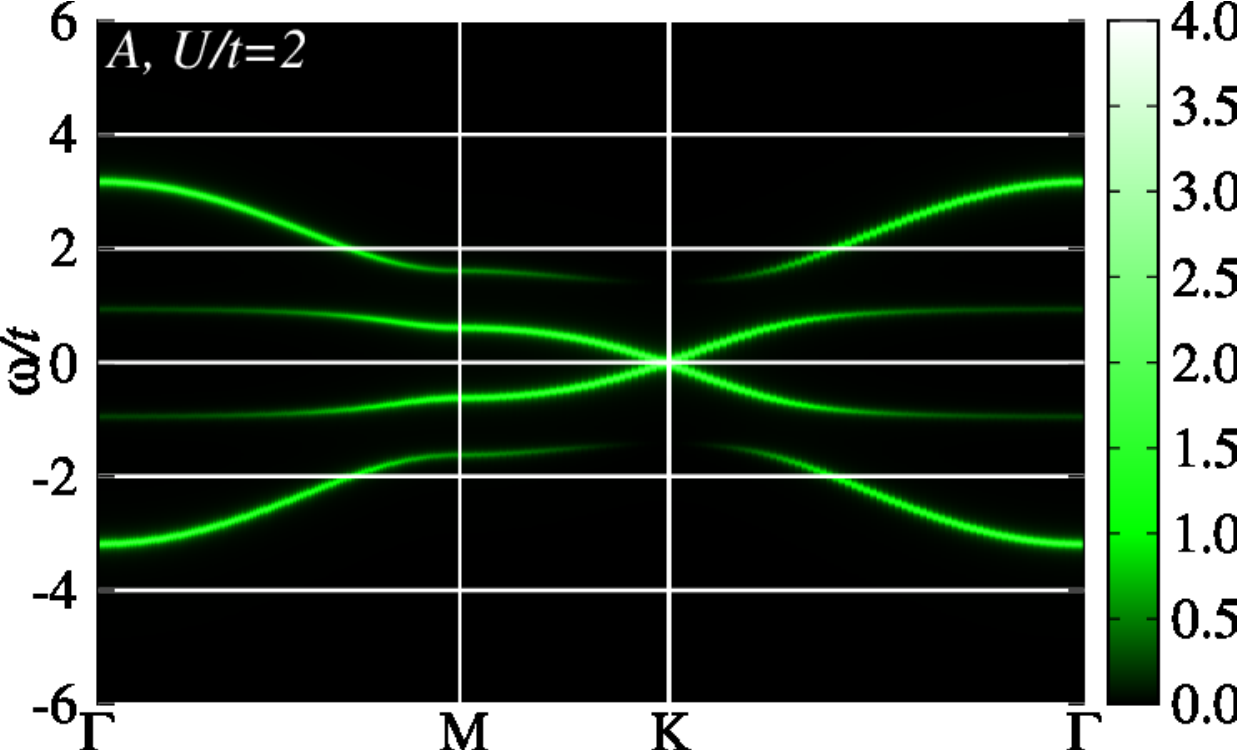}
  \includegraphics[width=4.3cm,angle=0]{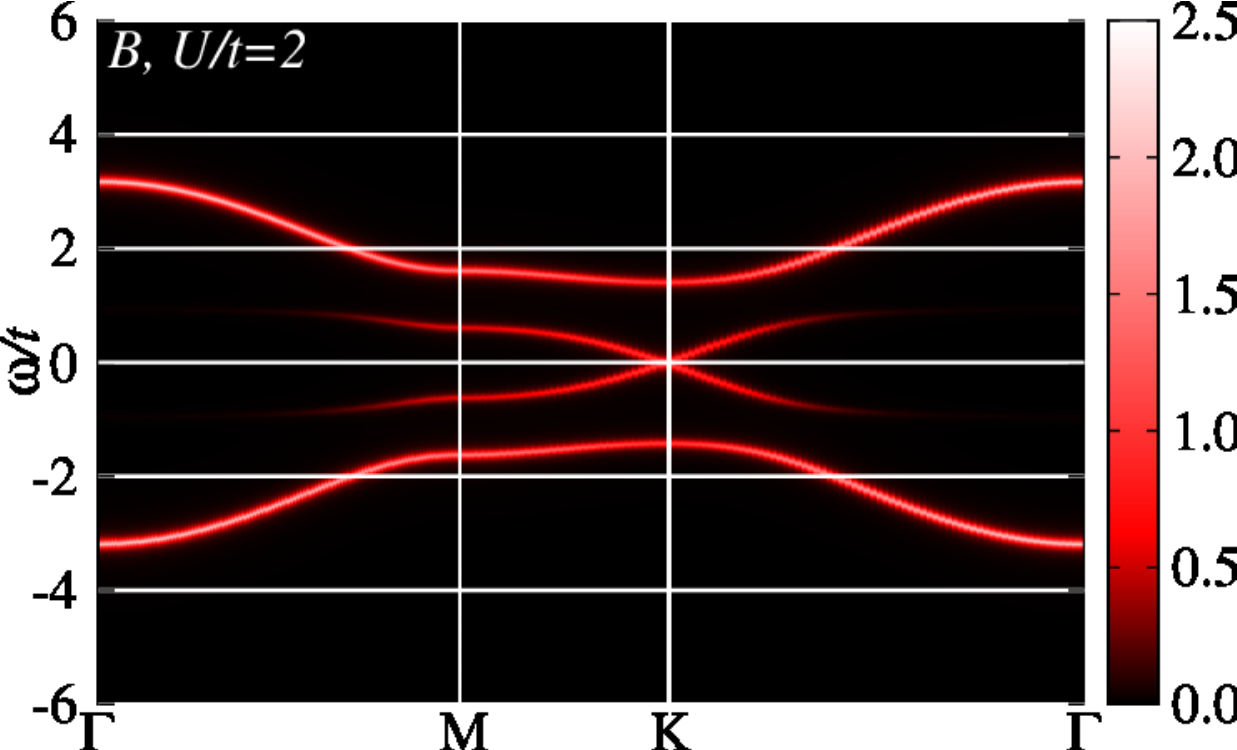}
  \includegraphics[width=4.3cm,angle=0]{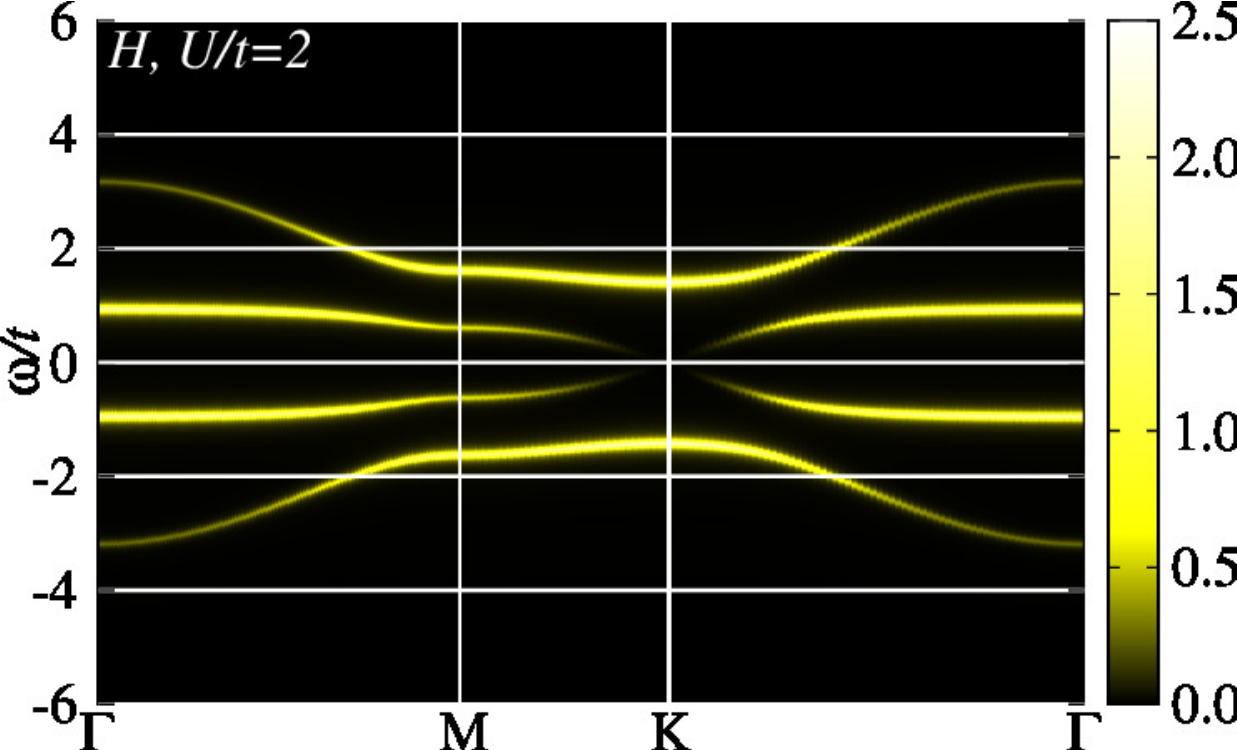}\\
  \includegraphics[width=4.3cm,angle=0]{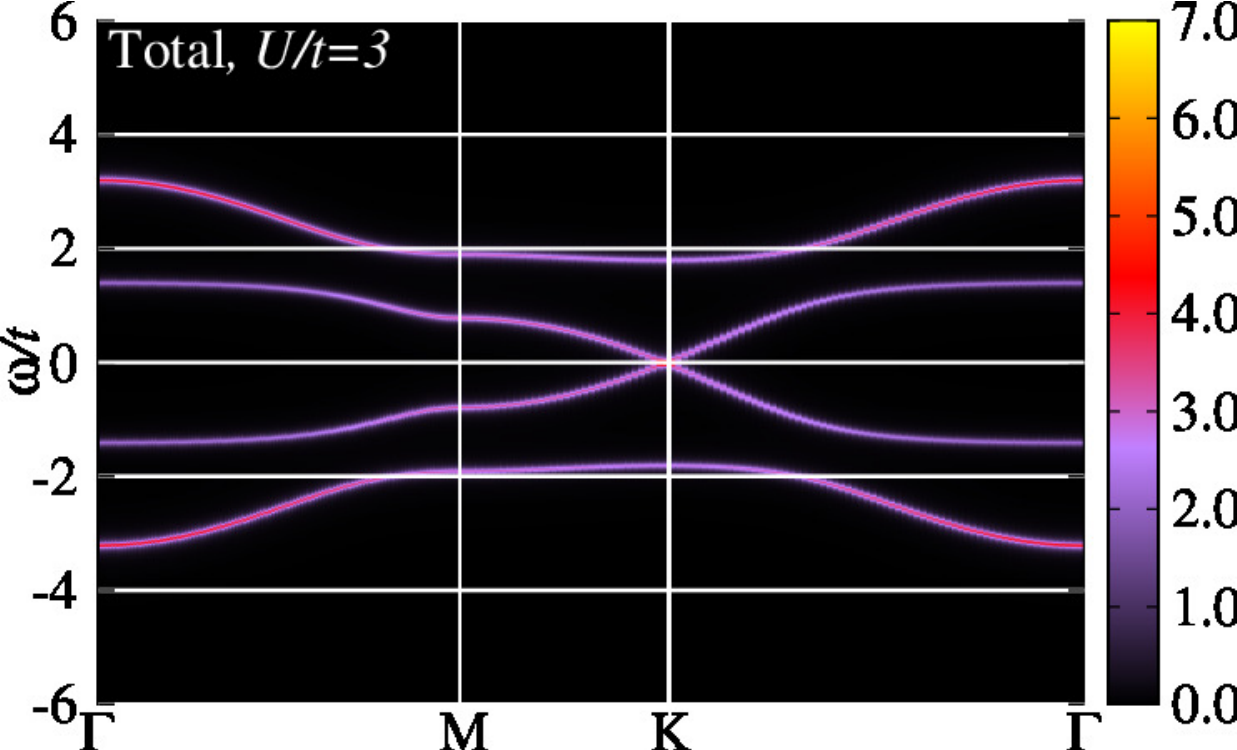}
  \includegraphics[width=4.3cm,angle=0]{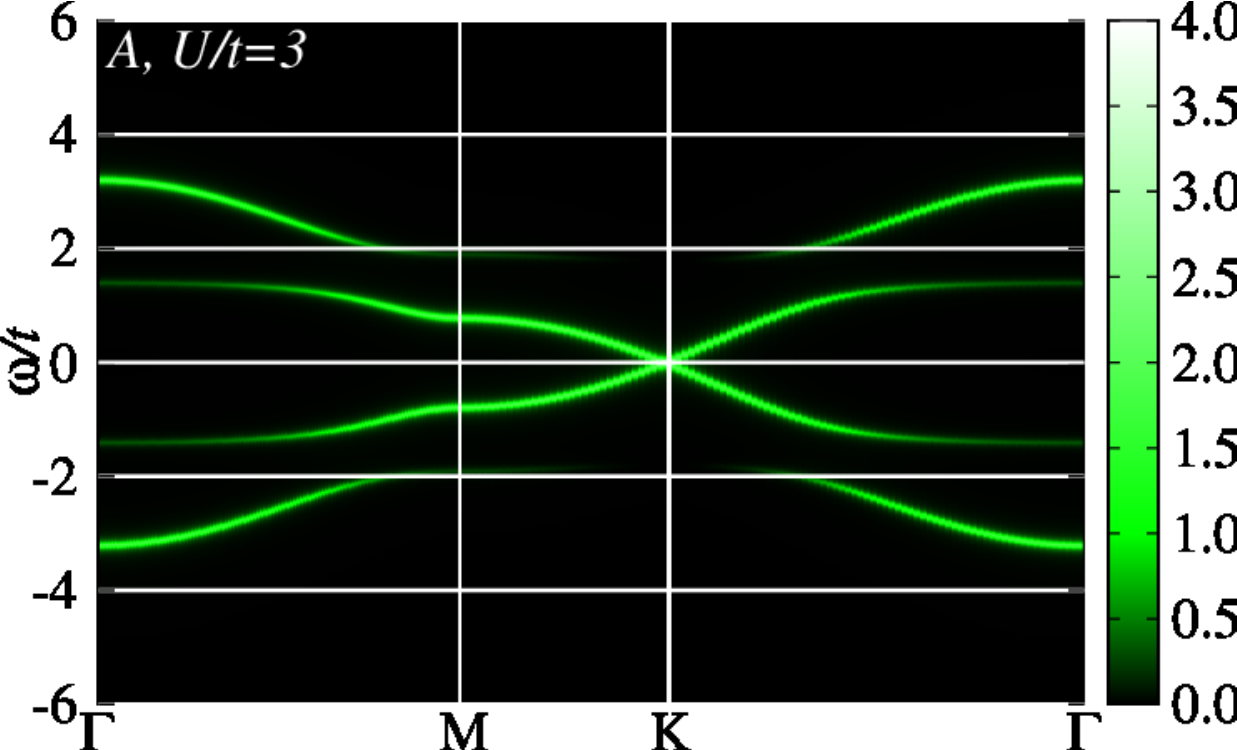}
  \includegraphics[width=4.3cm,angle=0]{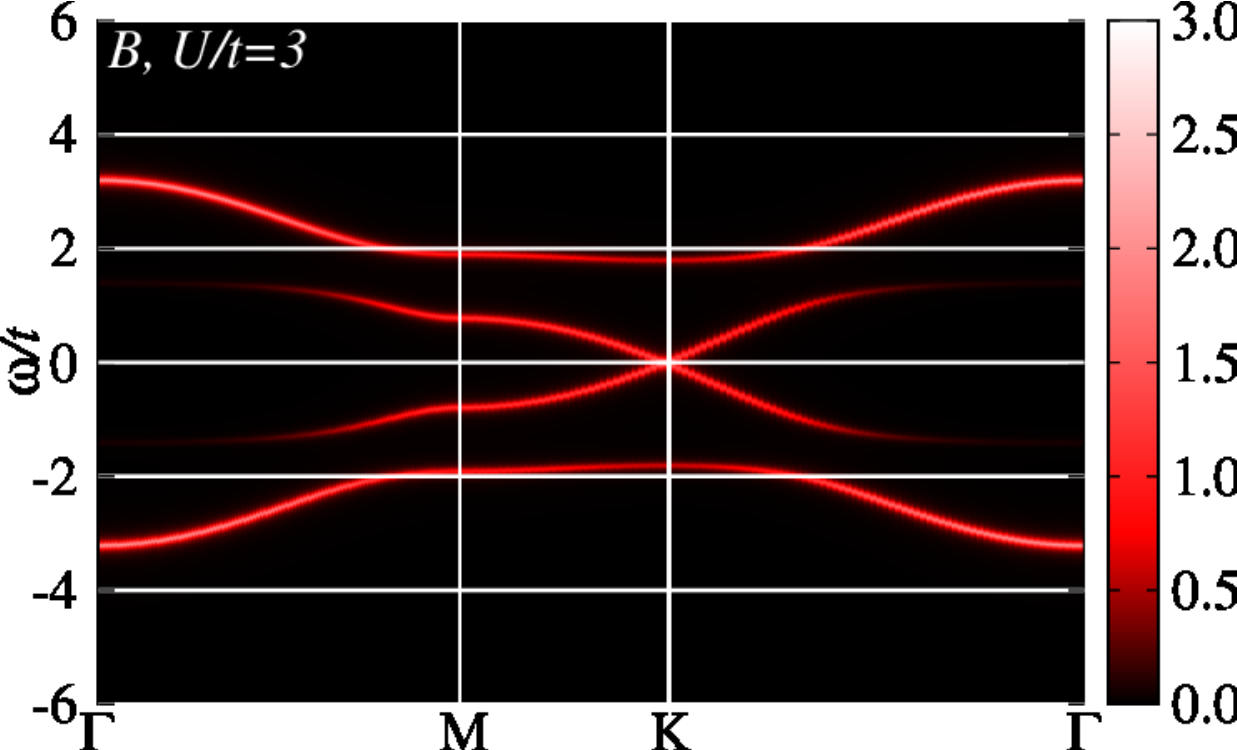}
  \includegraphics[width=4.3cm,angle=0]{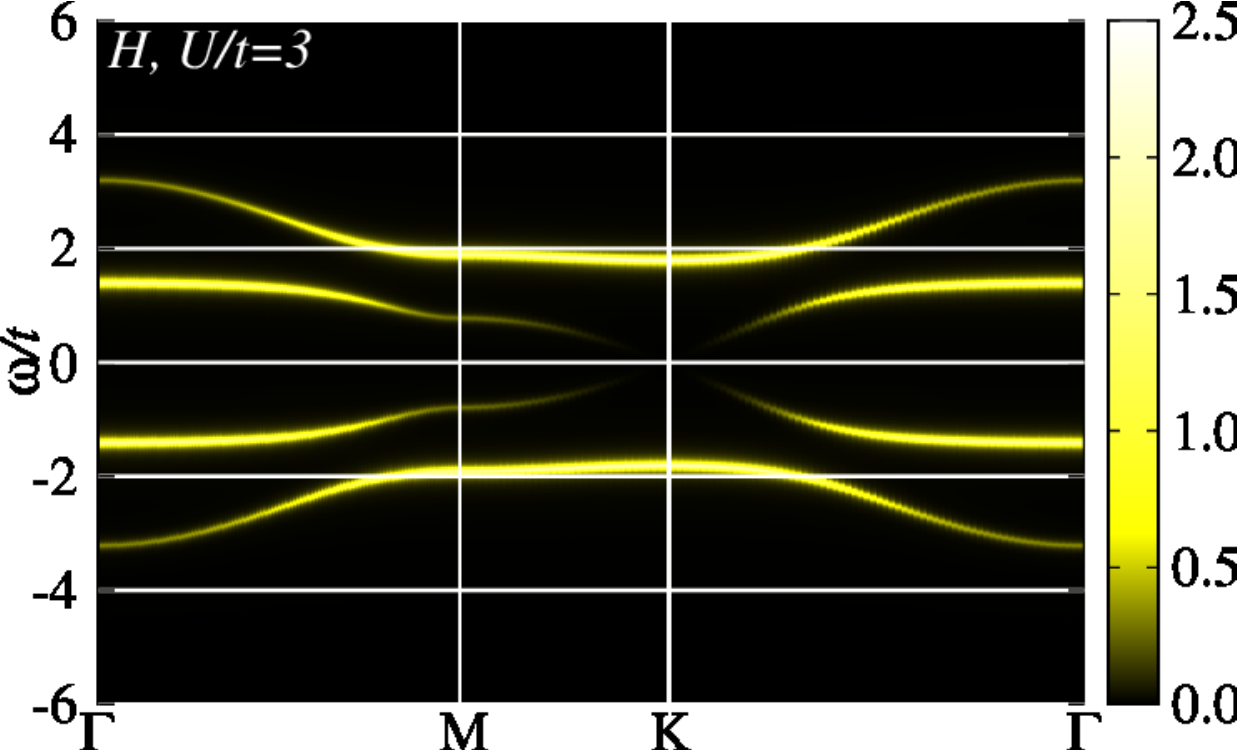}\\
  \includegraphics[width=4.3cm,angle=0]{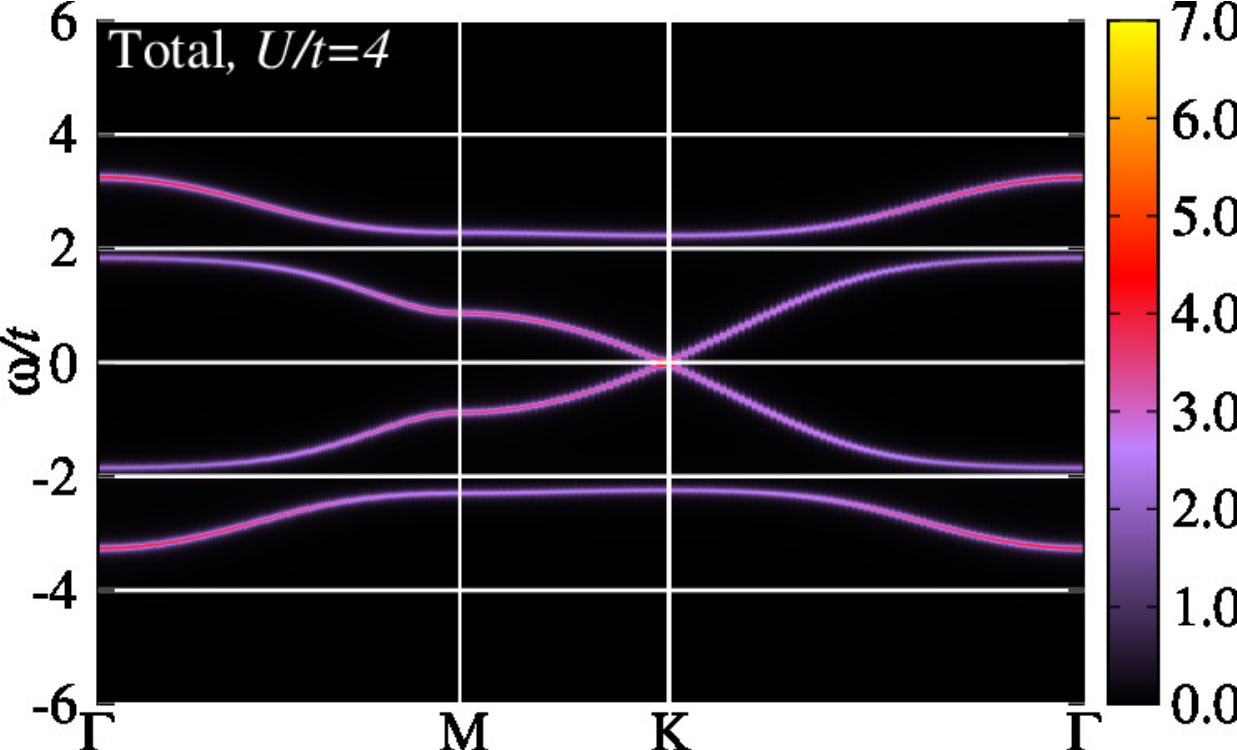}
  \includegraphics[width=4.3cm,angle=0]{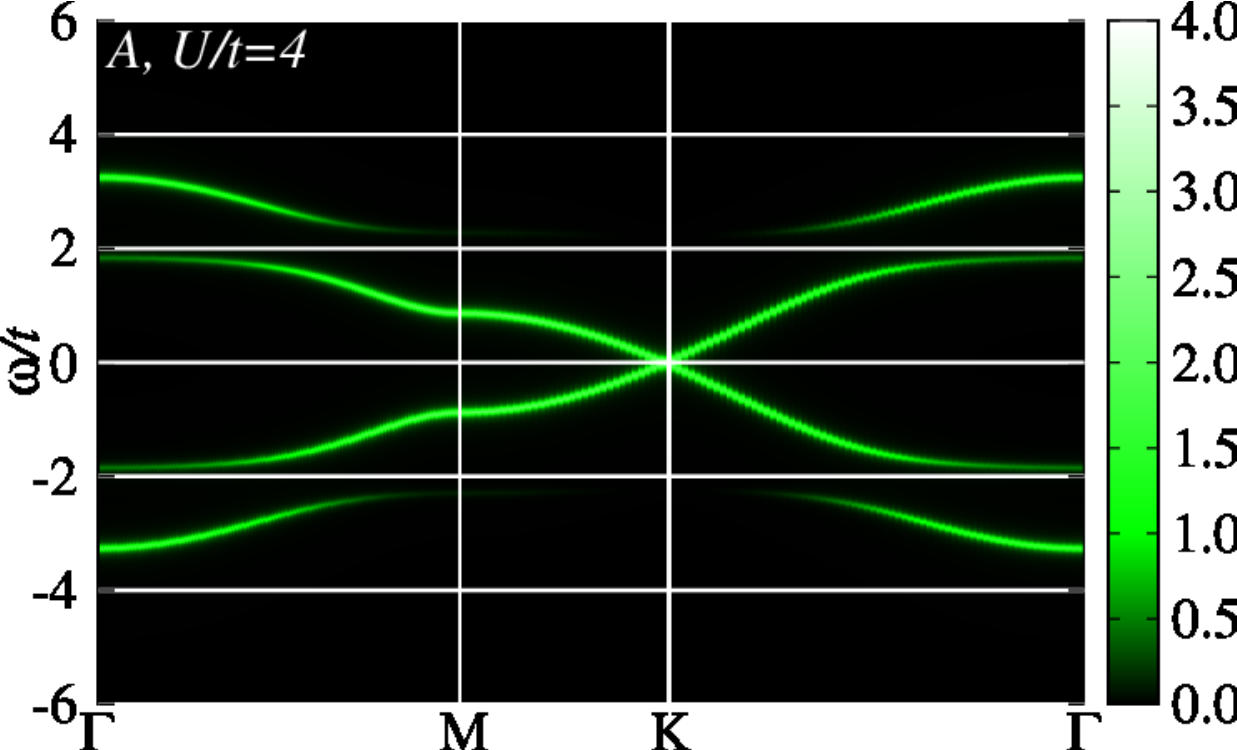}
  \includegraphics[width=4.3cm,angle=0]{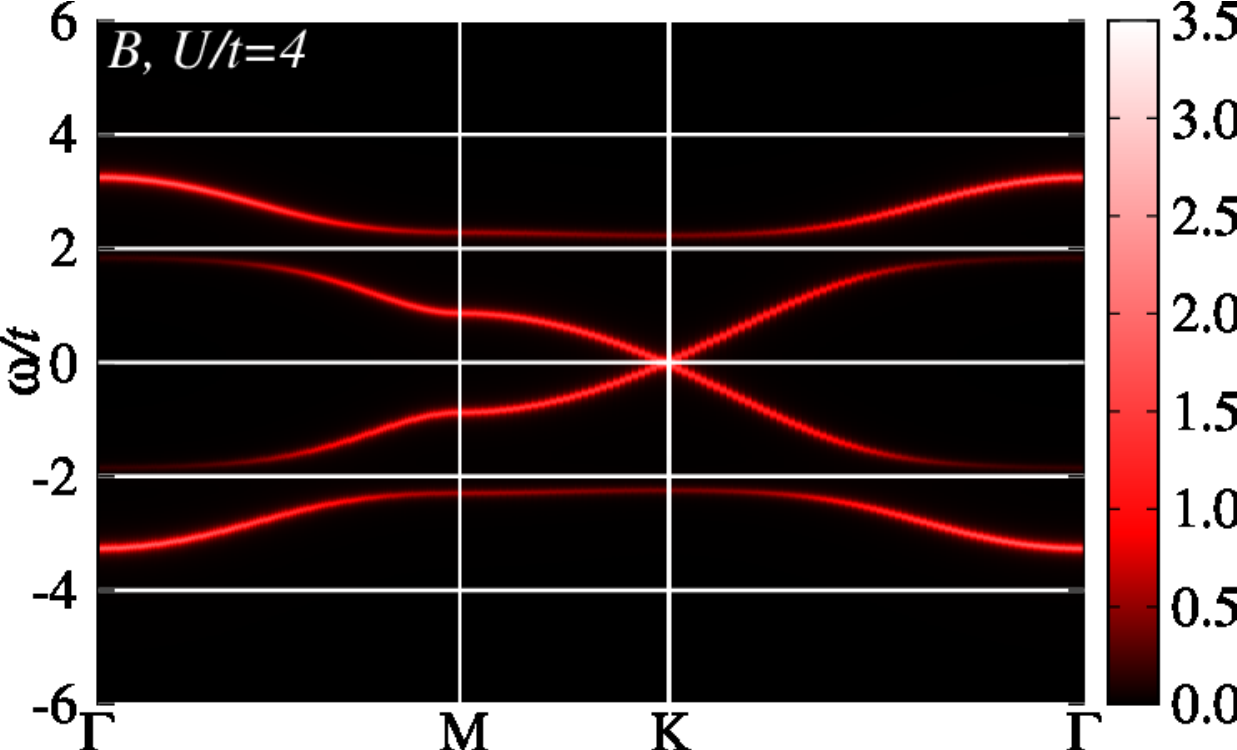}
  \includegraphics[width=4.3cm,angle=0]{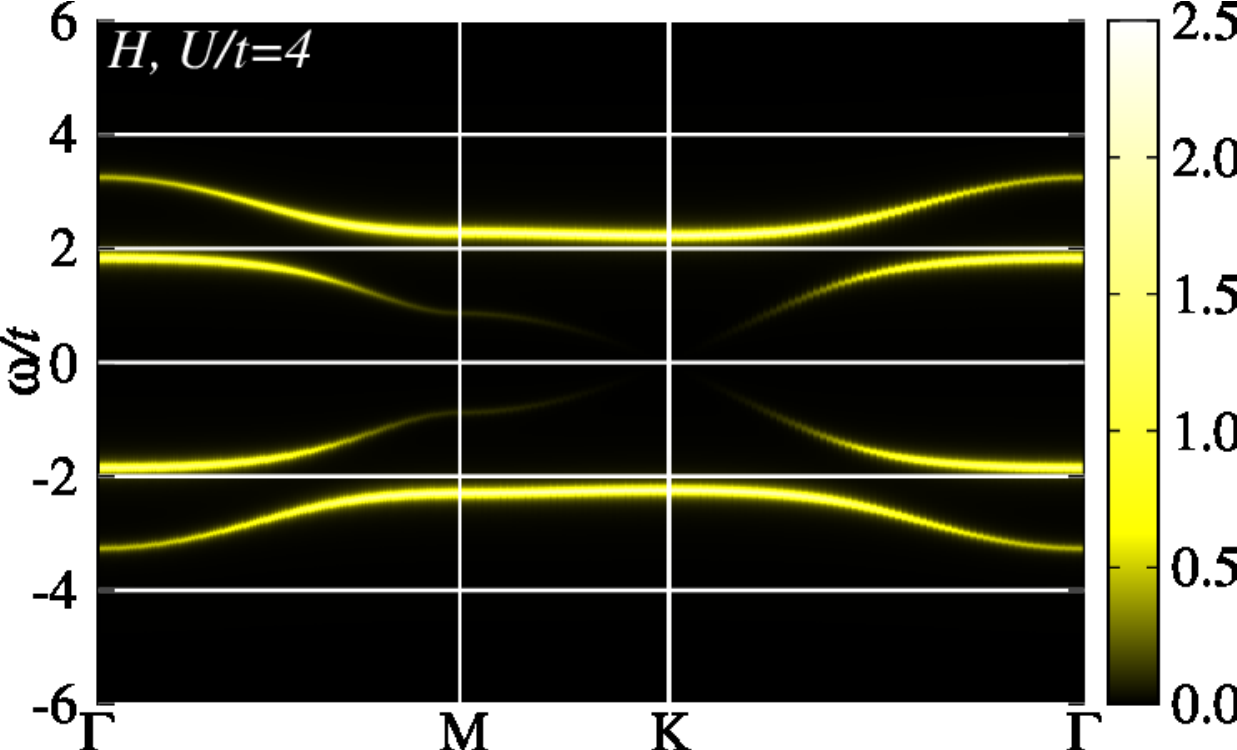}\\
  \includegraphics[width=4.3cm,angle=0]{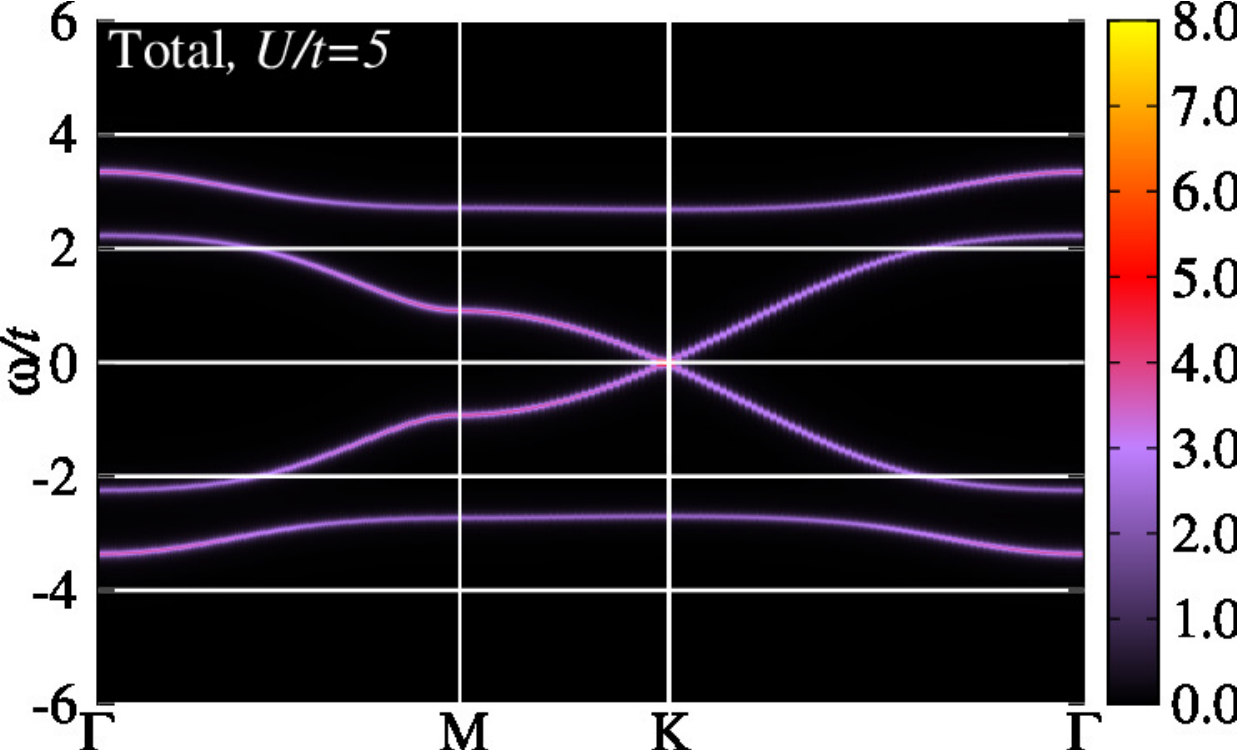}
  \includegraphics[width=4.3cm,angle=0]{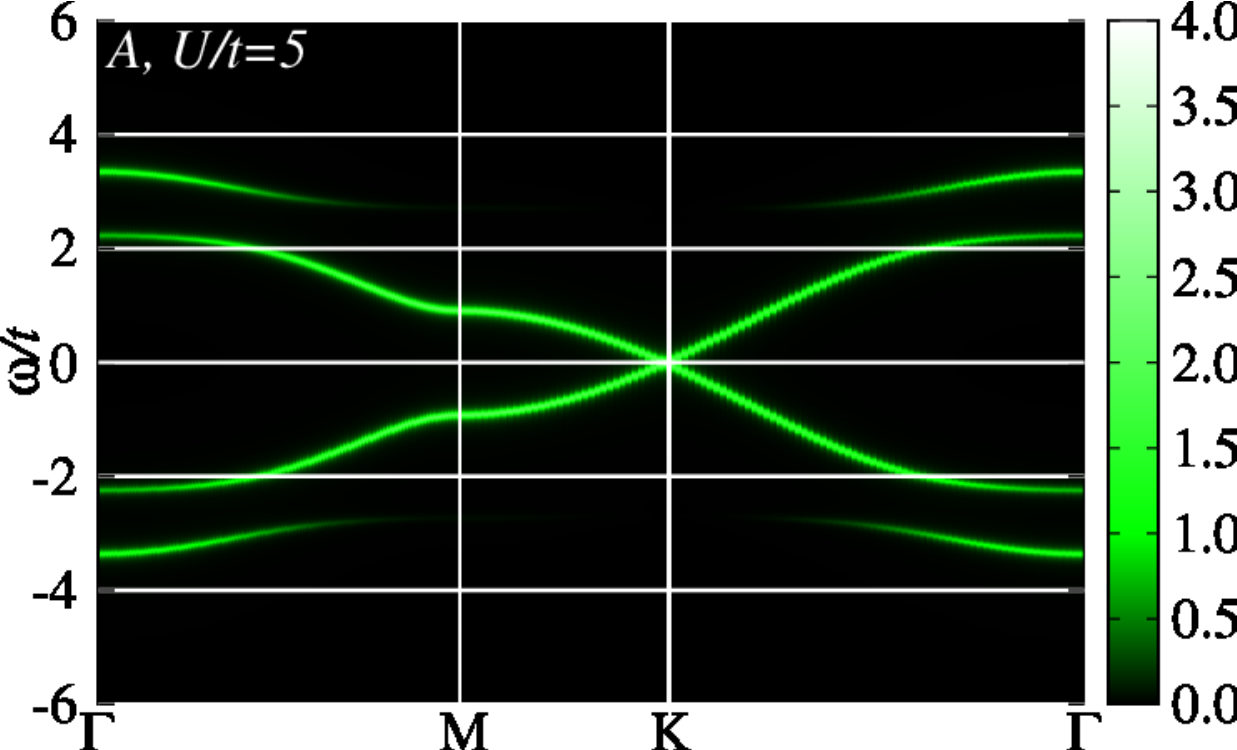}
  \includegraphics[width=4.3cm,angle=0]{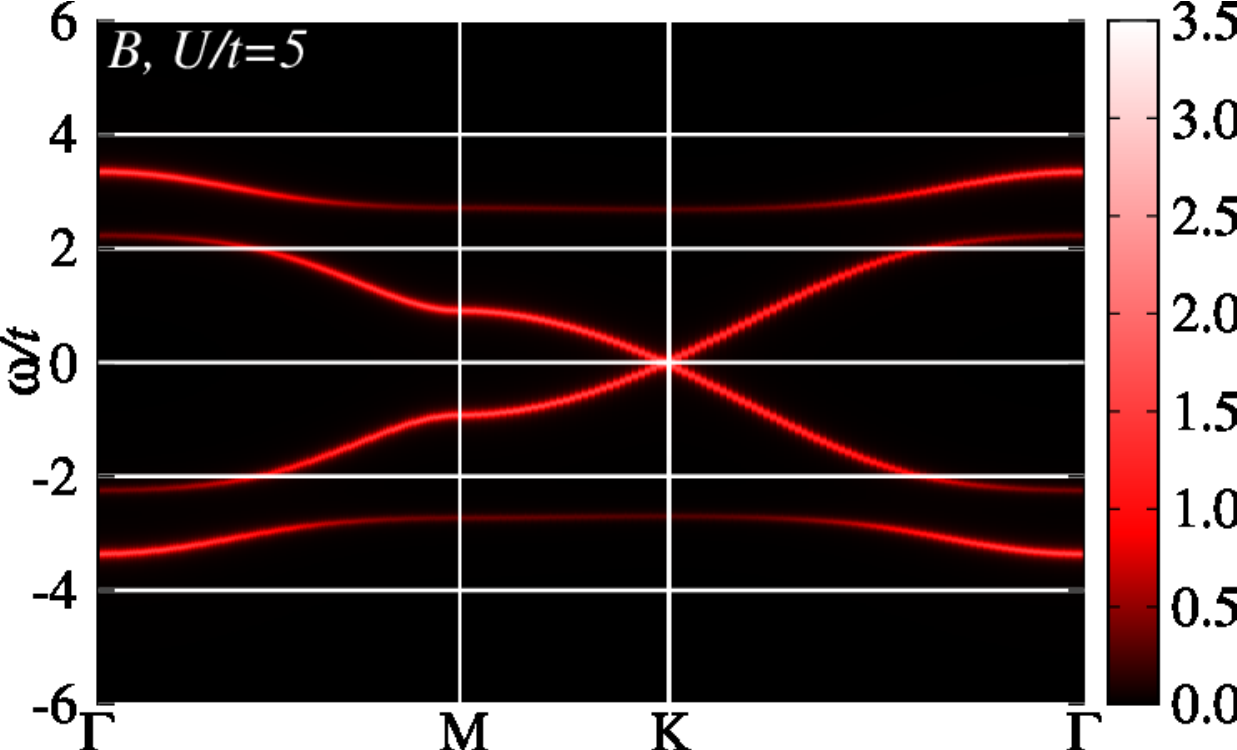}
  \includegraphics[width=4.3cm,angle=0]{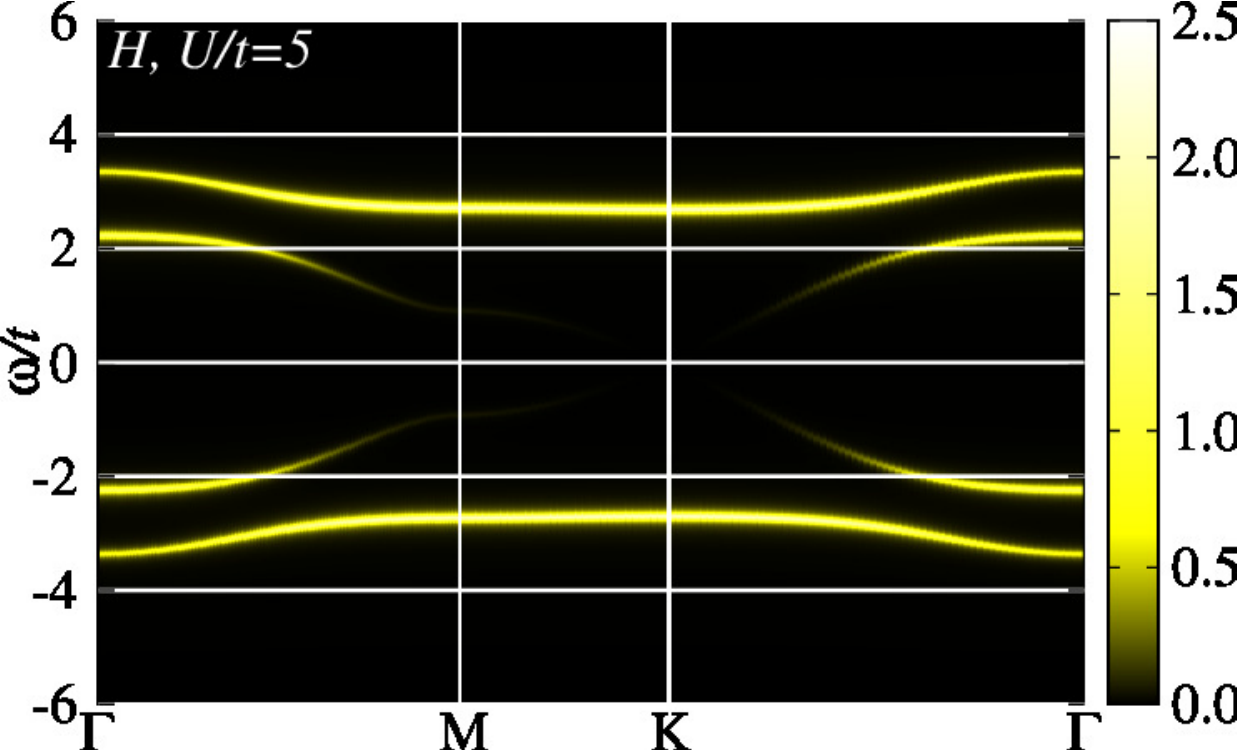}\\
  \includegraphics[width=4.3cm,angle=0]{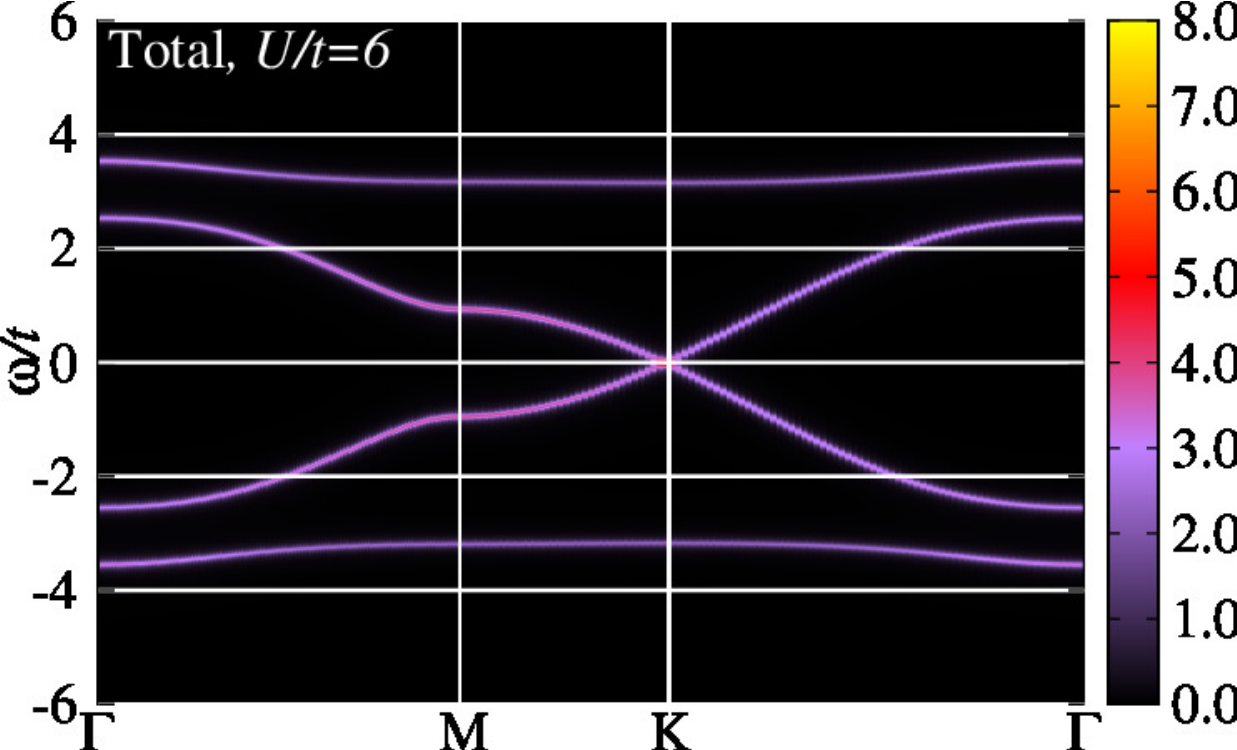}
  \includegraphics[width=4.3cm,angle=0]{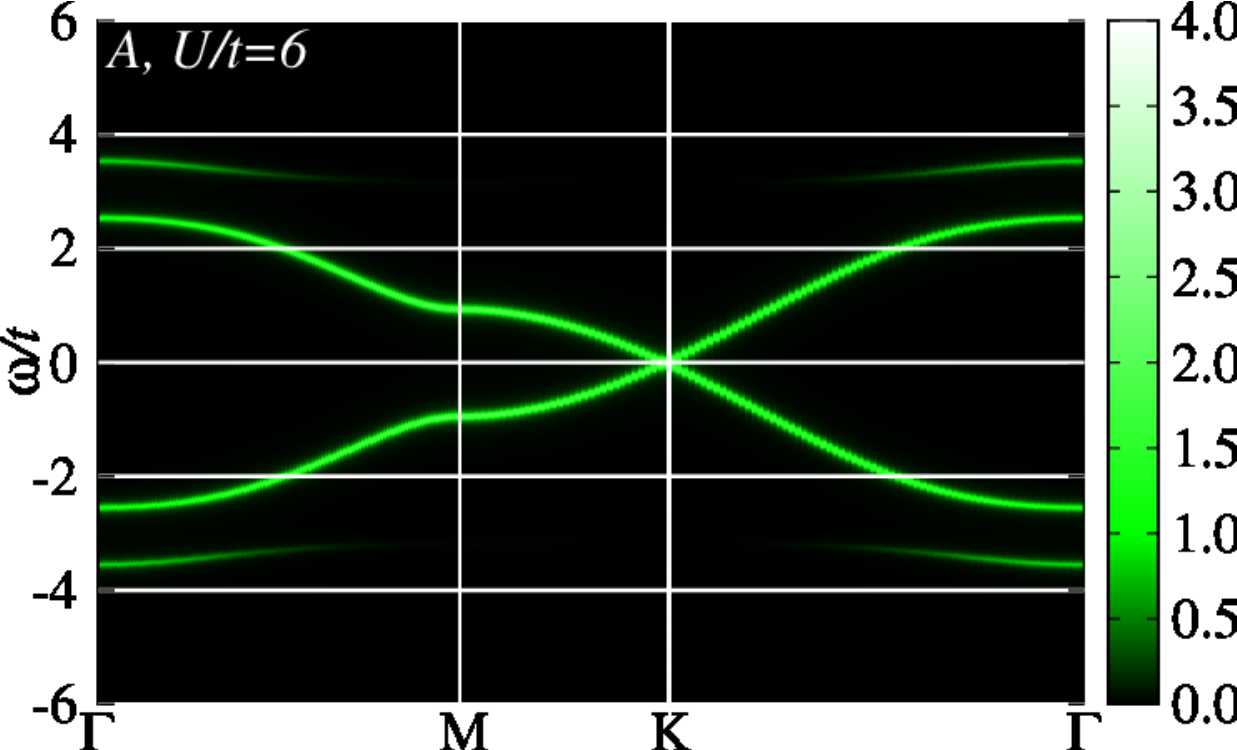}
  \includegraphics[width=4.3cm,angle=0]{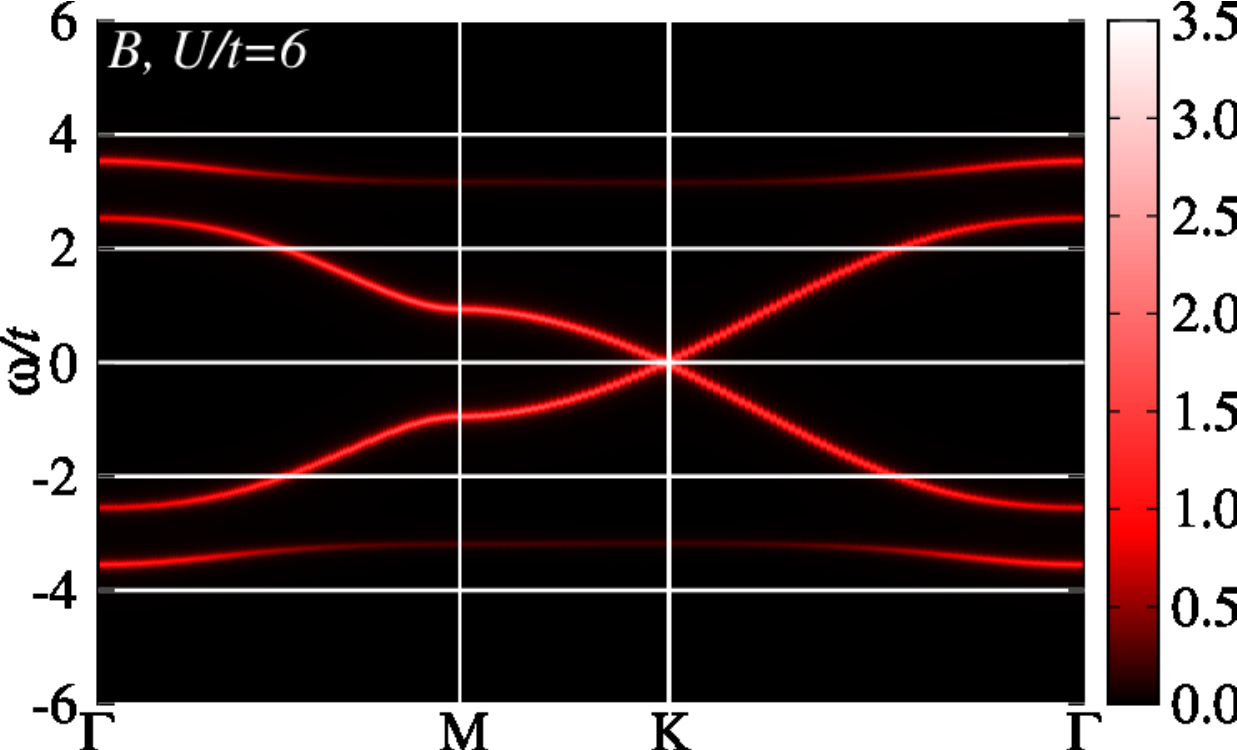}
  \includegraphics[width=4.3cm,angle=0]{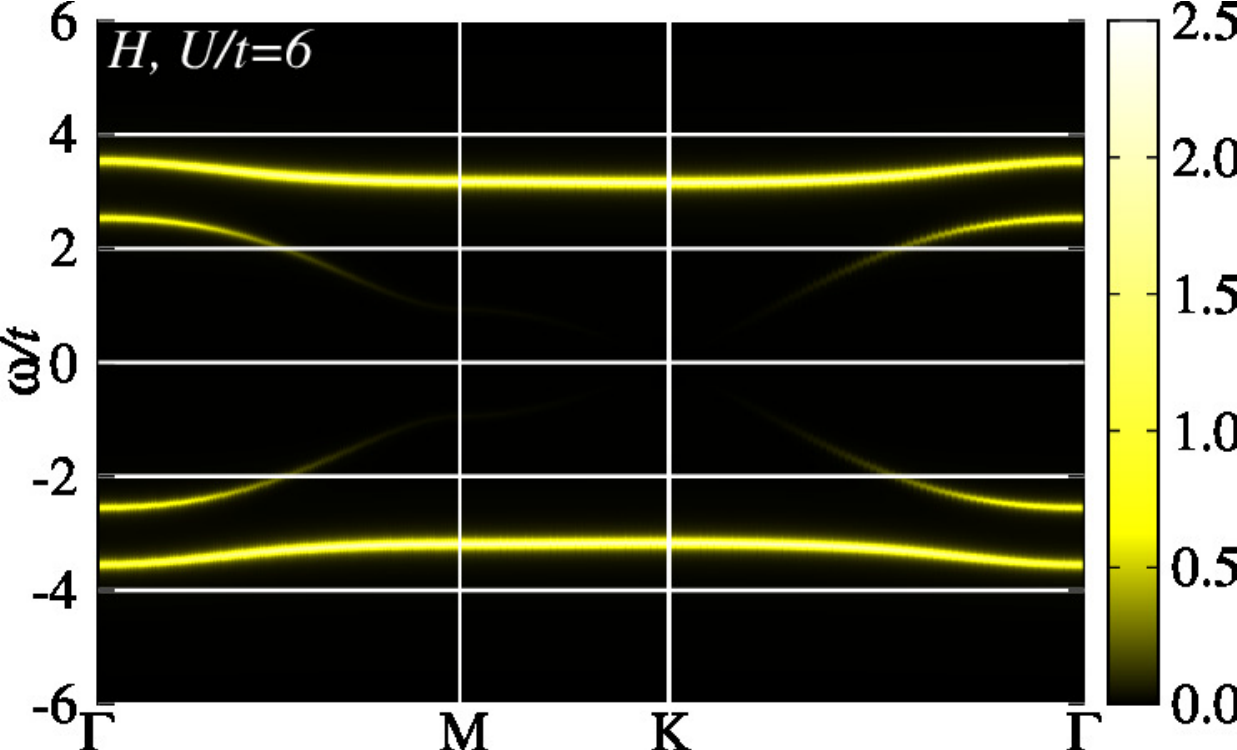}\\
  \includegraphics[width=4.3cm,angle=0]{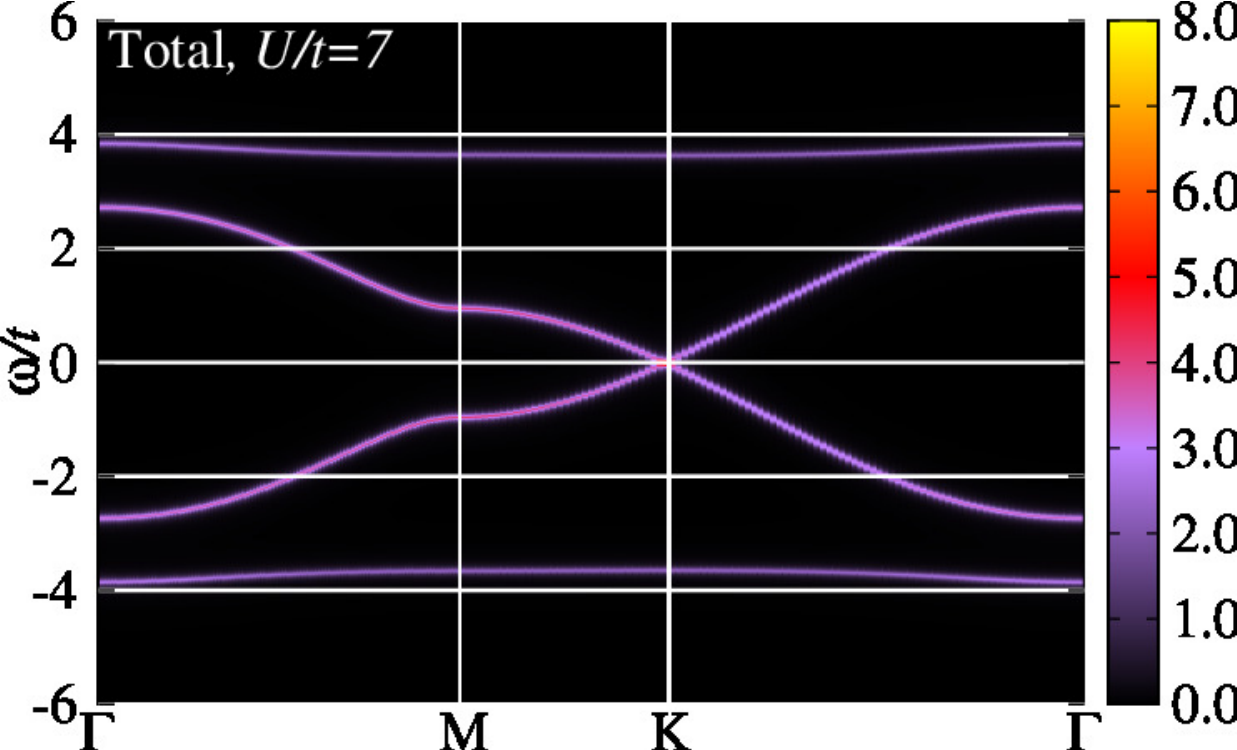}
  \includegraphics[width=4.3cm,angle=0]{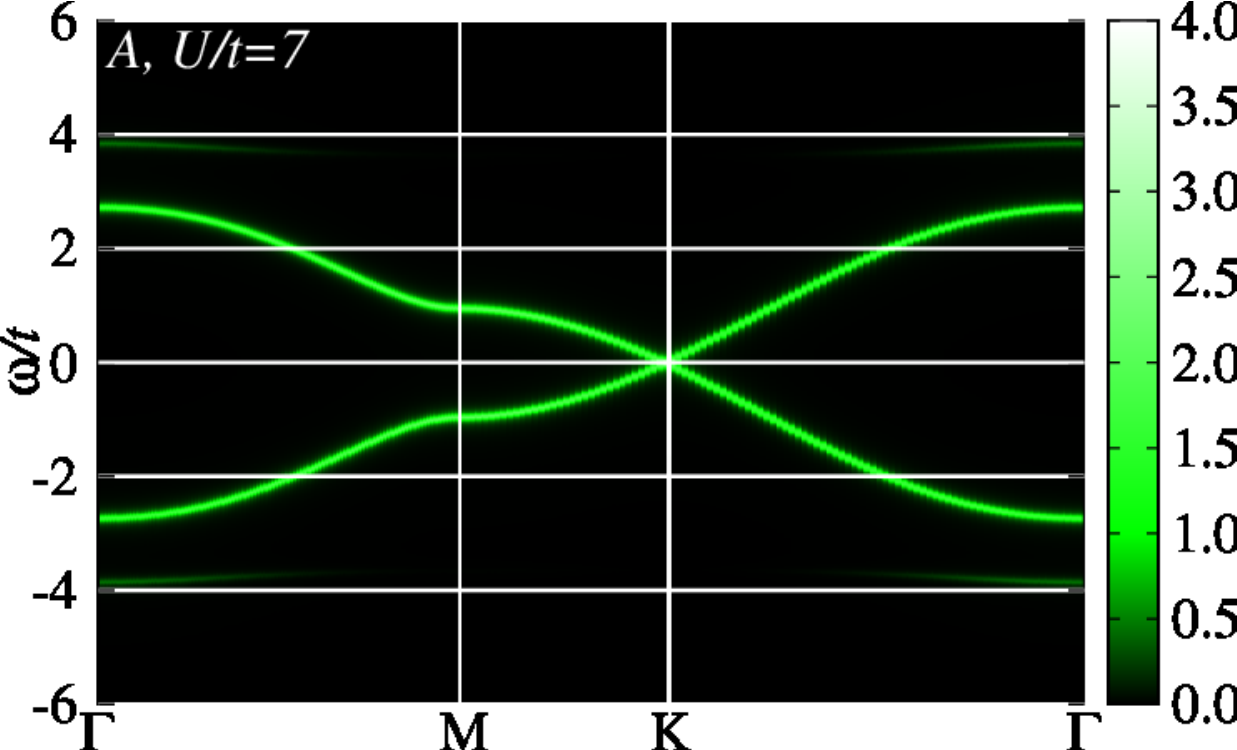}
  \includegraphics[width=4.3cm,angle=0]{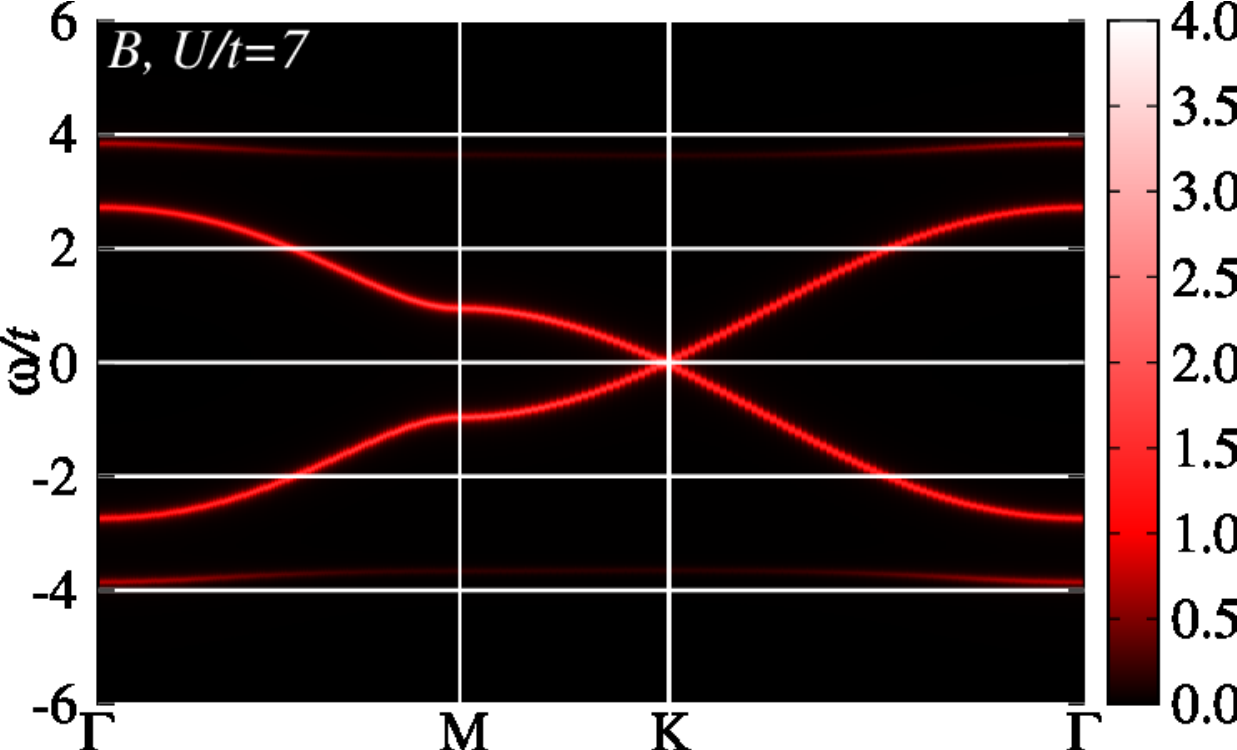}
  \includegraphics[width=4.3cm,angle=0]{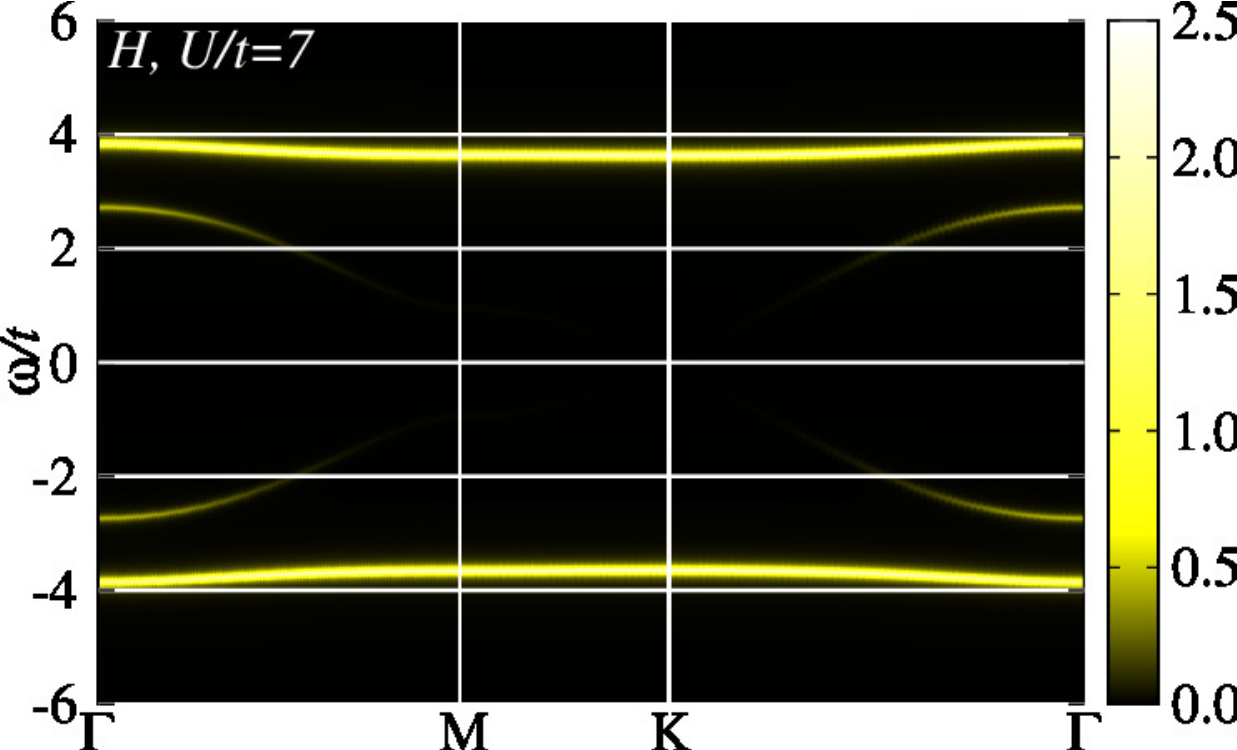}
  \caption{(color online) 
    Single-particle excitation spectra for $U/t=0,1, \cdots, 7$ (from top to bottom) 
    and $t_{sp}/t=1$ obtained by the Hubbard-I approximation. 
    $A_\sigma(\mb{ k},\w)=\sum_\alpha A^{\alpha\alpha}_\sigma(\mb{ k},\w)$, 
    $A^{AA}_\sigma(\mb{ k},\w)$, $A^{BB}_\sigma(\mb{ k},\w)$, and $A^{HH}_\sigma(\mb{ k},\w)$ are shown 
    from left to right panels. 
    Here, we only show the spectra $A^{\alpha\alpha}_\uparrow(\mb{k},\w)$ for up electrons,  
    which are exactly the same as $A^{\alpha\alpha}_\downarrow(\mb{k},\w)$.
    A Lorentzian broadening of $\eta/t=0.05$ is used.
    The spectral intensity is indicated by a color bar in each figure.
    Note that the different intensity scales are used for different figures.
    The Fermi energy $E_{\rm F}$ is located at $\w=0$. 
  } 
  \label{fig.akw.h1}
\end{figure*}

From Eq.~(\ref{eq.spectral}), the single-particle excitation spectrum $A^{\alpha\alpha}_\sigma(\mb{k},\w)$
for $\alpha$ orbital is given as  
\begin{equation}
  \label{Akw_HBI}
  A^{\alpha\alpha}_\sigma(\mb{k},\w) = \sum_{\nu=1}^4 \rho^{\alpha\alpha}_{\nu, \mb{k}} \delta(\w - \w_{\nu, \mb{k}}).
\end{equation}
Since the poles $\w_{\nu, \mb{k}}$ as well as the spectral weights $\rho^{\alpha\alpha}_{\nu, \mb{k}}$ are 
all known analytically in Eqs.~(\ref{eq.poles}), (\ref{eq.rho14}), and (\ref{eq.rho23}), the calculation of the 
single-particle excitation spectrum $A^{\alpha\alpha}_\sigma(\mb{k},\w)$ is straightforward and the results for 
various values of $U$ are shown in Fig.~\ref{fig.akw.h1}. 
We can clearly find in Fig.~\ref{fig.akw.h1} that 
(i) the flat band which is present only at $U=0$ evolves into 
the Dirac band with the massless Dirac dispersions emerging around $K$ and $K'$ points near 
the Fermi energy, (ii) the Dirac points are located exactly at the Fermi level and momentum ${\mb k}=K$ and $K'$, 
(iii) the contribution of $B$ orbital to the Dirac band becomes increasingly significant with increasing $U$, 
while $H$ orbital does not participate in the formation of the massless Dirac dispersion, and 
(iv) the highest and lowest bands which display the massive Dirac dispersions near $K$ and $K'$ points 
at $U=0$ 
evolve respectively into the upper and lower Hubbard 
bands in the high energy regions for $|\w|\gtrsim\Delta_c$. 
These characteristic features are in good qualitative agreement with those obtained by the CPT for 
the PM state shown in Fig.~\ref{fig:akw_pm}.

\begin{figure}
  \begin{center}
    \includegraphics[clip,width=6.0cm]{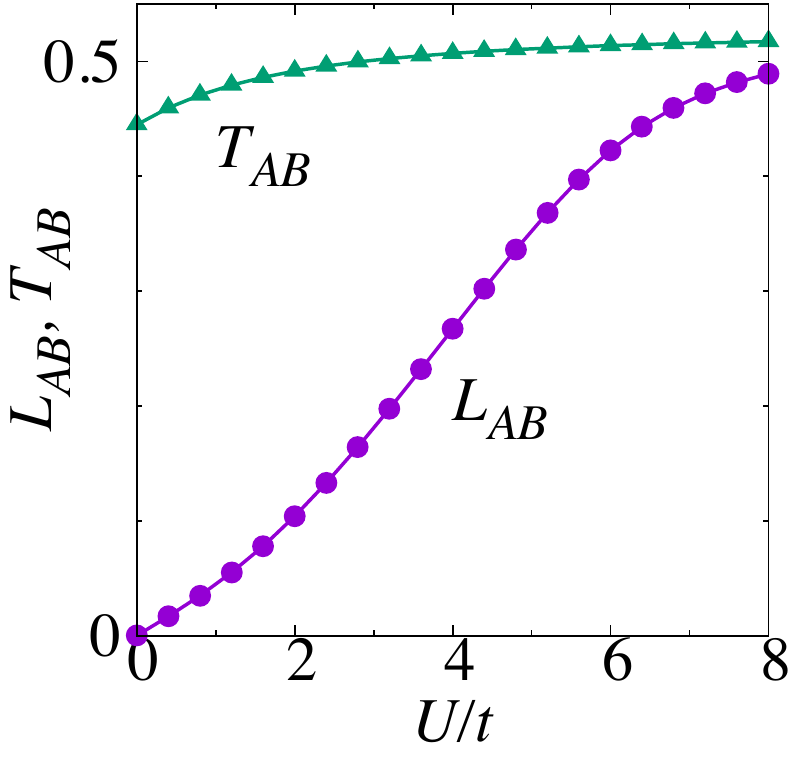}
    \caption{(color online) 
      $U$ dependence of the effective dynamical bonding strength 
      $L_{AB}$ and $T_{AB}$
      (see the text for definition) for $t_{sp}/t=1$ calculated using the Hubbard-I approximation. 
      \label{fig.dbs}
    }
  \end{center}
\end{figure}

    Although it is 
    already convincing that the contribution of $B$ orbital to the low energy 
    excitations is essential for the emergent massless Dirac quasiparticles, 
    here we quantify the low energy bonding character between $A$ and $B$ orbitals 
    and examine how this quantity evolves with increasing $U$ within the Hubbard-I approximation. 
    For this purpose, let us define the following effective dynamical bonding strength between 
    $A$ and $B$ orbitals 
  \begin{eqnarray}
    L_{AB} =  \frac{1}{N_a} \sum_{\mb{k},\sigma} \int_{-\Delta_c}^{0} d\w 
    \left|
    A^{AB}_\s(\mb{k},\w) + A^{BA}_\s(\mb{k},\w) 
    \right|,
  \end{eqnarray}
    where $-\Delta_c$ is the upper bound of the lower Hubbard band given in Eq.~(\ref{eq:delta_c}). 
    Note that 
    this quantity becomes $T_{AB}=\sum_\sigma|\langle c^\dag_{{\mb r}_i \s A}c_{{\mb r}_i \s B}+{\rm H.c.}\rangle|$ 
    if the lower bound of the integral is extended to $-\infty$. 
    As shown in Fig.~\ref{fig.dbs}, we find that 
    although $T_{AB}$ is almost constant and does not depend strongly on $U$,  
    $L_{AB}$ is rather sensitive to $U$ and monotonically increases from zero. 
    This clearly demonstrates that the low energy bonding between $A$ and $B$ orbitals becomes stronger 
    as the massless Dirac quasiparticles develops with increasing $U$, 
    which is accompanied by the large spectral weight redistribution from the upper and lower Hubbard 
    bands to the Dirac band. 

\subsection{Chiral symmetry in single-particle excitations} \label{sec:H-I_chiral}
  It is well known in the single-particle theory that for a bipartite system with no hopping between two 
  different sites on the same sublattice, there exist zero-energy states at least as many as the difference of 
  the number of sites on each sublattice~\cite{Lieb,Hatsugai2007,Hatsugai2013,Hatsugai2014}. 
  For example, the flat band found in the noninteracting limit of the periodic Anderson model $\mcal H$ 
  is a typical case because the number of sites on each sublattice is different 
  (i.e., sublattice imbalanced) by one per unit cell, thus leading to at least one zero-energy state at each 
  momentum, as discussed in Sec.~\ref{Sec:nonint}. 
  On the other hand, this theorem does not predict the existence 
  of the four Dirac cones with eight zero-energy states in the pure graphene model (including the spin degeneracy) 
  as the pure graphene model contains the same number of sites on each sublattice.  
  Instead, the four Dirac cones at $K$ and $K'$ points 
  in the pure graphene model are protected by the time-reversal symmetry, 
  120$^{\circ}$-rotational symmetry, and sublattice (or, equivalently, inversion) symmetry~\cite{Bernevig}. 

  Here, we argue that the quasiparticle excitations in the PM phase of the periodic Anderson 
  model dynamically recover the sublattice balance, thus eliminating a trivial zero-energy state,  
  and the Dirac cone like dispersions with point contacts at zero energy is protected by the electron correlation.  
  
  For this purpose, 
  we shall construct an effective Hamiltonian in a quadratic form of fermion quasiparticle operators, which 
  reproduces the single-particle excitations obtained by the Hubbard-I approximation, and 
  follow the chiral symmetry argument given in Refs.~\cite{Hatsugai2007, Hatsugai2013, Hatsugai2014}.  
  
\begin{figure}
  \begin{center}
    \includegraphics[width=7.5cm]{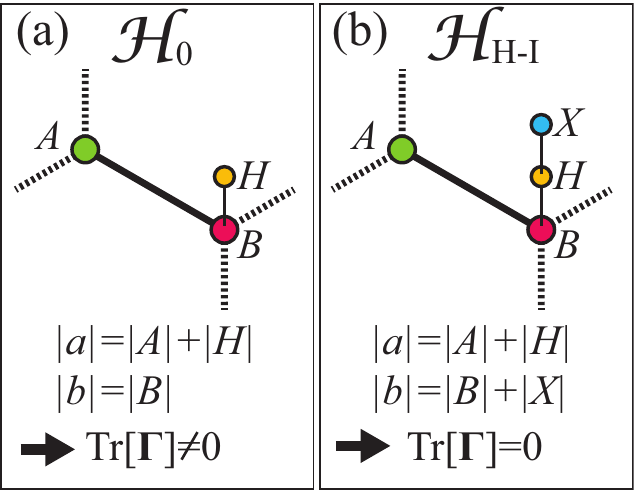}
    \caption{(color online) 
      Schematic representations for (a) the periodic Anderson model $\mcal H$ 
      in the noninteracting limit 
      and (b) the effective Hamiltonian $\mcal H_{\text{H-I}}$ 
      where the electron correlation $U$ is represented as the hybridization between 
      $H$ orbital and auxiliary $X$ orbital. 
      Both models are defined on bipartite lattices with the hopping between different sublattices, 
      and preserve the chiral symmetry. 
      The $A$-$B$ sublattice symmetry (the Hamiltonian is invariant under the exchange of $A$ and $B$ 
      orbitals) as in the pure graphene model is apparently broken in both models. 
      Although the sublattice balance is also broken in model (a), i.e., $|a| \not = |b|$, 
      it is preserved in model (b) due to the presence of auxiliary $X$ orbital. 
      \label{fig.rc}
    }
  \end{center}
\end{figure}

  By introducing an auxiliary orbital $X$, 
  we construct the following effective Hamiltonian: 
  \begin{equation} 
    \mcal{H}_{\text{H-I}} = \sum_{\mb{k}, \s} 
    \tilde{\bs{c}}_{\mb{k}\s}^\dag
    \left(
      \begin{array}{cccc}
        0 & \gamma_{\mb{k}} & 0 & 0  \\
        \gamma^*_{\mb{k}} & 0 & t_{sp} & 0  \\
        0 & t_{sp} & 0 & U/2  \\
        0 & 0 & U/2 & 0  
      \end{array}
      \right)
    \tilde{\bs{c}}_{\mb{k}\s},
    \label{eq.H_H-I}
  \end{equation}
  where fermion creation operators 
  $\tilde{\bs{c}}_{\mb{k}\s}^\dag =(
  \tilde{c}_{\mb{k} \s A}^\dag \, 
  \tilde{c}_{\mb{k} \s B}^\dag \, 
  \tilde{c}_{\mb{k} \s H}^\dag \, 
  \tilde{c}_{\mb{k} \s X}^\dag)$ 
  in the momentum space describe the 
  quasiparticles, not the bare electrons, in the Hubbard-I approximations. 
    Here, the dynamical electron correlation in the Hubbard-I approximation  
    is represented as the hybridization between the auxiliary orbital $X$ and the hydrogen orbital $H$ 
    with the hybridization strength $U/2$ (see Fig.~\ref{fig.rc}).  
  The Mott gap between the upper and lower Hubbard bands in the Hubbard-I approximation is then 
  interpreted as a single-particle hybridization gap generated by introducing $X$ orbital. 
  Indeed, we can show that the eigenvalues of 
  $\mcal{H}_{\text{H-I}}$ coincide with $\w_{\nu,\mb{k}}$ ($\nu=1,2,3$, and $4$) in the Hubbard-I approximation 
  (see Sec.~\ref{sec:HI_disp}). 
  The similar interpretation of the Mott gap 
  is recently emphasized by Sakai {\it et al.} in the context of 
  high-temperature cuprate superconductors~\cite{Sakai2014}. 
  The analysis of the effective Hamiltonian $\mcal{H}_{\text{H-I}}$ based on 
  the Brillouin-Wigner perturbation theory is given in Appendix~\ref{appsec:BW}.

    The spectral weight of the single-particle Green's function obtained by the Hubbard-I approximation 
    in Eq.~(\ref{eq.spectral}) 
    can also be reproduced from the eigenstates ${\bs\gamma}^\dag_{{\mb k}\s}$ of $\mcal{H}_{\text{H-I}}$, 
    i.e.,
  \begin{equation} 
    \mcal{H}_{\text{H-I}} = \sum_{\mb{k}, \s} 
    {\bs\gamma}^\dag_{{\mb k}\s}
    \left(
      \begin{array}{cccc}
        \w_{1,\mb{k}} & 0 & 0 & 0  \\
        0 & \w_{2,\mb{k}} & 0 & 0  \\
        0 & 0 & \w_{3,\mb{k}} & 0  \\
        0 & 0 & 0 & \w_{4,\mb{k}}  
      \end{array}
      \right)
    {\bs\gamma}_{{\mb k}\s},
  \end{equation}    
    where 
    \begin{equation}
    {\bs\gamma}^\dag_{{\mb k}\s} = \tilde{\bs{c}}_{\mb{k}\s}^\dag \bs{U}_{\mb{k}}
    \end{equation} 
    and 
    \begin{equation}
      \bs{U}_{\mb{k}} = 
      \left(
      \begin{array}{cccc}
        u_{1,\mb{k}}^{A} & u_{2,\mb{k}}^{A} & u_{3,\mb{k}}^{A} & u_{4,\mb{k}}^{A} \\
        u_{1,\mb{k}}^{B} & u_{2,\mb{k}}^{B} & u_{3,\mb{k}}^{B} & u_{4,\mb{k}}^{B} \\
        u_{1,\mb{k}}^{H} & u_{2,\mb{k}}^{H} & u_{3,\mb{k}}^{H} & u_{4,\mb{k}}^{H} \\
        u_{1,\mb{k}}^{X} & u_{2,\mb{k}}^{X} & u_{3,\mb{k}}^{X} & u_{4,\mb{k}}^{X}
      \end{array}
      \right),    
    \end{equation}
    by simply setting the $X$ components in $\bs{U}_{\mb{k}}$ to be zero. 
    The $\nu$-th band component of the spectral weight $\bs{\rho}_{\nu,\mb{k}}$ 
    in the Hubbard-I approximation is simply obtained as 
    \begin{equation}
      \rho_{\nu,\mb{k}}^{\alpha \beta} = (u_{\nu,\mb{k}}^{\alpha})^*  u_{\nu,\mb{k}}^{\beta}, 
    \end{equation}
    where $\alpha,\beta=A,B,$ and $H$. Notice that the unitarity of $\bs{U}_{\mb{k}}$ ensures the 
    spectral weight sum rule of the Hubbard-I approximation in Eq.~(\ref{sumrule}). 
  
  We now introduce the sublattice indexes $a$ and $b$ such that 
  $A$ and $H$ sites belong to $a$ sublattice, and $B$ and $X$ sites belong to $b$ sublattice. 
  By rearranging the rows and columns of the Hamiltonian matrix in Eq.~(\ref{eq.H_H-I}), 
  the effective Hamiltonian is  
  \begin{eqnarray}
    \label{eq.H_H-I2}
    \mcal{H}_{\text{H-I}} &=& 
    \sum_{\mb{k} \s} 
    \left(
      \tilde{c}_{\mb{k} \s A}^\dag \, 
      \tilde{c}_{\mb{k} \s H}^\dag \, 
      \tilde{c}_{\mb{k} \s B}^\dag \, 
      \tilde{c}_{\mb{k} \s X}^\dag 
      \right) \nonumber\\
    & & \quad\quad\quad \times \left(
      \begin{array}{cc}
        \bs{O} & \bs{T}_{\mb{\mb{k}}}  \\
        \bs{T}_{\mb{\mb{k}}}^\dag & \bs{O} 
      \end{array}
      \right)  
    \left(
      \begin{array}{c}
        \tilde{c}_{\mb{k} \s A} \\
        \tilde{c}_{\mb{k} \s H} \\ 
        \tilde{c}_{\mb{k} \s B} \\ 
        \tilde{c}_{\mb{k} \s X} 
      \end{array}
      \right),
  \end{eqnarray}
  where 
  \begin{equation}
    \label{eq.Tk}
    \bs{T}_{\mb{k}} = 
    \left(
      \begin{array}{cc}
        \gamma_{\mb{k}} & 0 \\
        t_{sp} & U/2
      \end{array}
      \right) 
  \end{equation}
  represents the hopping between sites on different sublattices. 

  We can now show that $\mcal{H}_{\text{H-I}}$ is chiral symmetric. 
  Let us define 
  $|a|=|A|+|X|$ and 
  $|b|=|B|+|H|$ 
  as the number of orbitals belonging to $a$ and $b$ sublattices, respectively, and 
  \begin{equation}
  \bs{H}_{\mb{k}} = \left(
      \begin{array}{cc}
        \bs{O} & \bs{T}_{\mb{\mb{k}}}  \\
        \bs{T}_{\mb{\mb{k}}}^\dag & \bs{O} 
      \end{array}
      \right)  \label{eq:H_k}
  \end{equation}
  is a $4 \times 4$ matrix in Eq.~(\ref{eq.H_H-I2}). 
  Then, following the argument given by Hatsugai {\it et al.}~\cite{Hatsugai2013}, 
  $\mcal{H}_{\text{H-I}}$ is said to be chiral symmetric if a 
  matrix $\bs{\Gamma}$ exists such that 
  \begin{eqnarray}
    \label{eq.chiral_1}
    &&\left\{\bs{H}_{\mb{k}},\bs{\Gamma} \right\} 
    = \bs{H}_{\mb{k}} \bs{\Gamma}  + \bs{\Gamma} \bs{H}_{\mb{k}} 
    = \bs{O}, \\
    \label{eq.chiral_2}
    &&(\bs{\Gamma})^2=\bs{I}, \, {\rm and}\\
    \label{eq.chiral_3}
    &&{\rm Tr}[ \bs{\Gamma} ] = |a|-|b|, 
  \end{eqnarray}
  where $\bs{O}$ ($\bs{I}$) is a null (unit) matrix. 
  Equations~(\ref{eq.chiral_1}) and (\ref{eq.chiral_2}) remind us 
  the Dirac matrices in relativistic quantum mechanics, 
  although the Dirac matrices must be traceless, 
  instead of Eq.~(\ref{eq.chiral_3}).
  We can easily find that for any $\mb{k}$ 
  \begin{equation}
    \bs{\Gamma}=
    \left(
      \begin{array}{cc}
        \bs{I} & \bs{O} \\
        \bs{O} & -\bs{I} 
      \end{array}
      \right)
    \label{eq.co}
  \end{equation} 
  is the matrix which defines the chiral symmetry of $\mcal{H}_{\text{H-I}}$.  
  The $\bs{\Gamma}$ matrix is the matrix representation of chiral operator and 
  represents the basis transformation,  
  $\tilde{c}_{\mb{k} \s A (H)} \to \tilde{c}_{\mb{k} \s A (H)}$ and 
  $\tilde{c}_{\mb{k} \s B (X)} \to -\tilde{c}_{\mb{k} \s B (X)}$.  
  Equation~(\ref{eq.chiral_1}) or equivalently $\bs{\Gamma} \bs{H}_{\mb{k}} \bs{\Gamma}^{-1} = - \bs{H}_{\mb{k}}$  
  thus implies that $\mcal{H}_{\text{H-I}}$ changes the sign by this transformation. 
  Notice also that $\bs{\Gamma}$ in Eq.~(\ref{eq.co}) is a traceless matrix, 
  i.e., ${\rm Tr}[\bs{\Gamma}] = |a| - |b| = 0$ in Eq.~(\ref{eq.chiral_3}). 

  It should be recalled here  that ${\rm Tr}[\bs{\Gamma}] = |a|-|b|$ represents 
  the difference of the number of sites belonging to $a$ and $b$ sublattices, 
  and gives the number of zero-energy states, 
  as first pointed out by Lieb~\cite{Lieb} (see also Ref.~\cite{Hatsugai2013}). 
  Indeed, we can find a $3 \times 3$ matrix $\bs{\Gamma}$ 
  even for the periodic Anderson model in the noninteracting limit, i.e., $\mcal{H}_0$ 
  in Eq.~(\ref{h_free}), which satisfies 
  Eqs.~(\ref{eq.chiral_1})--(\ref{eq.chiral_3}), but its trace is 
  ${\rm Tr}[\bs{\Gamma}] = |a| - |b| = 1$. 
  This immediately indicates the presence of the flat band due to the sublattice imbalance, 
  as already discussed in Sec.~\ref{Sec:nonint}. 
  The tracelessness of $\bs{\Gamma}$ in Eq.~(\ref{eq.co}) for $\mcal{H}_{\text{H-I}}$ 
  thus implies that 
  these {\it trivial} zero-energy states 
  are absent, which is similar to the cases of the pure graphene model and also 
  the relativistic particle in the Dirac equation, where the chiral symmetry is preserved. 

  It is now easy to show that the ``non-trivial'' zero-energy states exist only at $K$ and $K'$ points 
  for $\mcal{H}_{\text{H-I}}$ as long as the electron correlation $U$ is finite. 
  Since $t_{sp}$ and $U$ are independent of $\mb{k}$ 
  and $\gamma_{\mb{k}} = 0$ only at $K$ and $K'$ points, 
  we can readily find that 
  
  \begin{eqnarray}
    {\rm rank}\,\bs{T_{\mb{k}}} = \left\{
    \begin{array}{ll}
      1 \quad \text{($\mb{k} = K, K'$)}  \\
      2 \quad \text{(otherwise)} 
    \end{array}\right.
  \end{eqnarray} 
  provided that $t_{sp}$ and $U$ are both finite. 
  It is then immediately followed that 
  \begin{eqnarray}
    \label{rankH}
          {\rm rank}\,\bs{H_{\mb{k}}} = \left\{
          \begin{array}{ll}
            2 \quad \text{($\mb{k} = K, K'$)}  \\
            4 \quad \text{(otherwise).} 
          \end{array}\right.           
  \end{eqnarray} 
  Equation~(\ref{rankH}) therefore guarantees 
  the existence of two zero-energy modes at 
  $K$ ($K'$) point, 
  which represent the point contact of the single-particle excitations exactly 
  at Fermi energy. 
  In other words, 
  the finite electron correlation $U$ and the chiral symmetry of ${\mcal H}_{\text{H-I}}$ 
  with $\Tr[{\mb\Gamma}]=0$ 
  permit the point contacts of the single-particle excitations to appear only at $K$ and $K'$ points.  
  We should note that the argument given here is a direct extension of the pure graphene model~\cite{Hatsugai2014} 
  to the four orbital model ${\mcal H}_{\text{H-I}}$.

\section{Discussion} \label{sec:dis}

First, we should remark on the Hubbard-I approximation which has been repeatedly proved to 
successfully reproduce qualitatively and sometimes 
quantitatively the results obtained by the CPT for the PM state of the periodic Anderson 
model studied here. 
To understand the success of the Hubbard-I approximation, we should recall 
that there exists the flat band in the noninteracting limit, which 
is exactly half-filled. This flat band structure prohibits any perturbative 
treatment of $U$ since even a small $U$ should be regard as the strong correlation. This explains why the 
Hubbard-I approximation, which is usually a good approximation in the atomic 
(i.e., strong coupling) limit, gives the satisfactory results even for small $U$. 

Second, 
the emergent massless Dirac quasiparticles found here should be sharply contrasted to the 
recently discussed massless Dirac dispersion generated by band engineering~\cite{Ishizuka2012,Lin2013}. 
Our finding differs  from the previous reports in the following aspects: 
(i) while the electron correlation induces the massless Dirac quasiparticles in our case, 
breaking the spatial symmetry is essential to generate the massless Dirac dispersion in the band engineering, 
and  
(ii) the Dirac point appears exactly at the Fermi energy in our case, 
but it is generally away from the Fermi energy in the band engineered Dirac dispersion. 
To the best of our knowledge, this is the first example of 
the emergent massless Dirac quasiparticles due to dynamical electron correlations 
without breaking any spatial symmetry.

Third, let us briefly discuss the experimental implications of our results. 
We have studied the half-depleted periodic Anderson model on the 
honeycomb lattice at half filling, 
which can be considered as the simplest model for the single-side hydrogenated graphene. 
Recently, Ray {\it et al.}~\cite{Ray2014} 
reported a ferromagnetism in a partially hydrogenated graphene 
on the graphite substrate. 
Their observation of the ferromagnetism is consistent with our ground state
calculations. Although Lieb's theorem can not directly applied to the periodic 
Anderson model studied here, we have shown in Sec.~\ref{sec:pd} and Appendix~\ref{appsec:ed} that 
the FM ground state found in our calculations is smoothly connected to the Lieb-Mattis type ferromagnetism. 
Therefore, we attribute the ferromagnetism observed experimentally to the Lieb-Mattis type ferromagnetism. 

Another possible experiment which is relevant to our calculations is  
a graphene sheet on transition metal substrates. 
Varykhalov {\it et al.}~\cite{Varykhalov2012} reported 
angle-resolved photoemission spectroscopy (ARPES) experiments for 
graphene deposited on Ni(111) and Co(0001) surfaces. 
In these systems, the sublattice symmetry of graphene is apparently broken. 
This is because the carbon atoms on $B$ sublattice of the graphene sheet 
locate on top of the transition metal atoms of the substrate, whereas 
the carbon atoms on $A$ sublattice are placed on top of the interstitial sites of 
the transition metal atoms. 
Therefore, the carbon $2p_z$ orbitals on $B$ sublattice 
hybridize strongly with the transition metal $3d$ orbitals, but   
the hybridization between the carbon atoms on $A$ sublattice and the transition metal atoms 
is rather weak. 
In spite of the broken sublattice symmetry, 
they have observed in their ARPES experiments the linearly dispersing single-particle excitations 
with the Dirac point 
about 2.8 eV below the Fermi energy~\cite{Varykhalov2012}. 
The deviation of the Dirac point from the Fermi energy 
might be understood 
simply as a consequence of the electron transfer from the substrate.  
Since the most simplest model for these systems is the half-depleted periodic Anderson model studied here, 
their observation can be understood as the emergent massless Dirac quasiparticles induced by  
the electron correlation of transition metals. 
However, more detailed study is highly desired for quantitative comparison. 

\section{Summary} \label{sec:sum}

    Using the VCA and the CPT, we have studied the finite temperature phase diagram of the half-depleted 
    periodic Anderson model at half-filling for a model of graphone, i.e., the single-side hydrogenated graphene. 
    We have found that the ground state is FM as long as the electron correlation $U$ on the hydrogen 
    impurity sites is finite. 
    Although the single-particle excitations with the same spin are gapped, the quasiparticle dispersions 
    with the opposite spins touch at $K$ and $K'$ points. Therefore, this FM state is semi-metallic. 
    We have discussed the relevance of Lieb's theorem to the periodic Anderson model, 
    and shown, with the help of numerically exactly diagonalizing small clusters, 
    that the FM ground state found here is smoothly connected to the Lieb-Mattis type ferromagnetism. 
    We have also shown in the strong coupling limit that the FM state displays the linear spin wave dispersion at zero 
    temperature, rather than the quadratic spin wave dispersion often observed in the FM state. 
    This is simply because of the peculiar Dirac like electron energy dispersion of the conduction band. 
    However, we have found that the spin wave dispersion becomes quadratic at 
    finite temperatures, thus implying that the FM state is stable only at zero temperature, consistent
    with Mermin-Wagner theorem. 
    
    Indeed, we have found using the VCA that the FM state is fragile against thermal fluctuations,
    and the finite temperature phase diagram is dominated by the PM phase. 
    More surprisingly, our CPT calculations have revealed that the massless Dirac quasiparticles emerge 
    at $K$ and $K'$ points
    with the Dirac points exactly at the Fermi energy, once the electron correlation $U$ is introduced in the PM state. 
    This should be contrasted with the quadratic quasiparticle dispersions in the FM phase.     
    We have shown that the emergent massless Dirac quasiparticles in the PM phase 
    can be reproduced in the Hubbard-I approximation. Moreover, we have found that the formation of 
    the emergent massless Dirac quasiparticles is accompanied with the spectral weight redistribution of 
    the single-particle excitations, involving a large energy scale of $U$. 
    In fact, we have found in both CPT and Hubbard-I approximation that the 
    single-particle spectral weight $\rho_K^{BB}$ for $B$ orbital at the Dirac point is proportional to the square of the 
    Dirac Fermi velocity $v_{\rm F}$, i.e., $\rho_K^{BB}=(v_{\rm F}/v_0)^2$, where $v_{\rm F}$ is zero when $U=0$ 
    and monotonically increases with $U$. 
    This universal relation expresses that the involvement of $B$ orbital in the low-energy excitations 
    is essential for the formation of the emergent massless Dirac quasiparticles. 
    Constructing the effective quasiparticle Hamiltonian, we have argued that 
    the Dirac cones with the point contacts at $K$ and $K'$ points are protected by the electron correlation $U$. 
    Our finding therefore represents the first example of the emergence of massless Dirac quasiparticles induced 
    by the electron correlation without breaking any spatial symmetry. 
    
\acknowledgements
The computations have been done using the RIKEN Integrated Cluster of Clusters (RICC) facility and 
RIKEN supercomputer system (HOKUSAI GreatWave). 
This work has been supported in part by Grant-in-Aid for Scientific Research from MEXT Japan
under the grant Nos. 24740269 and 26800171, and by RIKEN iTHES Project and Molecular Systems. 
Q. Z. also acknowledges the National Natural Science Foundation of China (11204265, 11474246), 
the Natural Science Foundation of Jiangsu Province (BK2012248), 
and the College Natural Science Research Project of Jiangsu Province (13KJ430007). 

\appendix

\section{Linear spin wave analysis for the FM state in the strong coupling limit}\label{appsec:sw}

In this Appendix, we consider the large $U$ limit (i.e., Kondo limit), where a single electron is localized 
on each hydrogen impurity site, 
forming a localized spin with spin $S=1/2$. Recall here that the local electron density is always one at each site 
when the particle-hole symmetry is preserved, i.e., $\epsilon_H=-U/2$. 
In this limit, the periodic Anderson model is mapped onto an effective Kondo lattice model 
described by the following Hamiltonian: 
\begin{eqnarray}
  {\cal H}_{\rm KL} &=& -t\sum_{i}\sum_{\bs \delta} \sum_\sigma\left(c_{\mb{r}_i\sigma A}^\dag c_{\mb{r}_i+{\bs\delta}\sigma B} + {\rm H.c.}\right) \nonumber\\
  &+& 
  J_{\rm K}{\sum_i} {\mb S}_{\mb{r}_i}\cdot{\mb s}_{\mb{r}_i,B}, 
  \label{H_KL}
\end{eqnarray}
where 
$J_{\rm K}=8t^2_{sp}/U$, 
${\mb s}_{\mb{r}_i,B}=\frac{1}{2}\sum_{s,s'}c^\dag_{\mb{r}_isB}\left({\vec \sigma}\right)_{ss'}c_{\mb{r}_is'B}$ ($\vec\sigma$: Pauli matrix vector) 
is the spin operator of $B$ orbital, and ${\mb S}_{{\mb r}_i}$ is the spin-1/2 operator located 
at the hydrogen impurity site in the $i$-th unit cell (see Fig.~\ref{lattice}). 
We first analyze the RKKY interaction~\cite{RKKY}.  
Next, we analyze the magnetic excitations 
within the liner spin wave theory to 
discuss the stability of the FM state at finite temperatures. 

\subsection{RKKY interaction}\label{appsec:rkky}

By integrating out the conduction electron degrees of freedom, the magnetic coupling between the localized 
spins on the hydrogen impurity sites is described by the following spin Hamiltonian: 
\begin{equation}\label{h_rkky}
  {\cal H}_{\rm RKKY} = \frac{1}{2}{\sum_i}\sum_{\mb R} J_{\mb R} {\mb S}_{\mb{r}_i}\cdot{\mb S}_{{\mb r}_i+{\mb R}}, 
\end{equation}
where ${\mb R} = n_1{\mb d}_1+n_2{\mb d}_2$ ($n_1$ and $n_2$: integer) with 
${\mb d}_1$ and ${\mb d}_2$ being the primitive translational vectors (see Fig.~\ref{lattice}). 
The RKKY interaction $J_{\mb R}$ mediated by the conduction electrons is evaluated as
\begin{eqnarray}\label{rkky}
  J_{\mb{R}}=
  -J_{\mr{K}}^{2}\int_{0}^{\beta}d\tau 
  \left \langle s_{{\mb r}_i,B}^{-}(\tau)s_{{\mb r}_i+{\mb{R}},B}^{+}(0) \right \rangle,  
\end{eqnarray}
where $\beta=1/T$ is the inverse temperature, 
$s_{{\mb r}_i,B}^{-}=c_{{\mb r}_i\downarrow B}^{\dagger}c_{{\mb r}_i\uparrow B}$, and 
$s_{{\mb r}_i,B}^{+}=c_{{\mb r}_i\uparrow B}^{\dagger}c_{{\mb r}_i\downarrow B}$~\cite{RKKY,Saremi2007}. 
Given the following Hamiltonian for the conduction band, i.e., the pure graphene model,    
\begin{equation}
  {\cal H}_{\rm KL}^0=-t\sum_{i}\sum_{\bs \delta} \sum_\sigma\left(c_{\mb{r}_i\sigma A}^\dag c_{\mb{r}_i+{\bs\delta}\sigma B} + {\rm H.c.}\right), 
\end{equation} 
$s_{{\mb r}_i,B}^{\pm}(\tau)=e^{\tau {\cal H}_{\rm KL}^0}s_{{\mb r}_i,B}^{\pm}e^{-\tau {\cal H}_{\rm KL}^0}$ and 
$\langle\cdots\rangle$ in Eq.~(\ref{rkky}) represents ${\rm Tr}\left(e^{-\beta {\cal H}_{\rm KL}^0}\cdots\right)/{\rm Tr}e^{-\beta {\cal H}_{\rm KL}^0}$.  
Notice here that the chemical potential $\mu$ is zero for $n=1$. 

Applying Wick's theorem, the only non-zero term in Eq.~(\ref{rkky}) is 
$\langle c^\dag_{\mb{r}_i\dn B}(\tau) c_{\mb{r}_i \up B}(\tau) c^\dag_{\mb{r}_i+\mb{R} \up B} c_{\mb{r}_i+\mb{R} \dn B} \rangle = 
 \langle c^\dag_{\mb{r}_i\dn B}(\tau) c_{\mb{r}_i+\mb{R} \dn B}    \rangle 
 \langle c_{\mb{r}_i \up B}(\tau)    c^\dag_{\mb{r}_i+\mb{R} \up B}\rangle$ because the spin and the number of electrons 
 are conserved. 
 Therefore, the RKKY interaction is now written as
\begin{eqnarray}\label{j_r}
  J_{\mb{R}}&=&-\frac{J_{\mr{K}}^{2}}{N_a^{2}} \int_0^\beta d\tau
  \left( \sum_{\mb{k}} \langle c^\dag_{\mb{k} \dn B}(\tau) c_{\mb{k} \dn B} \rangle e^{i \mb{k} \cdot \mb{R}} \right)
  \nonumber \\
  & &\quad\quad\quad \times
  \left( \sum_{\mb{k}'} \langle c_{\mb{k}'\up B}(\tau) c^\dag_{\mb{k}'\up B} \rangle e^{-i \mb{k}' \cdot \mb{R}} \right), 
\end{eqnarray}
where 
\begin{equation}  
  c_{{\mb r}_i \s B} =  \frac{1}{\sqrt{N_a}}\sum_{\mb{k}} c_{\mb{k} \s B} e^{i \mb{k} \cdot \mb{r}_i}
\end{equation}
and we have used that 
$\langle c_{\mb{k}\s B}^\dag c_{\mb{k}'\s B} \rangle = \langle c_{\mb{k}\s B}^\dag c_{\mb{k}\s B} \rangle \delta_{\mb{k} \mb{k'}}$.

By introducing 
the canonical transformation 
\begin{equation}
  \label{unitary}
  \left(
    \begin{array}{c}
      \alpha_{\mb{k}\s} \\
      \beta_{\mb{k}\s} 
    \end{array}
  \right)
  = 
  \frac{1}{\sqrt{2}}
  \left(
    \begin{array}{cc}
      1 &  e^{i \theta_{\mb{k}}} \\
      1 & -e^{i \theta_{\mb{k}}} 
    \end{array}
  \right)  
  \left(
    \begin{array}{c}
      c_{\mb{k} \s A} \\
      c_{\mb{k} \s B} 
    \end{array}
  \right),
\end{equation}
where $e^{i\theta_{\mb{k}}} = \gamma_{\mb{k}}/|\gamma_{\mb{k}}|$ and $\gamma_{\mb{k}}$ is 
given in Eq.~(\ref{eq:gamma_k}), 
we can readily diagonalize ${\cal H}_{\rm KL}^0$ as 
\begin{equation}
  {\cal H}_{\rm KL}^0 = \sum_{\mb{k}\s} \left(|\gamma_{\mb{k}}| \alpha_{\mb{k}\s}^\dag \alpha_{\mb{k}\s} - |\gamma_{\mb{k}}| \beta_{\mb{k}\s}^\dag \beta_{\mb{k}\s} \right).
\end{equation}
Now the average of any operators composed of $c^\dag_{{\mb k}\sigma\alpha}$ and $c_{{\mb k}\sigma\alpha}$ 
can be expressed in terms of operators $\alpha_{{\mb k}\sigma}$ and $\beta_{{\mb k}\sigma}$, e.g.,
 $\langle c^\dag_{\mb{k}\s B} c_{\mb{k}\s B}\rangle 
= \langle \alpha_{\mb{k}\s}^\dag \alpha_{\mb{k}\s} + \beta_{\mb{k}\s}^\dag \beta_{\mb{k}\s} \rangle /2$.
This enables us to use the following equations: 
\begin{eqnarray}
  \langle \alpha_{\mb{k}\s}^\dag (\tau) \alpha_{\mb{k}\s} \rangle &=& e^{ \tau |\gamma_{\mb{k}}|} n_{\mr{F}}( |\gamma_{\mb{k}}|),\\
  \langle \beta_{\mb{k}\s}^\dag (\tau) \beta_{\mb{k}\s}   \rangle &=& e^{-\tau |\gamma_{\mb{k}}|} n_{\mr{F}}(-|\gamma_{\mb{k}}|),\\
  \langle \alpha_{\mb{k}\s} (\tau) \alpha^\dag_{\mb{k}\s} \rangle &=& \langle \beta_{\mb{k}\s}^\dag (\tau) \beta_{\mb{k}\s} \rangle \nonumber \\
  &=& e^{-\tau |\gamma_{\mb{k}}|} \left[1- n_{\mr{F}}( |\gamma_{\mb{k}}|) \right] ,\\
  \langle \beta_{\mb{k}\s} (\tau) \beta^\dag_{\mb{k}\s}   \rangle &=& \langle \alpha_{\mb{k}\s}^\dag (\tau) \alpha_{\mb{k}\s} \rangle \nonumber \\
  &=& e^{ \tau |\gamma_{\mb{k}}|} \left[1- n_{\mr{F}}(-|\gamma_{\mb{k}}|) \right] , 
\end{eqnarray}
where $n_{\mr{F}}(E) = \left( e^{\beta E} + 1 \right)^{-1}$ is the Fermi distribution function.

We can now explicitly perform the $\tau$ integral in Eq.~(\ref{j_r}) 
and finally obtain that 
\begin{eqnarray}
  J_{\mb{R}} &=&
  - \frac{J_{\mr{K}}^2}{2N_a^2} \sum_{\mb{k},\mb{k'}} 
  e^{i (\mb{k-k}')\cdot \mb{R}}
  \left(
    \frac{n_{\mr{F}}(-|\gamma_{\mb{k}}|) - n_{\mr{F}}( |\gamma_{\mb{k}'}|)} {|\gamma_{\mb{k}}| + |\gamma_{\mb{k}'}|} \right.\nonumber\\ 
    && \quad\quad\quad\quad\quad -
  \left.  \frac{n_{\mr{F}}( |\gamma_{\mb{k}}|) - n_{\mr{F}}( |\gamma_{\mb{k}'}|)} {|\gamma_{\mb{k}}| - |\gamma_{\mb{k}'}|}
  \right), 
  \label{eq.rkkyFT}
\end{eqnarray}
where we have used that $n_{\mr{F}}(-E)=1-n_{\mr{F}}(E)=e^{\beta E}n_{\mr{F}}(E)$. 
Notice here 
that the phase factor $e^{i \theta_{\mb{k}}}$ in Eq.~(\ref{unitary}) 
does not appear in Eq.~(\ref{eq.rkkyFT}) 
because the RKKY interaction considered here acts for spins only on the same $B$ sublattice.  
The phase factor $e^{i \theta_{\mb{k}}}$ becomes relevant when we consider the RKKY interaction 
for spins on different sublattices. 
At zero temperature, only the first term in Eq.~(\ref{eq.rkkyFT}) is finite and thus the RKKY interaction 
at zero temperature is given as 
\begin{eqnarray}
  \lim_{T \rightarrow 0} 
  J_{\mb{R}}=-\frac{J_{\mr{K}}^{2}}{2N_a^{2}}
  \sum_{\mb{k,k}'} e^{i(\mb{k-k}')\cdot{\mb{R}}} \frac{1}{|\gamma_{\mb{k}}|+|\gamma_{\mb{k}'}|}.
  \label{eq.rkky}
\end{eqnarray}
The RKKY interaction $J_{\mb R}$ at zero temperature is thus long ranged and 
it has been shown 
that (i) $J_\mb{ R}<0$ for all $\mb{ R}$, i.e., FM coupling,  
and (ii) the asymptotic behavior of $J_{\mb R}$ is $\sim |\mb{ R}|^{-3}$~\cite{Saremi2007}. 
The RKKY interaction $J_{\mb R}$ at zero temperature is thus long ranged and we can readily show

\subsection{Linear spin wave approximation}\label{appsec:tc-sw}

Let us now analyze the spin wave dispersion of the FM state for the effective spin Hamiltonian 
${\cal H}_{\rm RKKY}$ within the linear spin wave approximation. 
Introducing the Holstein-Primakoff transformation to the spin operators
\begin{eqnarray}
  S^z_{{\mb r}_i} &=& S - a_{{\mb r}_i}^\dag a_{{\mb r}_i}, \\
  S^-_{{\mb r}_i} &=& \sqrt{2S} a_{{\mb r}_i}^\dag \left(1-\frac{a_{{\mb r}_i}^\dag a_{{\mb r}_i}}{2S} \right)^{\frac{1}{2}}, \\
  S^+_{{\mb r}_i} &=& \sqrt{2S}       \left(1-\frac{a_{{\mb r}_i}^\dag a_{{\mb r}_i}}{2S} \right)^{\frac{1}{2}} a_{{\mb r}_i} ,
\end{eqnarray}
where $a_{{\mb r}_i}^\dag$ is a bosonic creation operator, i.e., $[a_{{\mb r}_i}^\dag,a_{{\mb r}_j}]=\delta_{{\mb r}_i{\mb r}_j}$,  
the spin Hamiltonian is now written in the linear spin wave approximation as 
\begin{equation}
  H_{\mr{RKKY}}^{\mr{sw}} = \sum_{i,\mb{R}} J_{\mb{R}} 
  \left( 
    \frac{S^2}{2} + S a_{{\mb r}_i}^\dag a_{{{\mb r}_i}+\mb{R}} - S a_{{\mb r}_i}^\dag a_{{\mb r}_i} 
  \right)  
\end{equation}
with keeping only up to quadratic terms in $a_{{\mb r}_i}^\dag$ and $a_{{\mb r}_i}$.  
This Hamiltonian is easily diagonalized in the momentum space as
\begin{equation}\label{h_sw}
  H_{\mr{RKKY}}^{\mr{sw}} = \frac{N_a J(\mb{0}) S^2}{2} + \sum_{\mb{q}} \Omega_{\mb{q}} a_{\mb{q}}^\dag a_{\mb{q}},
\end{equation} 
where $S=1/2$, $a_{\mb{q}} = \frac{1}{\sqrt{N_a}}\sum_i e^{i \mb{q}\cdot \mb{r}_i} a_{{\mb r}_i}$, and 
$J(\mb{q}) = \sum_{\mb{R}} J_{\mb{R}} e^{-i \mb{q} \cdot \mb{R}}$. 
The FM spin wave dispersion $\Omega_{\mb q}$ in Eq.~(\ref{h_sw}) is thus obtained as
\begin{equation}
  \Omega_{\mb{q}}= 
  S \left( J(\mb{q})-J(\mb{0}) \right) = 
  \frac{1}{2} \sum_{{\mb{R}}}
  J_{{\mb{R}}}\left(e^{-i\mb{q}\cdot{\mb{R}}} - 1\right).
  \label{eq.spinwave}
\end{equation} 
By substituting Eq.~(\ref{eq.rkkyFT}) into $J(\mb q)$ and $J(\mb 0)$ in Eq.~(\ref{eq.spinwave}), 
we explicitly obtain that 
\begin{eqnarray}
  J(\mb{q}) &=& -\frac{J_{\mr{K}}^{2}}{2N_a}\sum_{\mb{k}} 
  \left(
    \frac{n_{\mr{F}}(-|\gamma_{\mb{k}}|) - n_{\mr{F}}( |\gamma_{\mb{k+q}}|)} {|\gamma_{\mb{k}}| + |\gamma_{\mb{k+q}}|} \right. \nonumber \\
    &&\quad\quad\quad\quad - \left. 
    \frac{n_{\mr{F}}( |\gamma_{\mb{k}}|) - n_{\mr{F}}( |\gamma_{\mb{k+q}}|)} {|\gamma_{\mb{k}}| - |\gamma_{\mb{k+q}}|}
  \right)
  \label{eq.Jq}
\end{eqnarray}
and 
\begin{eqnarray}
  J(\mb{0}) &=& 
  - \frac{J_{\mr{K}}^{2}}{2N_a}\sum_{\mb{k}} 
  \left(
    \frac{n_{\mr{F}}(-|\gamma_{\mb{k}}|) - n_{\mr{F}}( |\gamma_{\mb{k}}|)} {2|\gamma_{\mb{k}}|} \right. \nonumber \\
    && \quad\quad\quad\quad 
    + \left. \beta n_{\mr{F}}(|\gamma_{\mb{k}}|) n_{\mr{F}}(-|\gamma_{\mb{k}}|)
  \right),
  \label{eq.J0}
\end{eqnarray}
 where 
 $J(\mb{0})$ is regarded as $J(\mb{0}) = \lim_{\mb{q}\rightarrow \mb{0}} J(\mb{q})$ and 
 we have used that $\partial n_{\mr{F}}(E)/\partial E = - \beta n_{\mr{F}}(E) n_{\mr{F}}(-E)$. 
 
\begin{figure}[htb]
  \includegraphics[width=8.0cm,angle=0]{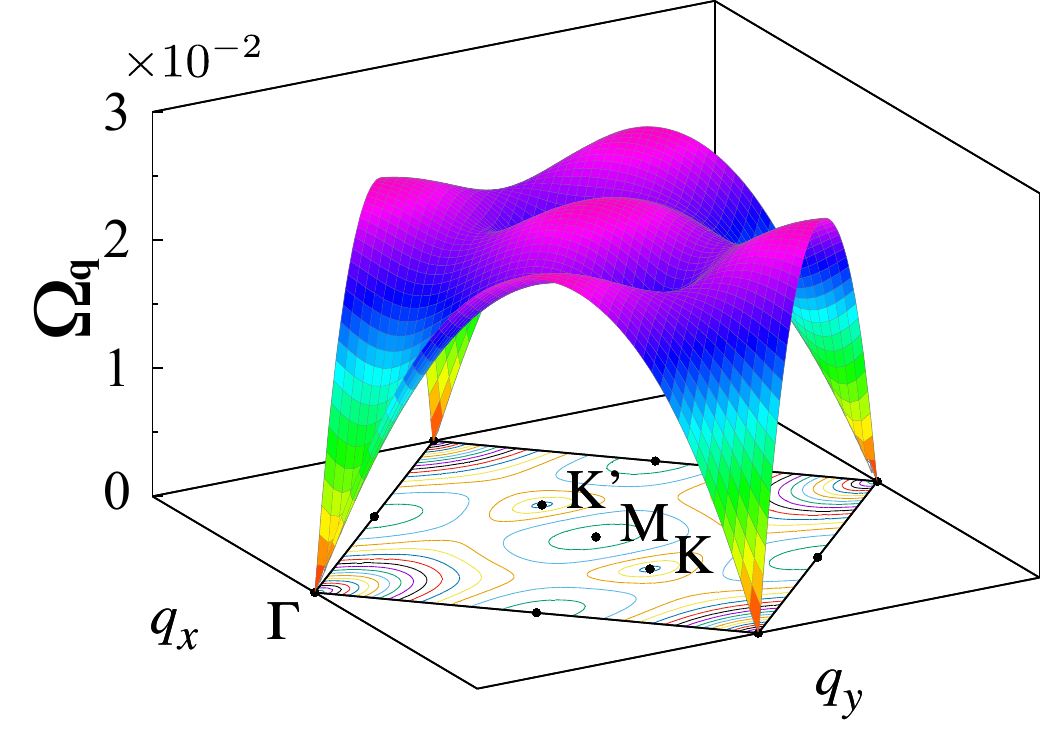}
  \caption{(color online)
    The FM spin wave dispersion for the effective spin Hamiltonian ${\cal H}_{\rm RKKY}$ obtained in 
    the linear spin wave approximation 
    at zero temperature. Here $J_{\mr{K}}$ is set to be one as the energy unit. 
    The black dots in the $\Omega_{\bm q}=0$ plane indicate high symmetric momenta such as 
    $\Gamma$, $M$, $K$, and $K'$ [see Fig.~\ref{disp_nonint}(b)]. 
  } \label{fig:spinwave}
\end{figure}

\begin{figure}[htb]
  \includegraphics[width=7cm,angle=0]{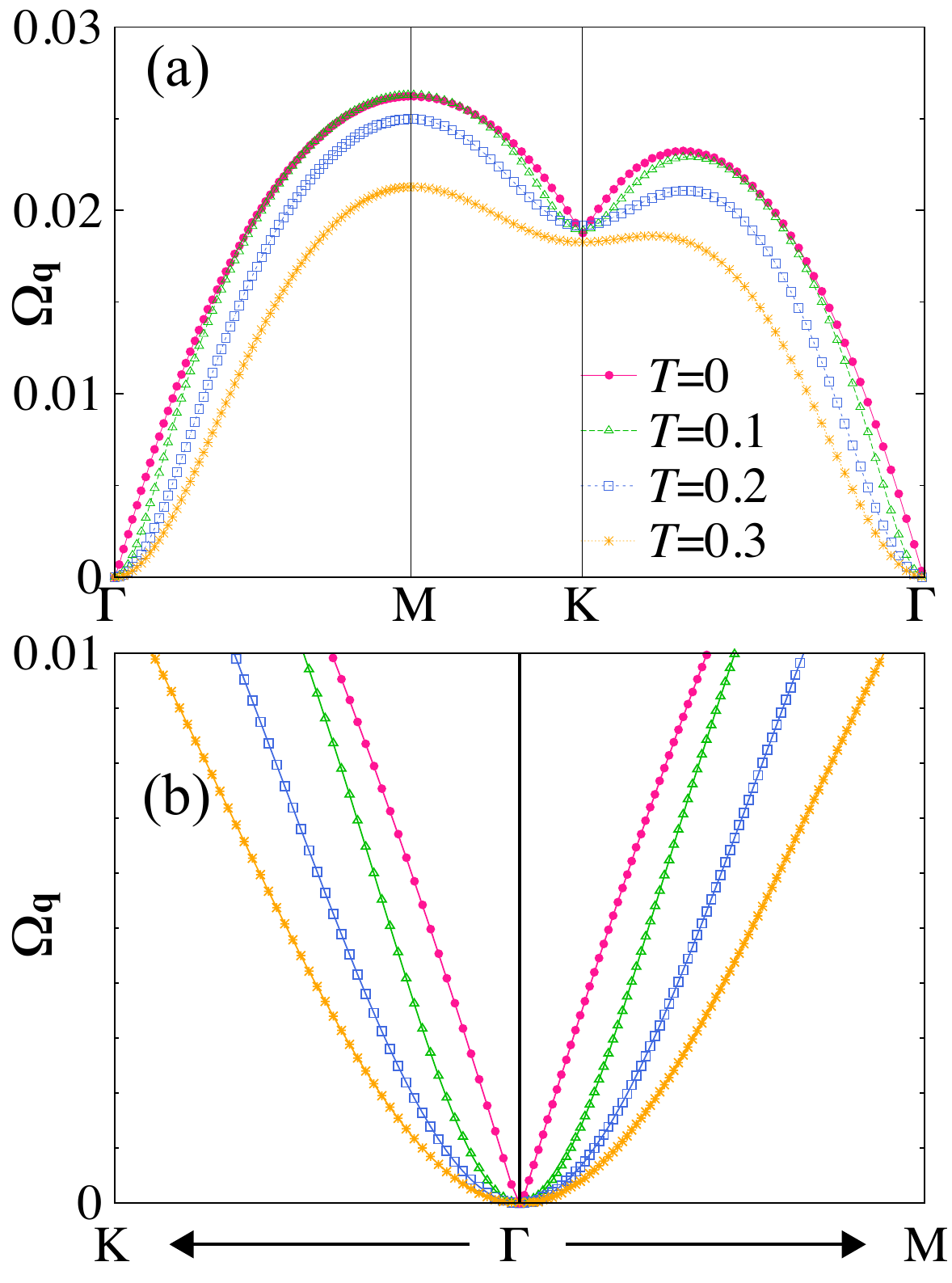}
  \caption{(color online)
    (a) Temperature dependence of the FM spin wave dispersion $\Omega_{\mb q}$ for the effective spin Hamiltonian 
    ${\cal H}_{\rm RKKY}$ obtained in the 
    linear spin wave approximation. Here the temperature dependence of the RKKY interaction $J_{\mb{ R}}$ 
    in Eq.~(\ref{eq.rkkyFT}) is explicitly considered. 
    (b) Same as (a) but the enlarged scale at the vicinity of $\Gamma$ point near zero energy. 
    The region of momenta taken in the horizontal axis is $0.5\pi$ in the $\Gamma$-$M$ ($\Gamma$-$K$) 
    direction from $\Gamma$. 
    Temperatures are indicated in (a) and $J_{\rm K}$ is set to be one as the energy unit. 
  } 
  \label{fig:spinwaveFT}
\end{figure}

In the zero temperature limit, the spin wave dispersion is therefore 
\begin{equation}\label{wq0}
  \lim_{T \rightarrow 0} 
  \Omega_{\mb{q}}=
  \frac{J_{\mr{K}}^{2}}{8N_a}\sum_{\mb{k}}
  \left(
    \frac{1}{|\gamma_{\mb{k}}|}-\frac{2}{|\gamma_{\mb{k}}|+|\gamma_{\mb{k+q}}|}
  \right).
\end{equation}
As shown in Fig.~\ref{fig:spinwave} (see also Fig.~\ref{fig:spinwaveFT}), we find that 
the spin wave dispersion $\Omega_{\mb{q}}$ is linear in the long 
wavelength limit, i.e., $|{\mb q}|\to0$, although the ground state is FM with no quantum
fluctuations. The linearity of the spin wave dispersion in the long wavelength limit is simply because of 
the massless Dirac dispersion of the conduction electrons, 
which induces the long range RKKY interaction. 
This should be contrasted to the spin wave dispersion of a FM Heisenberg model with 
a short range interaction, where the spin wave dispersion in the long wavelength limit is quadratic.    

The linear dispersion around $\Gamma$ point implies that the contribution from the thermal excitations 
of the spin wave is 
convergent even in two spatial dimensions. More explicitly, the magnetization $m(T) $ at temperature $T$ 
is evaluated as 
\begin{eqnarray}
m(T)&=&m(0)-\Delta m(T), 
\end{eqnarray}
where 
\begin{eqnarray}
  && \Delta m(T) = \frac{1}{N_a} \sum_{\mb{q}}\frac{1}{e^{\beta \Omega_{\mb{q}}} -1} \nonumber\\
  &\sim& \frac{1}{V_{\mr{BZ}}} \int_0^{2\pi} \dd \theta \int_0^\infty \frac{q \dd q }{e^{\beta c q} -1} = \frac{2 \pi \zeta(2)}{V_{\mr{BZ}}} \frac{T^2}{c^2} 
  \label{eq.dm}
\end{eqnarray}
with $\zeta(2) = \pi^2/6$ (the Riemann zeta function). 
Here, $m(0)=1/2$ is the magnetic moment for the fully polarized FM state at $T=0$ and  
the long wavelength approximation of the linear dispersion relation, i.e., $\Omega_{\mb{q}} \simeq c|{\mb q}|$, 
for the spin wave dispersion is used. 
It is thus tempting to conclude that the critical temperature $T_{\rm C}$ for the FM order is 
finite even in two dimensions and  is proportional to the magnon velocity $c$. 

However, it should be reminded that the RKKY interaction $J_{\mb{R}}$ in Eq.~(\ref{eq.rkkyFT}) itself 
is temperature dependent and the temperature 
dependence of $J_{\mb{R}}$ has to be considered explicitly when the spin wave dispersion is 
calculated at finite temperatures. 
The results of the finite temperature spin wave dispersion is summarized in Fig.~\ref{fig:spinwaveFT}. 
It is clearly found in Fig.~\ref{fig:spinwaveFT} that the spin wave dispersion $\Omega_{\mb q}$ 
is quadratic ($\Omega_{\mb q} \propto |{\mb q}|^2$) in the long wavelength limit around $\Gamma$ point 
at finite temperatures . 
It is now readily shown that the $q$ integral in $\Delta m(T)$ given in Eq.~(\ref{eq.dm}) is proportional to 
$\zeta(1)$, which is divergent.
Therefore, we can conclude that 
a finite temperature FM transition is impossible and $T_{\rm C}$ should be zero. 

\subsection{Summary and remark on the FM state}

Let us summarize the liner spin wave analysis for the effective spin Hamiltonian ${\cal H}_{\rm RKKY}$ 
and make a remark on the finite FM transition temperature $T_{\rm C}$ obtained by the VCA for the 
periodic Anderson model $\mcal H_{\rm }$ in Sec.~\ref{sec:pd}. 
Since the RKKY interaction is FM and 
long-ranged, i.e., $J_{\mb{R}} \propto |\mb{R}|^{-3}$, at zero temperature~\cite{Saremi2007}, 
the ground state of ${\cal H}_{\rm RKKY}$ is FM and 
the spin wave dispersion in the long wavelength limit is linear ($\sim|{\mb q}|$). 
However, this long range character of the RKKY interaction is true only at zero 
temperature and one can readily show that the RKKY interaction becomes short-ranged at finite temperatures. 
The resulting spin wave dispersion at finite temperatures is quadratic in the long wavelength limit 
($\sim|\mb q|^2$), and therefore 
$T_{\rm C}$ should be zero. After all, the model studied here, the periodic Anderson model,  
only includes short range interactions 
and thus Mermin-Wagner theorem~\cite{Mermin1966} guarantees that the FM instability 
should occur only at zero temperature. 

The finite $T_{\rm C}$ found in the VCA for the periodic Anderson model 
is due merely to a mean-filed like treatment of the electron correlation 
beyond the size of clusters and should be regarded as a temperature where the short range FM 
correlations are developed. Indeed, we have found that $T_{\rm C}$ decreases with increasing the size of 
clusters (see Fig.~\ref{fig:pd}). 
The considerably small $T_{\mr{C}}$ found in the VCA is due to 
the energy scale of the FM instability, namely, 
the exchange splitting, $\sim U( \langle n_{{\mb r}_i\up H} \rangle - \langle n_{{\mb r}_i\dn H}\rangle)$, 
for small $U$ and the RKKY interaction, $\sim t_{sp}^4/U^2 \times$ 
(static spin susceptibility of the conduction band),  
for large $U$.

\section{Application of Lieb's theorem}\label{appsec:lieb}

In this Appendix, we consider a Hubbard model $\mathcal{H}_{\rm HM}$ 
described by the same Hamiltonian 
for the half-depleted periodic Anderson model $\mcal H$ studied in the main text 
except that now the on-site Coulomb repulsion $U_C$ on the carbon conduction 
sites is incorporated, i.e., 
\begin{eqnarray}
  \mathcal{H}_{\rm HM} &=& 
  - t \sum_{i,{\mb\delta},\sigma} \left( c_{{\mb r}_i\sigma A}^{\dagger} c_{{\mb r}_i+{\mb\delta}\sigma B} + {\rm H.c.}\right)\nonumber \\ 
  &+& t_{sp} \sum_{i,\sigma} \left ( c_{{\mb r}_i\sigma B}^{\dagger} c_{{\mb r}_i\sigma H} + {\rm H.c.}\right) \nonumber \\ 
  &+& U \sum_{i} \left(n_{{\mb r}_i\uparrow H}- \frac{1}{2}\right) \left(n_{{\mb r}_i\downarrow H}- \frac{1}{2}\right) \nonumber \\
  &+& U_{C} \sum_{i,\alpha=A,B} \left(n_{{\mb r}_i\uparrow \alpha}-\frac{1}{2}\right) \left(n_{{\mb r}_i\downarrow \alpha} - \frac{1}{2} \right) 
\label{eq:ham1}
\end{eqnarray} 
defined on the lattice shown in Fig.~\ref{lattice}. 
In the following, we analyze the ground state of this Hubbard model $\mathcal{H}_{\rm HM}$ at half filling 
based on Lieb's theorem on a bipartite lattice~\cite{Lieb}. 

Following Lieb's argument in Ref.~\cite{Lieb}, we can show that the ground state of 
$\mathcal{H}_{\rm HM}$
has the following properties: (a) among the possibly degenerate ground states, 
there exists one state which has total spin $S=0$, when 
$U\le0$ and $U_C\le0$ and (b) the ground state is unique 
if $U < 0$ and $U_C < 0$. 
The details of the proof are found in Ref.~\cite{Lieb}. Here, we only note that in the proof 
the matrix $L_{{\mb r}_i} = \left< \bar\alpha \right| n_{{\mb r}_i\sigma} \left| \bar\beta \right>$ defined in Ref.~\cite{Lieb} 
should be replaced by $\left< \bar\alpha \right| (n_{{\mb r}_i\sigma\alpha} - 1/2) \left| \bar\beta \right> $, 
where $\left| \bar\alpha \right>$ and 
$\left| \bar\beta \right>$ represent sets of real space configurations of electrons with spin $\sigma$. 

Let us now map the Hubbard model $\mathcal{H}_{\rm HM}$ 
onto a negative $U$ Hubbard model by the particle-hole transformation 
\begin{eqnarray}
  \bar{c}_{{\mb r}_i\uparrow H} &=& c_{{\mb r}_i \uparrow H}, \quad \bar{c}_{{\mb r}_i\downarrow H} = c^\dag_{{\mb r}_i \downarrow H}, \nonumber \\
  \bar{c}_{{\mb r}_i\uparrow A} &=& c_{{\mb r}_i \uparrow A}, \quad \bar{c}_{{\mb r}_i\downarrow A} = c^\dag_{{\mb r}_i \downarrow A}, \nonumber \\
  \bar{c}_{{\mb r}_i\uparrow B} &=& c_{{\mb r}_i \uparrow B}, \quad \bar{c}_{{\mb r}_i\downarrow B} =-c^\dag_{{\mb r}_i \downarrow B}. \label{eq:trans}
\end{eqnarray}
With this transformation, the Hubbard model ${\mathcal H}_{\rm HM}$ is mapped onto 
\begin{eqnarray}
  \bar{\mathcal{H}}_{\rm HM} &=& 
  - t \sum_{i,{\mb\delta},\sigma} \left( \bar{c}_{{\mb r}_i\sigma A}^{\dagger} \bar{c}_{{\mb r}_i+{\mb \delta}\sigma B} + {\rm H.c.}\right) \nonumber\\
  &+& t_{sp} \sum_{i,\sigma} \left ( \bar{c}_{{\mb r}_i\sigma B}^{\dagger} \bar{c}_{{\mb r}_i\sigma H} + {\rm H.c.}\right)  \nonumber\\
  &-& U \sum_{i} \left(\bar{n}_{{\mb r}_i\uparrow H}- \frac{1}{2}\right) \left(\bar{n}_{{\mb r}_i\downarrow H}- \frac{1}{2}\right) \nonumber \\
  &-& U_{C} \sum_{i,\alpha=A,B} \left(\bar{n}_{{\mb r}_i\uparrow \alpha}-\frac{1}{2}\right) \left(\bar{n}_{{\mb r}_i\downarrow \alpha} - \frac{1}{2}\right),  
  \label{eq:ham2}
\end{eqnarray}
where $\bar{n}_{{\mb r}_i\sigma\alpha}=\bar{c}^\dag_{{\mb r}_i\sigma\alpha}\bar{c}_{{\mb r}_i\sigma\alpha}$. 

Applying Lieb's theorem to the negative $U$ Hubbard model 
${\bar{\mathcal H}}_{\rm HM}$, we can readily show 
that (i) the ground state of ${\mathcal H}_{\rm HM}$ at half filling is unique 
(apart from the trivial spin degeneracy) 
and it has total spin $S = \left( |H| + |A| - |B| \right)/2$ when $U > 0$ and $U_C > 0$ 
and (ii) one of the possibly degenerate ground states of ${\mathcal H}$ (i.e., ${\mathcal H}_{\rm HM}$ 
with $U_C=0$) at half filling has total spin $S = \left( |H| + |A| - |B| \right)/2$ when $U > 0$.

To prove statement (i), we should first notice that the corresponding spin-1/2 Heisenberg model obtained in the limit 
of $U$, $U_C \to \infty$ is defined on the bipartite lattice and thus Lieb-Mattis theorem guarantees 
the unique ground state of this spin-1/2 Heisenberg 
model with total spin $S = \left( |H| + |A| - |B| \right)/2 $~\cite{LM}. 
Applying Lieb's theorem (b) to ${\bar{\mathcal H}}_{\rm HM}$ for $U>0$ and $U_C>0$,  
we can now show that the ground state of ${\mathcal H}_{\rm HM}$ is unique, 
apart from the trivial degeneracy due to the 
spin rotational symmetry, for any finite value of $U_C$ until $U_C \to 0^+$ 
and the total spin of the ground state is $S = \left( |H| + |A| - |B| \right)/2 $~\cite{note10}.
 
When $U_C$ is exactly zero, 
the uniqueness of the ground state of ${\mathcal H}_{\rm HM}$ is no longer guaranteed. 
However, according to Lieb's theorem (a), the state with 
$S = \left( |H| + |A| - |B| \right)/2 $ must be the ground state or one of the possibly degenerate ground states, 
which thus proves statement (ii). In the next Appendix, 
we will show by numerically exactly diagonalizing 
small clusters that indeed the ground state of ${\mathcal H} $ is unique and it has total spin 
$S = \left( |H| + |A| - |B| \right)/2 $ when $U>0$.

\section{Numerically exact diagonalization study of the ground state phase diagram}\label{appsec:ed}

Although Lieb's theorem does not guarantee the unique ground state of ${\mathcal H}$ 
at half filling, here we 
perform numerically exact diagonalization calculations for small clusters to show that the ground state of 
${\mathcal H}_{\rm HM}$ at half filling for $U,\,U_{C}>0$ is smoothly connected to the one even for 
$U_C$ approaching exactly to zero, namely, 
the ground state of ${\mathcal H}$ at half filling is unique with its total spin 
$S = \left( |H| + |A| - |B| \right)/2 $. 

\begin{figure}[htb]
  \includegraphics[width=7.5cm,angle=0]{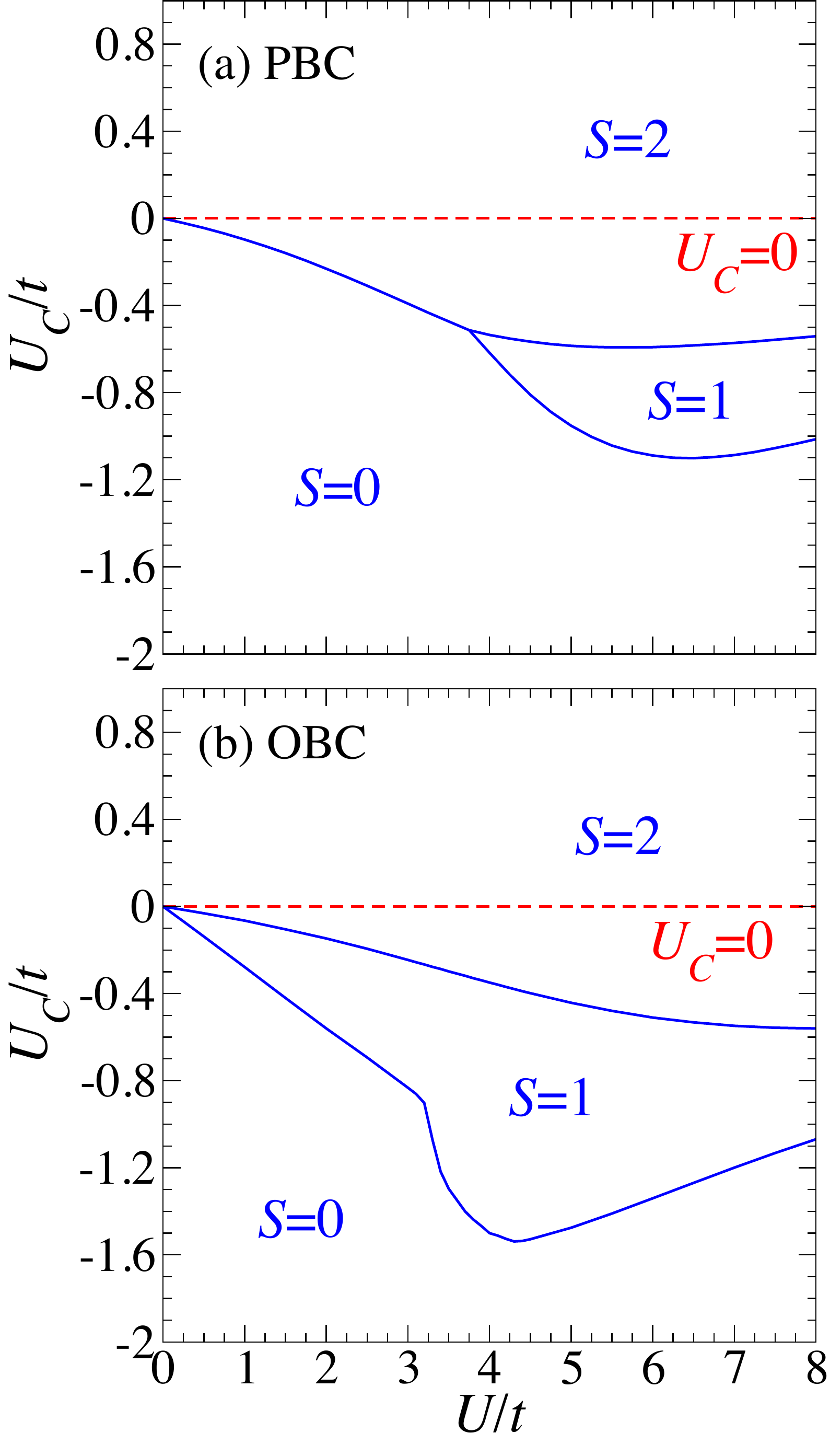}
  \caption{(color online)
    The ground state phase diagrams of ${\mathcal H}_{\rm HM}$ 
    for $t_{sp}=t$ at half filling. 
    The results are obtained by numerically exactly diagonalizing the 
    12-site clusters, containing 4 unit cells (see Fig~\ref{lattice}), with (a) periodic boundary conditions (PBC) and 
    (b) open boundary conditions (OBC). 
    The total spin $S$ of the ground state is indicated in the figures. 
    The ground state with $S=2$ is found all unique.  
    ${\mathcal H}_{\rm HM}$ with $U_C=0$ (indicated by red dashed lines) corresponds to the periodic 
    Anderson model $\mcal H$ studied in the min text. 
  } 
  \label{fig.pdg}
\end{figure}

Figure~\ref{fig.pdg} shows the ground state phase diagrams for ${\mathcal H}_{\rm HM}$ at hall filling 
obtained by numerically exactly diagonalizing the 12-site cluster (see Fig.~\ref{lattice}) with periodic 
and open boundary conditions. 
We find for both boundary conditions that the ground state for $U>0$ and $U_C>0$ is indeed 
unique and it has total spin $S = \left( |H| + |A| - |B| \right)/2 $, in good accordance with Lieb's theorem. 
We also find in Fig.~\ref{fig.pdg} that the ground state for $U>0$ and $U_C>0$ is smoothly connected to the 
unique ground state for $U_C=0$, where 
Lieb's theorem does not guarantee the uniqueness of the ground state. 
Therefore, we conclude that the ground state of ${\mathcal H}$ at half filling is 
unique and it has total spin $S = \left( |H| + |A| - |B| \right)/2 $ as long as $U>0$. 
This also proves that the FM ground state of ${\mathcal H}$ found in the main text is smoothly connected 
to the Lieb-Mattis type ferromagnetism.

\section{Single-particle excitations in the Hubbard-I approximation}\label{appsec:HubI}

The analytical forms of the single-particle excitation dispersion $\w_{\nu,\mb{k}}$ and the corresponding 
spectral weight $\bs{\rho}_{\nu,\mb{k}}$ obtained by the Hubbard-I approximation are provided 
in Eq.~(\ref{eq.poles}) and Eqs.~(\ref{eq.rho14}) and (\ref{eq.rho23}), respectively. 
Here, in this Appendix, we shall examine these quantities for several limiting cases.

\subsection{Noninteracting limit}

In the noninteracting limit, the energy dispersions are given as 
\begin{eqnarray}
\lim_{U \rightarrow 0} \w_{1/4,\mb{k}} &=& \pm \sqrt{|\gamma_{\mb{k}}|^2 + t_{sp}^2},\\
\lim_{U \rightarrow 0} \w_{2/3,\mb{k}} &=& 0, \label{flat_disp}
\end{eqnarray}
and the spectral weights are given as 
\begin{widetext}
\begin{eqnarray}
\lim_{U \rightarrow 0}\bs{\rho}_{1/4,\mb{k}} &=& \frac{1}{2(|\gamma_{\mb{k}}|^2 + t_{sp}^2)}
\left(
\begin{array}{ccc}
 |\gamma_{\mb{k}}|^2 &
 \pm \gamma_{\mb{k}}  \sqrt{|\gamma_{\mb{k}}|^2 + t_{sp}^2} &
 t_{sp} \gamma_{\mb{k}}  \\
 \pm \gamma_{\mb{k}}^* \sqrt{|\gamma_{\mb{k}}|^2 + t_{sp}^2} &
 |\gamma_{\mb{k}}|^2 + t_{sp}^2 &
 \pm  t_{sp} \sqrt{|\gamma_{\mb{k}}|^2 + t_{sp}^2}  \\
 t_{sp} \gamma_{\mb{k}}^*    &
 \pm  t_{sp} \sqrt{|\gamma_{\mb{k}}|^2 + t_{sp}^2}   &
 t_{sp}^2  
\end{array}
\right),\\
\lim_{U \rightarrow 0} \bs{\rho}_{2/3,\mb{k}} &=& \frac{-1}{2(|\gamma_{\mb{k}}|^2 + t_{sp}^2)}
\left(
\begin{array}{ccc}
  -t_{sp}^2               &
   0    & 
   t_{sp} \gamma_{\mb{k}}  \\
  0  &
  0  & 
  0  \\
  t_{sp} \gamma_{\mb{k}}^*    &
  0 &
 - |\gamma_{\mb{k}}|^2              
\end{array}
\right).  \label{flat_spec}
\end{eqnarray}
\end{widetext}
These results are also obtained directly by solving the noninteracting Hamiltonian ${\mathcal H}_0$. 
It is apparent in Eqs.~(\ref{flat_disp}) and (\ref{flat_spec}) that the flat band, corresponding to the inner 
two bands with $\nu=2$ and $3$, 
is composed solely of $A$ and $H$ orbitals, but not $B$ orbital, i.e., $B$ orbital being completely decoupled 
from the flat band (see the top panels of Fig.~\ref{fig.akw.h1}).

\subsection{Localized spin limit}

A single electron is localized on each hydrogen impurity site when $U$ is large enough, and 
eventually the hydrogen impurity sites are decoupled from 
the conduction band in the limit of $U\to\infty$.
In this limit, the energy dispersions are given as
\begin{eqnarray}
  \lim_{U \rightarrow \infty} \w_{1/4,\mb{k}} &\simeq& \pm \lim_{U \rightarrow \infty} \frac{U}{2}, \\
  \lim_{U \rightarrow \infty} \w_{2/3,\mb{k}} &=& \pm |\gamma_{\mb{k}}|,
\end{eqnarray}
and the corresponding spectral weights are 
\begin{eqnarray}
  \lim_{U \rightarrow \infty}\bs{\rho}_{1/4,\mb{k}} &=& \frac{1}{2}
  \left(
    \begin{array}{ccc}
      0 &
      0 &
      0 \\
      0 &
      0 &
      0 \\
      0 &
      0 &
      1
    \end{array}
  \right),\\
  \lim_{U \rightarrow \infty} \bs{\rho}_{2/3,\mb{k}} &=& \frac{1}{2}
  \left(
    \begin{array}{ccc}
      1               &
      \pm \frac{\gamma_{\mb{k}}}{|\gamma_{\mb{k}}|}   & 
      0 \\
      \pm \frac{\gamma_{\mb{k}}^*}{|\gamma_{\mb{k}}|}   &  
      1  & 
      0  \\
      0  &
      0 &
      0 
    \end{array}
  \right).
\end{eqnarray}
Thus, as expected, the hydrogen impurity sites are completely detached from the carbon conduction 
sites, forming the upper and lower Hubbard band in the atomic limit, which correspond to 
the outer two bands with $\nu=1$ and $4$, respectively. The inner two bands with $\nu=2$ and $3$ 
simply display the energy dispersion of the conduction band, i.e., the massless Dirac dispersion for the 
pure graphene model.

\subsection{Decoupling limit}

If $t_{sp}$ is zero, the hydrogen impurity sites are decoupled from the conduction band. 
In this limit, the energy dispersions are give as 
\begin{eqnarray}
  \lim_{t_{sp} \rightarrow 0} \w_{1/4,\mb{k}} &=& \pm \frac{U}{2},\\
  \lim_{t_{sp} \rightarrow 0} \w_{2/3,\mb{k}} &=& \pm |\gamma_{\mb{k}}|,
\end{eqnarray}
and the spectral weights are given as 
\begin{eqnarray}
  \lim_{t_{sp} \rightarrow 0}\bs{\rho}_{1/4,\mb{k}} &=& \frac{1}{2}
  \left(
    \begin{array}{ccc}
      0 &
      0 &
      0 \\
      0 &
      0 &
      0 \\
      0 &
      0 &
      1
    \end{array}
  \right),\\
  \lim_{t_{sp} \rightarrow 0} \bs{\rho}_{2/3,\mb{k}} &=& \frac{1}{2}
  \left(
    \begin{array}{ccc}
      1               &
      \pm \frac{\gamma_{\mb{k}}}{|\gamma_{\mb{k}}|}  & 
      0  \\
      \pm \frac{\gamma_{\mb{k}}^*}{|\gamma_{\mb{k}}|}  &  
      1  & 
      0  \\
      0  &
      0  &
      0 
    \end{array}
  \right).
\end{eqnarray}
It is interesting to notice that these are exactly the same as those in the limit of $U\to\infty$. 

\subsection{Strong bonding limit}

If $|t_{sp}|$ is large, it is expected that the bonding and anti-bonding ``molecular" orbitals are formed locally 
between the neighboring $B$ and $H$ orbitals, and as a result $A$ orbital is isolated. 
In this limit, the energy dispersions are given as 
\begin{eqnarray}
  \lim_{|t_{sp}| \rightarrow \infty} \w_{1/4,\mb{k}} &\simeq& \pm \lim_{|t_{sp}| \rightarrow \infty} |t_{sp}|  \\
  \lim_{|t_{sp}| \rightarrow \infty} \w_{2/3,\mb{k}} &=& 0,
\end{eqnarray}
and the spectral weights are given as 
\begin{eqnarray}
  \lim_{|t_{sp}| \rightarrow \infty}\bs{\rho}_{1/4,\mb{k}} &=& \frac{1}{2}
  \left(
    \begin{array}{ccc}
      0 &
      0 &
      0 \\
      0 &
      1 &
      \pm \frac{t_{sp}}{|t_{sp}|} \\
      0 &
      \pm \frac{t_{sp}}{|t_{sp}|} &
    1
    \end{array}
  \right),\\
  \lim_{|t_{sp}| \rightarrow \infty} \bs{\rho}_{2/3,\mb{k}} &=& \frac{1}{2}
  \left(
    \begin{array}{ccc}
      1  &
      0  & 
      0  \\
      0  &
      0  & 
      0  \\
      0  &
      0  &
      0 
    \end{array}
  \right).
\end{eqnarray}
It is apparent from these results that $B$ and $H$ orbitals are indeed tightly bound to form the bonding and 
anti-bonding ``molecular" orbitals and the isolated $A$ orbitals are completely localized.

\section{Brillouin-Wigner perturbation theory for the quasiparticle Hamiltonian $\mcal{H}_{\text{H-I}}$}\label{appsec:BW}

In this Appendix, we apply the Brillouin-Wigner (BW) perturbation theory~\cite{BW} to the quasiparticle Hamiltonian 
$\mcal{H}_{\text{H-I}}$ in Eq.~(\ref{eq.H_H-I}) and derive effective Hamiltonians for the Dirac band as well as 
for the upper and lower Hubbard bands. 

\subsection{BW perturbation theory}
Let us first divide the $4 \times 4$ quasiparticle Hamiltonian matrix $\bs{H}$, defining 
the quasiparticle Hamiltonian $\mcal{H}_{\text{H-I}}$ in Eq.~(\ref{eq.H_H-I}), into $2 \times 2$ submatrices, i.e., 
\begin{eqnarray}
\bs{H} &=& 
  \left(
  \begin{array}{cc|cc}
    0 & \gamma_{\mb{k}} & 0 & 0  \\
    \gamma^*_{\mb{k}} & 0 & t_{sp} & 0  \\
    \hline
    0 & t_{sp} & 0 & U/2  \\
    0 & 0 & U/2 & 0  
  \end{array}
  \right) \\
 &=& \left(
  \begin{array}{c|c}
    \bs{A} & \bs{B} \\
    \hline
    \bs{C} & \bs{D}
  \end{array}
  \right),
\end{eqnarray}
where $\bs A$, $\bs B$, $\bs C$, and $\bs D$ are the corresponding $2 \times 2$ matrices. 
Then, applying the BW perturbation theory, i.e.,  
the energy dependent perturbation theory, 
the energy ($\w$) dependent effective Hamiltonian matrix $\bs{H}_{\rm eff}(\w)$ is given as 
\begin{eqnarray}
  \bs{H}_{\rm eff}(\w) 
  &=& \bs{P} \bs{H} \sum_{n=0}^{\infty} \left[\bs{Q}\left(\w-\bs{H}_0\right)^{-1} \bs{H_1}\right]^n \notag \\
  &=& \bs{P} \bs{H} + \bs{P}\bs{H}\bs{Q}\left(\w-\bs{H}_0\right)^{-1}\bs{H_1} + \cdots, 
\end{eqnarray} 
where 
\begin{equation}
  \bs{H}_0=\left(
  \begin{array}{c|c}
    \bs{A} & \bs{0} \\
    \hline
    \bs{0} & \bs{D}
  \end{array}
  \right)
\end{equation}
is the unperturbed part and 
\begin{equation}
  \bs{H_1}=\left(
  \begin{array}{c|c}
    \bs{0} & \bs{B} \\
    \hline
    \bs{C} & \bs{0}
  \end{array}
  \right)
\end{equation}
is the perturbation. Here, $\bs{0}$ represents the $2 \times 2$ null matrix. 
$\bs P$ and $\bs Q$ are the projection matrices onto the target (i.e., effective model) space and 
the space orthogonal to the target space, respectively, and they satisfy that 
$\bs{P}^2=\bs{P}$ and $ \bs{P}+\bs{Q}=\bs{I}$, which lead to 
$\bs{Q}^2=\bs{Q}$ and $ \bs{PQ}=\bs{0}$. 

\subsection{Effective Hamiltonian for the Dirac band}

To obtain an effective Hamiltonian 
projected onto the carbon conduction sites, 
the projection matrices should be 
\begin{equation}
  \bs{P}={\rm diag}(1,1,0,0)
\end{equation}
and 
\begin{equation}
  \bs{Q}={\rm diag}(0,0,1,1).
\end{equation}
Then the effective Hamiltonian 
is given as 
\begin{equation}
  \bs{P}(\w - \bs{H}_{\rm eff}(\w))\bs{P} = 
  \left(
  \begin{array}{c|c}
    \bs{g}_{\rm eff,Gr}^{-1}(\w) & \bs{0} \\
    \hline
    \bs{0} & \bs{0}
  \end{array}
  \right),
\end{equation}
where, up to the second order of $t_{sp}$,  we obtain that 
\begin{eqnarray}
  \bs{g}_{\rm eff,Gr}^{-1}(\w)
&=& (\w-\bs{A}) - \bs{B} \left(\w-\bs{D}\right)^{-1} \bs{C} \notag \\
&=&  \left(
    \begin{array}{cc}
      \w & -\gamma \\
    -\gamma^* & \w 
    \end{array}
    \right)
    -
    \left(
    \begin{array}{cc}
      0 &  0\\
     0  & \displaystyle \frac{t^2_{sp}\w}{\w^2-(U/2)^2} 
    \end{array}
    \right) \notag \\
&=& 
    \left(
    \begin{array}{cc}
      \w &  -\gamma_{\mb{k}} \\
      -\gamma_{\mb{k}}^*  & \w \displaystyle - \frac{t^2_{sp}\w}{\w^2-(U/2)^2} 
    \end{array}
    \right).     
\end{eqnarray} 
Notice that this is a Schur's complement of $\w - \bs{H}$ with respect to $\w-\bs{D}$.
The effective Hamiltonian $\bs{h}_{\rm eff,Gr}(\w)$ of the target space is thus obtained as 
\begin{eqnarray} 
  \bs{h}_{\rm eff,Gr}(\w) :&=& \w - \bs{g}_{\rm eff,Gr}^{-1}(\w) \notag \\
  &=& 
  \left(
  \begin{array}{cc}
    0 &  \gamma_{\mb{k}} \\
    \gamma_{\mb{k}}^*  & \displaystyle \frac{t^2_{sp}\w}{\w^2-(U/2)^2} 
  \end{array}
  \right).
  \label{effGr}
\end{eqnarray} 
From Eq.~(\ref{effGr}), we find that 
(i) when $\w \to 0$ and  $|U|>0$, the effective Hamiltonian 
is the same as the pure graphene model, and  
(ii) when $U=0$, the effective on-site energy of $B$ orbital diverges in the limit of $\w \to 0$, 
implying that $B$ orbital does not involve the flat band formation. 

The eigenvalue problem of the 
target space in the BW perturbation theory is described as
\begin{equation}
  \bs{h}_{\rm eff,Gr}(\w) \bs{\psi}_{\rm eff, Gr} = \w \bs{\psi}_{\rm eff,Gr}, 
\end{equation}
where $ \bs{\psi}_{\rm eff, Gr}$ is the two dimensional eigenstate vector and 
the eigenvalues are given as the roots of the secular equation
\begin{equation}
\det\left[\w - \bs{h}_{\rm eff,Gr}(\w)\right] = \det \bs{g}_{\rm eff,Gr}^{-1}(\w)=0.  
\end{equation}
Noticing that the determinant formula for the block matrix 
\begin{equation}\label{formula}
  \det\bs{H} = \det\bs{D} \cdot \det(\bs{A} - \bs{B}\bs{D}^{-1} \bs{C}),  
\end{equation}
we find that the eigenvalues are given as the roots of 
\begin{equation}
 \det \bs{g}_{\rm eff,Gr}^{-1}(\w) = \frac{\det(\w-\bs{H})}{\det(\w-\bs{D})} = 0. 
\end{equation}
Therefore, the eigenvalues are 
identical to those obtained by the full eigenvalue problem of $\bs{H}$, i.e., 
\begin{equation}
  \w = \pm \w_{+,\mb{k}},\ {\rm and }\  \pm \w_{-,\mb{k}}
\end{equation}
in Eq.~(\ref{eq.poles}). 

We have obtained the exact eigenvalues from the effective Hamiltonian which is 
derived perturbatively only up to ${\mcal O}(t_{sp}^2)$. 
This is because the eigenvalues of the full Hamiltonian 
are determined by the roots of 
$\det(\w-\bs{H}) = \det(\w-\bs{D}) \cdot \det\left[\bs{A} - \bs{B}(\w-\bs{D})^{-1} \bs{C})\right] = 0$, 
which contains the term $\bs{B}(\w-\bs{D})^{-1}\bs{C}$, equivalent to 
the second order perturbation with respect to $t_{sp}$.

\subsection{Effective Hamiltonian for the upper and lower Hubbard bands}

An effective Hamiltonian projected onto the upper and lower Hubbard bands is obtained by 
considering the projection matrices 
\begin{equation} 
  \bs{P}={\rm diag}(0,0,1,1)
\end{equation}
and
\begin{equation}
  \bs{Q}={\rm diag}(1,1,0,0). 
\end{equation}
The effective Hamiltonian is then given as 
\begin{equation}
  \bs{P}(\w - \bs{H}_{\rm eff}(\w))\bs{P} = 
  \left(
  \begin{array}{c|c}
    \bs{0}  & \bs{0} \\
    \hline
    \bs{0} &    \bs{g}_{\rm eff,Hub}^{-1}(\w)  \\
  \end{array}
  \right), 
\end{equation}
where, up to $t_{sp}^2$, we obtain that 
\begin{eqnarray}
  \bs{g}_{\rm eff,Hub}^{-1}(\w)
  &=& (\w-\bs{D}) - \bs{C} \left(\w-\bs{A}\right)^{-1} \bs{B} \notag \\
  &=&  \left(
  \begin{array}{cc}
    \w & -U/2 \\
    -U/2 & \w 
  \end{array}
  \right)
  -
  \left(
  \begin{array}{cc}
    \displaystyle \frac{t^2_{sp}\w}{\w^2-|\gamma_{\mb{k}}|^2} &  0\\
    0  & 0 
  \end{array}
  \right) \notag \\ 
  &=& 
  \left(
  \begin{array}{cc}
    \w - \displaystyle \frac{t^2_{sp}\w}{\w^2-|\gamma_{\mb{k}}|^2}&  -U/2 \\
    -U/2  & \w 
  \end{array}
  \right). 
\end{eqnarray}
Notice that this is a Schur's complement of $\w - \bs{H}$ with respect to $\w - \bs{A}$. 
The effective Hamiltonian $\bs{h}_{\rm eff,Hub}(\w)$ is therefore obtained as 
\begin{eqnarray}
  \bs{h}_{\rm eff,Hub}(\w) :&=& \w - \bs{g}_{\rm eff,Hub}^{-1}(\w) \notag \\
  &=& 
  \left(
  \begin{array}{cc}
    \displaystyle \frac{t^2_{sp}\w}{\w^2-|\gamma_{\mb{k}}|^2}&  U/2 \\
    U/2  & 0 
  \end{array}
  \right). 
  \label{effHub}
\end{eqnarray}
From Eq.~(\ref{effHub}) we find that 
(i) when $\w \to 0$ and at momentum $\mb k$ away from $K$ and $K'$ points, 
the effective model 
simply describes the upper and lower Hubbard bands in the atomic limit, 
(ii) when $\w \to 0$ and at $\mb{k}=K (K')$, 
the effective on-site energy of $H$ orbital diverges, 
indicating that the contribution of $H$ orbital is absent at $K (K')$ in the Dirac band, and 
(iii) when $\w \to \pm |\gamma_{\mb{k}}|$, the effective on-site energy of 
$H$ orbital diverges, which is consistent with the ``dark spectral'' region found in both CPT 
and Hubbard-I approximation. 
Finally, we note that the eigenvalues of $\bs{h}_{\rm eff,Hub}(\w)$ are also 
identical to the ones obtained by the full eigenvalue problem of $\bs{H}$.

\end{document}